\documentclass[a4paper,11pt]{report}

\usepackage{epsfig}

\usepackage[english]{babel}

\usepackage{amsmath}

\usepackage{USCthesis2004}

          \renewcommand{\bibname}{BIBLIOGRAPHY}
        \renewcommand\contentsname{\centerline{\bf Table of Contents}}

\renewcommand{\thechapter}{\arabic{chapter}}

\title{TOPICS IN QUANTUM INFORMATION AND THE THEORY OF OPEN QUANTUM SYSTEMS}

\author{Ognyan Oreshkov}

\date{May 2008}

\renewcommand{\copyrightyear}{2008}

\renewcommand{\copyrightname}{Ognyan Oreshkov}

\DeclareGraphicsExtensions{.eps,.ps,.eps.gz,.ps.gz,.eps.Z}

\newcommand{\ra}{\rangle}
\newcommand{\la}{\langle}
\newcommand{\tr}{{\rm Tr}}

\newcommand{\be}{\begin{equation}}
\newcommand{\ee}{\end{equation}}
\newcommand{\ber}{\begin{eqnarray}}
\newcommand{\eer}{\end{eqnarray}}

\def\bra#1{{\langle#1|}}
\def\ket#1{{|#1\rangle}}

\def\expect#1{{\langle#1\rangle}}

\def\tr{{\rm Tr}}
\def\H{{ H}}

\def\U{{ U}}
\def\Udag{{ U}^\dagger}

\def\Op{{ O}}
\def\id{{ I}}
\def\bra#1{{\langle#1|}}
\def\ket#1{{|#1\rangle}}

\def\expect#1{{\langle#1\rangle}}

\def\tr{{\rm Tr}}
\def\H{{ H}}

\def\U{{ U}}
\def\Udag{{ U}^\dagger}

\def\Op{{ O}}

\def\Mhat{{ M}}
\def\Mdag{\Mhat^\dagger}

\def\epsop{{\varepsilon}}
\def\Xhat{{ X}}
\def\id{{ I}}

\makeatletter
\let\l@figureOLD \l@figure
\renewcommand{\l@figure}{\vspace{\baselineskip}\l@figureOLD}
\makeatother

\begin{document}

\renewcommand\contentsname{Table of Contents}

\maketitle

\begin{preface}

\pagebreak   

\chapter*{Dedication}

\begin{center}

{ \large To Iskra}

\end{center}


\addcontentsline{toc}{chapter}{Dedication}

\chapter*{Acknowledgements}

\addcontentsline{toc}{chapter}{Acknowledgements}

First and foremost, I would like to express my gratitude to Todd A.
Brun for his guidance and support throughout the years of our work
together. He has been a great advisor and mentor! I am deeply
indebted to him for introducing me to the field of quantum
information science and helping me advance in it. From him I learned
not only how to do research, but also numerous other skills
important for the career of a scientist, such as writing, giving
presentations, or communicating professionally. I highly appreciate
the fact that he was always supportive of any research direction I
wanted to undertake, never exerted pressure on my work, and was
available to give me advice or encouragement every time I needed
them. This provided for me the optimal environment to develop and
made my work with him a wonderful experience.

I am also greatly indebted to Daniel A. Lidar who has had an
enormous impact on my work. He provided inspiration for many of
the studies presented in this thesis. I have learned tons from my
discussions with him and from his courses on open quantum systems
and quantum error correction. His interest in what I do, the
lengthy email discussions we had, and his invitations to present
my research at his group meetings, have been a major stimulus for
my work.

I am also thankful to Igor Devetak to whom I owe a significant
part of my knowledge in quantum error correction and quantum
communication. It was a pleasure to have him in our research
group.

I want to thank Paolo Zanardi for encouraging me to complete my
work on holonomic quantum computation, and for the fruitful
discussions we had.

Special thanks are due to Stephan Haas for advising me about the
course of my Ph.D. studies from their very beginning. Back then,
he made me feel that I can rely on him for advice or help of any
sort, and he has been corroborating this ever since.

I thank Hari Krovi and Mikhail Ryazanov for our collaboration on
the project on the non-Markovian evolution of a qubit coupled to
an Ising spin bath. I also thank Alireza Shabani for stimulating
conversations regarding the measure of fidelity for encoded
information. Thanks to Martin Varbanov for sharing my excitement
about my research, and the numerous discussions we had.

Thanks are due also to all members of the Physics department that
I haven't mentioned explicitly but with whom I have interacted
during the course of my Ph.D. program, because I have learned a
lot from all of them.

Finally, I would like to acknowledge the people whose contribution
to my Ph.D. work has been indirect but of enormous significance. I
want thank my parents Katyusha and Viktor, who ensured I received
the best education when I was a child, gave me confidence in
myself, and supported all my endeavors throughout my life.

I thank my wife Iskra for her unconditional love, her belief in
me, and her constant support, without which this work would have
been impossible.


\begin{singlespace}

\newcommand*\oldhss{}
\let\oldhss\hss
\renewcommand*\hss{\oldhss\normalfont}
\tableofcontents
\let\hss\oldhss

\newpage

\addcontentsline{toc}{chapter}{List of Figures}

\addtocontents{lof}{\vspace*{-\baselineskip}}

\listoffigures


\end{singlespace}

\chapter*{Abstract}
\addcontentsline{toc}{chapter}{Abstract}

This thesis examines seven topics in quantum information and the
theory of open quantum systems. The first topic concerns weak
measurements and their universality as a means of generating quantum
measurements. It is shown that every generalized measurement can be
decomposed into a sequence of weak measurements which allows us to
think of measurements as resulting form continuous stochastic
processes. The second topic is an application of the decomposition
into weak measurements to the theory of entanglement. Necessary and
sufficient differential conditions for entanglement monotones are
derived and are used to find a new entanglement monotone for
three-qubit states. The third topic examines the performance of
different master equations for the description of non-Markovian
dynamics. The system studied is a qubit coupled to a spin bath via
the Ising interaction. The fourth topic studies continuous quantum
error correction in the case of non-Markovian decoherence. It is
shown that due to the existence of a Zeno regime in non-Markovian
dynamics, the performance of continuous quantum error correction may
exhibit a quadratic improvement if the time resolution of the
error-correcting operations is sufficiently high. The fifth topic
concerns conditions for correctability of subsystem codes in the
case of continuous decoherence. The obtained conditions on the
Lindbladian and the system-environment Hamiltonian can be thought of
as generalizations of the previously known conditions for noiseless
subsystems to the case where the subsystem is time-dependent. The
sixth topic examines the robustness of quantum error-correcting
codes against initialization errors. It is shown that operator codes
are robust against imperfect initialization without the need for
restriction of the standard error-correction conditions. For this
purpose, a new measure of fidelity for encoded information is
introduced and its properties are discussed.  The last topic
concerns holonomic quantum computation and stabilizer codes. A
fault-tolerant scheme for holonomic computations is presented,
demonstrating the scalability of the holonomic method. The scheme
opens the possibility for combining the benefits of error correction
with the inherent resilience of the holonomic approach.

\end{preface}

\chapter*{Chapter 1:  \hspace{1pt} Introduction}

\addcontentsline{toc}{chapter}{Chapter 1:\hspace{0.15cm}
Introduction}

\textbf{1.1 \hspace{2pt}  Quantum information and open quantum
systems}

\addcontentsline{toc}{section}{1.1 \hspace{0.15cm} Quantum
information and open quantum systems}

The field of quantum information and quantum computation has grown
rapidly during the last two decades \cite{NieChu00}. It has been
shown that quantum systems can be used for information processing
tasks that cannot be accomplished by classical means. Examples
include quantum algorithms that can outperform the best known
classical algorithms, such as Shor's factoring algorithm
\cite{Shor97} or Grover's search algorithm \cite{Grover97},
quantum communication protocols which use entanglement for
teleportation of quantum states \cite{BBC93} or superdense coding
\cite{BW92}, or quantum cryptographic protocols which offer
provably secure ways of confidential key distribution between
distant parties \cite{BB84}. This has triggered an immense amount
of research, leading to advances in many areas of quantum physics.

One area that has developed significantly as a result of the new
growing field is that of open quantum systems. This development has
been stimulated on one hand by the need to understand the full
spectrum of operations that can be applied to a quantum state, as
well as the information processing tasks that can be accomplished
with them. Except for unitary transformations, which generally
describe the dynamics of closed systems, the tools of quantum
information science involve also measurements, completely positive
(CP) maps  \cite{NieChu00}, and even non-CP operations
\cite{ShaLid07}. These more general operations result from
interactions of the system of interest with auxiliary systems, and
thus require knowledge of the dynamics of open quantum systems.

At the same time, it has been imperative to understand and find
means to overcome the effects of noise on quantum information.
Quantum superpositions, which are crucial for the workings of most
quantum information processing schemes, can be easily destroyed by
external interactions. This process, known as decoherence, has
presented a major obstacle to the construction of reliable quantum
information devices. This has prompted studies on the mechanisms
of information loss in open quantum systems and the invention of
methods to overcome them, giving rise to one of the pillars of
quantum information science---the theory of quantum error
correction \cite{Shor95, Ste96, Bennett96c, KL96, DG98, ZR97,
LCW98, LBKW01, KLV00, DeF00, KBLW01, YGB01, KLP05, KLPL06, BKK07}.

Quantum error correction studies the information-preserving
structures under open-system dynamics and the methods for encoding
and processing of information using these structures.  A major
result in the theory of error correction states that if the error
rate per information unit is below a certain value, by the use of
fault-tolerant techniques and concatenation, an arbitrarily large
information processing task can be implemented reliably with a
modest overhead of resources \cite{Sho96, DVS96, KL96, ABO98,
Kit97, KLZ98, Got97', Got97, Pre99}. This result, known as the
accuracy threshold theorem, is of fundamental significance for
quantum information science. It shows that despite the unavoidable
effects of noise, scalable quantum information processing is
possible in principle.

In this thesis, we examine topics from three intersecting areas in
the theory open quantum systems and quantum information---the
deterministic dynamics of open quantum systems, quantum
measurements, and quantum error correction.

\subsection*{1.1.1 \hspace{2pt} Deterministic dynamics of open quantum systems}

\addcontentsline{toc}{subsection}{1.1.1 \hspace{0.15cm}
Deterministic dynamics of open quantum systems}

All transformations in quantum mechanics, except for those that
result from measurements, are usually thought of as arising from
continuous evolution driven by a Hamiltonian that acts on the
system of interest and possibly other systems. These
transformations are therefore the result of the unitary evolution
of a larger system that contains the system in question.
Alternative interpretations are also possible---for example some
transformation can be thought of as resulting from measurements
whose outcomes are discarded. This description, however, can also
be understood as originating from unitary evolution of a system
which includes the measurement apparatus and all systems on which
the outcome has been imprinted.

Including the environment in the description is generally difficult
due to the large number of environment degrees of freedom. This is
why it is useful to have a description which involves only the
effective evolution of the reduced density operator of the system.
When the system and the environment are initially uncorrelated, the
effective evolution of the density operator of the system can be
described by a completely positive trace-preserving (CPTP) linear
map \cite{Kraus83}. CPTP maps are widely used in quantum information
science for describing noise processes and operations on quantum
states \cite{NieChu00}. They do not, however, describe the most
general form of transformation of the state of an open system, since
the initial state of the system and the environment can be
correlated in a way which gives rise to non-CP transformations.
Furthermore, the effective transformation by itself does not provide
direct insights into the process that drives the transformation. For
the latter one needs a description in terms of a generator of the
evolution, similar to the way the Schr\"{o}dinger equation describes
the evolution of a closed system as generated by a Hamiltonian. The
main difficulty in obtaining such a description for open systems is
that the evolution of the reduced density matrix of the system is
subject to non-trivial memory effects arising from the interaction
with the environment \cite{BrePet02}.

In the limit where the memory of the environment is short-lived, the
evolution of an open system can be described \cite{BrePet02} by a
time-local semi-group master equation in the Lindblad form
\cite{Lin76}. Such an equation can be thought of as corresponding to
a sequence of weak (infinitesimal) CPTP maps. When the memory of the
environment cannot be ignored and the effective transformation on
the initial state (which is not necessarily CP) is reversible, the
evolution can be described by a time-local master equation, known as
the \textit{time-convolutionless} (TCL) master equation
\cite{Shibata77,ShiAri80}. In contrast to the Lindblad equation,
this equation does not describe completely positive evolution.

The most general continuous deterministic evolution of an open
quantum system is described by the Nakajima-Zwanzig (NZ) equation
\cite{Nak58, Zwa60}. This equation involves convolution in time.
Both the TCL and NZ equations are quite complicated to obtain from
first principles and are usually used for perturbative descriptions.
Somewhere in between the Lindblad equation and the TCL or NZ
equations are the phenomenological post-Markovian master equations
such as the one proposed in Ref.~\cite{ShabaniLidar:05}.

In this thesis, we will examine the deterministic evolution of open
quantum systems both from the point of view of the full evolution of
the system and the environment and from the point of view of the
reduced dynamics of the system. We will study the performance of
different master equation for the description of the non-Markovian
evolution of a qubit coupled to a spin bath \cite{KORL07}, compare
Markovian and non-Markovian models in light of continuous quantum
error correction \cite{OB07}, and study the conditions for
preservation of encoded information under Markovian evolution of the
system and general Hamiltonian evolution of the system and the
environment \cite{OLB08}.

\subsection*{1.1.2 \hspace{2pt} Quantum measurements}

\addcontentsline{toc}{subsection}{1.1.2 \hspace{0.15cm} Quantum
measurements}

In addition to deterministic transformations, the state of an open
quantum system can also undergo stochastic transformations. These
are transformations for which the state may change in a number of
different ways with non-unit probability. Since according to the
postulates of quantum mechanics the only non-deterministic
transformations are those that result from measurements
\cite{vonNeumann,Lueders}, stochastic transformations most
generally result from measurements applied on the system and its
environment. Just like deterministic transformations, stochastic
transformations need not be completely positive. If the system of
interest is initially entangled with its environment and after
some joint unitary evolution of the system and the environment a
measurement is performed on the environment, the effective
transformation on the system resulting from this measurement need
not be CP.

The class of completely positive stochastic operations are
commonly referred to as generalized measurements \cite{Kraus83}.
Although this class includes standard projective measurements
\cite{vonNeumann,Lueders} as well as other operations whose
outcomes reveal information about the state, not all operations in
this category reveal information. Some operations simply consist
of deterministic operations applied with probabilities that do not
depend on the state, i.e., they amount to \textit{trivial}
measurements.

In recent years, a special type of generalized quantum measurements,
the so called \textit{weak} measurements \cite{AAV88, AV89, Leg89,
Per89, AV90}, have become of significant interest. A measurement is
called weak if all of its outcomes result in small (infinitesimal)
changes to the state. Weak measurements have been studied both in
the abstract, and as a means of understanding systems with
continuous monitoring. In the latter case, we can think of the
evolution as the limit of a sequence of weak measurements, which
gives rise to continuous stochastic evolutions called {\it quantum
trajectories} (see, e.g., \cite{Car93, DCM92, GPZ92, Gis84, Dio88,
GP92, PK98}). Such evolutions have been used also as models of
decoherence (see, e.g., \cite{Brun02}). Weak measurements have found
applications in feedback quantum control schemes \cite{DHJMT00} such
as state preparation \cite{Jac03, SJMBH04, SvHM04, WR06, CJ06} or
continuous quantum error correction \cite{ADL02, SarMil05g}.

In this thesis, we look at weak measurements as a means of
generating quantum transformations \cite{OB05}. We show that any
generalized measurement can be implemented as a sequence of weak
measurements, which allows us to use the tools of differential
calculus in studies concerning measurement-driven transformations.
We apply this result to the theory of entanglement, deriving
necessary and sufficient conditions for a function on quantum states
to be an entanglement monotone \cite{OB06}. We use these conditions
to find a new entanglement monotone for three-qubit pure states, a
subject of previously unsuccessful inquiries \cite{Gingrich02}. We
also discuss the use of weak measurements for continuous quantum
error correction.

\subsection*{1.1.3 \hspace{2pt} Quantum error correction}

\addcontentsline{toc}{subsection}{1.1.3 \hspace{0.15cm} Quantum
error correction}

Whether deterministic or stochastic, the evolution of a system
coupled to its environment is generally irreversible. This is
because the environment, by definition, is outside of the
experimenter's control. As irreversible transformations involve
loss of information, they could be detrimental to a quantum
information scheme unless an error-correcting method is employed.

A common form of error correction involves encoding the Hilbert
space of a single information unit, say a qubit, in a subspace of
the Hilbert space of a larger number of qubits \cite{Shor95,
Ste96, Bennett96c, KL96}. The encoding is such that if a single
qubit in the code undergoes an error, the original state can be
recovered by applying an appropriate operation. Clearly, there is
a chance that more than one qubit undergoes an error, but
according to the theory of fault tolerance \cite{Sho96, DVS96,
KL96, ABO98, Kit97, KLZ98, Got97', Got97, Pre99} this problem can
be dealt with by the use of fault-tolerant techniques and
concatenation.

Error correction encompasses a wide variety of methods, each
suitable for different types of noise, different tasks, or using
different resources. Examples include passive error-correction
methods which protect against correlated errors, such as
decoherence-free subspaces \cite{DG98, ZR97, LCW98, LBKW01} and
subsystems \cite{KLV00, DeF00, KBLW01, YGB01}, the standard active
methods \cite{Shor95, Ste96, Bennett96c, KL96} which are suitable
for fault-tolerant computation \cite{Got97}, entanglement assisted
quantum codes \cite{BDH06, BDH06'} useful in quantum communication,
or linear quantum error-correction codes \cite{ShaLid07} that
correct non-completely positive errors. Recently, a general
formalism called operator quantum error correction (OQEC)
\cite{KLP05, KLPL06, BKK07} was introduced, which unified in a
common framework all previously proposed error-correction methods.
This formalism employs the most general encoding of
information---encoding in subsystems \cite{Knill06, VKL01}. OQEC was
generalized to include entanglement-assisted error correction
resulting in the most general quantum error-correction formalism
presently known \cite{HDB07,GHWAC07}.

In the standard formulation of error correction, noise and the
error-correcting operations are usually represented by discrete
transformations \cite{KLP05, KLPL06, BKK07}. In practice, however,
these transformations result from continuous processes. The more
general situation where both the noise and the error-correcting
processes are assumed to be continuous, is the subject of continuous
quantum error correction \cite{PZ98, SarMil05, ADL02, SarMil05g}. In
the paradigm of continuous error correction, error-correcting
operations are generated by weak measurements, weak unitary
operations or weak completely positive maps. This approach often
leads to a better performance in the setting of continuous
decoherence than that involving discrete operations. In this thesis,
we will discuss topics concerning both the discrete formalism and
the continuous one. The topics we study include continuous quantum
error correction for non-Markovian decoherence \cite{OB07},
conditions for correctability of operator codes under continuous
decoherence \cite{OLB08}, the performance of OQEC under imperfect
encoding \cite{Ore08}, as well as fault-tolerant computation based
on holonomic operations \cite{OBL08}.

\section*{1.2 \hspace{2pt} Outline}

\addcontentsline{toc}{section}{1.2 \hspace{0.15cm} Outline}

This work examines seven topics in the areas of deterministic
open-quantum-system dynamics, quantum measurements, and quantum
error correction. Some of the topics concern all of these three
themes, while others concern only two or only one. As each of the
main results has a significance of its own, each topics has been
presented as a separate study in one of the following chapters. The
topics are ordered in view of the background material they introduce
and the logical relation between them.

We first study the theme of weak measurements and their applications
to the theory of entanglement. In Chapter 2, we show that every
generalized quantum measurement can be generated as a sequence of
weak measurements \cite{OB05}, which allows us to think of
measurements in quantum mechanics as generated by continuous
stochastic processes. In the case of two-outcome measurements, the
measurement procedure has the structure of a random walk along a
curve in state space, with the measurement ending when one of the
end points is reached. In the continuous limit, this procedure
corresponds to a quantum feedback control scheme for which the type
of measurement is continuously adjusted depending on the measurement
record. This result presents not only a practical prescription for
the implementation of any generalized measurement, but also reveals
a rich mathematical structure, somewhat similar to that of Lie
algebras, which allows us to study the transformations caused by
measurements by looking only at the properties of infinitesimal
stochastic generators. The result suggests the possibility of
constructing a unified theory of quantum measurement protocols.

Chapter 3 presents an application of the weak-measurement
decomposition to a study of entanglement. The theory of
entanglement concerns the transformations that are possible to a
state under local operations and classical communication (LOCC).
The universality of weak measurements allows us to look at LOCC as
the class of transformations generated by infinitesimal local
operations. We show that a necessary and sufficient condition for
a function of the state to be an entanglement monotone under local
operations that do not involve information loss is that the
function be a monotone under infinitesimal local operations.  We
then derive necessary and sufficient differential conditions for a
function of the state to be an entanglement monotone \cite{OB06}.
We first derive two conditions for local operations without
information loss, and then show that they can be extended to more
general operations by adding the requirement of convexity.  We
then demonstrate that a number of known entanglement monotones
satisfy these differential criteria. We use the differential
conditions to construct a new polynomial entanglement monotone for
three-qubit pure states.

In Chapter 4, we extend the scope of our studies to include the
deterministic dynamics of open quantum systems. We study the
analytically solvable Ising model of a single qubit system coupled
to a spin bath for a case for which the Markovian approximation of
short bath-correlation times cannot be applied \cite{KORL07}. The
purpose of this study is to analyze and elucidate the performance of
Markovian and non-Markovian master equations describing the dynamics
of the system qubit, in comparison to the exact solution. We find
that the time-convolutionless master equation performs particularly
well up to fourth order in the system-bath coupling constant, in
comparison to the Nakajima-Zwanzig master equation. Markovian
approaches fare poorly due to the infinite bath correlation time in
this model. A recently proposed post-Markovian master equation
performs comparably to the time-convolutionless master equation for
a properly chosen memory kernel, and outperforms all the
approximation methods considered here at long times. Our findings
shed light on the applicability of master equations to the
description of reduced system dynamics in the presence of spin
baths.

In Chapter 5, we investigate further the difference between
Markovian and non-Markovian decoherence---this time, form the point
of view of its implications for the performance of continuous
quantum error correction. We study the performance of a quantum-jump
error correction model in the case where each qubit in a codeword is
subject to a general Hamiltonian interaction with an independent
bath \cite{OB07}. We first consider the scheme in the case of a
trivial single-qubit code, which provides useful insights into the
workings of continuous error correction and the difference between
Markovian and non-Markovian decoherence. We then study the model of
a bit-flip code with each qubit coupled to an independent bath qubit
and subject to continuous correction, and find its solution. We show
that for sufficiently large error-correction rates, the encoded
state approximately follows an evolution of the type of a single
decohering qubit, but with an effectively decreased coupling
constant. The factor by which the coupling constant is decreased
scales quadratically with the error-correction rate. This is
compared to the case of Markovian noise, where the decoherence rate
is effectively decreased by a factor which scales only linearly with
the rate of error correction. The quadratic enhancement depends on
the existence of a Zeno regime in the Hamiltonian evolution which is
absent in purely Markovian dynamics. We analyze the range of
validity of this result and identify two relevant time scales.
Finally, we extend the result to more general codes and argue that
there the performance of continuous error correction will exhibit
the same qualitative characteristics. In the appendix of this
chapter, we discuss another application of weak measurements---we
show how the quantum-jump error-correction scheme can be implemented
using weak measurements and weak unitary operations.

In Chapter 6, we study the conditions under which a quantum code is
perfectly correctable during a time interval of continuous
decoherence for the most general type of encoding---encoding in
subsystems. We study the case of Markovian decoherence as well as
the general case of Hamiltonian evolution of the system and the
environment, and derive necessary and sufficient conditions on the
Lindbladian and the system-environment Hamiltonian \cite{OLB08},
respectively. Our approach is based on a result obtained in
Ref.~\cite{KS06} according to which a subsystem is correctable if
and only if it is unitarily recoverable. The conditions we derive
can be thought of as generalizations of the previously derived
conditions for decoherence-free subsystems to the case where the
subsystem is time-dependent. As a special case we consider
conditions for unitary correctability. In the case of Hamiltonian
evolution, the conditions for unitary correctability concern only
the effect of the Hamiltonian on the system, whereas the conditions
for general correctability concern the entire system-environment
Hamiltonian. We also derive conditions on the Hamiltonian which
depend on the initial state of the environment. We discuss possible
implications of our results for approximate quantum error
correction.

Chapter 7 also concerns subsystem codes. Here we study the
performance of operator quantum error correction (OQEC) in the
case of imperfect encoding \cite{Ore08}. In the OQEC, the notion
of correctability is defined under the assumption that states are
perfectly initialized inside a particular subspace, a factor of
which (a subsystem) contains the protected information. It was
believed that in the case of imperfect initialization, OQEC codes
would require more restrictive than the standard conditions if
they are to protect encoded information from subsequent errors. In
this chapter, we examine this requirement by looking at the errors
on the encoded state. In order to quantitatively analyze the
errors in an OQEC code, we introduce a measure of the fidelity
between the encoded information in two states for the case of
subsystem encoding. A major part of the chapter concerns the
definition of the measure and the derivation of its properties. In
contrast to what was previously believed, we obtain that more
restrictive conditions are not necessary neither for DFSs nor for
general OQEC codes. This is because the effective noise that can
arise inside the code as a result of imperfect initialization is
such that it can only increase the fidelity of an imperfectly
encoded state with a perfectly encoded one.

In Chapter 8, we present a scheme for fault-tolerant holonomic
computation on stabilizer codes \cite{OBL08}. In the holonomic
approach, logical states are encoded in the degenerate eigenspace of
a Hamiltonian and gates are implemented by adiabatically varying the
Hamiltonian along loops in parameter space. The result is a
transformation of purely geometric origin, which is robust against
various types of errors in the control parameters driving the
evolution. In the proposed scheme, single-qubit operations on
physical qubits are implemented by varying Hamiltonians that are
elements of the stabilizer, or in the case of subsystem
codes---elements of the gauge group. By construction, the geometric
transformations in each eigenspace of the Hamiltonian are
transversal, which ensures that errors do not propagate. We show
that for certain codes, such as the nine-qubit Shor code or its
subsystem versions, it is possible to realize universal
fault-tolerant computation using Hamiltonians of weight three. The
scheme proves that holonomic quantum computation is a scalable
method and opens the possibility for bringing together the benefits
of error correction and the operational robustness of the holonomic
approach. It also presents an alternative to the standard
fault-tolerant methods based on dynamical transformations, which
have been argued to be in a possible conflict with the assumption of
Markovian decoherence that often underlies the derivation of
threshold results.

Chapter 9 summarizes the results and discusses problems for future
research.

\chapter*{Chapter 2: \hspace{1pt} Generating quantum measurements using weak measurements}

\addcontentsline{toc}{chapter}{Chapter 2:\hspace{0.15cm}
Generating quantum measurements using weak measurements}

\section*{2.1 \hspace{2pt} Preliminaries}

\addcontentsline{toc}{section}{2.1 \hspace{0.15cm} Preliminaries}

In the original formulation of measurement in quantum mechanics,
measurement outcomes are identified with a set of orthogonal
projection operators, which can be thought of as corresponding to
the eigenspaces of a Hermitian operator, or {\it observable}
\cite{vonNeumann,Lueders}.  After a measurement, the state is
projected into one of the subspaces with a probability given by
the square of the amplitude of the state's component in that
subspace.

In recent years a more general notion of measurement has become
common:  the so called {\it generalized} measurement which
corresponds to a {\it positive operator valued measure} (POVM)
\cite{Kraus83}. This formulation can include many phenomena not
captured by projective measurements: detectors with non-unit
efficiency, measurement outcomes that include additional randomness,
measurements that give incomplete information, and many others.
Generalized measurements have found numerous applications in the
rapidly-growing field of quantum information processing
\cite{NieChu00}. Some examples include protocols for unambiguous
state discrimination \cite{Per88} and optimal entanglement
manipulation \cite{Nie99, Jonathan99a}.

Upon measurement, a system with density matrix $\rho$ undergoes a
random transformation
\begin{equation}
\rho \rightarrow \rho_j= {M}_j\rho {M}_j^{\dagger}/p_j,
\hspace{0.4cm}
\underset{j}{\sum}{M}_j^{\dagger}{M}_j={I},\label{genmeas}
\end{equation}
with probability $p_j=\textrm{Tr}({M}_j \rho {M}_j^{\dagger})$,
where the index $j$ labels the possible outcomes of the
measurement. Eq.~\eqref{genmeas} is not the most general
stochastic operation that can be applied to a state. For example,
one can consider the transformation
\begin{equation}
\rho \rightarrow \rho_j= \underset{i}{\sum}{M}_{ij}\rho
{M}_{ij}^{\dagger}/p_j, \hspace{0.4cm}
\underset{i,j}{\sum}{M}_{ij}^{\dagger}{M}_{ij}={I},
\end{equation}
where $p_j=\textrm{Tr}(\underset{i}{\sum}{M}_{ij} \rho
{M}_{ij}^{\dagger})$ is the probability for the $j^{\textrm{th}}$
outcome (see Chapter 3). The letter can be thought of as resulting
from a measurement of the type \eqref{genmeas} with measurement
operators $M_{ij}$ of which only the information about the index $j$
labeling the outcome is retained. In this thesis, when we talk about
generalized measurements, we will refer to the transformation
\eqref{genmeas}.

The transformation \eqref{genmeas} is commonly comprehended as a
spontaneous {\it jump}, unlike unitary transformations, for example,
which are thought of as resulting from {\it continuous} unitary
evolutions. Any unitary transformation can be implemented as a
sequence of {\it weak} (i.e., infinitesimal) unitary
transformations. One may ask if a similar decomposition exists for
generalized measurements. This would allow us to think of
generalized measurements as resulting from continuous stochastic
evolutions and possibly make use of the powerful tools of
differential calculus in the study of the transformations that a
system undergoes upon measurement.

In this chapter we show that any generalized measurement can be
implemented as a sequence of weak measurements and present an
explicit form of the decomposition. The main result was first
presented in Ref.~\cite{OB05}. We call a measurement {\it weak} if
all outcomes result in very small changes to the state. (There are
other definitions of weak measurements that include the
possibility of large changes to the state with low probability; we
will not be considering measurements of this type.) Therefore, a
weak measurement is one whose operators can be written as
\begin{equation}
{M}_j=q_j({I}+{\varepsilon}_j),\label{wmg}
\end{equation}
where $q_j\in {C}$, $0\leq |q_j| \leq 1$, and ${\varepsilon}$ is
an operator with small norm $\|{\varepsilon}\| \ll 1$.

\section*{2.2 \hspace{2pt} Decomposing projective measurements}
\addcontentsline{toc}{section}{2.2 \hspace{0.15cm} Decomposing
projective measurements}

It has been shown that any projective measurement can be implemented
as a sequence of weak measurements; and by using an additional {\it
ancilla} system and a joint unitary transformation, it is possible
to implement any generalized measurement using weak measurements
\cite{Bennett99}. This procedure, however, does not decompose the
operation on the original system into weak operations, since it uses
operations acting on a larger Hilbert space---that of the system
plus the ancilla.  If we wish to study the behavior of  a
function---for instance, an entanglement monotone---defined on a
space of a particular dimension, it complicates matters to add and
remove ancillas.  We will show that an ancilla is not needed, and
give an explicit construction of the weak measurement operators for
any generalized measurement that we wish to decompose.

It is easy to show that a measurement with any number of outcomes
can be performed as a sequence of measurements with two outcomes.
Therefore, for simplicity, we will restrict our considerations to
two-outcome measurements.  To give the idea of the construction,
we first show how every projective measurement can be implemented
as a sequence of weak generalized measurements. In this case the
measurement operators ${P}_1$ and ${P}_2$ are orthogonal
projectors whose sum ${P}_1+{P}_2={I}$ is the identity. We
introduce the operators
\begin{equation}
{P}(x)=\sqrt{\frac{1-\tanh(x)}{2}}{P}_1+\sqrt{\frac{1+\tanh(x)}{2}}{P}_2,
\hspace{0.5cm} x\in R. \label{measpro}
\end{equation}
Note that ${P}^2(x)+{P}^2(-x)={I}$ and therefore ${P}(x)$ and
${P}(-x)$ describe a measurement. If $x=\epsilon$, where
$|\epsilon| \ll 1$, the measurement is weak. Consider the effect
of the operators ${P}(x)$ on a pure state $|\psi\rangle$. The
state can be written as
$|\psi\rangle={P}_1|\psi\rangle+{P}_2|\psi\rangle=\sqrt{p_1}|\psi_1\rangle+\sqrt{p_2}|\psi_2\rangle$,
where $|\psi_{1,2}\rangle={P}_{1,2}|\psi\rangle/\sqrt{p_{1,2}}$
are the two possible outcomes of the projective measurement and
$p_{1,2}=\langle\psi|{P}_{1,2}|\psi\rangle$ are the corresponding
probabilities. If $x$ is positive (negative), the operator
${P}(x)$ increases (decreases) the ratio $\sqrt{p_2}/\sqrt{p_1}$
of the $|\psi_2\rangle$ and $|\psi_1\rangle$ components of the
state. By applying the same operator ${P}(\epsilon)$ many times in
a row for some fixed $\epsilon$, the ratio can be made arbitrarily
large or small depending on the sign of $\epsilon$, and hence the
state can be transformed arbitrarily close to $|\psi_1\rangle$ or
$|\psi_2\rangle$. The ratio of the $p_1$ and $p_2$ is the only
parameter needed to describe the state, since $p_1+p_2=1$.

Also note that ${P}(-x){P}(x)=(1-\tanh^2(x))^{1/2}{I}/2$ is
proportional to the identity. If we apply the same measurement
${P}(\pm\epsilon)$ twice and two opposite outcomes occur, the
system returns to its previous state. Thus we see that the
transformation of the state under many repetitions of the
measurement ${P}(\pm\epsilon)$ follows a random walk along a curve
$\ket{\psi(x)}$ in state space.  The position on this curve can be
parameterized by $x=\ln\sqrt{p_1/p_2}$. Then $\ket{\psi(x)}$ can
be written as $\sqrt{p_1(x)}\ket{\psi_1} +
\sqrt{p_2(x)}\ket{\psi_2}$, where $p_{1,2}(x) = (1/2)[1 \pm
\tanh(x)]$.

The measurement given by the operators ${P}(\pm\epsilon)$ changes
$x$ by $x\rightarrow x\pm\epsilon$, with probabilities
$p_{\pm}(x)=(1\pm\tanh(\epsilon)(p_1(x)-p_2(x)))/2$.  We continue
this random walk until $|x| \ge X$, for some $X$ which is
sufficiently large that $\ket{\psi(X)} \approx \ket{\psi_1}$ and
$\ket{\psi(-X)} \approx \ket{\psi_2}$ to whatever precision we
desire.  What are the respective probabilities of these two
outcomes?

Define $p(x)$ to be the probability that the walk will end at $X$
(rather than $-X$) {\it given} that it began at $x$.  This must
satisfy $p(x) = p_+(x) p(x+\epsilon) + p_-(x) p(x-\epsilon)$.
Substituting our expressions for the probabilities, this becomes
\begin{eqnarray}
p(x)= (p(x+\epsilon) + p(x-\epsilon))/2 \label{difference}
  +
\tanh(\epsilon)\tanh(x)(p(x+\epsilon)-p(x-\epsilon))/2.
\end{eqnarray}

If we go to the infinitesimal limit $\epsilon\rightarrow dx$, this
becomes a continuous differential equation
\begin{equation}
\frac{d^2p}{dx^2} + 2\tanh(x)\frac{dp}{dx} = 0 ,
\end{equation}
with boundary conditions $p(X)=1$, $p(-X)=0$.  The solution to
this equation is $p(x)=(1/2)[1 + \tanh(x)/\tanh(X)]$. In the limit
where $X$ is large, $\tanh(X)\rightarrow1$, so $p(x)=p_1(x)$. The
probabilities of the outcomes for the sequence of weak
measurements are exactly the same as those for a single projective
measurement. Note that this is also true for a walk with a step
size that is not infinitesimal, since the solution $p(x)$
satisfies \eqref{difference} for an arbitrarily large $\epsilon$.

Alternatively, instead of looking at the state of the system
during the process, we could look at an operator that effectively
describes the system's transformation to the current state. This
has the advantage that it is state-independent, and will lead the
way to decompositions of generalized measurements; it also becomes
obvious that the procedure works for mixed states, too.

We think of the measurement process as a random walk along a curve
${P}(x)$ in operator space, given by Eq.~(\ref{measpro}), which
satisfies ${P}(0)={I}/\sqrt{2}$, $\underset{x\rightarrow
-\infty}{\lim}{P}(x)={P}_1$, $\underset{x\rightarrow
\infty}{\lim}{P}(x)={P}_2$. It can be verified that
${P}(x){P}(y)\propto {P}(x+y)$, where the constant of
proportionality is $(\cosh(x+y)/2\cosh(x)\cosh(y))^{1/2}$. Due to
normalization of the state, operators which differ by an overall
factor are equivalent in their effects on the state.  Thus, the
random walk driven by weak measurement operators
${P}(\pm\epsilon)$ has a step size $|\epsilon|$.

\section*{2.3 \hspace{2pt} Decomposing generalized measurements}
\addcontentsline{toc}{section}{2.3 \hspace{0.15cm} Decomposing
generalized measurements}

Next we consider measurements where the measurement operators
${M}_1$ and ${M}_2$ are {\it positive} but not projectors. We use
the well known fact that a generalized measurement can be
implemented as joint unitary operation on the system and an
ancilla, followed by a projective measurement on the ancilla
\cite{NieChu00}.  (One can think of this as an {\it indirect}
measurement; one lets the system interact with the ancilla, and
then measures the ancilla.) Later we will show that the ancilla is
not needed. We consider two-outcome measurements and two-level
ancillas.  In this case ${M}_1$ and ${M}_2$ commute, and hence can
be simultaneously diagonalized.

Let the system and ancilla initially be in a state
$\rho\otimes|0\rangle\langle 0|$. Consider the unitary operation
\begin{equation}
{U}(0)= {M}_1\otimes {Z} +  {M}_2\otimes {X} ,\label{uzero}
\end{equation}
where ${X}=\sigma^x$ and ${Z}=\sigma^z$ are Pauli matrices acting
on the ancilla bit. By applying ${U}(0)$ to the extended system we
transform it to:
\begin{eqnarray}
{U}(0)(\rho\otimes|0\rangle\langle0|){U}^{\dagger}(0) &=&
{M}_1\rho {M}_1\otimes|0\rangle\langle 0|
+ {M}_1 \rho {M}_2\otimes|0\rangle\langle 1| \nonumber\\
 &+&{M}_2 \rho {M}_1\otimes|1\rangle\langle 0|  +
{M}_2 \rho {M}_2\otimes|1\rangle\langle 1|.
\end{eqnarray}
Then a projective measurement on the ancilla in the computational
basis would yield one of the possible generalized measurement
outcomes for the system. We can perform the projective measurement
on the ancilla as a sequence of weak measurements by the procedure
we described earlier. We will then prove that for this process,
there exists a corresponding sequence of generalized measurements
with the same effect acting solely on the system.  To prove this,
we first show that at any stage of the measurement process, the
state of the extended system can be transformed into the form
$\rho(x)\otimes |0\rangle\langle 0|$ by a  unitary operation which
does not depend on the state.

The net effect of the joint unitary operation ${U}(0)$, followed
by the effective measurement operator on the ancilla, can be
written in a block form in the computational basis of the ancilla:
\begin{eqnarray}
{\bar{M}}(x) &\equiv& ({I}\otimes{P}(x)){U}(0) =
\begin{pmatrix}
\sqrt{\frac{1-\tanh(x)}{2}}{M}_1&\sqrt{\frac{1-\tanh(x)}{2}}{M}_2\\
\sqrt{\frac{1+\tanh(x)}{2}}{M}_2&-\sqrt{\frac{1+\tanh(x)}{2}}{M}_1
\end{pmatrix}.
\end{eqnarray}
If the current state ${\bar{M}}(x)(\rho\otimes|0\rangle\langle
0|){\bar{M}}^\dagger$ can be transformed to
$\rho(x)\otimes|0\rangle\langle 0|$ by a unitary operator ${U}(x)$
which is independent of $\rho$, then the lower left block of
${U}(x){\bar{M}}(x)$ should vanish. We look for such a unitary
operator in block form, with each block being Hermitian and
diagonal in the same basis as ${M_1}$ and ${M_2}$. One solution
is:
\begin{equation}
{U}(x)=\begin{pmatrix} {A}(x)&{B}(x)\\
{B}(x)&-{A}(x)
\end{pmatrix},
\end{equation}
where
\begin{equation}
{A}(x)=\sqrt{1-\tanh(x)}{M}_1({I}+\tanh(x)({M}_2^2-{M}_1^2))^{-\frac{1}{2}},\label{A}
\end{equation}
\begin{equation}
{B}(x)=\sqrt{1+\tanh(x)}{M}_2({I}+\tanh(x)({M}_2^2-{M}_1^2))^{-\frac{1}{2}}.\label{B}
\end{equation}
(Since ${M}_1^2 + {M}_2^2 = {I}$, the operator
$({I}+\tanh(x)({M}_2^2-{M}_1^2))^{-\frac{1}{2}} $ always exists.)
Note that ${U}(x)$ is Hermitian, so ${U}(x)={U}^\dagger (x)$ is
its own inverse, and at $x=0$ it reduces to the operator
\eqref{uzero}.

After every measurement on the ancilla, depending on the value of
$x$, we apply the operation ${U}(x)$. Then, before the next
measurement, we apply its inverse ${U}^{\dagger}(x)={U}(x)$. By
doing this, we can think of the procedure as a sequence of
generalized measurements on the extended system that transform it
between states of the form $\rho(x)\otimes|0\rangle\langle 0|$ (a
generalized measurement preceded by a unitary operation and
followed by a unitary operation dependent on the outcome is again
a generalized measurement). The measurement operators are now
${\tilde{M}}(x,\pm\epsilon) \equiv {U}(x\pm\epsilon)({I} \otimes
{P}(\pm\epsilon)) {U}(x)$, and have the form
\begin{equation}
{\tilde{M}}(x,\pm\epsilon)=\begin{pmatrix}{M}(x,\pm\epsilon)&
{N}(x,\pm\epsilon)\\{0}&{O}(x,\pm\epsilon)
\end{pmatrix}.
\end{equation}
Here ${M},{N},{O}$ are operators acting on the system. Upon
measurement, the state of the extended system is transformed
\begin{equation}
\rho(x)\otimes|0\rangle\langle0|\rightarrow
\frac{{M}(x,\pm\epsilon)\rho(x){M}^{\dagger}(x,\pm\epsilon)}{p(x,\pm\epsilon)}\otimes|0\rangle\langle0|,
\end{equation}
with probability
\begin{equation}
p(x,\pm\epsilon) = \tr\left\{
{M}(x,\pm\epsilon)\rho(x){M}^{\dagger}(x,\pm\epsilon) \right\}.
\end{equation}
By imposing
${\tilde{M}}^\dagger(x,\epsilon){\tilde{M}}(x,\epsilon)+{\tilde{M}}(x,-\epsilon)^\dagger{\tilde{M}}(x,-\epsilon)={I}$,
we obtain that
\begin{equation}
{M}^\dagger(x,\epsilon){M}(x,\epsilon)+{M}^\dagger(x,-\epsilon){M}(x,-\epsilon)={I},
\end{equation}
where the operators in the last equation acts on the {\it system
space alone}. Therefore, the same transformations that the system
undergoes during this procedure can be achieved by the
measurements ${M}(x,\pm \epsilon)$ {\it acting solely on the
system}. Depending on the current value of $x$, we perform the
measurement ${M}(x,\pm\epsilon)$. Due to the one-to-one
correspondence with the random walk for the projective measurement
on the ancilla, this procedure also follows a random walk with a
step size $|\epsilon|$. It is easy to see that if the measurements
on the ancilla are weak, the corresponding measurements on the
system are also weak. Therefore we have shown that every
measurement with positive operators ${M}_1$ and ${M}_2$, can be
implemented as a sequence of weak measurements.  This is the main
result of this chapter.  From the construction above, one can find
the explicit form of the weak measurement operators:
\begin{eqnarray}
{M}(x,\epsilon) =
\sqrt{\frac{1-\tanh(\epsilon)}{2}}{A}(x){A}(x+\epsilon) +
\sqrt{\frac{1+\tanh(\epsilon)}{2}}{B}(x){B}(x+\epsilon).\label{measpos}
\end{eqnarray}

These expressions can be simplified further. The current state of
the system at any point during the procedure can be written as
\begin{equation}
\Mhat(x)\rho \Mhat(x)/\tr(\Mhat^2(x)\rho),
\end{equation}
where
\begin{equation}
\Mhat(x)=\sqrt{\frac{\id+\tanh(x)(\Mhat_2^2-\Mhat_1^2)}{2}},\ \
x\in R.
\end{equation}
The weak measurement operators can be written as
\begin{equation}
\Mhat(x,\pm\epsilon) = \sqrt{ C_\pm
  \frac{\id+\tanh(x\pm\epsilon)(\Mhat_2^2-\Mhat_1^2)}{\id+\tanh(x)(\Mhat_2^2-\Mhat_1^2)}},
\label{step_operator}
\end{equation}
where the weights $C_\pm$ are chosen to ensure that these
operators form a generalized measurement:
\begin{equation}
C_\pm = (1\pm\tanh(\epsilon)\tanh(x))/2  .
\end{equation}

Note that this procedure works even if the step of the random walk
is not small, since ${P}(x){P}(y)\propto {P}(x+y)$ for arbitrary
values of $x$ and $y$. So it is not surprising that the effective
operator which gives the state of the system at the point $x$ is
${M}(x)\equiv{M}(0,x)$.

In the limit when $\epsilon\rightarrow 0$, the evolution under the
described procedure can be described by a continuous stochastic
equation. We can introduce a time step $\delta t$ and a rate
\begin{equation}
\gamma=\epsilon^2/\delta t.
\end{equation}
Then we can define a mean-zero Wiener process $\delta W$ as
follows:
\begin{gather}
\delta W=(\delta x-M[\delta x])/\sqrt{\gamma},
\end{gather}
where $M[\delta x]$ is the mean of $\delta x$,
\begin{equation}
M[\delta x]=\epsilon (p_+(x)-p_-(x)).
\end{equation}
The probabilities $p_{\pm}(x)$ can be written in the form
\begin{equation}
p_{\pm}(x)=\frac{1}{2}(1\pm 2 \langle {Q}(x)\rangle \epsilon),
\end{equation}
where $\langle {Q}(x)\rangle$ denotes the expectation value of the
operator
\begin{equation}
Q(x)=\frac{1}{2}\frac{({M}_2^2-{M}_1^2)+\tanh(x){I}}{{I}+\tanh(x)({M}_2^2-{M}_1^2)}.
\end{equation}
Note that $M[(\delta W)^2]=\delta t+\textit{O}(\delta t^2)$.
Expanding the change of a state $|\psi\rangle$ upon the
measurement ${M}(x,\pm \epsilon)$ up to second order in $\delta W$
and taking the limit $\delta W\rightarrow 0$ averaging over many
steps, we obtain the following coupled stochastic differential
equations:
\begin{gather}
|d\psi\rangle=-\frac{\gamma}{2}({Q}(x)-\langle{Q}(x)\rangle)^2|\psi\rangle
dt+\sqrt{\gamma}({Q}(x)-\langle{Q}(x)\rangle)|\psi\rangle
dW,\\
dx=2\gamma\langle{Q}(x)\rangle dt+\sqrt{\gamma} dW.
\end{gather}
This process corresponds to a continuous measurement of an
observable ${Q}$ which is continuously changed depending on the
value of $x$. In other words, it is a feedback-control scheme where
depending on the measurement record, the type of measurement is
continuously adjusted.

Finally, consider the most general type of two-outcome generalized
measurement, with the only restriction being
${M}_1^{\dagger}{M}_1+{M}_2^{\dagger}{M}_2=I$. By polar
decomposition the measurement operators can be written
\begin{equation}
{M}_{1,2}={V}_{1,2}\sqrt{{M}_{1,2}^{\dagger}{M}_{1,2}},
\end{equation}
where ${V}_{1,2}$ are appropriate unitary operators.  One can
think of these unitaries as causing an additional disturbance to
the state of the system, in addition to the reduction due to the
measurement.  The operators $({M}_{1,2}^{\dagger}{M}_{1,2})^{1/2}$
are positive, and they form a measurement.  We could then measure
${M}_1$ and ${M}_2$ by first measuring these positive operators by
a sequence of weak measurements, and then performing either
${V}_1$ or ${V}_2$, depending on the outcome.

However, we can also decompose this measurement directly into a
sequence of weak measurements.  Let the weak measurement operators
for $({M}_{1,2}^{\dagger}{M}_{1,2})^{1/2}$ be
${M}_p(x,\pm\epsilon)$. Let  ${V}(x)$ be any continuous unitary
operator function satisfying ${V}(0)={I}$ and ${V}(\pm x)
\rightarrow {V}_{1,2}$ as $x\rightarrow\infty$.  We then define
\begin{equation}
{M}(x,y)\equiv{V}(x+y){M}_p(x,y){V}^{\dagger}(x).
\end{equation}
By construction ${M}(x,\pm y)$ are measurement operators. Since
${V}(x)$ is continuous, if $y=\epsilon$, where $\epsilon \ll 1$,
the measurements are weak. The measurement procedure is analogous
to the previous cases and follows a random walk along the curve
${M}(0,x)={V}(x){M}_p(0,x)$.

In summary, we have shown that for every two-outcome measurement
described by operators ${M}_1$ and ${M}_2$ acting on a Hilbert
space of dimension $d$, there exists a continuous two-parameter
family of operators ${M}(x,y)$ over the same Hilbert space with
the following properties:
\begin{gather}
 {M}(x,0)={I}/\sqrt{2},\\
 {M}(0,x)
\rightarrow {M}_1 \hspace{0.2cm} \textrm{as} \hspace{0.2cm}
x\rightarrow-\infty,\\
{M}(0,x) \rightarrow {M}_2\hspace{0.2cm} \textrm{as}
\hspace{0.2cm} x\rightarrow+\infty,\\
{M}(x+y,z){M}(x,y)\propto{M}(x,z+y),\\
{M}^{\dagger}(x,y){M}(x,y) + {M}^{\dagger}(x,-y){M}(x,-y) = {I}.
\end{gather}

We have presented an explicit solution for ${M}(x,y)$ in terms of
${M}_1$ and ${M}_2$. The measurement is implemented as a random
walk on the curve ${M}(0,x)$ by consecutive application of the
measurements ${M}(x,\pm\epsilon)$, which depend on the current
value of the parameter $x$.  In the case where $|\epsilon| \ll 1$,
the measurements driving the random walk are weak.  Since any
measurement can be decomposed into two-outcome measurements, weak
measurements are {\it universal}.

\section*{2.4 \hspace{2pt} Measurements with multiple
outcomes}

\addcontentsline{toc}{section}{2.4 \hspace{0.15cm} Measurements
with multiple outcomes}

Even though two-outcome measurements can be used to construct any
multi-outcome measurement, it is interesting whether a direct
decomposition similar to the one we presented can be obtained for
measurements with multiple outcomes as well. In Ref.~\cite{VB07}
it was shown that such a decomposition exists. For a measurement
with $n$ positive operators ${L}_j$, $j=1,...,n$,
$\overset{n}{\underset{j=1}{\sum}}{L}_j^2=I$, the effective
measurement operator $M(x)$ describing the state during the
procedure is given by \cite{VB07}
\begin{equation}
M(s)=\sqrt{f(s)}\sqrt{(\overset{n}{\underset{j=1}{\sum}}s^jL_j^2)},\label{multioutcomedecomposition}
\end{equation}
where
\begin{equation}
f(s)=1+n\overset{n}{\underset{j=1}{\sum}}s^j(1-s^j).
\end{equation}
Here the parameter $s$ is chosen such that
$\overset{n}{\underset{j=1}{\sum}}s^j=1$, $s\in[0,1]$, i.e., it
describes a simplex. The system of stochastic equations describing
the process in the case when the measurement operators ${L}_j$ are
commuting, can be written as
\begin{gather}
|d\psi\rangle=-\frac{\gamma}{8}g^{jk}(s)({Q}_j(s)-\langle{Q}_j(s)\rangle)({Q}_k(s)-\langle{Q}_k(s)\rangle)|\psi\rangle
dt+\notag\\
\frac{1}{2}\sqrt{\gamma}({Q_i}(s)-\langle{Q_i}(s)\rangle)|\psi\rangle
a^{i}_{\alpha}(s)
dW^{\alpha},\\
ds=\gamma g^{ij}(s)\langle{Q}_j(s)\rangle dt+\sqrt{\gamma}
a^{i}_{\alpha}(s) dW^{\alpha},
\end{gather}
where
\begin{equation}
{Q_i}(s)=\frac{L_i^2}{s^mL_m^2},
\end{equation}
\begin{equation}
g^{ij}(s)=\overset{n}{\underset{\alpha,\beta=1}{\sum}}s^i(\delta^i_{\alpha}-s^{\alpha})(\delta^{\alpha}_{\beta}-\frac{1}{n})s^j(\delta^j_{\beta}-s^{\beta}),
\end{equation}
$a(s)$ is the square root of $g(s)$,
\begin{equation}
g^{ij}(s)=\overset{n}{\underset{k=1}{\sum}}a_k^i(s)a_k^j(s),
\end{equation}
and we have assumed Einstein's summation convention.

The decomposition can be easily generalized to the case of
non-positive measurement operators in a way similar to the one we
described for the two-outcome case---by inserting suitable weak
unitaries between the weak measurements.

\section*{2.5 \hspace{2pt} Summary and outlook}

\addcontentsline{toc}{section}{2.5 \hspace{0.15cm} Summary and
outlook}

The result presented in this chapter may have important
implications for quantum control and the theory of quantum
measurements in general. It provides a practical prescription for
the implementation of any generalized measurement using weak
measurements which may be useful in experiments where strong
measurements are difficult to implement. The decomposition might
be experimentally feasible for some quantum optical or atomic
systems.

The result also reveals an interesting mathematical structure,
somewhat similar to that of Lie algebras, which allows us to think
of measurements as generated by infinitesimal stochastic
generators. One application of this is presented in the following
chapter, where we derive necessary and sufficient conditions for a
function on quantum states to be an entanglement monotone. An
entanglement monotone \cite{Vidal00b} is a function which does not
increasing on average under local operations. For pure states the
operations are unitaries and generalized measurements. Since all
unitaries can be broken into a series of infinitesimal steps and
all measurements can be decomposed into weak measurements, it
suffices to look at the behavior of a prospective monotone under
small changes in the state. Thus we can use this result to derive
differential conditions on the function.

These observations suggest that it may be possible to find a
unified description of quantum operations where every quantum
operations can be continuously generated. Clearly, measurements do
not form a group since they do not have inverse elements, but it
may be possible to describe them in terms of a semi-group. The
problem with using measurements as the elements of the semigroup
is that a strong measurement is not equal to a composition of weak
measurements, since the sequence of weak measurements that builds
up a particular strong measurement is not pre-determined---the
measurements depend on a stochastic parameter. It may be possible,
however, to use more general objects---\textit{measurement
protocols}---which describe measurements applied conditioned on a
parameter in some underlying manifold. If such a manifold exists
for the most general possible notion of a protocol, the basic
objects could be describable by stochastic matrices on this
manifold. Such a possibility is appealing since stochastic
processes are well understood and this may have important
implications for the study of quantum control protocols. In
addition, such a description could be useful for describing
general open-system dynamics. These questions are left open for
future investigation.

\chapter*{Chapter 3: \hspace{1pt} Applications of the decomposition into weak measurements to the theory of entanglement}

\addcontentsline{toc}{chapter}{Chapter 3:\hspace{0.15cm}
Applications of the decomposition into weak measurements to the
theory of entanglement}

\section*{3.1 \hspace{2pt} Preliminaries}

\addcontentsline{toc}{section}{3.1 \hspace{0.15cm} Preliminaries}

In this chapter we apply the result on the universality of weak
measurements to the theory of entanglement. The theory of
entanglement concerns the transformations that are possible to a
state under local operations with classical communication (LOCC).
The paradigmatic experiment is a quantum system comprising several
subsystems, each in a separate laboratory under control of a
different experimenter:  Alice, Bob, Charlie, etc.  Each
experimenter can perform any physically allowed operation on his or
her subsystem---unitary transformations, generalized measurements,
indeed any trace-preserving completely positive operation--and
communicate their results to each other without restriction.  They
are not, however, allowed to bring their subsystems together and
manipulate them jointly.  An LOCC protocol consists of any number of
local operations, interspersed with any amount of classical
communication; the choice of operations at later times may depend on
the outcomes of measurements at any earlier time.

The results of Bennett et al.
\cite{Bennett96a,Bennett96b,Bennett96c} and Nielsen \cite{Nie99},
among many others
\cite{Vidal99,Jonathan99a,Hardy99,Jonathan99b,Vidal00a}, have
given us a nearly complete theory of entanglement for {\it
bipartite} systems in pure states. Unfortunately, great
difficulties have been encountered in trying to extend these
results both to mixed states and to states with more than two
subsystems ({\it multipartite} systems).  The reasons for this are
many; but one reason is that the set LOCC is complicated and
difficult to describe mathematically \cite{Bennett99}.

One mathematical tool which has proven very useful is that of the
{\it entanglement monotone}:  a function of the state which is
invariant under local unitary transformations and always decreases
(or increases) on average after any local operation.  These
functions were described by Vidal \cite{Vidal00b}, and large
classes of them have been enumerated since then.

We will consider those protocols in LOCC that preserve pure states
as the set of operations generated by {\it infinitesimal local
operations}:  operations which can be performed locally and which
leave the state little changed including infinitesimal local
unitaries and weak generalized measurements. In Bennett et al.
\cite{Bennett99} it was shown that infinitesimal local operations
can be used to perform any local operation with the additional use
of local ancillary systems--extra systems residing in the local
laboratories, which can be coupled to the subsystems for a time
and later discarded. As we saw in the previous section, any local
generalized measurement can be implemented as a sequence of weak
measurements {\it without} the use of ancillas. This implies that
a necessary and sufficient condition for a function of the state
to be a monotone under local operations that preserve pure states
is the function to be a monotone under infinitesimal local
operations.

In this chapter we derive differential conditions for a function
of the state to be an entanglement monotone by considering the
change of the function on average under infinitesimal local
operations up to the lowest order in the infinitesimal parameter.
We thus obtain conditions that involve at most second derivatives
of the function.  We then prove that these conditions are both
necessary and sufficient. We show that the conditions are
satisfied by a number of known entanglement monotones and we use
them to construct a new polynomial entanglement monotone for
three-qubit pure states.

We hope that this approach will provide a new window with which to
study LOCC, and perhaps avoid some of the difficulties in the
theory of multipartite and mixed-state entanglement.  By looking
only at the differential behavior of entanglement monotones, we
avoid concerns about the global structure of LOCC or the class of
separable operations.

In Section 3.2, we define the basic concepts of this chapter: LOCC
operations, entanglement monotones, and infinitesimal operations.
In Section 3.3, we show how all local operations that preserve
pure states can be generated by a sequence of infinitesimal local
operations. In Section 3.4, we derive differential conditions for
a function of the state to be an entanglement monotone. There are
two such conditions for pure-state entanglement monotones: the
first guarantees invariance under local unitary transformations
(LU invariance), and involves only the first derivatives of the
function, while the second guarantees monotonicity under local
measurements, and involves second derivatives. For mixed-state
entanglement monotones we add a further condition, convexity,
which ensures that a function remains monotonic under operations
that lose information (and can therefore transform pure states to
mixed states). In Section 3.5, we look at some known
monotones--the norm of the state, the local purity, and the
entropy of entanglement--and show that they obey the differential
criteria. In Section 3.6, we use the differential conditions to
construct a new polynomial entanglement monotone for three-qubit
pure states which depends on the invariant identified by Kempe
\cite{Kempe99}. In Section 3.7 we conclude. In the Appendix
(Section 3.8), we show that higher derivatives of the function are
not needed to prove monotonicity.

\section*{3.2 \hspace{2pt} Basic definitions}

\addcontentsline{toc}{section}{3.2 \hspace{0.15cm} Basic
definitions}

\subsection*{3.2.1 \hspace{2pt} LOCC}

\addcontentsline{toc}{subsection}{3.2.1 \hspace{0.15cm} LOCC}

An operation (or protocol) in LOCC consists of a sequence of local
operations with classical communication between them.  Initially,
we will consider only those local operations that preserve pure
states: {\it unitaries}, in which the state is transformed
\begin{equation}
\rho \rightarrow \U\rho\Udag ,\ \ \Udag\U = \U\Udag= \id ,
\end{equation}
and {\it generalized measurements}, in which the state randomly
changes as in Eq.~\eqref{genmeas},
\begin{equation}
\rho \rightarrow \rho_j = \Mhat_j \rho \Mhat^{\dagger}_j /p_j ,\ \
\sum_j \Mdag_j\Mhat_j = \id ,\notag
\end{equation}
with probability $p_j = \tr\left\{\Mdag_j\Mhat_j\rho\right\}$,
where the index $j$ labels the possible outcomes of the
measurement. Note that we can think of a unitary as being a
special case of a generalized measurement with only one possible
outcome. One can think of this class of operations as being
limited to those which do not discard information.  Later, we will
relax this assumption to consider general operations, which can
take pure states to mixed states. Such operations do involve loss
of information. Examples include performing a measurement without
retaining the result, performing an unknown unitary chosen at
random, or entangling the system with an ancilla which is
subsequently discarded.

The requirement that an operation be local means that the
operators $\U$ or $\Mhat_j$ must have a tensor-product structure
$\U \equiv \U\otimes\id$, $\Mhat_j \equiv \Mhat_j \otimes \id$,
where they act as the identity on all except one of the
subsystems.  The ability to use classical communication implies
that the choice of later local operations can depend arbitrarily
on the outcomes of all earlier measurements.  One can think of an
LOCC operation as consisting of a series of ``rounds.''  In each
round, a single local operation is performed by one of the local
parties; if it is a measurement, the outcome is communicated to
all parties, who then agree on the next local operation.

\subsection*{3.2.2 \hspace{2pt} Entanglement monotones}

\addcontentsline{toc}{subsection}{3.2.2 \hspace{0.15cm}
Entanglement monotones}

For the purposes of this study, we define an entanglement monotone
to be a real-valued function of the state with the following
properties:  if we start with the system in a state $\rho$ and
perform a local operation which leaves the system in one of the
states $\rho_1,\cdots,\rho_n$ with probabilities $p_1,\ldots,p_n$,
then the value of the function must not increase on average:
\begin{subequations}
\label{eq:EM}
\begin{equation}
f(\rho) \ge \sum_j p_j f(\rho_j) . \label{monotonicity1}
\end{equation}

Furthermore, we can start with a state selected randomly from an
ensemble $\{\rho_k, p_k\}$. If we dismiss the information about
which particular state we are given (which can be done locally),
the function of the resultant state must not exceed the average of
the function we would have if we keep this information:
\begin{equation}
\sum_k p_k f(\rho_k) \geq f\left( \sum_k p_k \rho_k \right) .
\label{monotonicity2}
\end{equation}
\end{subequations}

Some functions may obey a stronger form of monotonicity, in which
the function cannot increase for any outcome:
\begin{equation}
f(\rho) \ge f(\rho_j),\  \forall j ,
\end{equation}
but this is not the most common situation.  Some monotones may be
defined only for pure states, or may only be monotonic for pure
states. In the latter case, monotonicity is defined as
non-increase on average under local operations that do not involve
information loss.

\subsection*{3.2.3 \hspace{2pt} Infinitesimal operations}

\addcontentsline{toc}{subsection}{3.2.3 \hspace{0.15cm}
Infinitesimal operations}

We call an operation {\it infinitesimal} if all outcomes result in
only very small changes to the state.  That is, if after an
operation the system can be left in states $\rho_1,\cdots,\rho_n$,
we must have
\begin{equation}
|| \rho - \rho_j || \ll 1, \ \forall j .
\end{equation}
For a unitary, this means that
\begin{equation}
\U = \exp(i{\epsop}) \approx \id + i {\epsop} ,
\end{equation}
where $\epsop$ is a Hermitian operator with small norm,
$||{\epsop}|| \ll 1$, ${\epsop}={\epsop}^\dagger$.  For a
generalized measurement, every measurement operator $\Mhat_j$ can
be written as in Eq.~\eqref{wmg},
\begin{equation}
\Mhat_j = q_j (\id + {\epsop}_j ) ,\notag
\end{equation}
where $0 \le q_j \le 1$ and ${\epsop}_j$ is an operator with small
norm $||{\epsop}_j|| \ll 1$.

\section*{3.3 \hspace{2pt} Local operations from infinitesimal local operations}

\addcontentsline{toc}{section}{3.3 \hspace{0.15cm} Local
operations from infinitesimal local operations}

In this section we show how any local operation that preserves
pure states can be performed as a sequence of infinitesimal local
operations. The operations that preserve pure states are unitary
transformations and generalized measurements.

\subsection*{3.3.1 \hspace{2pt} Unitary transformations}
\addcontentsline{toc}{subsection}{3.3.1 \hspace{0.15cm} Unitary
transformations}

Every local unitary operator has the representation
\begin{equation}
\U=e^{i\H},
\end{equation}
where $\H$ is a local hermitian operator. We can write
\begin{equation}
\U=\lim_{n\rightarrow\infty}(\id+i\H/n)^n,
\end{equation}
and define
\begin{equation}
\epsop=\H/n
\end{equation}
for a suitably large value of $n$.  Thus, in the limit
$n\rightarrow\infty$, any local unitary operation can be thought
of as an infinite sequence of infinitesimal local unitary
operations driven by operators of the form
\begin{equation}
\U_{\varepsilon}\approx \id+i\epsop,
\end{equation}
where $\epsop$ is a small ($\|{\epsop}\|\ll 1$) local hermitian
operator.

\subsection*{3.3.2 \hspace{2pt} Generalized measurements}
\addcontentsline{toc}{subsection}{3.3.2 \hspace{0.15cm}
Generalized measurements}

As was shown in Chapter 2, any measurement can be generated by a
sequence of weak measurements. Since a measurement with any number
of outcomes can be implemented as a sequence of two-outcome
measurements, it suffices to consider generalized measurements
with two outcomes. The form of the weak operators needed to
generate any measurement (Eq.~\eqref{step_operator}) is
\begin{equation}
\Mhat(x,\pm\epsilon) = \sqrt{ C_\pm
  \frac{\id+\tanh(x\pm\epsilon)(\Mhat_2^2-\Mhat_1^2)}{\id+\tanh(x)(\Mhat_2^2-\Mhat_1^2)}},\notag
\end{equation}
where
\begin{equation}
C_\pm = (1\pm\tanh(\epsilon)\tanh(x))/2  .\notag
\end{equation}
From these expressions it is easy to see that if $|\epsilon|\ll
1$, we have $\Mhat(x,\epsilon) = \sqrt{1/2}(\id+O(\epsilon))$,
i.e., the coefficients $q_j$ in Eq.~\eqref{wmg} are
$q_1=q_2=\frac{1}{\sqrt{2}}$. Furthermore, if the original
measurement is local, the weak measurements are also local.

Clearly, the fact that infinitesimal local operations are part of
the set of LO means that an entanglement monotone must be a
monotone under infinitesimal local operations. The discussion in
this section implies that if a function is a monotone under
infinitesimal local unitaries and generalized measurements, it is
a monotone under all local unitaries and generalized measurements
(the operations that do not involve information loss and preserve
pure states). Based on this result, in the next section we derive
necessary and sufficient conditions for a function to be an
entanglement monotone.

\section*{3.4 \hspace{2pt} Differential conditions for entanglement monotones}
\addcontentsline{toc}{section}{3.4 \hspace{0.15cm} Differential
conditions for entanglement monotones}

Let us now consider the change in the state under an infinitesimal
local operation. Without loss of generality, we assume that the
operation is performed on Alice's subsystem. In this case, it is
convenient to write the density matrix of the system as
\begin{equation}
{\rho}=\underset{i,j,l,m}\sum \rho_{ijlm} |i_A\rangle \langle
l_A|\otimes |j_{BC...}\rangle \langle m_{BC...}|,
\end{equation}
where the set $\{|i_A\rangle\}$ and the set
$\{|j_{BC...}\rangle\}$ are arbitrary orthonormal bases for
subsystem $A$ and the rest of the system, respectively. Any
function of the state $f(\rho)$ can be thought of as a function of
the coefficients in the above decomposition:
\begin{equation}
f(\rho) = f(\rho_{ijlm}).
\end{equation}

\subsection*{3.4.1 \hspace{2pt} Local unitary invariance}
\addcontentsline{toc}{subsection}{3.4.1 \hspace{0.15cm} Local
unitary invariance}

Unitary operations are invertible, and therefore the monotonicity
condition reduces to an invariance condition for LU
transformations. Under local unitary operations on subsystem $A$
the components of ${\rho}$ transform as follows:
\begin{equation}
\rho_{ijlm} \rightarrow \underset{k,p}\sum
U_{ik}\rho_{kjpm}U^*_{lp},
\end{equation}
where $U_{ik}$ are the components of the local unitary operator in
the basis $\{|i_A\rangle\}$. We consider infinitesimal local
unitary operations:
\begin{equation}
U_{lk} = \left(e^{i\epsop}\right)_{lk},
\end{equation}
where ${\epsop}$ is a local hermitian operator acting on subsystem
$A$, and
\begin{equation}
\|{\epsop}\|\ll 1.
\end{equation}
Up to first order in $\epsop$ the coefficients $\rho_{ijlm}$
transform as
\begin{equation}
\rho_{ijlm} \rightarrow \rho_{ijlm} + i[\epsop, \rho]_{ijlm}.
\end{equation}
Requiring LU-invariance of $f(\rho)$, we obtain that the function
must satisfy
\begin{equation}
\underset{i,j,l,m}\sum\frac{\partial f}{\partial
\rho_{ijlm}}[\epsop, \rho]_{ijlm}=0. \label{three}
\end{equation}
Analogous equations must be satisfied for arbitrary hermitian
operators $\epsop$ acting on the other parties' subsystems. In a
more compact form, the condition can be written as
\begin{equation}
\tr\left\{ \frac{\partial f}{\partial\rho}[\epsop , \rho] \right\}
= 0, \label{mixed1}
\end{equation}
where $\epsop$ is an arbitrary local hermitian operator.

\subsection*{3.4.2 \hspace{2pt} Non-increase under infinitesimal local measurements}
\addcontentsline{toc}{subsection}{3.4.2 \hspace{0.15cm}
Non-increase under infinitesimal local measurements}

As mentioned earlier, a measurement with any number of outcomes
can be implemented as a sequence of measurements with two
outcomes, and a general measurement can be done as a measurement
with positive operators, followed by a unitary conditioned on the
outcome; therefore, it suffices to impose the monotonicity
condition for two-outcome measurements with positive measurement
operators. Consider local measurements on subsystem $A$ with two
measurement outcomes, given by operators ${\Mhat}_1^2+{\Mhat}_2^2
= \id$. Without loss of generality, we assume
\begin{eqnarray}
{\Mhat}_1 &=&
\sqrt{ ({\id}+{\epsop})/2 },\nonumber\\
{\Mhat}_2 &=& \sqrt{ ({\id}-{\epsop})/2 }, \label{mm}
\end{eqnarray}
where ${\epsop}$ is again a small local hermitian operator acting
on $A$ (in the previous section we saw that any two-outcome
measurement with positive operators can be generated by weak
measurements of this type). Upon measurement, the state undergoes
one of two possible transformations
\begin{eqnarray}
\rho &\rightarrow& \frac{\Mhat_{1,2}\rho \Mhat_{1,2}}{p_{1,2}},
\end{eqnarray}
with probabilities $ p_{1,2} = \tr\left\{ {\Mhat_{1,2}}^2 \rho
\right\}$. Since $\epsop$ is small, we can expand
\begin{eqnarray}
{\Mhat}_1 &=& \frac{1}{\sqrt{2}}({\id} + {\epsop}/2 - {\epsop}^2/8 - \cdots), \label{m1} \\
{\Mhat}_2 &=&\frac{1}{\sqrt{2}}({\id} - {\epsop}/2 - {\epsop}^2/8
- \cdots).\label{m2}
\end{eqnarray}
The condition for non-increase on average of the function $f$
under infinitesimal local measurements is
\begin{equation}
p_1 f(\Mhat_1\rho \Mhat_1/p_1) + p_2 f(\Mhat_2\rho \Mhat_2/p_2)
\le f(\rho) . \label{nonincrease4}
\end{equation}
Expanding (\ref{nonincrease4}) in powers of $\epsop$ up to second
order, we obtain
\begin{equation}
\frac{1}{4}\tr\left\{ \frac{\partial f}{\partial\rho}[[\epsop,
\rho],\epsop] \right\}
  + \tr\left\{ \frac{\partial^2 f}{\partial\rho^{\otimes2}}
  \left( \tr(\epsop\rho)\rho - \frac{1}{2}\{\epsop,\rho\} \right)^{\otimes 2} \right\}
  \leq 0 ,
\label{mixed3}
\end{equation}
where $\{\epsop, \rho \}$ is the anti-commutator of $\epsop$ and
$\rho$. The inequality must be satisfied for an arbitrary local
hermitian operator $\epsop$.

So long as (\ref{mixed3}) is satisfied by a strict inequality, it
is obvious that we need not consider higher-order terms in
$\epsop$. But what about the case when the condition is satisfied
by equality?  In the appendix we will show that even in the case
of equality, (\ref{mixed3}) is still the necessary and sufficient
condition for monotonicity under local generalized measurements.
There we also prove the sufficiency of the LU-invariance condition
\eqref{mixed1}. This allows us to state the following
\\
\textbf{Theorem 1}: A twice-differentiable function $f(\rho)$ of
the density matrix is a monotone under local unitary operations
and generalized measurements, if and only if it satisfies
\eqref{mixed1} and \eqref{mixed3}.
\\
We point out that from the condition of LU invariance applied up
to second-order in $\epsop$, one obtains
\begin{equation}
\tr\left\{ \frac{\partial f}{\partial\rho}[[\epsop, \rho],\epsop]
\right\}=-\tr\left\{ \frac{\partial^2 f}{\partial\rho^{\otimes2}}
  \left( i[\epsop,\rho] \right)^{\otimes 2}
  \right\}.\label{LUequiv}
\end{equation}
Therefore, in the case when both Eq.~\eqref{mixed1} and
Eq.~\eqref{mixed3} are satisfied, condition \eqref{mixed3} can be
written equivalently in the form
\begin{equation}
\tr\left\{ \frac{\partial^2 f}{\partial\rho^{\otimes2}}
 \left[ \left( \tr(\epsop\rho)\rho - \frac{1}{2}\{\epsop,\rho\} \right)^{\otimes 2}- \left( \frac{i}{2}[\epsop,\rho] \right)^{\otimes 2} \right] \right\}
  \leq 0.
\end{equation}

Unitary operations and generalized measurements are the operations
that preserve pure states. Other operations (which involve loss of
information), such as positive maps, would in general cause pure
states to evolve into mixed states. A measure of pure-state
entanglement need not be defined over the entire set of density
matrices, but only over pure states. Thus a measure of pure-state
entanglement, when expressed as a function of the density matrix,
may have a significantly simpler form than its generalizations to
mixed states. For example, the entropy of entanglement for
bipartite pure states can be written in the well-known form
$S_A(\rho)=-\tr(\rho_A\log \rho_A)$, where $\rho_A$ is the reduced
density matrix of one of the parties' subsystems. When directly
extended over mixed states, this function is not well justified,
since $S_A(\rho)$ may have a different value from $S_B(\rho)$.
Moreover, $S_A(\rho)$ by itself is not a mixed-state entanglement
monotone, since it may increase under local positive maps on
subsystem A (these properties of the entropy of entanglement will
be discussed further in Section 3.5). One generalization of the
entropy of entanglement to mixed states is the entanglement of
formation \cite{Bennett96c}, which is defined as the minimum of
$\sum_i p_i S_A(\rho_i)$ over all ensembles of bipartite pure
states $\{\rho_i, p_i\}$ realizing the mixed state: $\rho=\sum_i
p_i \rho_i$. This quantity is a mixed-state entanglement monotone.
As a function of $\rho$, it has a much more complicated form than
the above expression for the entropy of entanglement. In fact,
there is no known analytic expression for the entanglement of
formation in general. The problem of extending pure-state
entanglement monotones to mixed states is an important one, since
every mixed-state entanglement monotone can be thought of as an
extension of a pure-state entanglement monotone. Note, however,
that a pure-state entanglement monotone may have many different
mixed-state generalizations. The relation between the entanglement
of formation and the entropy of entanglement presents one way to
perform such an extension (convex-roof extension). For every
pure-state entanglement monotone $m(\rho)$, one can define a
mixed-state extension $M(\rho)$ as the minimum of $\sum_i p_i
m(\rho_i)$ over all ensembles of pure states $\{\rho_i, p_i\}$
realizing the mixed state: $\rho=\sum_i p_i \rho_i$. It is easy to
verify that $M(\rho)$ is an entanglement monotone for mixed
states. On the set of pure states the function $M(\rho)$ reduces
to $m(\rho)$. As the example with the entropy of entanglement
suggests, not every form of a pure-state entanglement monotone
corresponds to a mixed-state entanglement monotone when trivially
extended to all states---there are additional conditions that a
mixed-state entanglement monotone must satisfy. On the basis of
the above considerations, it makes sense to consider separate sets
of differential conditions for pure-state and mixed-state
entanglement monotones.
\\
\textbf{Corollary 1:} A twice-differentiable function $f(\rho)$ of
the density matrix is a pure-state entanglement monotone, if and
only if it satisfies \eqref{mixed1} and \eqref{mixed3} for pure
$\rho$.
\\
For pure states $\rho = \ket\psi\bra\psi$, the elements of $\rho$
are $\rho_{ij\ell m} = \alpha_{ij}\alpha^*_{\ell m}$, where the
$\{\alpha_{ij}\}$ are the state amplitudes: $ |\psi\rangle =
\underset{i,j}{\sum}\alpha_{ij}|i_A\rangle |j_{BC...}\rangle $.
Any function on pure states $f(\rho)\equiv f(|\psi\rangle)$ is
therefore a function of the state amplitudes and their complex
conjugates:
\begin{equation}
f(|\psi\rangle) = f(\{\alpha_{ij}\}, \{\alpha^{\ast}_{ij}\}).
\end{equation}
By making the substitution $\rho_{ij\ell m} =
\alpha_{ij}\alpha^*_{\ell m}$ into (\ref{mixed1}) and
(\ref{mixed3}), we can (after considerable algebra) derive
alternative forms of the differential conditions for functions of
the state vector:
\begin{equation}
\underset{i,j,k}\sum\frac{\partial f}{\partial \alpha_{ij}}
\varepsilon_{ik} \alpha_{kj}=\underset{i,j,k}\sum\frac{\partial
f}{\partial
\alpha^{\ast}_{ij}}\varepsilon^{\ast}_{ik}\alpha^{\ast}_{kj},
\label{LU}
\end{equation}
\begin{equation}
\underset{i,j,k,l,m,n}\sum\frac{\partial^2f}{\partial
\alpha_{ij}\partial\alpha_{mn}}
  \left( \varepsilon_{ik}\alpha_{kj}-\expect{\epsop}\alpha_{ij} \right)
  \left( \varepsilon_{m\ell}\alpha_{\ell n}-\expect{\epsop}\alpha_{mn} \right)
+ c.c. \leq 0. \label{nonincrease3}
\end{equation}
Here $\epsop$ is a local hermitian operator acting on subsystem A.
Analogous conditions must be satisfied for $\epsop$ acting on the
other parties' subsystems.

\subsection*{3.4.3 \hspace{2pt} Monotonicity under operations with information
loss}
\addcontentsline{toc}{subsection}{3.4.3 \hspace{0.15cm}
Monotonicity under operations with information loss}

Besides monotonicity under local unitaries and generalized
measurements, an entanglement monotone for mixed states should
also satisfy monotonicity under local operations which involve
{\it loss of information}. The most general transformation that
involves loss of information has the form
\begin{equation}
\rho \rightarrow \rho_k = \frac{1}{p_k} \sum_j \Mhat_{k,j} \rho
\Mdag_{k,j} ,\label{mostgeneral}
\end{equation}
where
\begin{equation}
p_k = \tr\left\{ \sum_j \Mhat_{k,j} \rho \Mdag_{k,j} \right\}
\end{equation}
is the probability for outcome $k$.  The operators
$\{\Mhat_{k,j}\}$ must satisfy
\begin{equation}
\sum_{k,j} \Mdag_{k,j} \Mhat_{k,j} = \id .
\end{equation}
We can see that this includes unitary transformations, generalized
measurements, and completely positive trace-preserving maps as
special cases.

It occasionally makes sense to consider even more general
transformations, where the operators need not sum to the identity:
\begin{equation}
\sum_{k,j} \Mdag_{k,j} \Mhat_{k,j} \le \id .
\end{equation}
This corresponds to a situation where only certain outcomes are
retained, and others are discarded; the probabilities add up to
less than 1 due to these discarded outcomes.  We say such a
transformation involves {\it postselection}.

With or without postselection, we are concerned with the case
where all operations are done locally, so that all the operators
$\{\Mhat_{k,j}\}$ act on a single subsystem. Every such
transformation can be implemented as a sequence of local
generalized measurements (possibly discarding some of the
outcomes) and local completely positive maps. In operator-sum
representation \cite{Kraus83}, a completely positive map can be
written
\begin{equation}
\rho \rightarrow \sum_k \Mhat_k \rho
\Mhat_k^{\dagger},\label{firstCPTPmap}
\end{equation}
where
\begin{equation}
\sum_k \Mhat_k^{\dagger}\Mhat_k \leq \id. \label{positivity}
\end{equation}
Therefore, in addition to \eqref{mixed1} and \eqref{mixed3} we
must impose the condition
\begin{equation}
f(\rho) \geq f\left( \sum_k \Mhat_k\rho \Mhat_k^{\dagger} \right)
. \label{maps}
\end{equation}
for all sets of local operators $\{\Mhat_k\}$ satisfying
(\ref{positivity}).

Suppose the parties are supplied with a state $\rho_k$ taken from
an ensemble $\{\rho_k, p_k\}$.  Discarding the information of the
actual state amounts to the transformation
\begin{equation}
\{\rho_k, p_k\} \rightarrow \rho '=\underset{k}{\sum} p_k \rho_k.
\end{equation}
As pointed out in \cite{Vidal00b}, discarding information should
not increase the entanglement of the system on average. Therefore,
for any ensemble $\{\rho_k, p_k\}$, an entanglement monotone on
mixed states should be {\it convex}:
\begin{equation}
\sum_k p_k f(\rho_k) \geq f\left( \sum_k p_k \rho_k \right) .
\label{convex}
\end{equation}
Condition \eqref{convex}, together with condition (\ref{mixed3})
for monotonicity under local generalized measurements, implies
monotonicity under local completely positive maps:
\begin{equation}
f\left( \sum_k \Mhat_k\rho \Mhat_k^{\dagger} \right)
  \leq \sum_k p_k f \left( \frac{\Mhat_k\rho \Mhat^{\dagger}_k}{p_k} \right)
  \leq f(\rho).
\end{equation}
It is easy to see that if this inequality holds without
postselection, it must also hold with postselection.

It follows that a function of the density matrix is an
entanglement monotone for mixed states if and only if it is (1) a
convex function on the set of density matrices and (2) a monotone
under local unitaries and generalized measurements. Fortunately,
there are also simple differential conditions for convexity.  A
necessary and sufficient condition for a twice-differentiable
function of multiple variables to be convex on a convex set is
that its Hessian matrix be positive on the interior of the convex
set (in this case, the set of density matrices). Therefore, in
addition to \eqref{mixed1} and \eqref{mixed3} we add the
differential condition
\begin{equation}
\tr\left\{ \frac{\partial^2 f(\rho)}{\partial\rho^{\otimes
2}}\sigma^{\otimes 2} \right\} \geq 0, \label{mixed4}
\end{equation}
which must be satisfied at every $\rho$ on the interior of the set
of density matrices for an arbitrary traceless hermitian matrix
$\sigma$.
\\\\
\textbf{Corollary 2:} A twice-differentiable function $f(\rho)$ of
the density matrix is a mixed-state entanglement monotone, if and
only if it satisfies \eqref{mixed1}, \eqref{mixed3} and
\eqref{mixed4}.

\section*{3.5 \hspace{2pt} Examples}
\addcontentsline{toc}{section}{3.5 \hspace{0.15cm} Examples}

In this section we demonstrate how conditions \eqref{mixed1},
\eqref{mixed3} and \eqref{mixed4} can be used to verify if a
function is an entanglement monotone. We show this for three well
known entanglement monotones: the norm of the state of the system,
the trace of the square of the reduced density matrix of any
subsystem, and the entropy of entanglement. In the next section we
will use some of the observations made here to construct a new
polynomial entanglement monotone for three-qubit pure states.

\subsection*{3.5.1 \hspace{2pt} Norm of the state}
\addcontentsline{toc}{subsection}{3.5.1 \hspace{0.15cm} Norm of
the state}
The most trivial example is the norm or the trace of
the density matrix of the system:
\begin{equation}
I_1=\tr\{\rho\}.
\end{equation}
Clearly $I_1$ is a monotone under LOCC, since all operations that
we consider either preserve or decrease the trace. But for the
purpose of demonstration, let us verify that $I_1$ satisfies the
differential conditions.

The LU-invariance condition \eqref{mixed1} reads
\begin{equation}
\tr\left\{ \frac{\partial I_1}{\partial\rho}[\epsop , \rho]
\right\}
  = \tr\left\{ [\epsop , \rho ] \right\} = 0.
\end{equation}
The second equality follows from the cyclic invariance of the
trace.

Since the trace is linear, the second term in condition
\eqref{mixed3} vanishes, and we consider only the first term:
\begin{equation}
\tr\left\{ \frac{\partial I_1}{\partial\rho}[[\epsop,
\rho],\epsop] \right\}
  = \tr\left\{ [[\epsop, \rho],\epsop] \right\} = 0.
\end{equation}
The condition is satisfied with equality, again due to the cyclic
invariance of the trace, implying that the norm remains invariant
under local measurements. The convexity condition \eqref{mixed4}
is also satisfied by equality.

\subsection*{3.5.2 \hspace{2pt} Local purity}
\addcontentsline{toc}{subsection}{3.5.2 \hspace{0.15cm} Local
purity}

The second example is the purity of the reduced density matrix:
\begin{equation}
I_2=\tr\left\{ \rho_A^2 \right\},
\end{equation}
where $\rho_A$ is the reduced density matrix of subsystem $A$
(which in general need not be a one-party subsystem).  Note that
this is an {\it increasing} entanglement monotone for pure
states---the purity of the local reduced density matrix can only
increase under LOCC.

It has been shown in \cite{Brun04} that every $m$-th degree
polynomial of the components of the density matrix $\rho$ can be
written as an expectation value of an observable $\Op$ on $m$
copies of $\rho$:
\begin{equation}
f(\rho)=\tr\left\{ \Op \rho^{\otimes m} \right\} .
\end{equation}
Here we have
\begin{equation}
\tr\left\{ \rho_A^2 \right\}
  = \tr\left\{ {C} \rho^{\otimes 2} \right\} ,
\end{equation}
where the components of ${C}$ are
\begin{equation}
C_{lpsnkjqm}=\delta_{jp}\delta_{mn}\delta_{lq}\delta_{ks}.
\end{equation}
Therefore
\begin{eqnarray}
\tr\left\{ \frac{\partial I_2}{\partial\rho}[\epsop , \rho]
\right\}
  &=&  \tr\left\{ {C}\left( [\epsop , \rho ]\otimes\rho
    + \rho\otimes[\epsop , \rho ] \right) \right\} \nonumber\\
  &=& \tr_A \left\{ [\epsop , \rho ]_A \rho_A + \rho_A [\epsop,\rho]_A \right\} \nonumber\\
  &=& 2 \tr_A \left\{ \rho_A [\epsop , \rho]_A \right\} ,
\label{ifour}
\end{eqnarray}
where by $\Op_A$ we denote the partial trace of an operator $\Op$
over all subsystems except $A$. If $\epsop$ does not act on
subsystem $A$, then $[\epsop,\rho ]_A = 0$ and the above
expression vanishes. If it acts on subsystem $A$, then $[\epsop ,
\rho ]_A =[\epsop,\rho_A]$ and the expression vanishes due to the
cyclic invariance of the trace.

Now consider condition \eqref{mixed3}. If $\epsop$ does not act on
subsystem $A$, then
\begin{equation}
[[\epsop, \rho],\epsop]_A=0.\label{doublecom}
\end{equation}
From \eqref{mixed3} we get
\begin{eqnarray}
0 &\le& \frac{1}{4}\tr\left\{ \frac{\partial I_2}{\partial\rho}
[[\epsop, \rho],\epsop] \right\}
  + \tr\left\{ \frac{\partial^2 I_2}{\partial\rho^{\otimes2}} \left( \tr\left\{ \epsop\rho \right\} \rho
  - \frac{1}{2} \{\epsop,\rho\} \right)^{\otimes 2} \right\} \nonumber\\
&& = 2\tr\left\{ \left( \tr\{\epsop\rho\} \rho
  - \frac{1}{2}\{\epsop,\rho\} \right)_A^2 \right\}.
\end{eqnarray}
The inequality follows from the fact that
$\left(\tr\{\epsop\rho\}\rho - (1/2) \{\epsop,\rho\}\right)_A^2$
is a positive operator.

If $\epsop$ acts on $A$, we can use the fact that for pure states
\begin{equation}
\tr\left\{ \rho_A^2 \right\}= \tr\left\{ \rho_B^2 \right\},
\end{equation}
where $B$ denotes the subsystem complementary to $A$. Then we can
apply the same argument as before for the function $\tr\left\{
\rho_B^2 \right\}$. Therefore $I_2$ does not \emph{decrease} on
average under local generalized measurements, and is an
entanglement monotone for pure states.

What about mixed states?  For \emph{increasing} entanglement
monotones the convexity condition \eqref{mixed4} becomes a
\emph{concavity} condition---the direction of the inequality is
inverted. In the case of $I_2$, however, we have
\begin{equation}
\tr\left\{ \frac{\partial^2 I_2(\rho)}{\partial\rho^{\otimes
2}}\sigma^{\otimes 2} \right\}
  = 2\tr\left\{ \sigma_A^2 \right\} \geq 0,
\end{equation}
i.e., the function is convex. This means that $\tr\{\rho_A^2\}$ is
\emph{not} a good measure of entanglement for mixed states.
Indeed, when extended to mixed states, $I_2$ cannot distinguish
between entanglement and classical disorder.

\subsection*{3.5.3 \hspace{2pt} Entropy of entanglement}
\addcontentsline{toc}{subsection}{3.5.3 \hspace{0.15cm} Entropy of
entanglement}

Finally consider the von~Neumann entropy of entanglement:
\begin{equation}
S_A=-\tr(\rho_A\log \rho_A).
\end{equation}
Expanding around $\rho_A=\id$, we get
\begin{equation}
S_A=-\tr[(\rho_A-\id)+\frac{1}{2}(\rho_A-\id)^2-\frac{1}{6}(\rho_A-\id)^3
+ ...].
\end{equation}
The LU-invariance follows from the fact that every term in this
expansion satisfies \eqref{mixed1}. If we substitute the $n$-th
term in the condition, we obtain
\begin{equation}
\tr([\epsop,\rho]_A (\rho_A-\id)^{n-1})=0.
\end{equation}
This is true either because $[\epsop,\rho]_A=0$ when $\epsop$ does
not act on $A$, or because otherwise $[\epsop,\rho]_A=
[\epsop,\rho_A]$ and the equation follows from the cyclic
invariance of the trace.

Now to prove that $S_A$ satisfies \eqref{mixed3}, we will first
assume that $\rho_A^{-1}$ exists. Then we can formally write
\begin{equation}
\frac{\partial}{\partial \rho}\log{\rho_A} =
\frac{\partial\rho_A}{\partial \rho}
\frac{\partial}{\partial\rho_A}\log{\rho_A} = \frac{\partial
\rho_A}{\partial\rho}\rho_A^{-1}.
\end{equation}
Consider the case when $\epsop$ does not act on $A$. Substituting
$S_A$ in \eqref{mixed3}, we get
\begin{gather}
\frac{1}{4}\tr\left\{ \frac{\partial S_A}{\partial\rho}[[\epsop,
\rho],\epsop] \right\}
  + \tr\left\{ \frac{\partial^2 S_A}{\partial\rho^{\otimes2}}
  \left( \tr\{\epsop\rho\}\rho - \frac{1}{2}\{\epsop,\rho\}\right)^{\otimes 2} \right\} \nonumber\\
= 0 + \tr\left\{ \left( \frac{\partial}{\partial\rho}\otimes
\left( -\log{\rho_A}\frac{\partial \rho_A}{\partial\rho}
  -\frac{\partial \rho_A}{\partial \rho} \right) \right)
\left( \tr\{\epsop\rho\}\rho - \frac{1}{2}\{\epsop,\rho\} \right)^{\otimes 2} \right\} \nonumber\\
= -\tr\left\{ \left( \rho_A^{-1}\frac{\partial \rho_A}{\partial
\rho}\frac{\partial \rho_A}{\partial \rho} \right)
  \left( \tr\{\epsop\rho\}\rho - \frac{1}{2}\{\epsop,\rho\} \right)^{\otimes 2} \right\} \nonumber\\
= - \tr_A \left\{ \rho_A^{-1}\left(  \tr\{\epsop\rho\}\rho -
\frac{1}{2}\{\epsop,\rho\} \right)_A
  \left( \tr\{\epsop\rho\}\rho - \frac{1}{2}\{\epsop,\rho\} \right)_A \right\} \nonumber\\
= - \tr_A \Biggl\{ \left|\rho_A^{-1/2}\left(\tr\{\epsop\rho\}\rho
- \frac{1}{2}\{\epsop,\rho\} \right)_A\right|^2 \Biggr\} \leq 0.
\label{ent}
\end{gather}
If $\rho_A^{-1}$ does not exist, it is only on a subset of measure
zero---where one or more of the eigenvalues of $\rho_A$ vanish.
Therefore, we can always find an arbitrarily close vicinity in the
parameters describing $\rho_A$, where $\rho_A^{-1}$ is regular and
where \eqref{mixed3} is satisfied. Since the condition is
continuous, it cannot be violated on this special subset.

If $\epsop$ acts on $A$, we can use an equivalent definition of
the entropy of entanglement:
\begin{equation}
S_A=S_B=-\tr\{\rho_B\log \rho_B\},
\end{equation}
and apply the same arguments. Therefore $S_A$ is an entanglement
monotone for pure states.

The convexity condition is not satisfied, since
\begin{equation}
\tr\left\{ \frac{\partial^2
S_A}{\partial\rho^{\otimes2}}\sigma^{\otimes 2} \right\}
  = - \tr\{\rho_A^{-1}\sigma_A^2 \} \leq 0.
\end{equation}
This reflects the fact that the entropy of entanglement, like
$I_2$, does not distinguish between entanglement and classical
randomness.

\section*{3.6 \hspace{2pt} A new entanglement monotone}
\addcontentsline{toc}{section}{3.6 \hspace{0.15cm} A new
entanglement monotone}

It has been shown \cite{Gingrich02} that the set of all
entanglement monotones for a multipartite pure state uniquely
determine the orbit of the state under the action of the group of
local unitary transformations. For three-qubit pure states the
orbit is uniquely determined by 5 independent continuous
invariants (not counting the norm) and one discrete invariant
\cite{Acin00,Carteret00}. Therefore, for pure states of three
qubits there must exist five independent continuous entanglement
monotones that are functions of the five independent continuous
invariants.

Any polynomial invariant in the amplitudes of a state
\[
|\psi\rangle =
\underset{i,j,k\ldots}{\sum}\alpha_{ijk\ldots}|i_A\rangle
|j_B\rangle |k_C\rangle \cdots
\]
is a sum of homogenous polynomials of the form \cite{Sudbery01}
\begin{equation}
P_{\sigma\tau\cdots}(|\psi\rangle)= \alpha_{i_1 j_1 k_1 \ldots}
\alpha^*_{i_1 j_{\sigma(1)} k_{\tau(1)} \ldots} \cdots \alpha_{i_n
j_n k_n \ldots} \alpha^*_{i_n j_{\sigma(n)} k_{\tau(n)} \ldots},
\label{inv}
\end{equation}
where $\sigma, \tau, \ldots$ are permutations of (1,2,\ldots,n),
and repeated indices indicate summation. A set of five independent
polynomial invariants for three-qubit pure states is
\cite{Sudbery01}
\begin{eqnarray}
I_1&=&P_{e,(12)}\\
I_2&=&P_{(12),e}\\
I_3&=&P_{(12),(12)}\\
I_4&=&P_{(123),(132)}\\
I_5&=&| \alpha_{i_1j_1k_1} \alpha_{i_2j_2k_2} \alpha_{i_3j_3k_3}
\alpha_{i_4j_4k_4} \epsilon_{i_1i_2} \epsilon_{i_3i_4}
\epsilon_{j_1j_2} \epsilon_{j_3j_4} \epsilon_{k_1k_3}
\epsilon_{k_2k_4} |^2.
\end{eqnarray}
In the last expression $\epsilon_{ij}$ is the antisymmetric tensor
in 2 dimensions. The first three invariants are the local purities
of subsystems C, B and A, $I_4$ is the invariant identified by
Kempe \cite{Kempe99} and $I_5$ is (up to a factor) the square of
the 3-tangle identified by Coffman, Kundu and Wootters
\cite{Coffman00}. According to \cite{Gingrich02} the four known
independent continuous entanglement monotones that do not require
maximization over a multi-dimensional space are
\begin{gather}
\tau_{(AB)C}=2(1-I_1)\\
\tau_{(AC)B}=2(1-I_2)\\
\tau_{(BC)A}=2(1-I_3)\\
\tau_{ABC}= 2\sqrt{I_5},
\end{gather}
and any fifth independent entanglement monotone must depend on
$I_4$. Numerical evidence suggested that the tenth order
polynomial $\sigma_{ABC}= 3-(I_1+I_2+I_3)I_4$ might be such an
entanglement monotone. However, no rigorous proof of monotonicity
was given. Here, we will use conditions \eqref{mixed1} and
\eqref{mixed3} to construct a different independent entanglement
monotone, which is of sixth order in the amplitudes of the state
and their complex conjugates.

Observe that in \eqref{inv} the amplitudes have been combined in
such a way that subsystem A is manifestly traced out. By
appropriate rearrangement, one can write the same expression in a
form where an arbitrary subsystem is manifestly traced out.
Therefore, any polynomial invariant can be written entirely in
terms of the components of $\tr_A\left\{\rho\right\}$ or
$\tr_B\left\{\rho\right\}$, etc. This immediately implies that the
LU-invariance condition \eqref{mixed1} is satisfied, since if
$\epsop$ acts on subsystem A, we can consider the expression in
terms of $\rho_{BC...}$, which, when substituted in
\eqref{mixed1}, would yield zero because
$[\epsop,\rho]_{BC...}=0$. It also implies that in order to prove
monotonicity under local measurements we can only consider the
second term in \eqref{mixed3}, since when $\epsop$ acts on
subsystem A, we can again consider the expression for the function
only in terms of $\rho_{BC...}$ and the first term would vanish
according to \eqref{doublecom}.

We will aim at constructing a polynomial function of three-qubit
pure states $\rho$ which has the same form when expressed in terms
of $\rho_{AB}$, $\rho_{AC}$, or $\rho_{BC}$, in order to avoid the
necessity for separate proofs of monotonicity under measurements
on the different subsystems. It has been shown in \cite{Sudbery01}
that
\begin{eqnarray}
I_4&=& 3\tr\left\{\rho_{AB}(\rho_A\otimes \rho_B)\right\} -
\tr\left\{\rho_A^3\right\} - \tr\left\{\rho_B^3\right\}\nonumber\\
&=&3\tr\left\{\rho_{AC}(\rho_A\otimes \rho_C)\right\} -
\tr\left\{\rho_A^3\right\} - \tr\left\{\rho_C^3\right\}\nonumber\\
&=&3\tr\left\{\rho_{BC}(\rho_B\otimes \rho_C)\right\} -
\tr\left\{\rho_B^3\right\} -
\tr\left\{\rho_C^3\right\}.\label{sudbery}
\end{eqnarray}
For local measurements on subsystem C it is convenient to use the
first of the above expressions for $I_4$. The terms
$\tr\left\{\rho_A^3\right\}$ and $\tr\left\{\rho_B^3\right\}$ are
entanglement monotones by themselves. This can be easily seen by
plugging them in condition \eqref{mixed3}:
\begin{gather}
\frac{1}{4}\tr\left\{ \frac{\partial
\tr\left\{\rho_{A,B}^3\right\}}{\partial\rho}[[\epsop,
\rho],\epsop] \right\}
  + \tr\left\{ \frac{\partial^2 \tr\left\{\rho_{A,B}^3\right\}}{\partial\rho^{\otimes2}}
  \left( \tr(\epsop\rho)\rho - \frac{1}{2}\{\epsop,\rho\} \right)^{\otimes 2} \right\}
\nonumber\\
  =0+6 \tr\left\{\rho_{A,B} \left( \tr\{\epsop\rho\} \rho
  - \frac{1}{2}\{\epsop,\rho\} \right)_{A,B}^2 \right\}\geq 0.
\end{gather}

These terms, however, are not independent of the invariants $I_2$
and $I_3$. The term which is independent of the other polynomial
invariants is $\tr\left\{\rho_{AB}(\rho_A\otimes\rho_B)\right\}$.
When we plug this term into condition \eqref{mixed3} we obtain an
expression which is not manifestly positive or negative. Is it
possible to construct a function dependent on this term, which
similarly to $\tr\left\{\rho_{A,B}^3\right\}$ would yield a trace
of a manifestly positive operator when substituted in
\eqref{mixed3}?

It is easy to see that if the function has the form
$\tr\left\{\Xhat^3\right\}$, where the operator $\Xhat(\rho_{AB})$
is a positive operator linearly dependent on $\rho_{AB}$, it will
be an increasing monotone under local measurements on C (for
simplicity we assume $\Xhat(0)={0}$):
\begin{gather}
\frac{1}{4}\tr\left\{ \frac{\partial\tr\left\{ \Xhat^3(\rho_{AB})
\right\}}{\partial\rho}[[\epsop,\rho],\epsop] \right\}
  + \tr\left\{ \frac{\partial^2 \tr\left\{ \Xhat^3(\rho_{AB}) \right\}}{\partial\rho^{\otimes2}}
  \left( \tr(\epsop\rho)\rho - \frac{1}{2}\{\epsop,\rho\} \right)^{\otimes 2} \right\} \nonumber\\
= 0 + 6 \tr\left\{\Xhat(\rho_{AB}) \Xhat^2(( \tr\{\epsop\rho\}
\rho - \frac{1}{2}\{\epsop,\rho\} )_{AB}) \right\} \geq 0.
\end{gather}
Since we want the function to depend on
$\tr\left\{\rho_{AB}(\rho_A\otimes \rho_B)\right\}$, we choose
$\Xhat(\rho_{AB}) = 2\rho_{AB} + \rho_A\otimes I_B + I_A\otimes
\rho_B$.  This is clearly positive for positive $\rho_{AB}$.
Expanding the trace, we obtain:
\begin{gather}
\tr\left\{ \Xhat^3(\rho_{AB}) \right\} =
12\tr\left\{\rho_{AB}(\rho_A\otimes\rho_B) \right\}
  +12\tr\left\{ \rho_{AB}^2(I_A\otimes\rho_B)\right\} \nonumber\\
+  12\tr\left\{ \rho_{AB}^2(\rho_A\otimes I_B) \right\}
  + 6\tr\left\{\rho_{AB}(I_A\otimes\rho_B)^2 \right\}
  + 6\tr\left\{\rho_{AB}(\rho_A\otimes I_B)^2 \right\}\nonumber\\
  + 3\tr\left\{\rho_A\otimes\rho_B^2\right\}
  + 3\tr\left\{ \rho_A^2\otimes\rho_B \right\}
  + \tr\left\{ I_A\otimes\rho_B^3 \right\}
  + \tr\left\{ \rho_A^3\otimes I_B \right\}
  + 8\tr\left\{\rho_{AB}^3 \right\} \nonumber\\
= 12\tr\left\{\rho_{AB}(\rho_A\otimes\rho_B) \right\}
  + 12\tr\left\{ \rho_{AB}^2(I_A\otimes\rho_B) \right\}
  + 12\tr\left\{\rho_{AB}^2(\rho_A\otimes I_B)\right\} \nonumber\\
 +8\tr\left\{\rho_A^3\right\}
  + 8\tr\left\{\rho_B^3\right\}+8\tr\left\{\rho_{AB}^3 \right\}
  + 3\tr\left\{\rho_A^2\right\}+3\tr\left\{\rho_B^2\right\} .
\end{gather}

One can show that \begin{gather} \tr\left\{
\rho_{AB}^2(I_A\otimes\rho_B) \right\} = \tr\left\{
\rho_{BC}(\rho_B\otimes\rho_C) \right\},\\
 \tr\left\{
\rho_{AB}^2(\rho_A\otimes I_B) \right\}= \tr\left\{
\rho_{AC}(\rho_A\otimes\rho_C) \right\}.
\end{gather}
We also have that $\tr\left\{ \rho_{AB}^3 \right\} = \tr\left\{
\rho_C^3 \right\}$. Using this and \eqref{sudbery}, we obtain
\begin{equation}
\tr\left\{ \Xhat^3(\rho_{AB}) \right\} = 12 I_4 + 16
\left(\tr\left\{ \rho_A^3 \right\} + \tr\left\{ \rho_B^3 \right\}
  +\tr\left\{ \rho_C^3\right\} \right) +3\tr\left\{ \rho_A^2 \right\} + 3\tr\left\{ \rho_B^2 \right\} .
\end{equation}
This expression is an increasing monotone under local measurements
on C. If we add to it $3\tr\left\{ \rho_{AB}^2\right\} =
3\tr\left\{ \rho_C^2 \right\}$, it becomes invariant under
permutations of the subsystems. Since $\tr\left\{ \rho_C^2
\right\}$ is an increasing entanglement monotone, the whole
expression will be a monotone under operations on any subsystem.
We can define the closely related quantity
\begin{equation}
\phi_{ABC}=69-\tr\left\{(2\rho_{AB} + \rho_A\otimes I_B +
I_A\otimes \rho_B)^3\right\}-3\tr\left\{ \rho_{AB}^2\right\}.
\end{equation}
This is a {\it decreasing} entanglement monotone that vanishes for
product states, which is more standard for a measure of
entanglement.  It depends on the invariant identified by Kempe and
is therefore independent of the other known monotones for
three-qubit pure states.

\section*{3.7 \hspace{2pt} Summary and outlook}
\addcontentsline{toc}{section}{3.7 \hspace{0.15cm} Summary and
outlook}

We have derived differential conditions for a twice-differentiable
function on quantum states to be an entanglement monotone.  There
are two such conditions for pure-state entanglement
monotones---invariance under local unitaries and diminishing under
local measurements---plus a third condition (overall convexity of
the function) for mixed-state entanglement monotones. We have
shown that these conditions are both necessary and sufficient.  We
then verified that the conditions are satisfied by a number of
known entanglement monotones and we used them to construct a new
polynomial entanglement monotone for three-qubit pure states.

It is our hope that this approach to the study of entanglement may
circumvent some of the difficulties that arise due the
mathematically complicated nature of LOCC. It may be possible to
find new classes of entanglement monotones, for both pure and
mixed states, and to look for functions with particularly
desirable properties (such as additivity).

There may also be other areas of quantum information theory where
it will prove advantageous to consider general quantum operations
as continuous processes. This seems a very promising new direction
for research.

\section*{3.8 \hspace{1pt} Appendix: Proof of sufficiency}
\addcontentsline{toc}{section}{3.8 \hspace{0.15cm} Appendix: Proof
of sufficiency}

The LU-invariance condition can be written as
\begin{equation}
F(\rho,\epsop)=0,
\end{equation}
where we define
\begin{equation}
F(\rho,\epsop)=f(e^{i\epsop}\rho e^{-i\epsop})-f(\rho)
\end{equation}
with $\epsop$ being a local hermitian operator. This condition has
to be satisfied for every $\rho$ and every $\epsop$. By expanding
up to first order in $\epsop$ we obtained condition
\eqref{mixed1}, which is equivalent to
\begin{equation}
\tr\left\{ \left.\frac{\partial F(\rho,\epsop)}{\partial\epsop}
  \right|_{\epsop={0}}\epsop\right\} = 0.
\end{equation}
This is a linear form of the components of $\epsop$ and the
requirement that it vanishes for every $\epsop$ implies that
\begin{equation}
\left.\frac{\partial F(\rho,\epsop)}{\partial\varepsilon_{ij}}
\right|_{\epsop={0}}=0.
\end{equation}
This has to be satisfied for every $\rho$. Consider the first
derivative of $F(\rho,\epsop)$ with respect to $\varepsilon_{ij}$,
taken at an arbitrary point $\epsop_0$. We have
\begin{equation}
\left.\frac{\partial F(\rho,\epsop)}{\partial\varepsilon_{ij}}
  \right|_{\epsop=\epsop_0}
= \left.\frac{\partial
F(\rho,\epsop_0+\epsop)}{\partial\varepsilon_{ij}}
  \right|_{\epsop={0}}.
\end{equation}
But from the form of $F(\rho,\epsop)$ one can see that
$F(\rho,\epsop_0+\epsop)=F(\rho',\epsop)$, where
$\rho'=e^{i\epsop_0}\rho e^{-i\epsop_0}$. Therefore

\begin{equation}
\left.\frac{\partial F(\rho,\epsop)}{\partial\varepsilon_{ij}}
  \right|_{\epsop=\epsop_0}
= \left.\frac{\partial F(\rho',\epsop)}{\partial\varepsilon_{ij}}
  \right|_{\epsop={0}} =0,
\end{equation}
i.e., the first derivatives of $F(\rho,\epsop)$ with respect to
the components of $\epsop$ vanish identically. This means that
$F(\rho,\epsop)=F(\rho,{0})=0$ for every $\epsop$ and condition
\eqref{mixed1} is sufficient.

The condition for non-increase on average under local generalized
measurements \eqref{nonincrease4} can be written as
\begin{equation}
G(\rho,\epsop) \leq 0, \label{nonincrease5}
\end{equation}
where
\begin{equation}
G(\rho,\epsop) =p_1 f(\Mhat_1\rho \Mhat_1/p_1) + p_2 f(\Mhat_2\rho
\Mhat_2/p_2) - f(\rho).
\end{equation}
The operators $\Mhat_1$ and $\Mhat_2$ in terms of $\epsop$ are
given by \eqref{mm}, and the probabilities $p_1$ and $p_2$ are
defined as before. As we have argued in Section 3.3, it is
sufficient that this condition is satisfied for infinitesimal
$\epsop$. By expanding the condition up to second order in
$\epsop$ we obtained condition \eqref{mixed3}, which is equivalent
to
\begin{equation}
\tr\left\{
\left.\frac{\partial^2G(\rho,\epsop)}{\partial\epsop^{\otimes2}}
  \right|_{\epsop={0}}\epsop^{\otimes 2} \right\} \leq 0.
\end{equation}
Clearly, if this condition is satisfied by a strict inequality, it
is sufficient, since corrections of higher order in $\epsop$ can
be made arbitrarily smaller in magnitude by taking $\epsop$ small
enough. Concerns about the contribution of higher-order
corrections may arise only if the second-order correction to
$G(\rho, \epsop)$ vanishes in some open vicinity of $\rho$ and
some open vicinity of $\epsop$ (we have assumed that the function
$f(\rho)$ is continuous). But the second-order correction is a
real quadratic form of the components of $\epsop$ and it can
vanish in an open vicinity of $\epsop$, only if it vanishes for
every $\epsop$, i.e., if
\begin{equation}
\left.\frac{\partial^2G(\rho,\epsop)}{\partial\varepsilon_{ij}\partial\varepsilon_{kl}}
  \right|_{\epsop={0}}=0.
\label{zero}
\end{equation}
We will now show that if \eqref{zero} is satisfied in an open
vicinity of $\rho$, there exists an open vicinity of $\epsop={0}$
in which all second derivatives of $G(\rho,\epsop)$ with respect
to $\epsop$ vanish identically. This means that all higher-order
corrections to $G(\rho,\epsop)$ vanish in this vicinity and
\eqref{nonincrease5} is satisfied with equality.

Consider the two terms of $G(\rho,\epsop)$ that depend on
$\epsop$:
\begin{equation}
G_1(\rho,\epsop)=p_1 f(\Mhat_1\rho \Mhat_1/p_1),
\end{equation}
\begin{equation}
G_2(\rho,\epsop)=p_2 f(\Mhat_2\rho \Mhat_2/p_2).
\end{equation}
They differ only by the sign of $\epsop$, i.e. $G_1(\rho,\epsop)
=G_2(\rho,-\epsop)$, and therefore
\begin{equation}
\left.\frac{\partial^2
G_1(\rho,\epsop)}{\partial\varepsilon_{ij}\partial\varepsilon_{kl}}\right|_{\epsop={0}}=\left.\frac{\partial^2
G_2(\rho,\epsop)}{\partial\varepsilon_{ij}\partial\varepsilon_{kl}}\right|_{\epsop={0}}=\frac{1}{2}\left.\frac{\partial^2
G(\rho,\epsop)}{\partial\varepsilon_{ij}\partial\varepsilon_{kl}}\right|_{\epsop={0}}.
\end{equation}
If \eqref{zero} is satisfied in an open vicinity of $\rho$, we
have
\begin{equation}
\left.\frac{\partial^2
G_1(\rho,\epsop)}{\partial\varepsilon_{ij}\partial\varepsilon_{kl}}\right|_{\epsop={0}}=\left.\frac{\partial^2
G_2(\rho,\epsop)}{\partial\varepsilon_{ij}\partial\varepsilon_{kl}}\right|_{\epsop={0}}=0
\label{cond}
\end{equation}
in this vicinity. Consider the second derivatives of
$G(\rho,\epsop)$ with respect to the components of $\epsop$, taken
at a point $\epsop_0$:
\begin{equation}
\begin{split}
\left.\frac{\partial^2G(\rho,\epsop)}{\partial\varepsilon_{ij}\partial\varepsilon_{kl}}
  \right|_{\epsop=\epsop_0}
=
\left.\frac{\partial^2G_1(\rho,\epsop)}{\partial\varepsilon_{ij}\partial\varepsilon_{kl}}
  \right|_{\epsop=\epsop_0}
+
\left.\frac{\partial^2G_2(\rho,\epsop)}{\partial\varepsilon_{ij}\partial\varepsilon_{kl}}
  \right|_{\epsop=\epsop_0}  \\
=
\left.\frac{\partial^2G_1(\rho,\epsop_0+\epsop)}{\partial\varepsilon_{ij}\partial\varepsilon_{kl}}
  \right|_{\epsop={0}}
+
\left.\frac{\partial^2G_2(\rho,\epsop_0+\epsop)}{\partial\varepsilon_{ij}\partial\varepsilon_{kl}}
  \right|_{\epsop={0}}.
\end{split}
\end{equation}

From the expression for $G_1(\rho,\epsop)$ one can see that
$\epsop$ occurs in $G_1(\rho,\epsop)$ only in the combination
$\sqrt{\frac{\id-\epsop}{2}}\rho\sqrt{\frac{\id-\epsop}{2}}$. In
$G_1(\rho,\epsop_0+\epsop)$ it will appear only in
$\sqrt{\frac{\id-\epsop_0-\epsop}{2}}\rho\sqrt{\frac{\id-\epsop_0-\epsop}{2}}$.
But
\begin{equation}
\sqrt{\frac{\id-\epsop_0-\epsop}{2}}=\sqrt{\frac{\id-\epsop'}{2}}\sqrt{\id-\epsop_0},
\end{equation}
where
\begin{equation}
\epsop'=\epsop(\id-\epsop_0)^{-1}.
\end{equation}
So we can write
\begin{equation}
\sqrt{\frac{\id-\epsop_0-\epsop}{2}}\rho\sqrt{\frac{\id-\epsop_0-\epsop}{2}}
= p'\sqrt{\frac{\id-\epsop'}{2}}\rho'\sqrt{\frac{\id-\epsop'}{2}},
\end{equation}
where
\begin{equation}
\rho' = \left( \sqrt{\id-\epsop_0}\rho\sqrt{\id-\epsop_0}
\right)/p' \label{rhoprime}
\end{equation}
and
\begin{equation}
p'=\tr\left\{\sqrt{\id-\epsop_0}\rho\sqrt{\id-\epsop_0}\right\}.
\end{equation}
Then one can verify that
\begin{equation}
G_1(\rho,\epsop_0+\epsop)=p'G_1(\rho',\epsop').
\end{equation}
Similarly
\begin{equation}
G_2(\rho,\epsop_0+\epsop)=p''G_2(\rho'',\epsop''),
\end{equation}
where
\begin{equation}
\epsop''=\epsop(\id+\epsop_0)^{-1},
\end{equation}
\begin{equation}
\rho'' = \left(\sqrt{\id+\epsop_0}\rho\sqrt{\id+\epsop_0}
\right)/p'', \label{rhodoubleprime}
\end{equation}
\begin{equation}
p''=\tr\left\{\sqrt{\id+\epsop_0}\rho\sqrt{\id+\epsop_0}\right\}.
\end{equation}
Note that $\partial\varepsilon'_{pq}/\partial\varepsilon_{ij}$ and
$\partial\varepsilon''_{pq}/\partial\varepsilon_{ij}$ have no
dependence on $\epsop$.  Nor do $p'$ and $p''$.  Therefore we
obtain
\begin{equation}
\begin{split}
\left.\frac{\partial^2
G(\rho,\epsop)}{\partial\varepsilon_{ij}\partial\varepsilon_{kl}}\right|_{\epsop=\epsop_0}
=p'\left.\frac{\partial^2
G_1(\rho',\epsop')}{\partial\varepsilon_{ij}\partial\varepsilon_{kl}}\right|_{\epsop={0}}+
p''\left.\frac{\partial^2
G_2(\rho'',\epsop'')}{\partial\varepsilon_{ij}\partial\varepsilon_{kl}}\right|_{\epsop={0}} \\
=
\sum_{p,q,r,s}\frac{\partial\varepsilon'_{pq}}{\partial\varepsilon_{ij}}
\frac{\partial\varepsilon'_{rs}}{\partial\varepsilon_{kl}}
p'\left.\frac{\partial^2
G_1(\rho',\epsop')}{\partial\varepsilon'_{pq}\partial\varepsilon'_{rs}}\right|_{\epsop'={0}}
 +\\
\sum_{p,q,r,s}\frac{\partial\varepsilon''_{pq}}{\partial\varepsilon_{ij}}
\frac{\partial\varepsilon''_{rs}}{\partial\varepsilon_{kl}}
p''\left.\frac{\partial^2
G_2(\rho'',\epsop'')}{\partial\varepsilon''_{pq}\partial\varepsilon''_{rs}}\right|_{\epsop''={0}}.
\end{split}
\end{equation}
We assumed that \eqref{cond} is satisfied in an open vicinity of
$\rho$. If $\rho'$ and $\rho''$ are within this vicinity, the
above expression will vanish. But from \eqref{rhoprime} and
\eqref{rhodoubleprime} we see that as $\|\epsop_0\|$ tends to
zero, the quantities $\|\rho'-\rho\|$ and $\|\rho''-\rho\|$ also
tend to zero. Therefore there exists an open vicinity of
$\epsop_0={0}$, such that for every $\epsop_0$ in this vicinity,
the corresponding $\rho'$ and $\rho''$ will be within the vicinity
of $\rho$ for which \eqref{cond} is satisfied and
\begin{equation}
\left.\frac{\partial^2
G(\rho,\epsop)}{\partial\varepsilon_{ij}\partial\varepsilon_{kl}}\right|_{\epsop=\epsop_0}=0.
\end{equation}
This means that higher derivatives of $G(\rho,\epsop)$ with
respect to the components of $\epsop$ taken at points in this
vicinity will vanish, in particular derivatives taken at
$\epsop={0}$. So higher order corrections in $\epsop$ to
$G(\rho,\epsop)$ will also vanish. Therefore $G(\rho,\epsop)=0$ in
the vicinity of $\rho$ for which we assumed that \eqref{mixed3} is
satisfied with equality, which implies that condition
\eqref{mixed3} is sufficient.

\chapter*{Chapter 4: \hspace{1pt} Non-Markovian dynamics of a qubit coupled to a spin bath
via the Ising interaction} \addcontentsline{toc}{chapter}{Chapter
4:\hspace{0.15cm} Non-Markovian dynamics of a qubit coupled to a
spin bath via the Ising interaction}

In this chapter, we turn our attention to the deterministic
dynamics of open quantum systems. The chapter is based on a study
made in collaboration with Hari Krovi, Mikhail Ryazanov and Daniel
Lidar \cite{KORL07}.

\section*{4.1 \hspace{2pt} Preliminaries}
\addcontentsline{toc}{section}{4.1 \hspace{0.15cm} Preliminaries}

As we pointed out in Chapter 1, a major conceptual as well as
technical difficulty in the practical implementation of quantum
information processing schemes is the unavoidable interaction of
quantum systems with their environment. This interaction can
destroy quantum superpositions and lead to an irreversible loss of
information, a process known as decoherence. Understanding the
dynamics of open quantum systems is therefore of considerable
importance. The Schr\"{o}dinger equation, which describes the
evolution of closed systems, is generally inapplicable to open
systems, unless one includes the environment in the description.
This is, however, generally difficult, due to the large number of
environment degrees of freedom. An alternative is to develop a
description for the evolution of only the subsystem of interest. A
multitude of different approaches have
been developed in this direction, exact as well as approximate \cite%
{Alicki:87,BrePet02}. Typically the exact approaches are of
limited practical usefulness as they are either phenomenological
or involve complicated integro-differential equations. The various
approximations lead to regions of validity that have some overlap.
Such techniques have been studied for many different models, but
their performance in general, is not fully understood.

In this work we consider an exactly solvable model of a single
qubit coupled to an environment of qubits. We are motivated by the
physical importance of such spin bath models \cite{Prokofev:00} in
the description of decoherence in solid state quantum information
processors,
such as systems based on the nuclear spin of donors in semiconductors \cite%
{Kane:98,Vrijen:00}, or on the electron spin in quantum dots
\cite{Loss:98}. Rather than trying to accurately model decoherence
due to the spin bath in such systems (as in, e.g., Refs.
\cite{sousa:115322,Witzel:06}), our goal in this work is to
compare the performance of different master equations which have
been proposed in the literature. Because the model we consider is
exactly solvable, we are able to accurately assess the performance
of the approximation techniques that we study. In particular, we
study the Born-Markov and Born master equations, and the
perturbation expansions of the Nakajima-Zwanzig (NZ)
\cite{Nak58,Zwa60} and the time-convolutionless (TCL) master
equations \cite{Shibata77,ShiAri80} up to fourth order in the
coupling constant. We also study the post-Markovian (PM) master
equation proposed in \cite{ShabaniLidar:05}.

The dynamics of the system qubit in the model we study is highly
non-Markovian and hence we do not expect the traditional Markovian
master equations commonly used, e.g., in quantum optics \cite{Car99}
and nuclear magnetic resonance \cite{Slichter:book}, to be accurate.
This is typical of spin baths, and was noted, e.g., by Breuer et al.
\cite{BBP04}. As we will see in Chapter 5, the non-Markovian
character of the dynamics can be used to our advantage in
error-correction schemes, hence understanding these models is of
special significance. The work by Breuer et al. (as well as by other
authors in a number of
subsequent publications \cite%
{Palumbo:06,Burgarth:06,Hamdouni:06,Yuan:07,Camalet:07,Jing:07})
is conceptually close to ours in that in both cases an
analytically solvable spin-bath model is considered and the
analytical solution for the open system dynamics is compared to
approximations. However, there are also important differences,
namely, in Ref. \cite{BBP04} a so-called spin-star system was
studied, where the system spin has equal couplings to all the bath
spins, and these are of the XY exchange-type. In contrast, in our
model the system spin interacts via Ising couplings with the bath
spins, and we allow for arbitrary coupling constants. As a result
there are also important
differences in the dynamics. For example, unlike the model in Ref. \cite%
{BBP04}, for our model we find that the odd order terms in the
perturbation expansions of Nakajima-Zwanzig and
time-convolutionless master equations are
non-vanishing. This reflects the fact that there is a coupling between the $%
x $ and $y$ components of the Bloch vector which is absent in
\cite{BBP04}. In view of the non-Markovian behavior of our model,
we also discuss the relation between a representation of the
analytical solution of our model in terms of completely positive
maps, and the Markovian limit obtained via a coarse-graining
method introduced in \cite{Lidar:CP01}, and the performance of the
post-Markovian master equation \cite{ShabaniLidar:05}.

This chapter is organized as follows. In Section 4.2, we present
the model, derive the exact solution and discuss its behavior in
the limit of small times and large number of bath spins, and in
the cases of discontinuous spectral density co-domain and
alternating sign of the system-bath coupling constants. In Section
4.3, we consider second order approximation methods such as the
Born-Markov and Born master equations, and a coarse-graining
approach to the Markovian semigroup master equation. Then we
derive solutions to higher order corrections obtained from the
Nakajima-Zwanzig and time-convolutionless projection techniques as
well as derive the optimal approximation achievable through the
post-Markovian master equation. In Section 4.4, we compare these
solutions for various parameter values in the model and plot the
results. Finally in Section 4.5, we present our conclusions.

\section*{4.2 \hspace{2pt} Exact dynamics}
\addcontentsline{toc}{section}{4.2 \hspace{0.15cm} Exact dynamics}

\subsection*{4.2.1 \hspace{2pt} The model}
\addcontentsline{toc}{subsection}{4.2.1 \hspace{0.15cm} The model}

We consider a single spin-$\frac{1}{2}$ system (i.e., a qubit with
a
two-dimensional Hilbert space $\mathcal{H}_{S}$) interacting with a bath of $%
N$ spin-$\frac{1}{2}$ particles (described by an $N$-fold tensor
product of two-dimensional Hilbert spaces denoted
$\mathcal{H}_{B}$). The observables describing the spin of a
spin-$\frac{1}{2}$ particle in each of the three spatial
directions are described by the Pauli operators
\begin{equation}
\sigma^x=\begin{pmatrix} 0&1\\
1&0
\end{pmatrix},
\sigma^y=\begin{pmatrix} 0&-i\\
i&0
\end{pmatrix},
\sigma^z=\begin{pmatrix} 1&0\\
0&-1
\end{pmatrix}.
\end{equation}
We model the interaction between the system qubit and the bath by
the Ising Hamiltonian
\begin{equation}
H_{I}^{\prime }=\alpha \sigma ^{z}\otimes
\sum_{n=1}^{N}g_{n}\sigma _{n}^{z}, \label{eq:HI}
\end{equation}%
where $g_{n}$ are dimensionless real-valued coupling constants in
the interval $[-1,1]$ ($n$ labels the different qubits in the
bath), and $\alpha >0$ is a parameter having the dimension of
frequency (we work in units in which $\hbar =1$), which describes
the
coupling strength and will be used below in conjunction with time ($\alpha t$%
) for perturbation expansions.

The system and bath Hamiltonians are
\begin{equation}
H_{S}=\frac{1}{2}\omega _{0}\sigma ^{z}  \label{SystemHam}
\end{equation}%
and
\begin{equation}
H_{B}=\sum_{n=1}^{N}\frac{1}{2}\Omega _{n}\sigma _{n}^{z}.
\end{equation}%
For definiteness, we restrict the frequencies $\omega _{0}$ and
$\Omega _{n}$ to the interval $[-1,1]$, in inverse time units.
Even though the units of time can be arbitrary, by doing so we do
not lose generality, since we will be working in the interaction
picture where only the frequencies $\Omega _{n} $ appear in
relation to the state of the bath [Eq.~(\ref{eq:rhoB0})]. Since
the ratios of these frequencies and the temperature of the bath
occur in the equations, only their values relative to the
temperature are of interest. Therefore, henceforth we will omit
the units of frequency and temperature and will treat these
quantities as dimensionless.

The interaction picture is defined as the transformation of any
operator
\begin{equation}
A\mapsto A(t)=\exp (iH_{0}t)A\exp (-iH_{0}t),
\end{equation}%
where $H_{0}=H_{S}+H_{B}$. The interaction Hamiltonian $H_{I}$
chosen here is invariant under this transformation since it
commutes with $H_{0}$. [Note that in the next subsection, to
simplify our calculations we redefine $H_{S}$ and $H_{I}^{\prime
}$ (whence $H_{I}^{\prime }$ becomes $H_{I}$), but this does not
alter the present analysis.] All the quantities discussed in the
rest of this article are assumed to be in the interaction picture.

The dynamics can be described using the superoperator notation for
the Liouville operator
\begin{equation}
\mathcal{L}\rho (t)\equiv -i[H_{I}^{\prime },\rho (t)],
\end{equation}%
where $\rho (t)$ is the density matrix for the total system in the
Hilbert space $\mathcal{H}_{S}\otimes \mathcal{H}_{B}$. The
dynamics is governed by the von Neumann equation
\begin{equation}
\frac{d}{dt}\rho (t)=\alpha \mathcal{L}\rho (t)
\end{equation}%
and the formal solution of this equation can be written as
follows:
\begin{equation}
\rho (t)=\exp (\alpha \mathcal{L}t)\rho (0).  \label{VNeqnSoln}
\end{equation}%
The state of the system is given by the reduced density operator
\begin{equation}
\rho _{S}(t)=\mathrm{Tr}_{B}\{\rho (t)\},
\end{equation}%
where $\mathrm{Tr}_{B}$ denotes a partial trace taken over the
bath Hilbert space $\mathcal{H}_{B}$. This can also be written in
terms of the Bloch sphere vector
\begin{equation}
\vec{v}(t)=%
\begin{pmatrix}
v_{x}(t) \\
v_{y}(t) \\
v_{z}(t)%
\end{pmatrix}%
=\mathrm{Tr}\{\vec{\sigma}\rho _{S}(t)\},
\end{equation}%
where $\vec{\sigma}\equiv (\sigma ^{x},\sigma ^{y},\sigma ^{z})$
is the vector of Pauli matrices. In the basis of $\sigma ^{z}$
eigenstates this is equivalent to
\begin{eqnarray}
\rho _{S}(t) =\frac{1}{2}(I+\vec{v}\cdot \vec{\sigma})
=\frac{1}{2}%
\begin{pmatrix}
1+v_{z}(t) & v_{x}(t)-iv_{y}(t) \\
v_{x}(t)+iv_{y}(t) & 1-v_{z}(t)%
\end{pmatrix}%
.  \label{I+vs}
\end{eqnarray}%
We assume that the initial state is a product state, i.e.,
\begin{equation}
\rho (0)=\rho _{S}(0)\otimes \rho _{B},
\end{equation}%
and that the bath is initially in the Gibbs thermal state at a
temperature $T $
\begin{equation}
\rho _{B}=\exp (-H_{B}/kT)/\mathrm{Tr}[\exp (-H_{B}/kT)],
\label{eq:rhoB0}
\end{equation}%
where $k$ is the Boltzmann constant. Since $\rho _{B}$ commutes
with the interaction Hamiltonian $H_{I}$, the bath state is
stationary throughout the dynamics:\ $\rho _{B}(t)=\rho _{B}$.
Finally, the bath spectral density
function is defined as usual as%
\begin{equation}
J(\Omega )=\sum_{n}|g_{n}|^{2}\delta (\Omega -\Omega _{n}).
\label{eq:J}
\end{equation}

\subsection*{4.2.2 \hspace{2pt} Exact solution for the evolution of the system qubit}
\addcontentsline{toc}{subsection}{4.2.2 \hspace{0.15cm} Exact
solution for the evolution of the system qubit}

We first shift the system Hamiltonian in the following way:
\begin{eqnarray}
H_{S} \mapsto H_{S}+\theta I, \hspace{0.5cm}  \theta \equiv
\mathrm{Tr}\{\sum_{n}g_{n}\sigma _{n}^{z}\rho _{B}\}.
\label{theta}
\end{eqnarray}%
As a consequence the interaction Hamiltonian is modified from Eq. (\ref%
{eq:HI}) to
\begin{equation}
H_{I}^{\prime }\mapsto H_{I}=\alpha \sigma ^{z}\otimes B,
\end{equation}%
where
\begin{equation}
B\equiv \sum_{n}g_{n}\sigma _{n}^{z}-\theta I_{B}.  \label{Bcomp}
\end{equation}%
This shift is performed because now $\mathrm{Tr}_{B}[H_{I},\rho
(0)]=0$, or equivalently
\begin{equation}
\mathrm{Tr}_{B}\{B\rho _{B}\}=0.
\end{equation}%
This property will simplify our calculations later when we
consider approximation techniques in Section 4.3. Now, we derive
the exact solution for the reduced density operator $\rho _{S}$
corresponding to the system. We do this in two different ways. The
Kraus operator sum representation is a standard description of the
dynamics of a system initially decoupled from its environment and
it will also be helpful in studying the coarse-graining approach
to the quantum semigroup master equation. The second method is
computationally more effective and is helpful in obtaining
analytical expressions for $N\gg 1$.

\subsubsection*{4.2.2.1 \hspace{2pt} Exact Solution in the Kraus Representation}
\addcontentsline{toc}{subsubsection}{4.2.2.1 \hspace{0.15cm} Exact
Solution in the Kraus Representation}

In the Kraus representation the system state at any given time can
be written as
\begin{equation}
\rho _{S}(t)=\sum_{i,j}K_{ij}(t)\rho _{S}(0)K_{ij}(t)^{\dag },
\label{KrausForm}
\end{equation}%
where the Kraus operators satisfy $\sum_{ij}K_{ij}(t)^{\dag }K_{ij}(t)=I_{S}$ \cite%
{Kraus83}. These operators can be expressed easily in the
eigenbasis of the initial state of the bath density operator as
\begin{equation}
K_{ij}(t)=\sqrt{\lambda _{i}}\langle j|\exp (-iH_{I}t)|i\rangle ,
\end{equation}%
where the bath density operator at the initial time is $\rho
_{B}(0)=\sum_{i}\lambda _{i}|i\rangle \langle i|$. For the Gibbs
thermal
state chosen here, the eigenbasis is the $N$-fold tensor product of the $%
\sigma ^{z}$ basis. In this basis
\begin{equation}
\rho _{B}=\sum_{l}\frac{\exp (-\beta E_{l})}{Z}|l\rangle \langle
l|,
\end{equation}%
where $\beta =1/kT$. Here
\begin{equation}
E_{l}=\sum_{n=1}^{N}\frac{1}{2}\hbar \Omega _{n}(-1)^{l_{n}},
\label{eq:El}
\end{equation}%
is the energy of each eigenstate $|l\rangle $, where
$l=l_{1}l_{2}\dots l_{n} $ is the binary expansion of the integer
$l$, and the partition function is $Z=\sum_{l}\exp (-\beta
E_{l})$. Therefore, the Kraus operators become
\begin{equation}
K_{ij}(t)=\sqrt{\lambda _{i}}\exp (-it\alpha \tilde{E}_{i}\sigma
^{z})\delta _{ij},
\end{equation}%
where
\begin{equation}
\tilde{E}_{i}=\langle i|B|i\rangle =\sum_{n=1}^{N}g_{n}(-1)^{i_{n}}-\mathrm{%
Tr}\{\sum_{n}g_{n}\sigma _{n}^{z}\rho _{B}\},  \label{eq:Eitilde}
\end{equation}%
and $\lambda _{i}=\exp (-\beta E_{i})/Z$. Substituting this expression for $%
K_{ij}$ into Eq. (\ref{KrausForm}) and writing the system state in
the Bloch vector form given in Eq. (\ref{I+vs}), we obtain
\begin{eqnarray}
v_{x}(t) &=&v_{x}(0)C(t)-v_{y}(0)S(t),  \notag \\
v_{y}(t) &=&v_{x}(0)S(t)+v_{y}(0)C(t),  \label{ExactSoln} \\
v_{z}(t) &=&v_{z}(0),  \notag
\end{eqnarray}%
where%
\begin{eqnarray}
C(t) &=&\sum_{i}\lambda _{i}\cos 2\alpha \tilde{E}_{i}t,  \notag \\
S(t) &=&\sum_{i}\lambda _{i}\sin 2\alpha \tilde{E}_{i}t.
\label{CSeqn}
\end{eqnarray}

The equations (\ref{ExactSoln}) are the exact solution to the
system dynamics of the above spin bath model. We see that the
evolution of the Bloch vector is a linear combination of rotations
around the $z$ axis. This evolution reflects the symmetry of the
interaction Hamiltonian which is diagonal in the $z$ basis. By
inverting Eqs. (\ref{ExactSoln}) for $v_{x}(0)$
or $v_{y}(0)$, we see that the Kraus map is irreversible when $%
C(t)^{2}+S(t)^{2}=0$. This will become important below, when we
discuss the validity of the time-convolutionless approximation.

\subsubsection*{4.2.2.2 \hspace{2pt} Alternative Exact Solution}
\addcontentsline{toc}{subsubsection}{4.2.2.2 \hspace{0.15cm}
Alternative Exact Solution}

Another way to derive the exact solution which is computationally
more useful is the following. Since all $\sigma _{n}^{z}$ commute,
the initial bath density matrix factors and can be written as
\begin{eqnarray}
\rho _{B} =\bigotimes\limits_{n=1}^{N}\frac{\exp \left( -\frac{\Omega _{n}%
}{2kT}\sigma _{n}^{z}\right) }{\mathrm{Tr}\left[ \exp \left(
-\frac{\Omega _{n}}{2kT}\sigma _{n}^{z}\right) \right] }
=\bigotimes\limits_{n=1}^{N}\frac{1}{2}\left( I+\beta _{n}\sigma
_{n}^{z}\right) \equiv \prod_{n=1}^{N}\rho _{n},
\label{eq_rho_B_inter}
\end{eqnarray}%
where
\begin{equation}
\beta _{n}=\tanh \left( -\frac{\Omega _{n}}{2kT}\right) ,
\end{equation}%
and $-1\leq \beta _{n}\leq 1$. Using this, we obtain an expression for $%
\theta $ defined in Eq. (\ref{theta})
\begin{eqnarray}
\theta  &=&\mathrm{Tr}\{\sum_{n=1}^{N}g_{n}\sigma
_{n}^{z}\bigotimes\limits_{m=1}^{N}\frac{1}{2}(I+\beta _{m}\sigma
_{m}^{z})\}
\notag \\
&=&\sum_{n=1}^{N}g_{n}\mathrm{Tr}\{\frac{1}{2}(\sigma
_{n}^{z}+\beta _{n}I)\}\prod\limits_{m\neq
n}\mathrm{Tr}\{\frac{1}{2}(I+\beta _{m}\sigma
_{m}^{z})\}  \notag \\
&=&\sum_{n=1}^{N}g_{n}\beta _{n}.  \label{eq:theta}
\end{eqnarray}%
The evolution of the system density matrix in the interaction
picture is
\begin{equation}
\rho _{S}(t)=\mathrm{Tr}_{B}\{e^{-iH_{I}t}\rho (0)e^{iH_{I}t}\}.
\end{equation}%
In terms of the system density matrix elements in the computational basis $%
\{|0\rangle ,|1\rangle \}$ (which is an eigenbasis of $\sigma ^{z}$ in $%
H_{I}=\alpha \sigma ^{z}\otimes B$), we have
\begin{eqnarray}
\langle j|\rho _{S}(t)|k\rangle  &=&\langle
j|\mathrm{Tr}_{B}\{e^{-iH_{I}t} \rho
_{S}(0)\bigotimes\limits_{m=1}^{N}\rho
_{m}e^{iH_{I}t}\}|k\rangle   \notag \\
&=&\mathrm{Tr}_{B}\{e^{-i\alpha \langle j|\sigma ^{z}|j\rangle Bt}
\langle j|\rho _{S}(0)|k\rangle \bigotimes\limits_{m=1}^{N}\rho
_{m}e^{+i\alpha \langle k|\sigma ^{z}|k\rangle Bt}\}.  \notag
\end{eqnarray}%
Let us substitute $\langle j|\sigma ^{z}|j\rangle =(-1)^{j}$ and
rewrite
\begin{eqnarray}
e^{-i\alpha \langle j|\sigma ^{z}|j\rangle Bt} =e^{-i\alpha
(-1)^{j}\left( \sum_{l=1}^{N}g_{l}\sigma _{l}^{z}-\theta I\right)
t}  =\bigotimes\limits_{l=1}^{N}e^{-i(-1)^{j}\alpha \left(
g_{l}\sigma _{l}^{z}-\frac{\theta }{N}I\right) t}.  \notag
\end{eqnarray}%
Since all the matrices are diagonal, they commute and we can
collect the terms by qubits:
\begin{eqnarray}
\langle j|\rho _{S}(t)|k\rangle  =\langle j|\rho _{S}(0)|k\rangle
\mathrm{Tr}\{\bigotimes\limits_{m=1}^{N}e^{-i\left[
(-1)^{j}-(-1)^{k}\right] \alpha \left( g_{l}\sigma _{l}^{z}-\frac{\theta }{N}%
I\right) t}\rho _{n}\}.  \notag
\end{eqnarray}%
Let us denote $(-1)^{j}-(-1)^{k}=2\epsilon _{jk}$. The trace can
be easily computed to be
\begin{eqnarray*}
&\prod_{n=1}^{N}&\mathrm{Tr}\{e^{-i2\epsilon _{jk}\alpha \left(
g_{n}\sigma _{n}^{z}-\frac{\theta }{N}I\right)
t}\tfrac{1}{2}(I+\beta _{n}\sigma
_{n}^{z})\} \\
&=&\prod_{n=1}^{N}e^{i2\epsilon _{jk}\alpha \frac{\theta
}{N}t}\left[ \cos
(2\epsilon _{jk}\alpha g_{n}t)-i\beta _{n}\sin (2\epsilon _{jk}\alpha g_{n}t)%
\right] .
\end{eqnarray*}%
Thus the final expression for the system density matrix elements
is
\begin{eqnarray}
\langle j|\rho _{S}(t)|k\rangle  &=&\langle j|\rho
_{S}(0)|k\rangle e^{i2\epsilon _{jk}\alpha \theta t}
\prod_{n=1}^{N}\left[ \cos (2\epsilon _{jk}\alpha g_{n}t)-i\beta
_{n}\sin (2\epsilon _{jk}\alpha g_{n}t)\right] . \notag
\end{eqnarray}%
Notice that $\epsilon _{00}=\epsilon _{11}=0$, hence the diagonal
matrix elements do not depend on time as before:
\begin{gather*}
\langle 0|\rho _{S}(t)|0\rangle =\langle 0|\rho _{S}(0)|0\rangle , \\
\langle 1|\rho _{S}(t)|1\rangle =\langle 1|\rho _{S}(0)|1\rangle .
\end{gather*}%
For the off-diagonal matrix elements $\epsilon _{01}=1$, $\epsilon _{10}=-1$%
, and the evolution is described by
\begin{eqnarray}
\langle 0|\rho _{S}(t)|1\rangle  &=&\langle 0|\rho
_{S}(0)|1\rangle f(t),
\notag \\
\langle 1|\rho _{S}(t)|0\rangle  &=&\langle 1|\rho
_{S}(0)|0\rangle f^{\ast }(t),  \label{ExactSoln2}
\end{eqnarray}%
where
\begin{equation}
f(t)=e^{i2\alpha \theta t}\prod_{n=1}^{N}\left[ \cos (2\alpha
g_{n}t)-i\beta _{n}\sin (2\alpha g_{n}t)\right] .  \label{prod}
\end{equation}%
In terms of the Bloch vector components, this can be written in
the form of Eq. (\ref{ExactSoln}), where
\begin{eqnarray}
C(t) &=&(f(t)+f^{\ast }(t))/2,  \notag \\
S(t) &=&(f(t)-f^{\ast }(t))/2i.  \label{eq:CS}
\end{eqnarray}

\subsection*{4.2.3 \hspace{2pt} Limiting cases}
\addcontentsline{toc}{subsection}{4.2.3 \hspace{0.15cm} Limiting
cases}

\subsubsection*{4.2.3.1 \hspace{2pt} Short Times}
\addcontentsline{toc}{subsubsection}{4.2.3.1 \hspace{0.15cm} Short
Times}

Consider the evolution for short times where $\alpha t\ll 1$. Then
\begin{eqnarray}
\lefteqn{\left\vert \prod_{n=1}^{N}\left[ \cos (2\alpha g_{n}t)\pm
i\beta
_{n}\sin (2\alpha g_{n}t)\right] \right\vert }  \notag \\
&=&\prod_{n=1}^{N}\sqrt{1-(1-\beta _{n}^{2})\sin ^{2}(2\alpha
g_{n}t)}
\notag \\
&\approx &\prod_{n=1}^{N}[1-2(1-\beta _{n}^{2})(\alpha
g_{n}t)^{2}]  \notag
\\
&\approx &1-2\left[ \alpha ^{2}\sum_{n=1}^{N}g_{n}^{2}(1-\beta _{n}^{2})%
\right] t^{2}  \notag \\
&\approx &\exp [-2(\alpha t)^{2}Q_{2}],  \label{eq:f-approx}
\end{eqnarray}%
where (see Appendix A at the end of this chapter)
\begin{eqnarray}
Q_{2} \equiv \mathrm{Tr}\{B^{2}\rho
_{B}\}=\sum_{n=1}^{N}g_{n}^{2}(1-\beta _{n}^{2})
=\int_{-\infty }^{\infty }\frac{2J(\Omega )}{1+\cosh (\frac{\Omega }{kT})}%
\mathrm{d}\Omega .  \label{Q_2}
\end{eqnarray}%
Note that for the above approximation to be valid, we need
$2(\alpha t)^{2}Q_{2}\ll 1$. The total phase of $f(t)$ in Eq.
(\ref{prod}) is
\begin{eqnarray}
\phi  \approx 2\theta \alpha t+\sum_{n=1}^{N}(-\beta _{n}2\alpha
g_{n}t) =2\theta \alpha t-2\alpha \left( \sum_{n=1}^{N}g_{n}\beta
_{n}\right) t=0,
\end{eqnarray}%
where we have used Eq. (\ref{eq:theta}). Thus, the off-diagonal
elements of the system density matrix become
\begin{eqnarray}
\rho _{S}^{01}(t) &\approx &\rho _{S}^{01}(0)e^{-2(\alpha
t)^{2}Q_{2}},
\notag \\
\rho _{S}^{10}(t) &\approx &\rho _{S}^{10}(0)e^{-2(\alpha
t)^{2}Q_{2}}.
\end{eqnarray}

Finally, the dynamics of the Bloch vector components are:
\begin{eqnarray}
v_{x,y}(t) &\approx &v_{x,y}(0)e^{-2(\alpha t)^{2}Q_{2}},  \notag \\
v_{z}(t) &=&v_{z}(t).  \label{shorttimes}
\end{eqnarray}%
This represents the well known behavior \cite{NNP96} of the
evolution of an open quantum system in the Zeno regime. In this
regime coherence does not decay exponentially but is initially
flat, as is the case here due to the vanishing time derivative of
$\rho _{S}^{01}(t)$ at $t=0$. As we will see in Section 4.3, the
dynamics in the Born approximation (which is also the second order
time-convolutionless approximation) exactly matches the last
result.

\subsubsection*{4.2.3.2 \hspace{2pt} Large $N$}
\addcontentsline{toc}{subsubsection}{4.2.3.2 \hspace{0.15cm} Large
$N$}

When $N\gg 1$ and the values of $g_{n}$ are random, then the
different terms in the product of Eq. (\ref{prod}) are smaller
than $1$ most of the time and have recurrences at different times.
Therefore, we expect the function $f(t)$ to be close to zero in
magnitude for most of the time and full recurrences,
if they exist, to be extremely rare. When $g_{n}$ are equal and so are $%
\Omega _{n}$, then partial recurrences occur periodically, independently of $%
N$. Full recurrences occur with a period which grows at least as fast as $N$%
. This can be argued from Eq. (\ref{ExactSoln}) by imposing the
condition that the arguments of all the cosines and sines are
simultaneously equal to an integer multiple of $2\pi $. When
$J(\Omega )$ has a narrow high peak, e.g., one $g_{n}$ is much
larger than the others, then the corresponding terms in the
products in Eq. (\ref{prod}) oscillate faster than the rate at
which the whole product decays. This is effectively a modulation
of the decay.

\subsubsection*{4.2.3.3 \hspace{2pt} Discontinuous spectral density co-domain}
\addcontentsline{toc}{subsubsection}{4.2.3.3 \hspace{0.15cm}
Discontinuous spectral density co-domain}

As can be seen from Eq.~(\ref{prod}), the coupling constants $g_{n}$
determine the oscillation periods of the product terms, while the
temperature factors $\beta _{n}$ determine their modulation depths.
If the codomain of spectral density is not continuous, i.e. it can
be split into non-overlapping intervals $G_{j}$, $j=1,...,J$, then
Eq.~(\ref{prod}) can be represented in the following form:
\begin{equation}
f(t)=e^{i2\alpha \theta t}P_{1}(t)P_{2}(t)\dots P_{J}(t),
\end{equation}%
where
\begin{equation}
P_{j}(t)=\prod_{g_{n}\in G_{j}}\big[\cos (2\alpha g_{n}t)-i\beta
_{n}\sin (2\alpha g_{n}t)\big].
\end{equation}%
In this case, if $G_{j}$ are separated by large enough gaps, the
evolution rates of different $P_{j}(t)$ can be significantly
different. This is particularly noticeable if one $P_{j}(t)$
undergoes partial recurrences while another $P_{j^{\prime }}(t)$
slowly decays.

For example, one can envision a situation with two intervals such
that one term shows frequent partial recurrences that slowly decay
with time, while the other term decays faster, but at times larger
than the recurrence time. The overall evolution then consists in a
small number of fast partial recurrences. In an extreme case, when
one $g_{n}$ is much larger then the others, this results in an
infinite harmonic modulation of the decay with depth dependent on
$\beta _{n}$, i.e., on temperature.

\subsubsection*{4.2.3.4 \hspace{2pt} Alternating signs}
\addcontentsline{toc}{subsubsection}{4.2.3.4 \hspace{0.15cm}
Alternating signs}

If the bath has the property that every bath qubit $m$ has a pair
$-m$ with the same frequency $\Omega _{-m}=\Omega _{m}$, but
opposite coupling constant $g_{-m}=-g_{m}$, the exact solution can
be simplified. First, $\beta _{-m}=\beta _{m}$, and $\theta =0$.
Next, Eq.~(\ref{prod}) becomes
\begin{eqnarray}
f(t) &=&\prod_{m=1}^{N/2}\big[\cos (2\alpha g_{m}t)-i\beta
_{m}\sin (2\alpha g_{m}t)\big]\big[\cos (2\alpha g_{-m}t)-i\beta
_{-m}\sin (2\alpha g_{-m}t)\big]  \notag
\\
&=&\prod_{m=1}^{N/2}\big[\cos ^{2}(2\alpha g_{m}t)+\beta
_{m}^{2}\sin ^{2}(2\alpha g_{m}t)\big].
\end{eqnarray}%
This function is real, thus Eq.~(\ref{eq:CS}) becomes
$C(t)=f(t),S(t)=0$, so that $v_{x}(t)=v_{x}(0)f(t)$ and
$v_{y}(t)=v_{y}(0)f(t)$. The exact solution is then symmetric
under the interchange $v_{x}\leftrightarrow v_{y}$, a property
shared by all the second order approximate solutions considered
below, as well as the post-Markovian master equation. The limiting case Eq.~(%
\ref{eq:f-approx}) remains unchanged, and since $Q_{2}$ depends on
$g_{n}^{2} $, but not $g_{n}$, it and all second order
approximations also remain unchanged. In the special case
$|g_{m}|=g$, the exact solution exhibits full recurrences with
period $T=\pi /\alpha g$.

\section*{4.3 \hspace{2pt} Approximation methods}
\addcontentsline{toc}{section}{4.3 \hspace{0.15cm} Approximation
methods}

In this section we discuss the performance of different
approximation
methods developed in the open quantum systems literature \cite%
{Alicki:87,BrePet02}. The corresponding master equations for the
system density matrix can be derived explicitly and since the
model considered here is exactly solvable, we can compare the
approximations to the exact dynamics. We use the Bloch vector
representation and since the $z$ component has no dynamics, a fact
which is reflected in all the master equations, we omit it from
our comparisons.

\subsection*{4.3.1 \hspace{2pt} Born and Born-Markov approximations}
\addcontentsline{toc}{subsection}{4.3.1 \hspace{0.15cm} Born and
Born-Markov approximations}

Both the Born and Born-Markov approximations are second order in
the coupling strength $\alpha $.

\subsubsection*{4.3.1.1 \hspace{2pt} Born approximation}
\addcontentsline{toc}{subsubsection}{4.3.1.1 \hspace{0.15cm} Born
approximation}

The Born approximation is equivalent to a truncation of the
Nakajima-Zwanzig projection operator method at the second order,
which is discussed in detail in Section 4.3.2. The Born
approximation is given by the following
integro-differential master equation:%
\begin{equation}
\dot{\rho}_{S}(t)=-\int_{0}^{t}\mathrm{Tr}_{B}\{[H_{I}(t),[H_{I}(s),\rho
_{S}(s)\otimes \rho _{B}]]\}\text{d}s.
\end{equation}%
Since in our case the interaction Hamiltonian is time-independent,
the integral becomes easy to solve. We obtain
\begin{equation}
\dot{\rho}_{S}(t)=-2\alpha ^{2}Q_{2}\int_{0}^{t}(\rho
_{S}(s)-\sigma ^{z}\rho _{S}(s)\sigma ^{z})\text{d}s,
\end{equation}%
where $Q_{2}$ is the second order bath correlation function in Eq. (\ref{Q_2}%
). Writing $\rho _{S}(t)$ in terms of Bloch vectors as $(I+\vec{v}\cdot \vec{%
\sigma})/2$ [Eq. (\ref{I+vs})], we obtain the following
integro-differential equations:
\begin{eqnarray}
\dot{v}_{x,y}(t) &=&-4\alpha
^{2}Q_{2}\int_{0}^{t}v_{x,y}(s)\text{d}s. \label{BornApprox}
\end{eqnarray}

These equations can be solved by taking the Laplace transform of
the variables. The equations become
\begin{equation}
sV_{x,y}(s)-v_{x,y}(0)=-4\alpha ^{2}Q_{2}\frac{V_{x,y}(s)}{s},
\end{equation}%
where $V_{x,y}(s)$ is the Laplace transform of $v_{x,y}(t)$. This
gives
\begin{equation}
V_{x,y}(s)=\frac{v_{x,y}(0)s}{s^{2}+4Q_{2}\alpha ^{2}},
\end{equation}%
which can be readily solved by taking the inverse Laplace
transform. Doing so, we obtain the solution of the Born master
equation for our model:
\begin{eqnarray}
v_{x,y}(t) &=&v_{x,y}(0)\cos (2\alpha \sqrt{Q_{2}}t).
\label{Born}
\end{eqnarray}%
Note that this solution is symmetric under the interchange $%
v_{x}\leftrightarrow v_{y}$, but the exact dynamics in Eq.
(\ref{ExactSoln})
does not have this symmetry. The exact dynamics respects the symmetry: $%
v_{x}\rightarrow v_{y}$ and $v_{y}\rightarrow -v_{x}$, which is a
symmetry of the Hamiltonian. This means that higher order
corrections are required to break the symmetry
$v_{x}\leftrightarrow v_{y}$ in order to approximate the exact
solution more closely.

One often makes the substitution $v_{x,y}(t)$ for $v_{x,y}(s)$ in Eq. (\ref%
{BornApprox}) since the integro-differential equation obtained in
other models may not be as easily solvable. This approximation,
which is valid for short times, yields
\begin{eqnarray}
\dot{v}_{x,y}(t) &=&-4\alpha ^{2}Q_{2}tv_{x,y}(t),
\end{eqnarray}%
which gives
\begin{eqnarray}
v_{x,y}(t) &=&v_{x,y}(0)\exp (-2Q_{2}\alpha ^{2}t^{2}),
\label{TCL2}
\end{eqnarray}%
i.e., we recover Eq. (\ref{shorttimes}). This is the same solution
obtained in the second order approximation using the
time-convolutionless (TCL) projection method discussed in Section
4.3.2.

\subsubsection*{4.3.1.2 \hspace{2pt} Born-Markov approximation}
\addcontentsline{toc}{subsubsection}{4.3.1.2 \hspace{0.15cm}
Born-Markov approximation}

In order to obtain the Born-Markov approximation, we use the
following quantities \cite{BrePet02}[Ch.3]:
\begin{eqnarray}
R(\omega ) &=&\sum_{E_{2}-E_{1}=\omega }P_{E_{1}}\sigma
^{z}P_{E_{2}},
\notag \\
\Gamma (\omega ) &=&\alpha ^{2}\int_{0}^{\infty }e^{i\omega
s}Q_{2}\text{d}s,
\notag \\
H_{L} &=&\sum_{\omega }T(\omega )R(\omega )^{\dag }R(\omega ),
\end{eqnarray}%
where $T(\omega )=(\Gamma (\omega )-\Gamma (\omega )^{\ast })/2i$,
$E_{i}$ is an eigenvalue of the system Hamiltonian $H_{S}$, and
$P_{E_{i}}$ is the
projector onto the eigenspace corresponding to this eigenvalue. In our case $%
H_{S}$ is diagonal in the eigenbasis of $\sigma ^{z}$, and only
$\omega =0$ is relevant. This leads to $R(0)=\sigma ^{z}$ and
$\Gamma (0)=\alpha
^{2}\int_{0}^{\infty }Q_{2}\text{d}t$. Since $\Gamma (0)$ is real, we have $%
T(0)=0.$ Hence the Lamb shift Hamiltonian $H_{L}=0$, and the
Lindblad form of the Born-Markov approximation is
\begin{equation}
\dot{\rho}_{S}(t)=\gamma (\sigma ^{z}\rho _{S}\sigma ^{z}-\rho
_{S}), \label{Lindblad}
\end{equation}%
where $\gamma =\Gamma (0)+\Gamma (0)^{\ast }=2\alpha
^{2}\int_{0}^{\infty }Q_{2}$d$t$. But note that
$Q_{2}=\mathrm{Tr}_{B}\{B^{2}\rho _{B}\}$ does not depend on time.
This means that $\Gamma $ and hence $\gamma $ are both infinite.
Thus the Born-Markov approximation is not valid for this model and
the main reason for this is the time independence of the bath
correlation functions. The dynamics is inherently non-Markovian.

A different approach to the derivation of a Markovian semigroup
master equation was proposed in \cite{Lidar:CP01}. In this
approach, a Lindblad equation is derived from the Kraus
operator-sum representation by a coarse-graining procedure defined
in terms of a phenomenological coarse-graining time scale $\tau $.
The general form of the equation is:
\begin{eqnarray}
\frac{\partial \rho (t)}{\partial t} =-i[\langle \dot{Q}\rangle
_{\tau },\rho (t)] +\frac{1}{2}\sum_{\alpha ,\beta =1}^{M}\langle
\dot{\chi}_{\alpha ,\beta }\rangle _{\tau }([A_{\alpha },\rho
(t)A_{\beta }^{\dagger }]+[A_{\alpha }\rho (t),A_{\beta }^{\dagger
}]),
\end{eqnarray}%
where the operators $A_{0}=I$ and $A_{\alpha },\alpha =1,...,M$
form an
arbitrary fixed operator basis in which the Kraus operators (\ref{KrausForm}%
) can be expanded as
\begin{equation}
K_{i}=\sum_{\alpha =0}^{M}b_{i\alpha }A_{\alpha }.
\end{equation}%
The quantities $\chi _{\alpha ,\beta }(t)$ and $Q(t)$ are defined
through
\begin{equation}
\chi _{\alpha ,\beta }(t)=\sum_{i}b_{i\alpha }(t)b_{i\beta }^{\ast
}(t),
\end{equation}%
\begin{equation}
Q(t)=\frac{i}{2}\sum_{\alpha =1}^{M}(\chi _{\alpha 0}(t)K_{\alpha
}-\chi _{0\alpha }(t)K_{\alpha }^{\dagger }),
\end{equation}%
and
\begin{equation}
\langle X\rangle _{\tau }=\frac{1}{\tau }\int_{0}^{\tau }X(s)ds.
\end{equation}%
For our problem we find
\begin{equation}
\frac{\partial \rho (t)}{\partial t}=-i\tilde{\omega}[\sigma _{Z},\rho (t)]+%
\tilde{\gamma}(\sigma _{Z}\rho (t)\sigma _{Z}-\rho (t)),
\label{coarsegrain}
\end{equation}%
where
\begin{equation}
\tilde{\omega}=\frac{1}{2\tau }S(\tau )
\end{equation}%
and
\begin{equation}
\tilde{\gamma}=\frac{1}{2\tau }(1-C(\tau ))
\end{equation}%
with $C(t)$ and $S(t)$ defined in Eq. (\ref{CSeqn}). In order for
this approximation to be justified, it is required that the
coarse-graining time scale $\tau $ be much larger than any
characteristic time scale of the bath \cite{Lidar:CP01}. However,
in our case the bath correlation time is infinite which, once
again, shows the inapplicability of the Markovian approximation.
This is further supported by the performance of the optimal
solution that one can achieve by varying $\tau $, which is
discussed in Section 4.4. There we numerically examine the average
trace-distance between the
solution to Eq. (\ref{coarsegrain}) and the exact solution as a function of $%
\tau $. The average is taken over a time $T$, which is greater
than the decay time of the exact solution. We determine an optimal
$\tau $ for which the average trace distance is minimum and then
determine the approximate solution. The solution of Eq.
(\ref{coarsegrain}) for a particular $\tau $ in terms of the Bloch
vector components is
\begin{eqnarray}
v_{x}(t) &=&v_{x}(0)\tilde{C}_{\tau }(t)+v_{y}(0)\tilde{S}_{\tau
}(t)  \notag
\\
v_{y}(t) &=&v_{y}(0)\tilde{C}_{\tau }(t)-v_{x}(0)\tilde{S}_{\tau
}(t),
\end{eqnarray}%
where $\tilde{C}_{\tau }(t)=e^{-\tilde{\gamma}(\tau )t}\cos (\tilde{\omega}%
(\tau )t)$ and $\tilde{S}_{\tau }(t)=e^{-\tilde{\gamma}(\tau )t}\sin (\tilde{%
\omega}(\tau )t)$. The average trace distance as a function of
$\tau $ is given by,
\begin{eqnarray}
&\bar{D}&(\rho _{\mathrm{exact}},\rho _{\mathrm{CG}})\equiv \frac{1}{2}%
\mathrm{Tr}|\rho _{\mathrm{exact}}-\rho _{\mathrm{CG}}|  \notag \\
&=&\frac{1}{2T}\sum_{t=0}^{T}\sqrt{(C(t)-\tilde{C}(t))^{2}+(S(t)-\tilde{S}%
(t))^{2}} \sqrt{v_{x}(0)^{2}+v_{y}(0)^{2}},
\end{eqnarray}%
where $\rho _{\mathrm{CG}}$ represents the coarse-grained solution
and where $|X|=\sqrt{X^{\dag }X}$. The results are presented in
Section 4.4. Next we consider the Nakajima-Zwanzig (NZ) and the
time-convolutionless (TCL) master equations for higher order
approximations.

\subsection*{4.3.2 \hspace{2pt} NZ and TCL master equations}
\addcontentsline{toc}{subsection}{4.3.2 \hspace{0.15cm} NZ and TCL
master equations}

Using projection operators one can obtain approximate
non-Markovian master equations to higher orders in $\alpha t$. A
projection is defined as follows,
\begin{equation}
\mathcal{P}\rho=\mathrm{Tr}_B\{\rho\}\otimes\rho_B ,
\end{equation}
and serves to focus on the ``relevant dynamics'' (of the system)
by removing the bath (a recent generalization is discussed in Ref.
\cite{Breuer:07}).
The choice of $\rho_B$ is somewhat arbitrary and can be taken to be $%
\rho_B(0)$ which significantly simplifies the calculations. Using
the notation introduced in \cite{BBP04}, define
\begin{equation}
\left\la \mathcal{S}\right\ra\equiv
\mathcal{P}\mathcal{S}\mathcal{P}
\end{equation}
for any superoperator $\mathcal{S}$. Thus
$\left\la\mathcal{S}^n\right\ra$ denote the moments of the
superoperator. Note that for the Liouvillian
superoperator, $\left\la\mathcal{L}\right\ra=0$ by virtue of the fact that $%
\mathrm{Tr}_B\{B\rho_B(0)\}=0$ (see \cite{BrePet02}). Since we
assume that the initial state is a product state, both the NZ and
TCL equations are homogeneous equations. The NZ master equation is
an integro-differential equation with a memory kernel
$\mathcal{N}(t,s)$ and is given by
\begin{equation}
\dot{\rho}_S(t)\otimes\rho_B=\int_0^t\mathcal{N}(t,s)\rho_S(s)\otimes\rho_B
\text{d}s .
\end{equation}
The TCL master equation is a time-local equation given by
\begin{equation}
\dot{\rho}_S(t)\otimes\rho_B=\mathcal{K}(t)\rho_S(t)\otimes\rho_B
.
\end{equation}
When these equations are expanded in $\alpha t$ and solved we
obtain the higher order corrections. When the interaction
Hamiltonian is time independent (as in our case), the above
equations simplify to
\begin{equation}  \label{NZ}
\int_0^t\mathcal{N}(t,s)\rho_S(s)\otimes\rho_B
\text{d}s=\sum_{n=1}^\infty \alpha^n
\mathcal{I}_n(t,s)\left\la\mathcal{L}^n \right\ra_{pc}\rho_S(s)
\end{equation}
and
\begin{equation}  \label{TCL}
\mathcal{K}(t)=\sum_{n=1}^\infty \alpha^n
\frac{t^{n-1}}{(n-1)!}\left\la \mathcal{L}^n\right\ra_{oc}
\end{equation}
for the NZ and TCL equations, respectively, where the time-ordered
integral operator $\mathcal{I}_n(t,s)$ is defined as
\begin{equation}
\mathcal{I}_n(t,s)\equiv \int_0^t\text{d}t_1\int_0^{t_1}\text{d}%
t_2\cdots\int_0^{t_{n-2}}\text{d}s .
\end{equation}
The definitions of the partial cumulants
$\left\la\mathcal{L}\right\ra_{pc}$ and the ordered cumulants
$\left\la\mathcal{L}\right\ra_{oc}$ are given in Refs.
\cite{ShiAri80,Royer:72,Kam74}. For our model we have
\begin{equation}
\left\la\mathcal{L}\right\ra_{pc}=\left\la\mathcal{L}\right\ra_{oc}=0
,
\end{equation}
and
\begin{eqnarray}
\left\la\mathcal{L}^2\right\ra_{pc}=\left\la\mathcal{L}^2\right\ra  \notag \\
\left\la\mathcal{L}^2\right\ra_{oc}=\left\la\mathcal{L}^2\right\ra  \notag \\
\left\la\mathcal{L}^3\right\ra_{pc}=\left\la\mathcal{L}^3\right\ra  \notag \\
\left\la\mathcal{L}^3\right\ra_{oc}=\left\la\mathcal{L}^3\right\ra  \notag \\
\left\la\mathcal{L}^4\right\ra_{pc}=\left\la\mathcal{L}^4\right\ra - \left\la%
\mathcal{L}^2\right\ra^2  \notag \\
\left\la\mathcal{L}^4\right\ra_{oc}=\left\la\mathcal{L}^4\right\ra
- 3\left\la\mathcal{L}^2\right\ra^2 .  \label{eq:cumulants}
\end{eqnarray}
Explicit expressions for these quantities are given in Appendix
\ref{app:B} at the end of this chapter. Substituting these into
the NZ and TCL equations (\ref{NZ}) and (\ref{TCL}), we obtain
what we refer to below as the NZ$n$ and TCL$n$ master equations,
with $n=2,3,4$. These approximate master equations are,
respectively, second, third and fourth order in the coupling
constant $\alpha$, and they can be solved analytically. The second
order solution of the NZ equation
(NZ2) is exactly the Born approximation and the solution is given in Eq. (%
\ref{Born}). The third order NZ master equation is given by
\begin{eqnarray}
\dot{\rho}_S(t)&=&-2\alpha^2Q_2 \mathcal{I}_2(t,s)
(\rho_S(s)-\sigma^z\rho_S(s)\sigma^z)  \notag \\
&+&
i4\alpha^3Q_3\mathcal{I}_3(t,s)(\sigma^z\rho_S(s)-\rho_S(s)\sigma^z)
,
\end{eqnarray}
and the fourth order is
\begin{eqnarray}
\dot{\rho}_S(t)&=&-2\alpha^2Q_2 \mathcal{I}_2(t,s)
(\rho_S(s)-\sigma^z\rho_S(s)\sigma^z)  \notag \\
&+&
i4\alpha^3Q_3\mathcal{I}_3(t,s)(\sigma^z\rho_S(s)-\rho_S(s)\sigma^z)
\notag \\
&+& 8\alpha^4(Q_4-Q_2^2)\mathcal{I}_4(t,s)(\rho_S(s)-\sigma^z\rho_S(s)%
\sigma^z).  \notag \\
\end{eqnarray}
These equations are equivalent to, respectively, 6th and 8th order
differential equations (with constant coefficients) and are
difficult to solve analytically. The results we present in the
next section were therefore obtained numerically.

The situation is simpler in the TCL approach. The second order TCL
equation is given by
\begin{eqnarray}
\dot{\rho}_S(t)&=&-\alpha^2t\mathrm{Tr}_B\{[H_I,[H_I,\rho_S(t)\otimes%
\rho_B(0)]]\}  \notag \\
&=& -2\alpha^2tQ_2(\rho_S(t)-\sigma^z\rho_S(t)\sigma^z) ,
\end{eqnarray}
whose solution is as given in Eq. (\ref{TCL2}) in terms of Bloch
vector components. For TCL3 we find
\begin{eqnarray}
\dot{\rho}_S(t)=-2\alpha^2
tQ_2(\rho_S(t)-\sigma_z\rho_S(t)\sigma_z) +
4iQ_3\alpha^3\frac{t^2}{2}(\sigma_z\rho_S(t)-\rho_S(t)\sigma_z),
\end{eqnarray}
and for TCL4 we find
\begin{eqnarray}
\dot{\rho}_S(t) &=& [-2\alpha^2 tQ_2 +
(8Q_4-24Q_2^2)\alpha^4\frac{t^3}{6}]
(\rho_S(t)-\sigma_z\rho_S(t)\sigma_z)  \notag \\
&+&
4iQ_3\alpha^3\frac{t^2}{2}(\sigma_z\rho_S(t)-\rho_S(t)\sigma_z) .
\end{eqnarray}
These equation can be solved analytically, and the solutions to
the third and fourth order TCL equations are given by
\begin{eqnarray}  \label{TCL3}
v_x(t)&=&f_n(\alpha t)\left
[v_x(0)\cos(g(t))+v_y(0)\sin(g(t))\right ],
\notag \\
v_y(t)&=&f_n(\alpha t)\left
[v_y(0)\cos(g(t))-v_x(0)\sin(g(t))\right ] .
\notag \\
\end{eqnarray}
where $g(t)=4Q_3\alpha^3t^3/3$, $f_3(\alpha
t)=\exp(-2Q_2\alpha^2t^2)$ (TCL3) and $f_4(\alpha
t)=\exp(-2Q_2\alpha^2t^2\\ +(2Q_4-6Q_2^2)\alpha^4t^4/3)$ (TCL4).
It is interesting to note that the second order expansions of the
TCL and NZ master equations exhibit a $v_x\leftrightarrow v_y$
symmetry between the components of the Bloch vector, and only the
third order correction breaks this symmetry. Notice that the
coefficient of $\alpha^3$ does not vanish in this model unlike in
the one considered in \cite{BBP04}
because both $\left\la\mathcal{L}^3\right\ra_{pc}\neq 0$ and $\left\la%
\mathcal{L}^3\right\ra_{oc}\neq 0$ and hence the third order (and
other odd order) approximations exist.

\subsection*{4.3.3 \hspace{2pt} Post-Markovian (PM) master equation}
\addcontentsline{toc}{subsection}{4.3.3 \hspace{0.15cm}
Post-Markovian (PM) master equation}

In this section we study the performance of the post-Markovian
master equation recently proposed in \cite{ShabaniLidar:05}:
\begin{equation}
\frac{\partial \rho (t)}{\partial
t}=\mathcal{D}\int_{0}^{t}dt^{\prime }k(t^{\prime })\exp
(\mathcal{D}t^{\prime })\rho (t-t^{\prime })\mathrm{.}
\label{PMAL}
\end{equation}%
This equation was constructed via an interpolation between the
exact dynamics and the dynamics in the Markovian limit. The
operator $\mathcal{D}$ is the dissipator in the Lindblad equation
(\ref{Lindblad}), and $k(t)$ is a phenomenological memory kernel
which must be found by fitting to data or guessed on physical
grounds. As was discussed earlier, the Markovian approximation
fails for our model, nevertheless, one can use the form of the
dissipator we obtained in Eq. (\ref{Lindblad})
\begin{equation}
\mathcal{D}\rho =\sigma ^{z}\rho \sigma ^{z}-\rho .
\label{dissipator}
\end{equation}%
It is interesting to examine to what extent Eq. (\ref{PMAL}) can
approximate the exact dynamics. As a measure of the performance of
the post-Markovian
equation, we will take the trace-distance between the exact solution $\rho _{%
\mathrm{exact}}(t)$ and the solution to the post-Markovian
equation $\rho _{1}(t)$. The general solution of Eq. (\ref{PMAL})
can be found by expressing $\rho (t)$ in the damping basis
\cite{Briegel:93} and applying a
Laplace transform \cite{ShabaniLidar:05}. The solution is%
\begin{equation}
\rho (t)=\sum_{i}\mu _{i}(t)R_{i}=\sum_{i}{\text{Tr}}(L_{i}\rho
(t))R_{i},
\end{equation}%
where
\begin{equation}
\mu _{i}(t)=\mathrm{Lap}^{-1}\left[ \frac{1}{s-\lambda _{i}\tilde{k}%
(s-\lambda _{i})}\right] \mu _{i}(0)\equiv \xi _{i}(t)\mu _{i}(0),
\end{equation}%
($\mathrm{Lap}^{-1}$ is the inverse Laplace transform) with
$\tilde{k}$ being the Laplace transform of the kernel $k$,
$\{L_{i}\}$ and $\{R_{i}\}$ being the left and right eigenvectors
of the superoperator $\mathcal{D}$, and $\lambda _{i}$ the
corresponding eigenvalues. For our dissipator the
damping basis is $\{L_{i}\}=\{R_{i}\}=\{\frac{I}{\sqrt{2}},\frac{\sigma ^{x}%
}{\sqrt{2}},\frac{\sigma ^{y}}{\sqrt{2}},\frac{\sigma
^{z}}{\sqrt{2}}\}$ and the eigenvalues are $\{0,-2,-2,0\}$.
Therefore, we can immediately write the formal solution in terms
of the Bloch vector components:
\begin{gather}
v_{x,y}(t)={\text{Lap}}^{-1}\left[
\frac{1}{s+2\tilde{k}(s+2)}\right] v_{x,y}(0)\equiv \xi
(t)v_{x,y}(0).  \label{eq:vy-post}
\end{gather}

We see that $v_{x}(t)$ has no dependence on $v_{y}(0)$, and neither does $%
v_{y}(t)$ on $v_{x}(0)$, in contrast to the exact solution. The
difference
comes from the fact that the dissipator $\mathcal{D}$ does not couple $%
v_{x}(t)$ and $v_{y}(t)$. This reveals an inherent limitation of
the post-Markovian master equation:\ it inherits the symmetries of
the Markovian dissipator $\mathcal{D}$, which may differ from
those of the generator of the exact dynamics. In order to
rigorously determine the optimal performance, we use the trace
distance between the exact solution and a solution to the
post-Markovian equation:

\begin{eqnarray}
D(\rho _{\mathrm{exact}}(t),\rho _{1}(t))
=\frac{1}{2}\sqrt{(C(t)-\xi (t))^{2}+S(t)^{2}}
\sqrt{v_{x}(0)^{2}+v_{y}(0)^{2}}.
\end{eqnarray}%
Obviously this quantity reaches its minimum for $\xi
(t)=C(t),\forall t$ independently of the initial conditions. The
kernel for which the optimal performance of the post-Markovian
master equation is achieved, can thus be formally expressed, using
Eq. (\ref{eq:vy-post}), as:
\begin{equation}
k_{\mathrm{opt}}(t)=\frac{1}{2}e^{2t}\mathrm{Lap}^{-1}\left\{ \frac{1}{%
\mathrm{Lap}(C(t))}-s\right\} .  \label{kopt}
\end{equation}%
It should be noted that the condition for complete positivity of
the map generated by Eq. (\ref{PMAL}), $\sum_{i}\xi
_{i}(t)L_{i}^{T}\otimes
R_{i}\geq 0$ \cite{ShabaniLidar:05}, amounts here to $|\xi (t)|=|C(t)|\leq 1$%
, which holds for all $t$. Thus the minimum achievable
trace-distance between the two solutions is given by
\begin{equation}
D_{\mathrm{min}}(\rho _{\mathrm{exact}}(t),\rho _{1}(t))=\frac{1}{2}S(t)%
\sqrt{v_{x}(0)^{2}+v_{y}(0)^{2}}.
\end{equation}%
The optimal fit is plotted in Section 4.4.

Finding a simple analytical expression for the optimal kernel Eq. (\ref{kopt}%
) seems difficult due to the complicated form of $C(t)$. One way
to approach this problem is to expand $C(t)$ in powers of $\alpha
t$ and consider terms which give a valid approximation for small
times $\alpha t\ll 1$. For example, Eq. (\ref{eq:f-approx}) yields
the lowest non-trivial order as:
\begin{equation}
C_{2}(t)=1-2Q_{2}\alpha ^{2}t^{2}+\mathcal{O}(\alpha ^{4}t^{4}).
\end{equation}%
Note that this solution violates the complete positivity condition
for times larger than $t=1/\alpha \sqrt{2Q_{2}}$. The
corresponding kernel is:
\begin{equation}
k_{2}(t)=2\alpha ^{2}Q_{2}e^{2t}\cosh (2\sqrt{Q_{2}}\alpha t).
\end{equation}%
Alternatively we could try finding a kernel that matches some of
the approximate solutions discussed so far. For example, it turns
out that the kernel
\begin{equation}
k_{\mathrm{NZ2}}(t)=2\alpha ^{2}Q_{2}e^{2t}
\end{equation}%
leads to an exact match of the NZ2 solution. Finding a kernel
which gives a good description of the evolution of an open system
is an important but in general, difficult question which remains
open for further investigation. We note that this question was
also taken up in the context of the PM\ in the recent study
\cite{ManPet06}, where the PM\ was applied to an exactly solvable
model describing a qubit undergoing spontaneous emission and
stimulated absorption. No attempt was made to optimize the memory
kernel and hence the agreement with the exact solution was not as
impressive as might be possible with optimization.

\section*{4.4 \hspace{2pt} Comparison of the analytical solution and the different
approximation techniques}
\addcontentsline{toc}{section}{4.4
\hspace{0.15cm} Comparison of the analytical solution and the
different approximation techniques}

In the results shown below, all figures express the evolution in
terms of the dimensionless parameter $\alpha t$ (plotted on a
logarithmic scale). We
choose the initial condition $v_{x}(0)=v_{y}(0)=1/\sqrt{2}$ and plot only $%
v_{x}(t)$ since the structure of the equations for $v_{x}(t)$ and
$v_{y}(t)$ is similar. In order to compare the different methods of
approximation, we consider various choices of parameter values in
our model. Among these choices we consider both low and high
temperature cases. We note that in a spin bath model it is assumed
that the environment degrees of freedom are localized and this is
usually the case at low temperatures. At higher temperatures one may
need to consider delocalized environment degrees of freedom in order
to account for such environment modes such as phonons, magnons etc.
A class of models known as oscillator bath models (e.g.,
Ref.~\cite{Weiss:Book}) consider such effects. In this study, we
restrict attention to the spin bath model described here for both
low and high temperatures.

\subsection*{4.4.1 \hspace{2pt} Exact solution}
\addcontentsline{toc}{subsection}{4.4.1 \hspace{0.15cm} Exact
solution}

\begin{figure}[h]
\begin{center}
\includegraphics[scale=0.45]{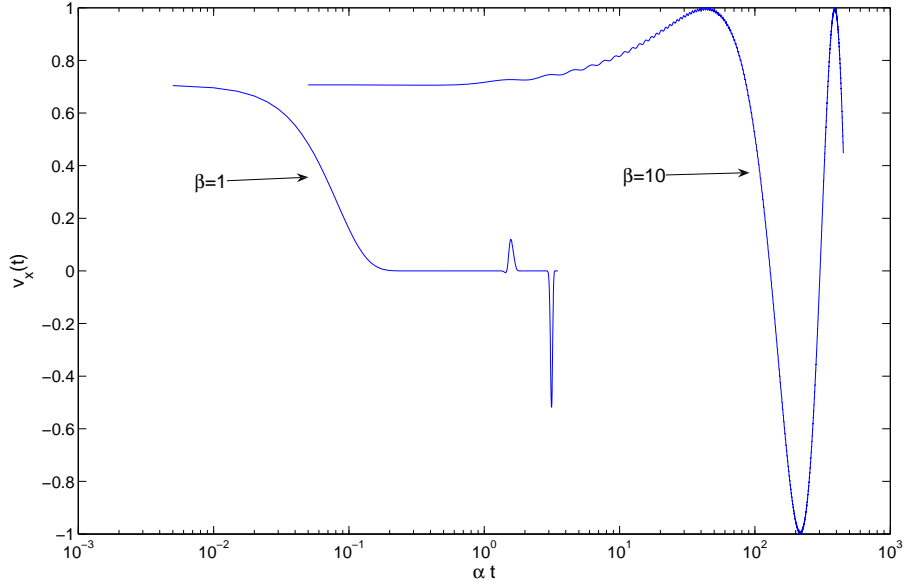}
\end{center}
\caption{Comparison of the exact solution at $\protect\beta =1 $
and $\protect\beta =10$ for $N=100$.} \label{N100_exact}
\end{figure}

\begin{figure}[h]
\begin{center}
\includegraphics[scale=0.45]{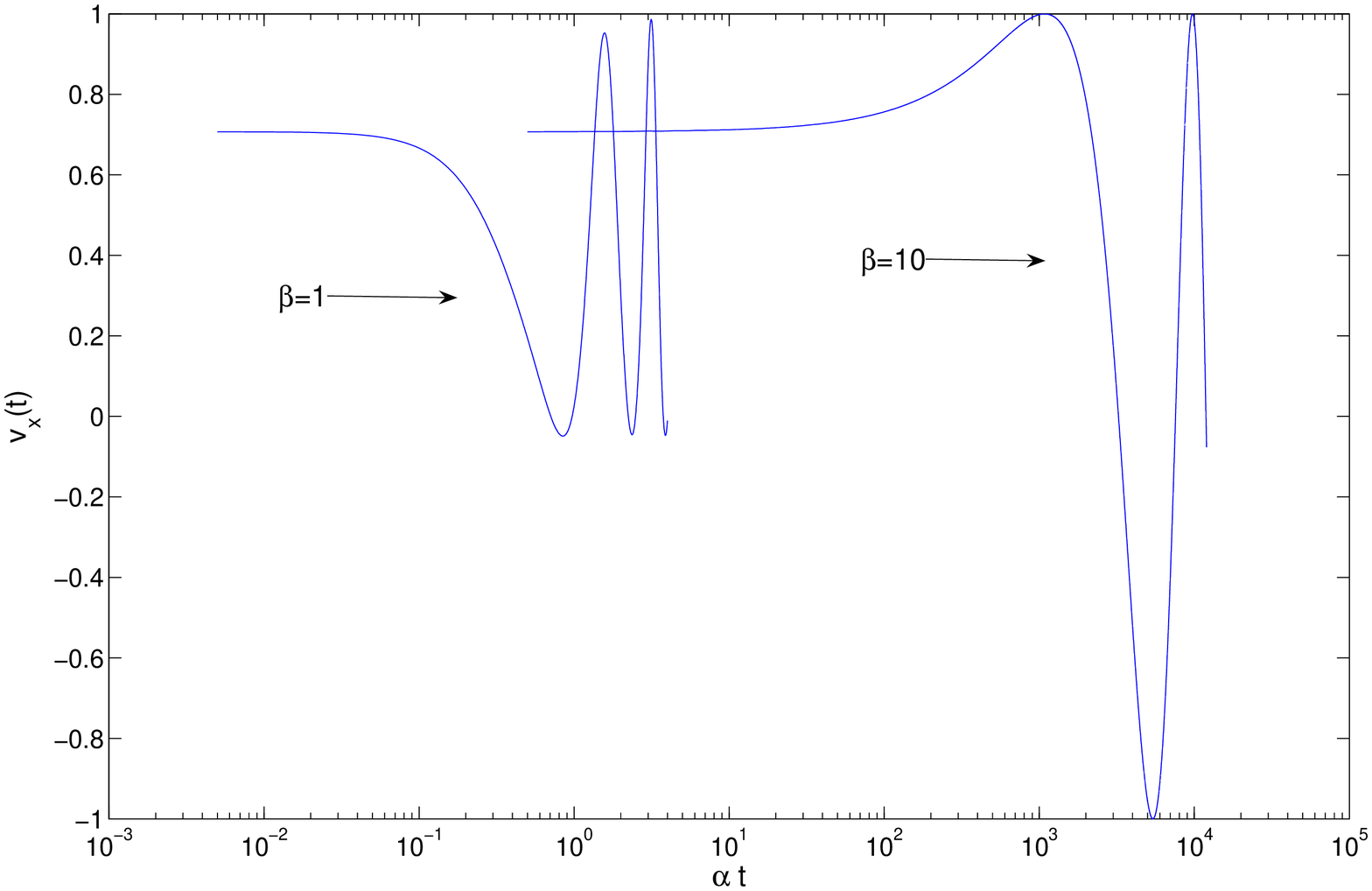}
\end{center}
\caption{Comparison of the exact solution at $\protect\beta =1 $
and $\protect\beta =10$ for $N=4$.} \label{N4_exact}
\end{figure}

We first assume that the frequencies of the qubits in the bath are equal ($%
\Omega _{n}=1$, $\forall n$), and so are the coupling constants ($g_{n}=1$, $%
\forall n$). In this regime, we consider large and small numbers
of bath
spins $N=100$ and $N=4$, and two different temperatures $\beta =1$ and $%
\beta =10$. Figs.~\ref{N100_exact} and \ref{N4_exact} show the
exact solution for $N=100$ and $N=4$ spins, respectively, up to
the second recurrence time. For each $N$, we plot the exact
solution for $\beta =1$ and $\beta =10$.

\begin{figure}[h]
\begin{center}
\includegraphics[scale=0.45]{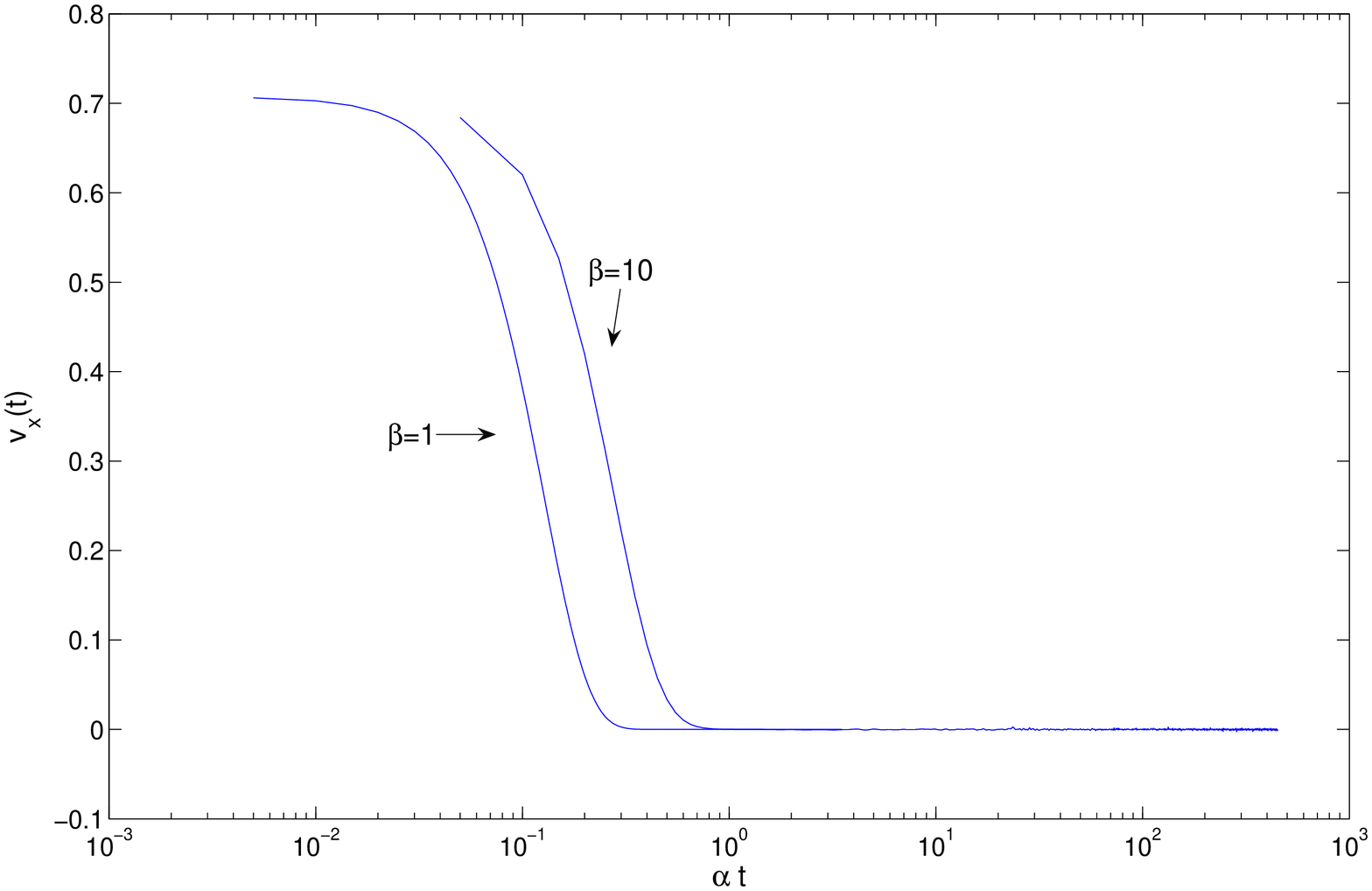}
\end{center}
\caption{Comparison of the exact solution at $\protect\beta =1
$ and $\protect\beta =10$ for $N=100$ for randomly generated $g_{n}$ and $%
\Omega _{n}$.} \label{N100_exact_random_long}
\end{figure}

We also consider the case where the frequencies $\Omega _{n}$ and
the coupling constants $g_{n}$ can take different values. We
generated uniformly distributed random values in the interval
$[-1,1]$ for both $\Omega _{n}$
and $g_{n}$. In Figs.~(\ref{N100_exact_random_long}) and (\ref%
{N4_exact_random}) we plot the ensemble average of the solution
over 50 random ensembles. The main difference from the solution
with equal $\Omega _{n}$ and $g_{n}$ is that the partial
recurrences decrease in size, especially as $N$ increases. We
attribute this damping partially to the fact that we look at the
ensemble average, which amounts to averaging out the positive and
negative oscillations that arise for different values of the
parameters. The main reason, however, is that for a generic
ensemble of random $\Omega _{n}$ and $g_{n}$ the positive and
negative oscillations in the sums \eqref{CSeqn} tend to average
out. This is particularly true for large $N$, as reflected in
Fig.~\ref{N100_exact_random_long}. We looked at a few individual
random cases for $N=100$ and recurrences were not present there.
For $N=20$ (not shown here), some small recurrences were still
visible.

We also looked at the case where one of the coupling constants,
say $g_i$, has a much larger magnitude than the other ones (which
were made equal). The behavior was similar to that for a bath
consisting of only a single spin.

\begin{figure}[h]
\begin{center}
\includegraphics[scale=0.45]{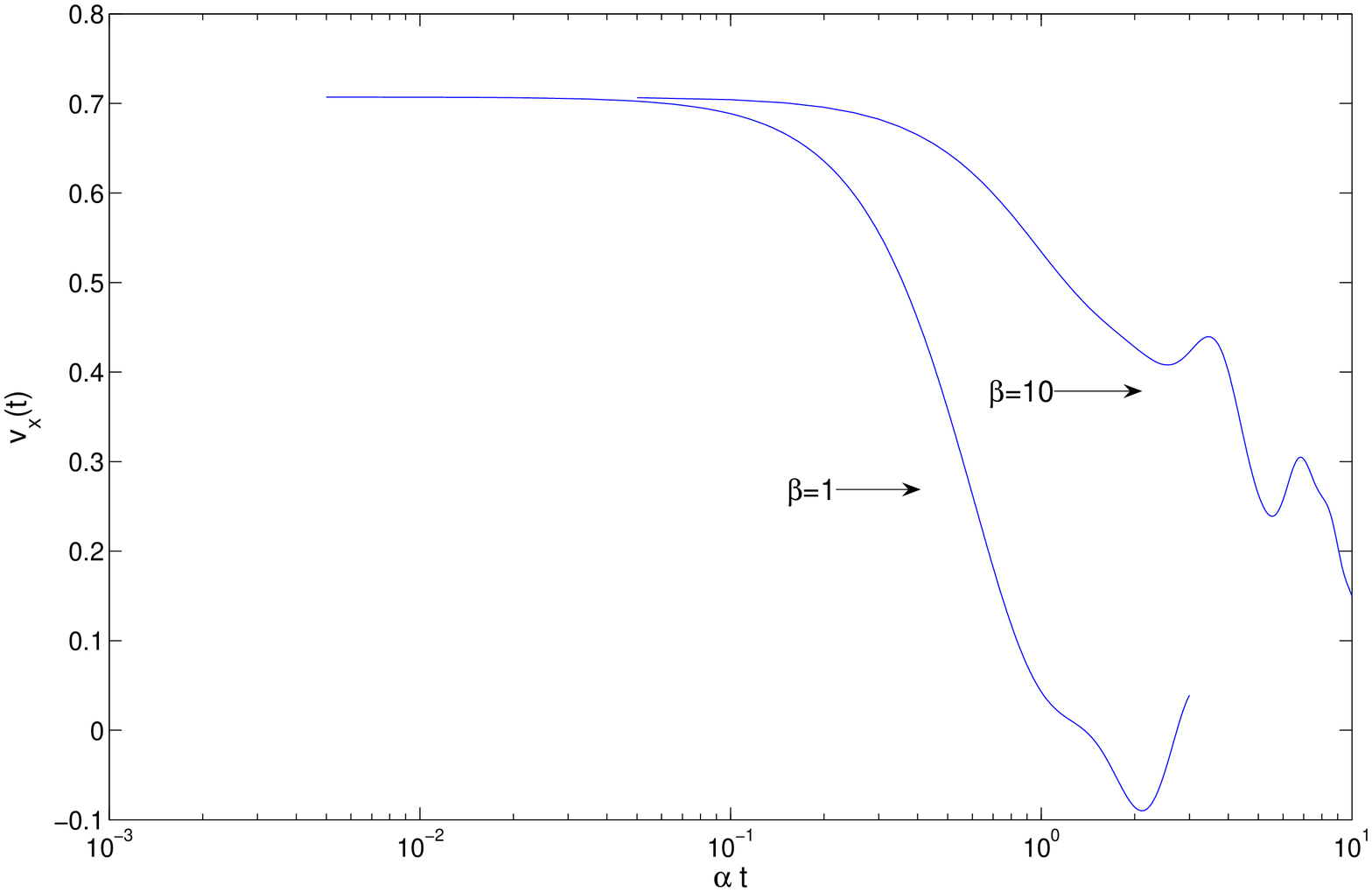}
\end{center}
\caption{Comparison of the exact solution at $\protect\beta =1
$ and $\protect\beta =10$ for $N=4$ for randomly generated $g_{n}$ and $%
\Omega _{n}$.} \label{N4_exact_random}
\end{figure}

In the following, we plot the solutions of different orders of the
NZ, TCL and PM master equations and compare them for the same
parameter values.

\subsection*{4.4.2 \hspace{2pt} NZ}
\addcontentsline{toc}{subsection}{4.4.2 \hspace{0.15cm} NZ}

\begin{figure}[h]
\begin{center}
\includegraphics[scale=0.45]{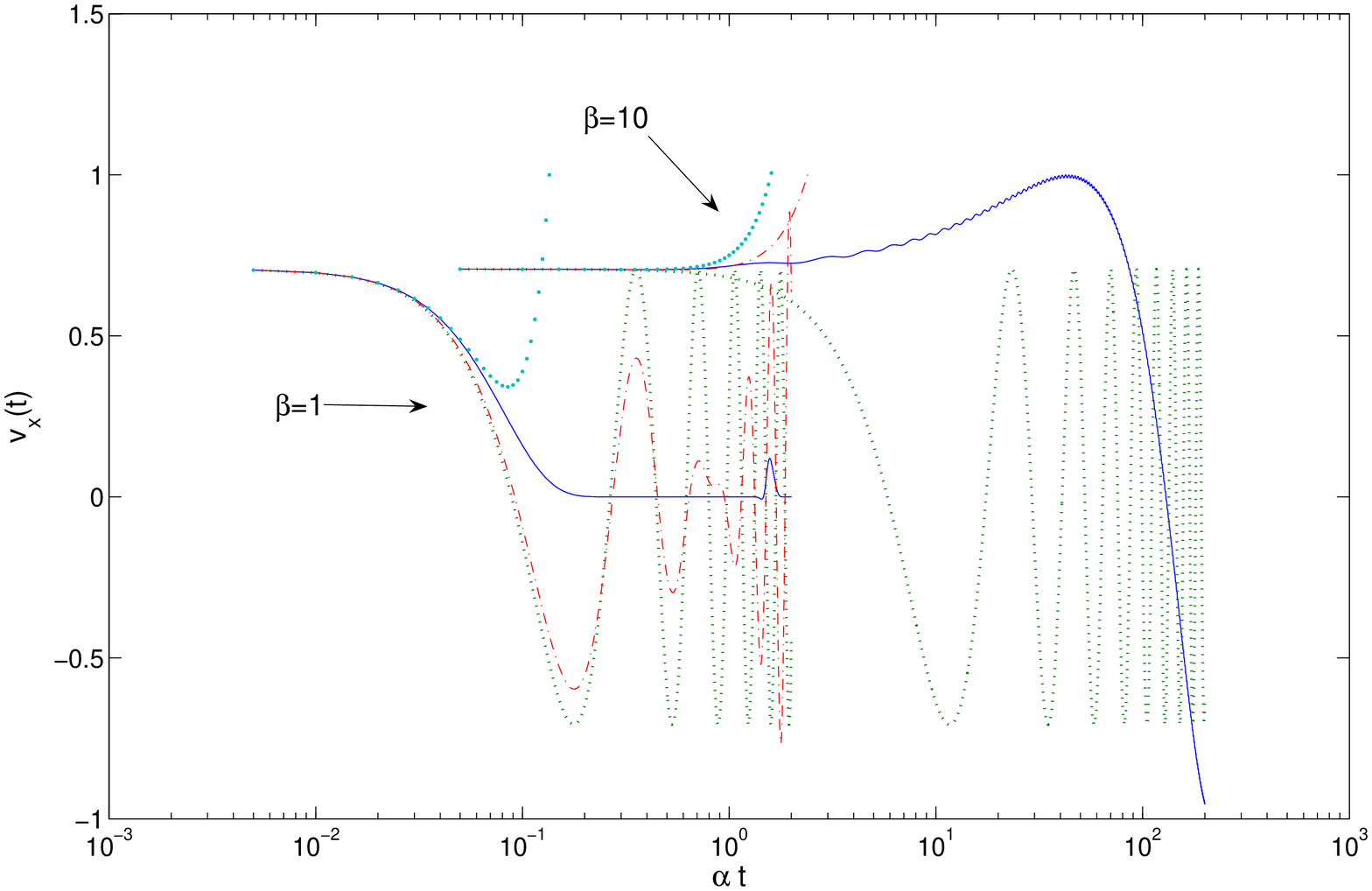}
\end{center}
\caption{Comparison of the exact solution, NZ2, NZ3 and NZ4 at
$\protect\beta =1$ and $\protect\beta =10$ for $N=100$. The exact
solution is the solid (blue) line, NZ2 is the dashed (green) line,
NZ3 is the dot-dashed (red) line and NZ4 is the dotted (cyan)
line.} \label{N100_NZ}
\end{figure}

\begin{figure}[h]
\begin{center}
\includegraphics[scale=0.45]{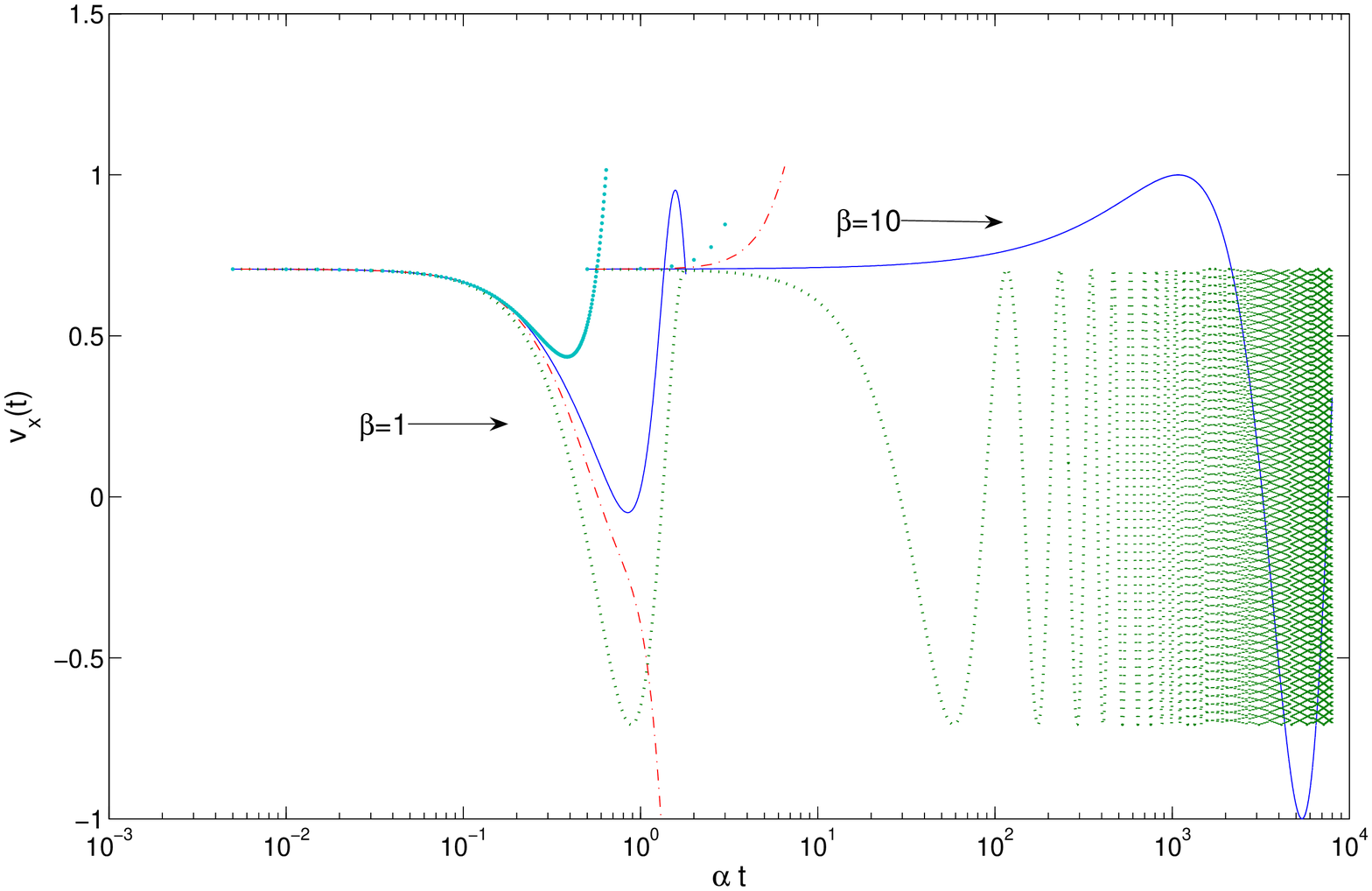}
\end{center}
\caption{Comparison of the exact solution, NZ2, NZ3 and NZ4 at
$\protect\beta =1$ and $\protect\beta =10$ for $N=4$. The exact
solution is the solid (blue) line, NZ2 is the dashed (green) line,
NZ3 is the dot-dashed (red) line and NZ4 is the dotted (cyan)
line.} \label{N4_NZ}
\end{figure}

In this subsection, we compare the solutions of different orders
of the NZ master equation for $\Omega _{n}=g_{n}=1$.
Fig.~(\ref{N100_NZ}) shows the solutions to NZ2, NZ3, NZ4 and the
exact solution for $\beta =1$ and $\beta =10$ up to the first
recurrence time of the exact solution. For short times NZ4 is the
better approximation. It can be seen that while NZ2 and NZ3 are
bounded, NZ4 leaves the Bloch sphere. But note that the
approximations under which these solutions have been obtained are
valid for $\alpha t\ll 1$. The NZ4 solution leaves the Bloch
sphere in a regime where the approximation is not valid. For
$\beta =10$, NZ2 again has a periodic behavior (which is
consistent with the solution), while the NZ3 and NZ4 solutions
leave the Bloch sphere after small times. Fig.~(\ref{N4_NZ}) shows
the same graphs for $N=4$. In this case both NZ3 and NZ4 leave the
Bloch sphere for $\beta =1$ and $\beta =10$, while NZ2 has a
periodic behavior. A clear conclusion from these plots is that the
NZ\ approximation is truly a short-time one:\ it becomes
completely unreliable for times longer than $\alpha t\ll 1$.

\subsection*{4.4.3 \hspace{2pt} TCL}
\addcontentsline{toc}{subsection}{4.4.3 \hspace{0.15cm} TCL}

\begin{figure}[h]
\begin{center}
\includegraphics[scale=0.45]{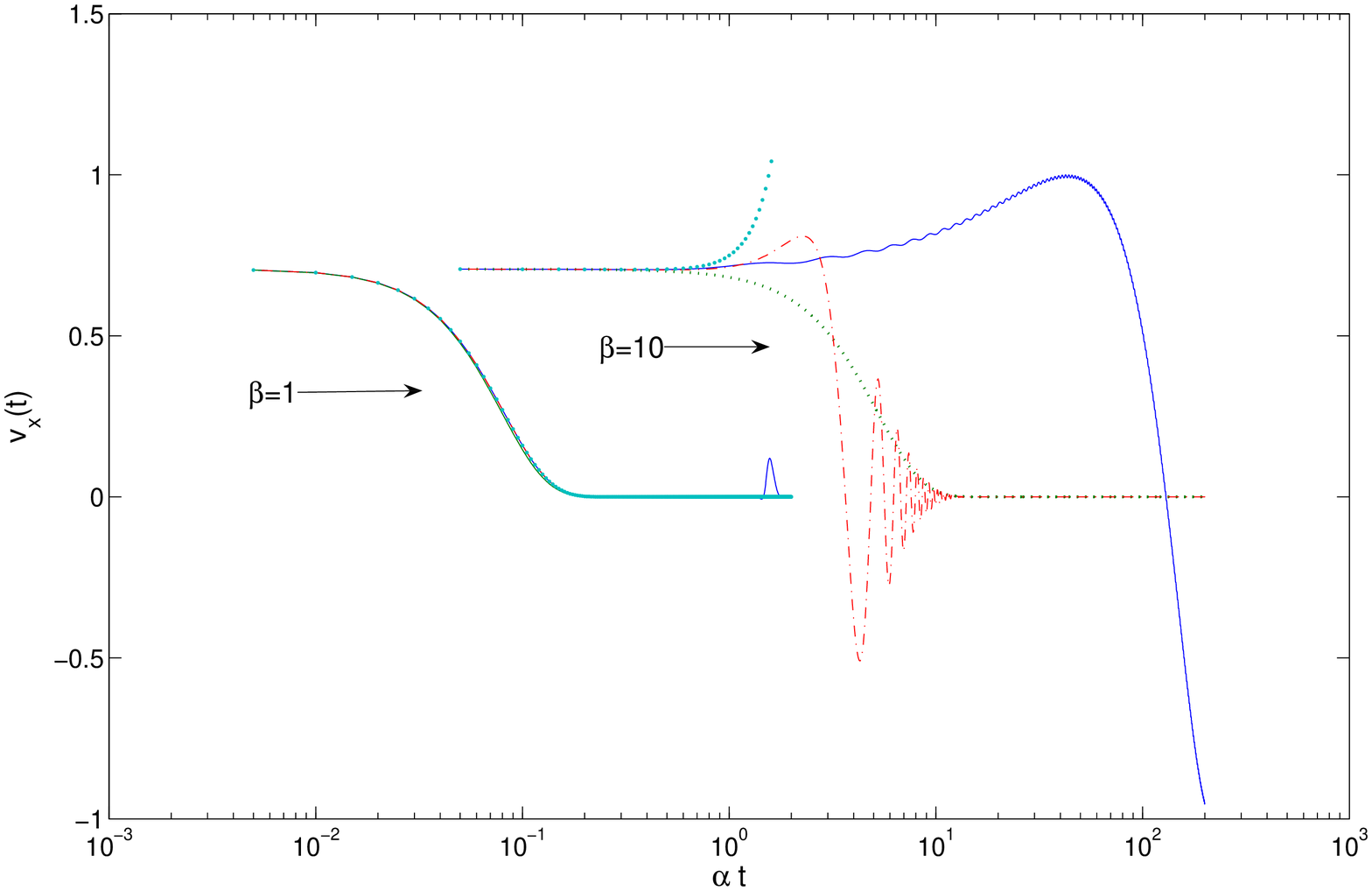}
\end{center}
\caption{Comparison of the exact solution, TCL2, TCL3 and TCL4 at
$\protect\beta =1$ and $\protect\beta =10$ for $N=100$. The exact
solution is the solid (blue) line, TCL2 is the dashed (green)
line, TCL3 is
the dot-dashed (red) line and TCL4 is the dotted (cyan) line. Note that for $%
\protect\beta =1$, the curves nearly coincide.} \label{N100_TCL}
\end{figure}

\begin{figure}[h]
\begin{center}
\includegraphics[scale=0.45]{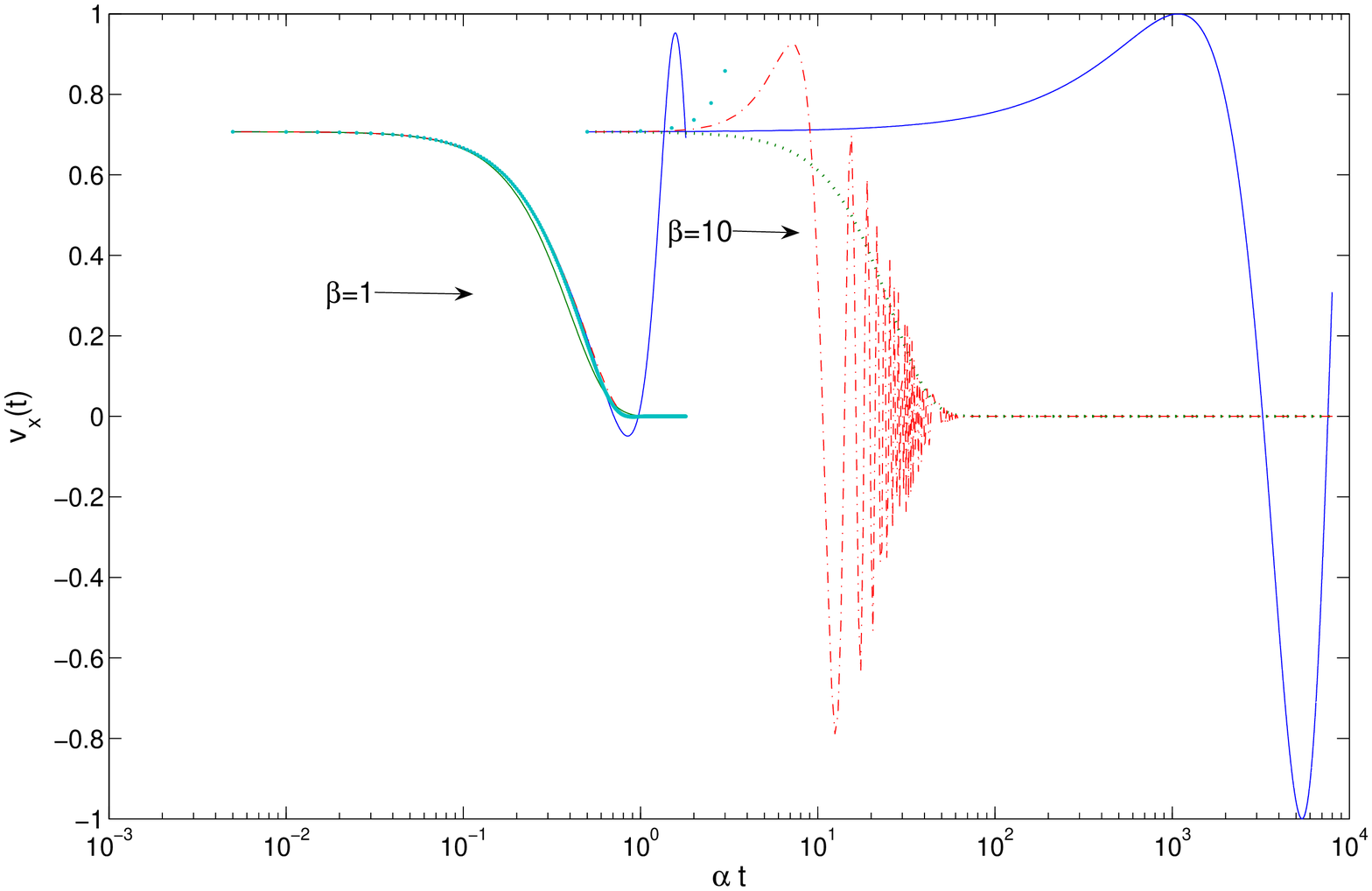}
\end{center}
\caption{Comparison of the exact solution, TCL2, TCL3 and TCL4 at
$\protect\beta =1$ and $\protect\beta =10$ for $N=4$. The exact
solution is the solid (blue) line, TCL2 is the dashed (green)
line, TCL3 is
the dot-dashed (red) line and TCL4 is the dotted (cyan) line. Note that for $%
\protect\beta =1$, TCL3, TCL4 and the exact solution nearly
coincide.} \label{N4_TCL}
\end{figure}

Fig.~(\ref{N100_TCL}) plots the exact solution, TCL2, TCL3 and TCL4 at $%
\beta =1$ and $\beta =10$ for $N=100$ spins and $\Omega
_{n}=g_{n}=1$. It can be seen that for $\beta =1$, the TCL
solution approximates the exact solution well even for long times.
However, the TCL solution cannot reproduce the recurrence behavior
of the exact solution (also shown in the figure.)
Fig.~(\ref{N4_TCL}) shows the same graphs for $N=4$. In this case,
while TCL2 and TCL3 decay, TCL4 increases exponentially and leaves
the Bloch sphere after a short time. This is because the exponent
in the solution of TCL4 in Eq.~(\ref{TCL3}) is positive. Here
again the approximations under which the solutions have been
obtained are valid only for small time scales and the graphs
demonstrate the complete breakdown of the perturbation expansion
for large values of $\alpha t$. Moreover, the graphs reveal the
sensitivity of the approximation to temperature:\ the TCL\ fares
much better at high temperatures.

\begin{figure}[h]
\begin{center}
\includegraphics[scale=0.45]{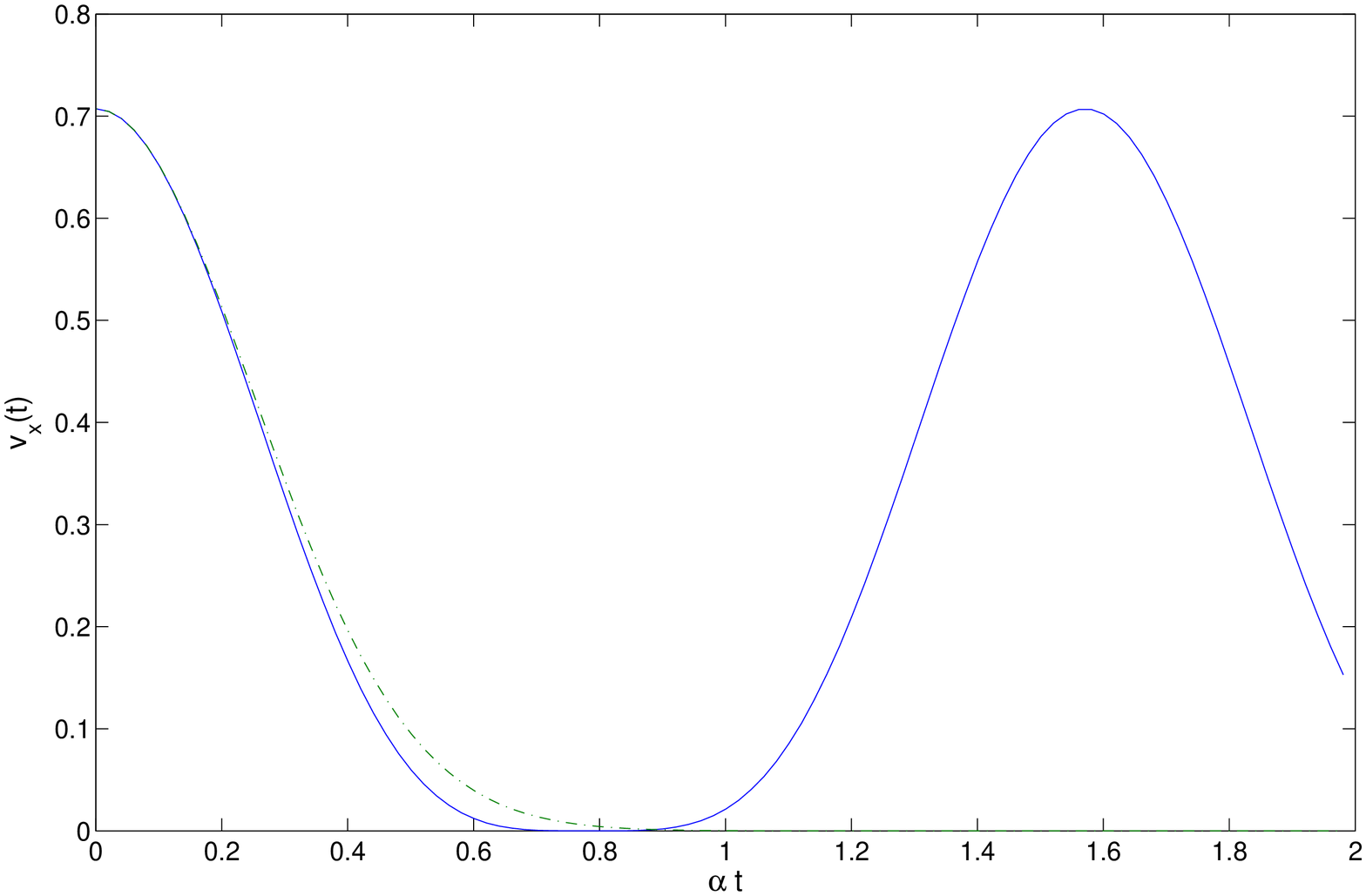}
\end{center}
\caption{Comparison of TCL2 and the exact solution to
demonstrate the validity of the TCL approximation for $N=4$ and $\protect%
\beta =1$. The solid (blue) line denotes the exact solution and
the dashed (green) line is TCL2. Note that the time axis here is
on a linear scale. TCL2 breaks down at $\protect\alpha t\approx
0.9$, where it remains flat, while the exact solution has a
recurrence.} \label{N4B0TCL2}
\end{figure}

In order to determine the validity of the TCL approximation, we
look at the invertibility of the Kraus map derived in Eq.
(\ref{KrausForm}) or equivalently Eq. (\ref{CSeqn}). As mentioned
earlier, this map is
non-invertible if $C(t)^{2}+S(t)^{2}=0$ for some $t$ (or equivalently $%
v_{x}(t)=0$ and $v_{y}(t)=0$). This will happen if and only if at
least one of the $\beta _{n}$ is zero. This can occur when the
bath density matrices of some of the bath spins are maximally
mixed or in the limit of a very high bath temperature. Clearly,
when the Kraus map is non-invertible, the TCL approach becomes
invalid since it relies on the assumption that the information
about the initial state is contained in the current state. This
fact has also been observed for the spin-boson model with a damped
Jaynes-Cummings Hamiltonian \cite{BrePet02}. At the point where
the Kraus map becomes non-invertible, the TCL solution deviates
from the exact solution (see Fig.~ \ref{N4B0TCL2}). We verified
that both $v_{x}$ and $v_{y} $ vanish at this point.

\subsection*{4.4.4 \hspace{2pt} NZ, TCL, and PM}
\addcontentsline{toc}{subsection}{4.4.4 \hspace{0.15cm} NZ, TCL,
and PM}

\begin{figure}[h]
\begin{center}
\includegraphics[scale=0.45]{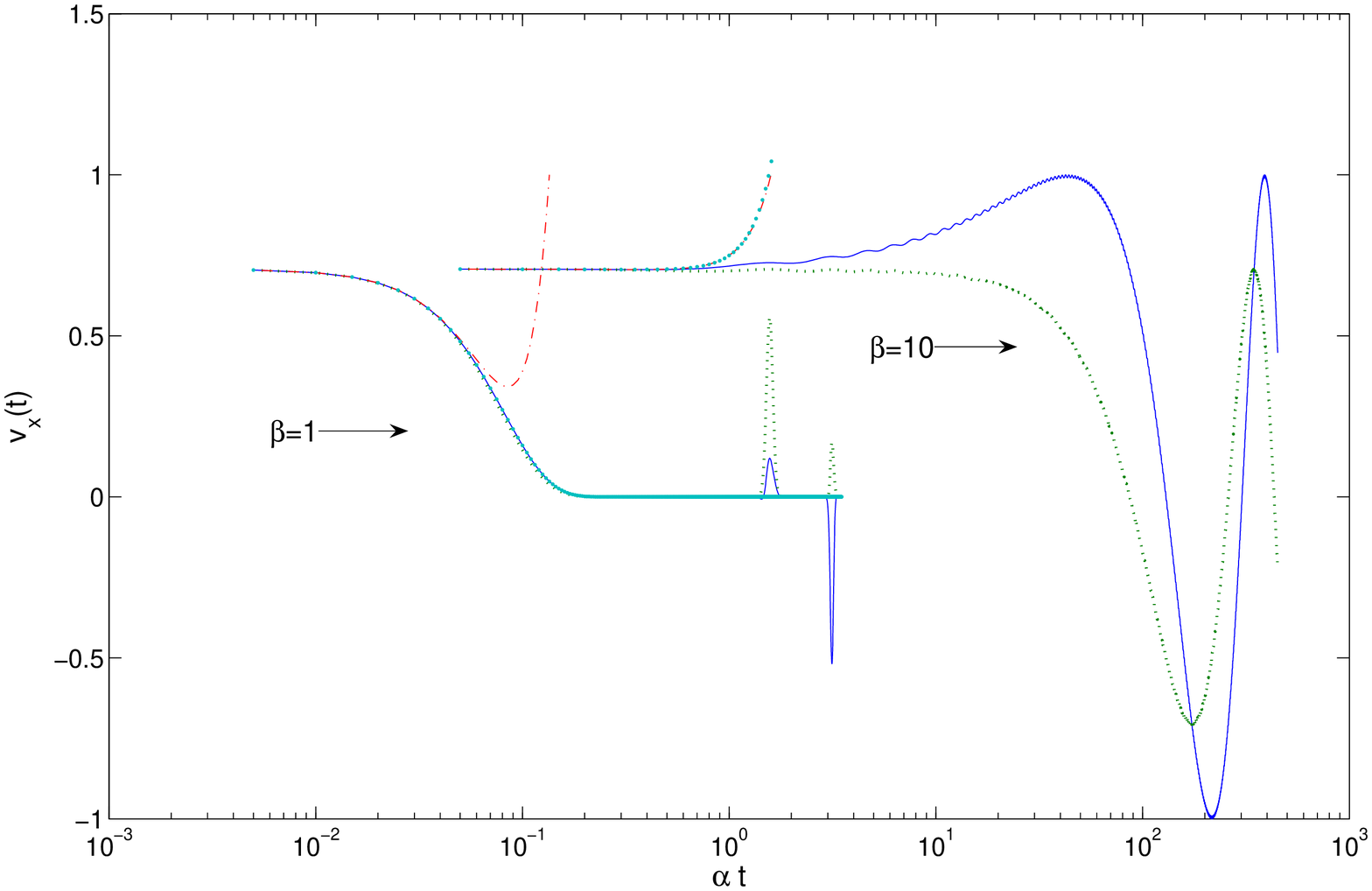}
\end{center}
\caption{Comparison of the exact solution, NZ4, TCL4 and PM at
$\protect\beta =1$ and $\protect\beta =10$ for $N=100$. The exact
solution is the solid (blue) line, PM is the dashed (green) line,
NZ4 is the
dot-dashed (red) line and TCL4 is the dotted (cyan) line. Note that for $%
\protect\beta =1$, TCL4, PM and the exact solution nearly coincide
for short and medium times. Only PM\ captures the recurrences of
the exact solution at long times.} \label{N100_best}
\end{figure}

In this subsection, we compare the exact solution to TCL4, NZ4 and
the solution of the optimal PM master equation.
Fig.~(\ref{N100_best}) shows these solutions for $N=100$ and
$\beta =1$ and $\beta =10$ when $\Omega _{n}=g_{n}=1$. Here we
observe that while the short-time behavior of the exact solution
is approximated well by all the approximations we consider, the
long-time behavior is approximated well only by PM.

\begin{figure}[h]
\begin{center}
\includegraphics[scale=0.45]{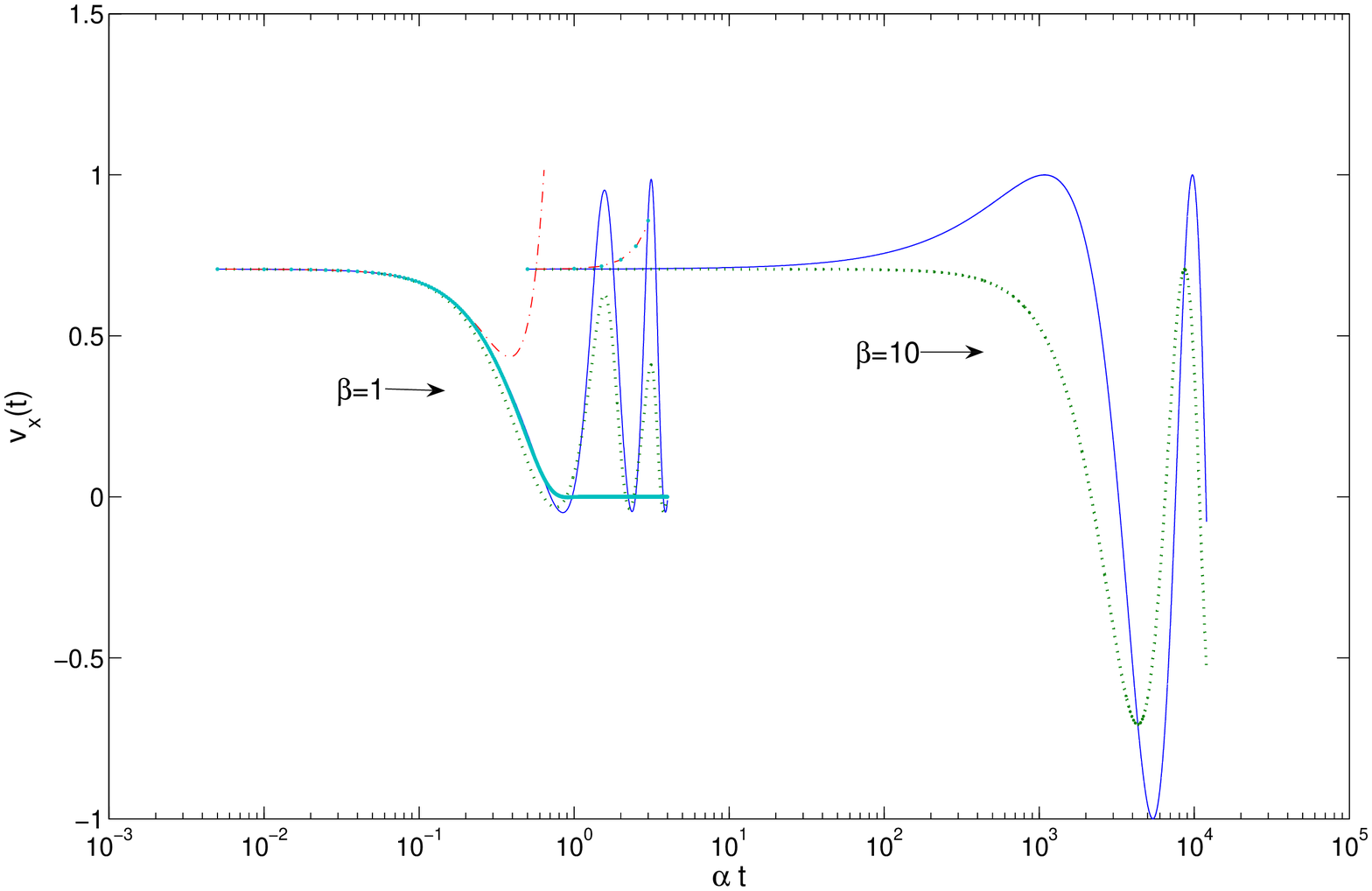}
\end{center}
\caption{Comparison of the exact solution, NZ4, TCL4 and PM at
$\protect\beta=1$ and $\protect\beta=10$ for $N=4$. The exact
solution is the solid (blue) line, PM is the dashed (green) line,
NZ4 is the dot-dashed (red) line and TCL4 is the dotted (cyan)
line. Note that for $\protect\beta=1 $, TCL4 and the exact
solution nearly coincide for short and medium times.}
\label{N4_best}
\end{figure}

For $\beta =1$, NZ4 leaves the Bloch sphere after a short time
while TCL4 decays with the exact solution. But as before, the TCL
solution cannot reproduce the recurrences seen in the exact
solution. The optimal PM solution, by contrast, is capable of
reproducing both the decay and the
recurrences. TCL4 and NZ4 leave the Bloch sphere after a short time for $%
\beta =10$, while PM again reproduces the recurrences in the exact
solution. Fig.~\ref{N4_best} shows the corresponding graphs for
$N=4$ and it can be
seen that again PM can outperform both TCL and NZ for long times. Figs.~\ref%
{N100_beta} and \ref{N4_beta} show the performance of TCL4, NZ4
and PM compared to the exact solution at a fixed time (for which
the approximations are valid) for different temperatures ($\beta
\in \lbrack 0.01,10]$). It can be seen that both TCL4 and the
optimal PM solution perform better than NZ4 at medium and high
temperatures, with TCL4 outperforming PM at medium temperatures.
The performance of NZ4 is enhanced at low temperatures, where
it performs similarly to TCL4 (see also Figs.~\ref{N100_best} and \ref%
{N4_best}). This can be understood from the short-time
approximation to the exact solution given in Eq.
(\ref{shorttimes}), which up to the precision for which it was
derived is also an approximation of NZ2 [Eq. (\ref{Born})]. As
discussed above, this approximation (which also coincides with
TCL2) is valid when $2Q_{2}(\alpha t)^{2}\ll 1$. As temperature
decreases, so does
the magnitude of $Q_{2}$, which leads to a better approximation at fixed $%
\alpha t$. Since NZ2 gives the lowest-order correction, this
improvement is reflected in NZ4 as well.

In Figs. \ref{N100_best_random} and \ref{N4_best_random} we plot
the
averaged solutions over 50 ensembles of random values for $\Omega _{n}$ and $%
g_{n}$ in the interval $[-1,1]$. We see that on average TCL4, NZ4
and the optimal PM solution behave similarly to the case when
$\Omega _{n}=g_{n}=1$. Due to the damping of the recurrences,
especially when $N=100$, the TCL4 and the PM solutions match the
exact solution closely for much longer times than in the
deterministic case. Again, the PM solution is capable of
qualitatively matching the behavior of the exact solution at long
times.

\begin{figure}[h]
\begin{center}
\includegraphics[scale=0.45]{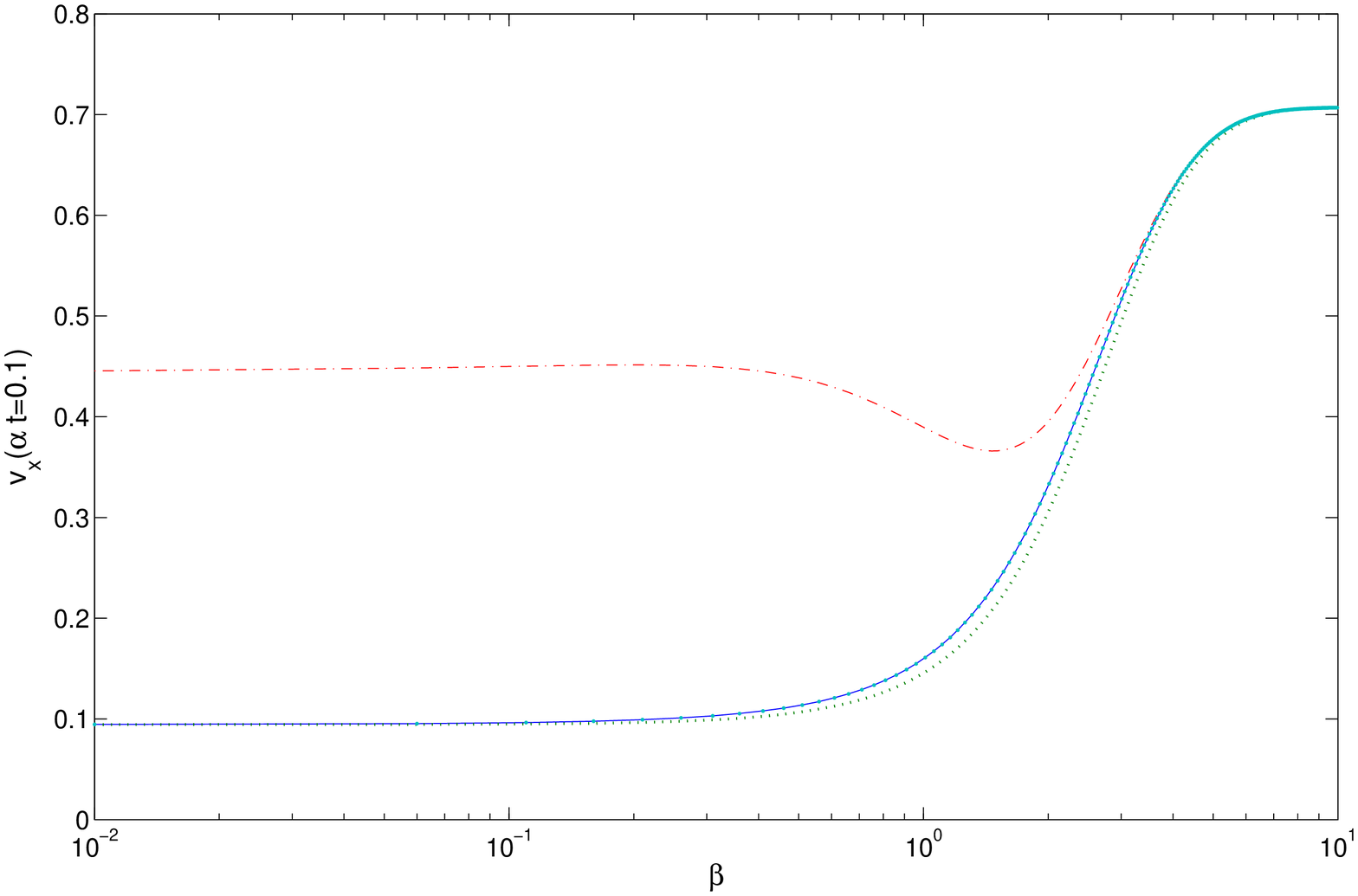}
\end{center}
\caption{Comparison of the exact solution, NZ4, TCL4 and PM at
$\protect\alpha t=0.1$ for $N=100$ for different $\protect\beta\in
[0.01,10]$. The exact solution is the solid (blue) line, PM is the
dashed (green) line, NZ4 is the dot-dashed (red) line and TCL4 is
the dotted (cyan) line.} \label{N100_beta}
\end{figure}

\begin{figure}[h]
\begin{center}
\includegraphics[scale=0.45]{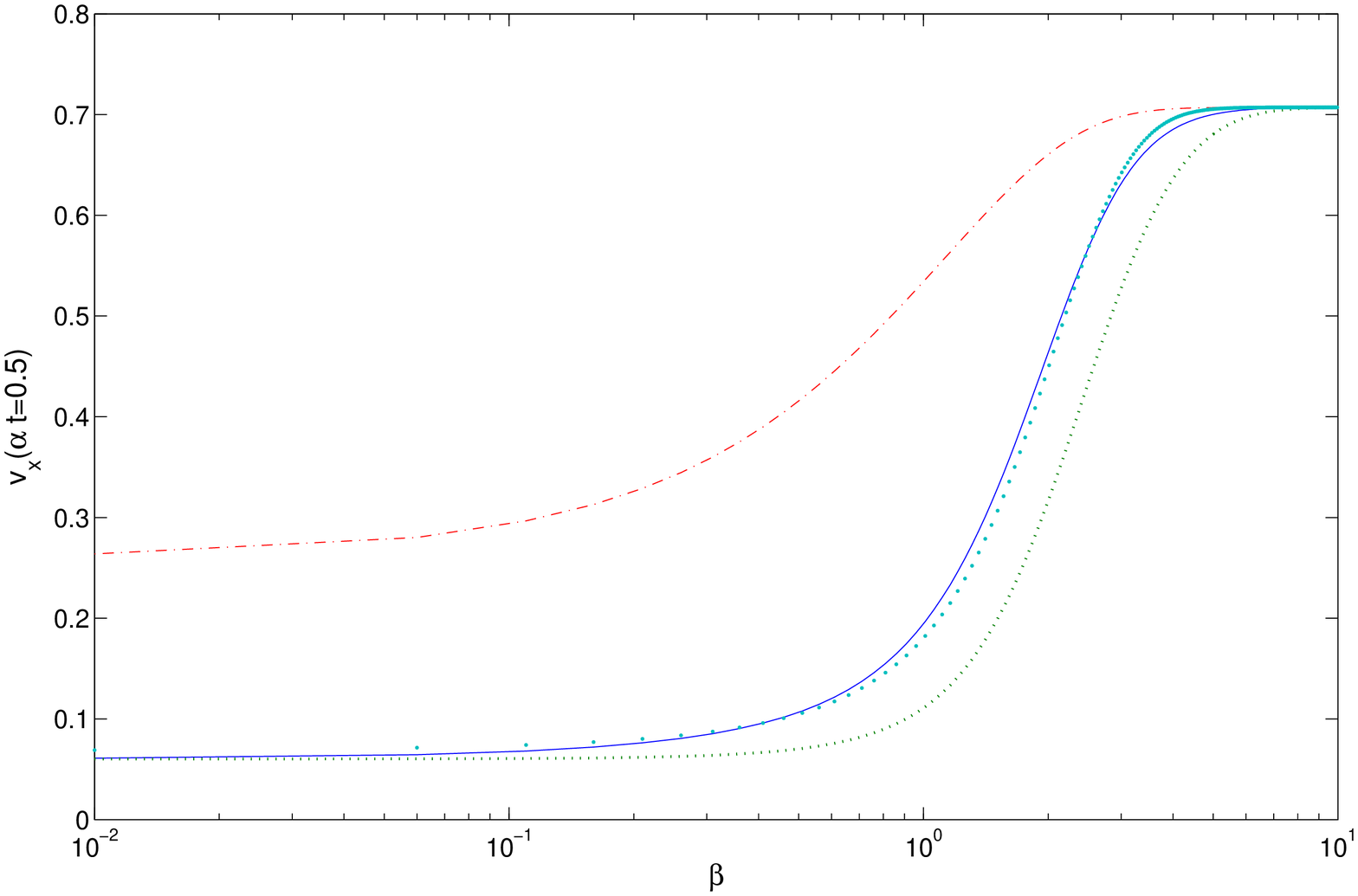}
\end{center}
\caption{Comparison of the exact solution, NZ4, TCL4 and PM at
$\protect\alpha t=0.5$ for $N=4$ for different $\protect\beta\in
[0.01,10] $. The exact solution is the solid (blue) line, PM is
the dashed (green) line, NZ4 is the dot-dashed (red) line and TCL4
is the dotted (cyan) line.} \label{N4_beta}
\end{figure}

\begin{figure}[h]
\begin{center}
\includegraphics[scale=0.45]{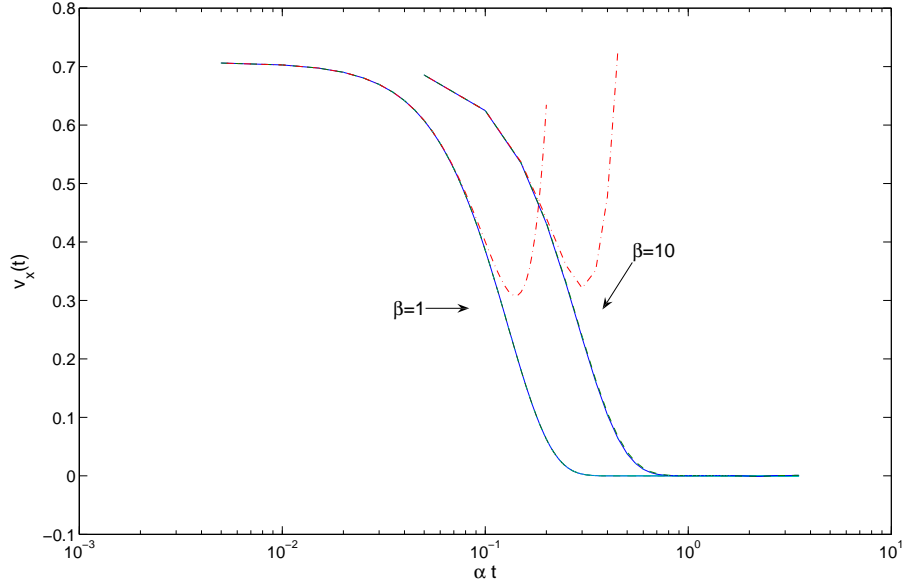}
\end{center}
\caption{Comparison of the exact solution, NZ4, TCL4 and PM at
$\protect\beta=1$ and $\protect\beta=10$ for $N=100$ for random
values of $g_n$ and $\Omega_n$. The exact solution is the solid
(blue) line, PM is the dashed (green) line, NZ4 is the dot-dashed
(red) line and TCL4 is the dotted (cyan) line. Note that for
$\protect\beta=1$ and $\protect\beta=10$, TCL4, PM and the exact
solution nearly coincide.} \label{N100_best_random}
\end{figure}

\begin{figure}[h]
\begin{center}
\includegraphics[scale=0.45]{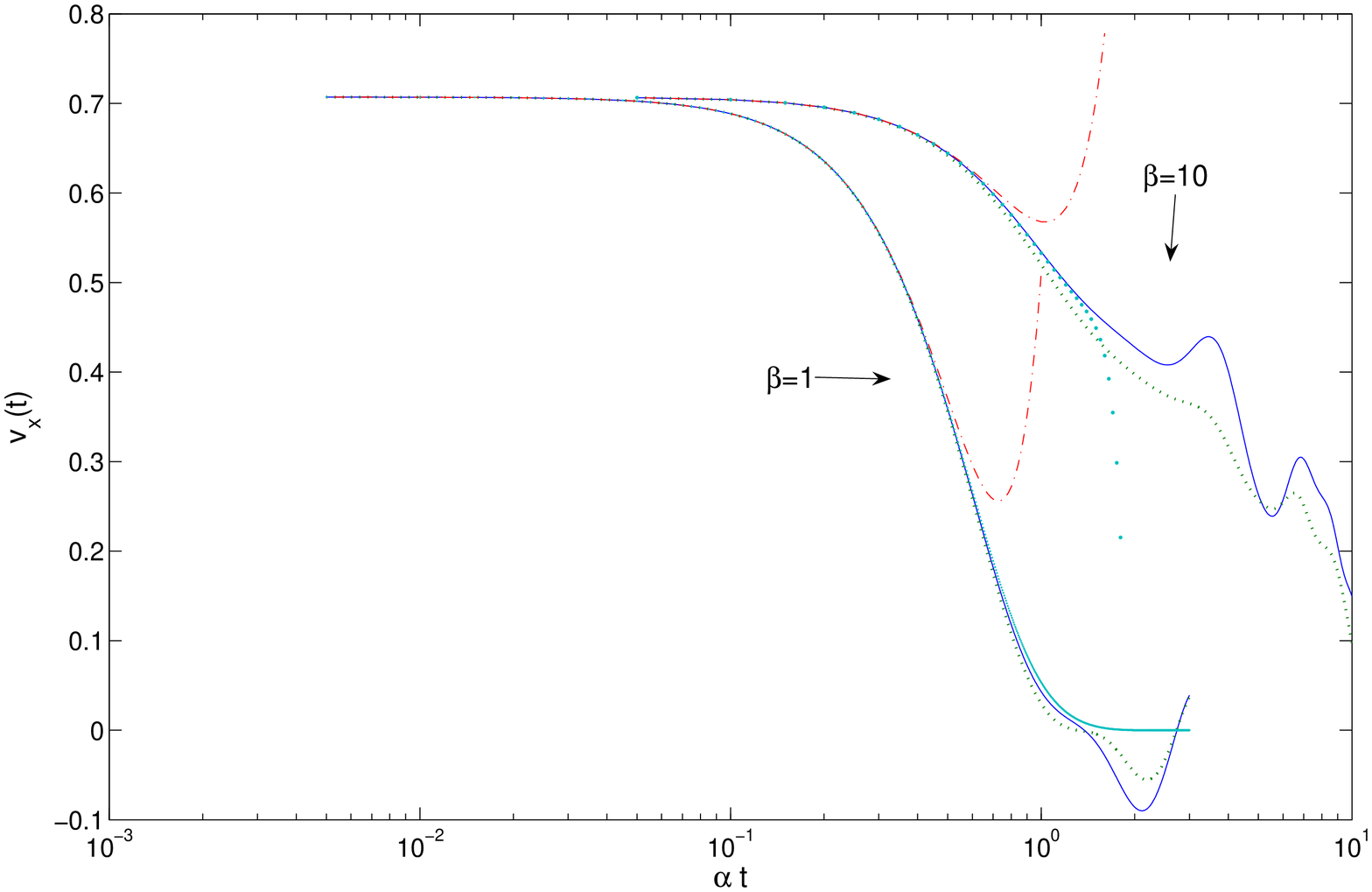}
\end{center}
\caption{Comparison of the exact solution, NZ4, TCL4 and PM
at $\protect\beta=1$ and $\protect\beta=10$ for $N=4$ for random values of $%
g_n$ and $\Omega_n$. The exact solution is the solid (blue) line,
PM is the dashed (green) line, NZ4 is the dot-dashed (red) line
and TCL4 is the dotted (cyan) line. Note that for
$\protect\beta=1$, TCL4, PM and the exact solution nearly coincide
for short and medium times.} \label{N4_best_random}
\end{figure}

\subsection*{4.4.5 \hspace{2pt} Coarse-graining approximation}
\addcontentsline{toc}{subsection}{4.4.5 \hspace{0.15cm}
Coarse-graining approximation}

\begin{figure}[h]
\begin{center}
\includegraphics[scale=0.45]{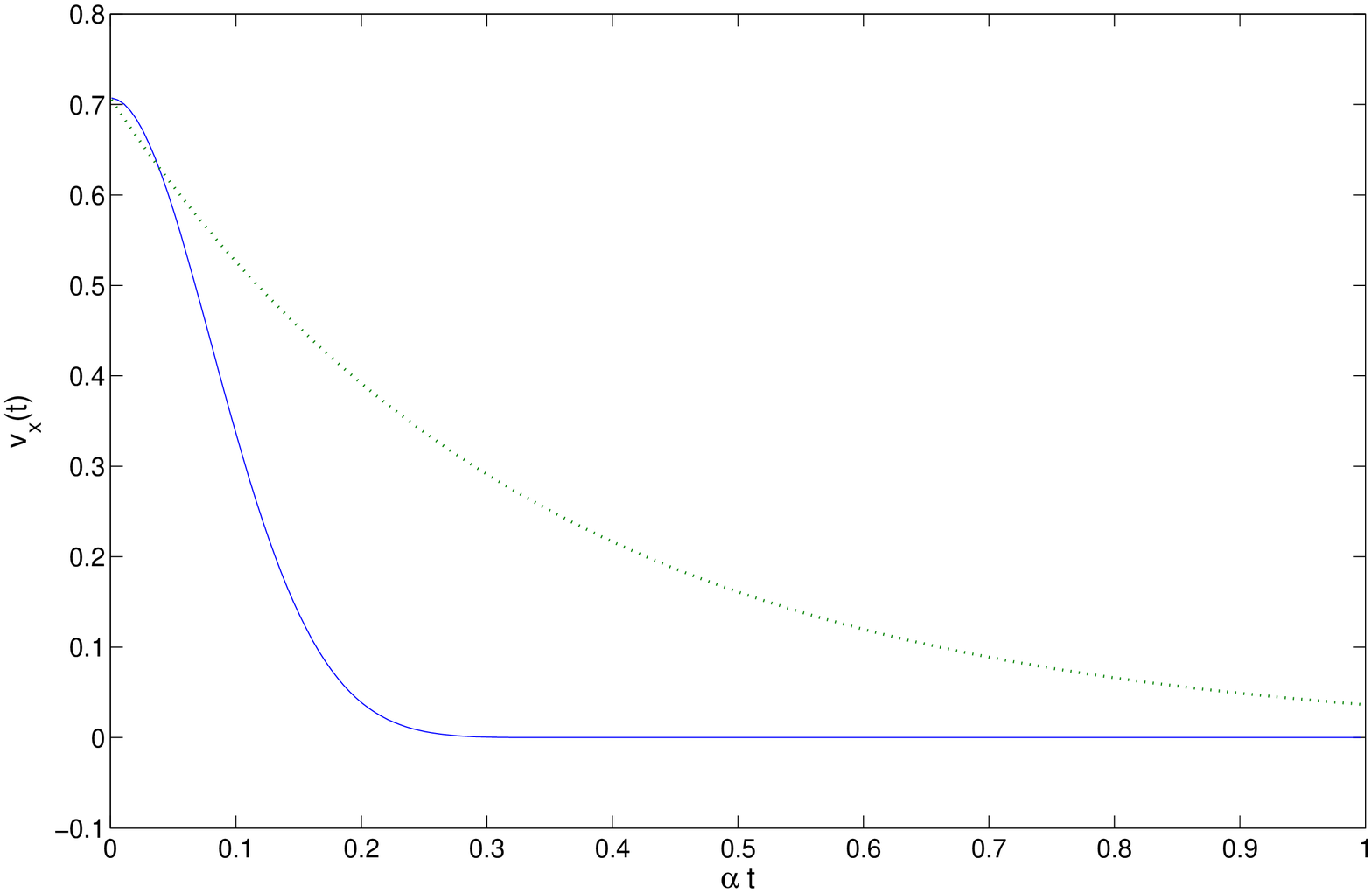}
\end{center}
\caption{Comparison of the exact solution and the optimal
coarse-graining approximation for $N=50$ and $\protect\beta =1$.
The exact solution is the solid (blue) line and the
coarse-graining approximation is the dashed (green) line. Note the
linear scale time axis.} \label{N50B1CG}
\end{figure}

Finally, we examine the coarse-graining approximation discussed in
Section 4.3.1. We choose the time over which the average trace
distance is calculated to be the time where the exact solution
dies down. In Fig. \ref{N50B1CG} we plot the coarse-grained
solution for the value of $\tau $ for which the trace distance to
the exact solution is minimum. As can be seen, the coarse-graining
approximation does not help since the Markovian assumption is not
valid for this model. In deriving the coarse-graining
approximation \cite{Lidar:CP01} one makes the assumption that the
coarse-graining time scale is greater than any characteristic bath
time scale. But the characteristic time scale of the bath is
infinite in this case.

\section*{4.5 \hspace{2pt} Summary and conclusions}
\addcontentsline{toc}{section}{4.5 \hspace{0.15cm} Summary and
conclusions}

We studied the performance of various methods for approximating the
evolution of an Ising model of an open quantum system for a qubit
system coupled to a bath bath consisting of $N$ qubits. The high
symmetry of the model allowed us to derive the exact dynamics of the
system as well as find analytical solutions for the different master
equations. We saw that the Markovian approximation fails for this
model due to the time independence of the bath correlation
functions. This is also reflected in the fact that the
coarse-graining method \cite{Lidar:CP01} does not approximate the
exact solution well. We discussed the performance of these solutions
for various parameter regimes. Unlike other spin bath models
discussed in literature (e.g., Ref.~\cite{BBP04}), the odd-order
bath correlation functions do not vanish, leading to the existence
of odd-order terms in the solution of TCL and NZ equations. These
terms describe the rotation around the $z$ axis of the Bloch sphere,
a fact which is reflected in the exact solution. We showed that up
to fourth order TCL performs better than NZ at medium and high
temperatures. For low temperatures we demonstrated an enhancement in
the performance of NZ and showed that NZ and TCL perform equally
well. We showed that the TCL approach breaks down for certain
parameter choices and related this to the non-invertibility of the
Kraus map describing the system dynamics. We also studied the
performance of the post-Markovian master equation obtained in
\cite{ShabaniLidar:05} with an optimal memory kernel. We discussed
possible ways of approximating the optimal kernel for short times
and derived the kernel which leads to an exact fit to the NZ2
solution. It turns out that PM master equation performs as well as
the TCL2 for a large number of spins and outperforms all orders of
NZ\ and TCL\ considered here at long times, as it captures the
recurrences of the exact solution.

Our study reveals the limitations of some of the best known master
equations available in the literature, in the context of a spin
bath. In general, perturbative approaches such as low-order NZ\
and TCL do well at short times (on a time scale set by the
system-bath coupling constant) and fare very poorly at long times.
These approximations are also very sensitive to temperature and do
better in the high temperature limit. The PM\ does not do as well
as TCL4 at short times but has the distinct advantage of retaining
a qualitatively correct character for long times. This conclusion
depends heavily on the proper choice of the memory kernel; indeed,
when the memory kernel is not optimally chosen the PM can yield
solutions which are not as satisfactory \cite{ManPet06}.

\section*{4.6 \hspace{2pt} Appendix A: Bath correlation functions}

\label{app:A}

\addcontentsline{toc}{section}{4.6 \hspace{0.15cm} Appendix A:
Bath correlation functions}

Here we show how to calculate the bath correlation functions used
in our simulations. The $k^{\mathrm{th}}$ order bath correlation
function is defined as
\begin{equation*}
Q_{k}=\mathrm{Tr}\{B^{k}\rho _{B}\},
\end{equation*}%
where $B$ and $\rho _{B}$ were given in Eqs. (\ref{Bcomp})\ and (\ref%
{eq:rhoB0}), respectively.
This yields:%
\begin{eqnarray*}
Q_{k} &=&\mathrm{Tr}\{(\sum_{n}g_{n}\sigma _{n}^{z}-\theta I_{B})^{k}\sum_{l}%
\frac{\exp (-\beta E_{l})}{Z}|l\rangle \langle l|\} \\
&=&\sum_{l}\frac{\exp (-\beta E_{l})}{Z}\langle
l|(\sum_{n}g_{n}\sigma
_{n}^{z}-\theta I_{B})^{k}|l\rangle \\
&=&\sum_{l,l^{\prime },...,l^{\prime \prime \prime }}\frac{\exp
(-\beta E_{l})}{Z}\langle l|(\sum_{n}g_{n}\sigma _{n}^{z}-\theta
I_{B})|l^{\prime }\rangle \langle l^{\prime }|(\sum_{n^{\prime
}}g_{n^{\prime }}\sigma _{n^{\prime }}^{z}-\theta I_{B})|l^{\prime
\prime }\rangle \langle l^{\prime \prime }|\cdots\notag\\
&& \cdots |l^{\prime \prime \prime }\rangle \langle l^{\prime
\prime \prime }|(\sum_{n^{\prime \prime \prime}}g_{n^{\prime
\prime \prime}}\sigma _{n^{\prime \prime \prime}}^{z}-\theta I_{B})|l\rangle \\
&=&\sum_{l,l^{\prime },...,l^{\prime \prime \prime }}\frac{%
\exp (-\beta E_{l})}{Z}(\sum_{n}g_{n}\langle l|\sigma
_{n}^{z}|l^{\prime }\rangle -\theta )\delta _{ll^{\prime
}}(\sum_{n^{\prime }}g_{n^{\prime }}\langle l^{\prime }|\sigma
_{n^{\prime }}^{z}|l^{\prime \prime }\rangle -\theta )\delta
_{l^{\prime }l^{\prime \prime }}\cdots \notag\\
&& \cdots(\sum_{n^{\prime \prime \prime }}g_{n^{\prime \prime
\prime}}\langle l^{\prime \prime \prime }|\sigma _{n^{\prime
\prime \prime }}^{z}|l\rangle -\theta )\delta
_{l^{\prime \prime \prime }l} \\
&=&\sum_{l}\frac{\exp (-\beta E_{l})}{Z}(\sum_{n}g_{n}\langle
l|\sigma _{n}^{z}|l\rangle -\theta )(\sum_{n^{\prime
}}g_{n^{\prime }}\langle l|\sigma _{n^{\prime }}^{z}|l\rangle
-\theta )\cdots \notag\\
&&\cdots(\sum_{n^{\prime \prime \prime }}g_{n^{\prime \prime
\prime }}\langle l|\sigma _{n^{\prime \prime \prime
}}^{z}|l\rangle
-\theta ) \\
&=&\sum_{l}\frac{\exp (-\beta E_{l})}{Z}(\sum_{n}g_{n}\langle
l|\sigma _{n}^{z}|l\rangle -\theta )^{k},
\end{eqnarray*}

or
\begin{equation}
Q_{k}=\frac{1}{Z}\sum_{l}(\tilde{E}_{l})^{k}\exp (-\beta E_{l}),
\end{equation}%
where $Z=\sum_{l}\exp (-\beta E_{l})$ and the expressions for $E_{l}$ and $%
\tilde{E}_{l}$ were given in Eqs. (\ref{eq:El})\ and
(\ref{eq:Eitilde}), respectively.

The above formulas are useful when the energy levels $E_{l}$ and $\tilde{E}%
_{l}$ are highly degenerate, which is the case for example when
$g_{n}\equiv g$ and $\Omega _{n}\equiv \Omega $ for all $n$. For a
general choice of these parameters, it is computationally more
efficient to consider $\theta $ in the form (\ref{eq:theta}) and
the initial bath density matrix in the form
(\ref{eq_rho_B_inter}). For example, the second order bath
correlation function is
  \begin{eqnarray}
Q_{2} &=&\mathrm{Tr}\{(\sum_{m=1}^{N}g_{m}\sigma _{m}^{z}-\theta
I)(\sum_{n=1}^{N}g_{n}\sigma _{n}^{z}-\theta I)\rho _{B}\}  \notag \\
&=&\mathrm{Tr}\{\sum_{n,m=1}^{N}g_{n}g_{m}\sigma _{n}^{z}\sigma
_{m}^{z}\rho _{B}\} -2\theta
\underbrace{\mathrm{Tr}\{\sum_{n=1}^{N}g_{n}\sigma
_{n}^{z}\rho _{B}\}}_{\theta }+\theta ^{2}  \notag \\
&=&\mathrm{Tr}\{\sum_{n,m=1}^{N}g_{n}g_{m}\sigma _{n}^{z}\sigma
_{m}^{z}\bigotimes\limits_{n=1}^{N}\frac{1}{2}(I+\beta _{n}\sigma
_{n}^{z})\}-\theta ^{2}  \notag \\
&=&\sum_{n\neq m}^{N}\mathrm{Tr}\{g_{m}\frac{1}{2}(\sigma
_{m}^{z}+\beta _{m}I)\}\mathrm{Tr}\{g_{n}\frac{1}{2}(\sigma
_{n}^{z}+\beta _{n}I)\}\prod\limits_{j\neq
m,n}\mathrm{Tr}\{\frac{1}{2}(I+\beta _{j}\sigma
_{j}^{z})\}\notag\\
&& +\mathrm{Tr}\{\sum_{n=1}^{N}g_{n}^{2}\rho _{B}\}-\theta ^{2}
\notag \\
&=&\underbrace{\sum_{n,m=1}^{N}g_{m}\beta _{m}g_{n}\beta
_{n}}_{\theta ^{2}}-\sum_{n=1}^{N}g_{n}^{2}\beta
_{n}^{2}+\sum_{n=1}^{N}g_{n}^{2}-\theta
^{2}  \notag \\
&=&\sum_{n=1}^{N}g_{n}^{2}(1-\beta _{n}^{2}).
  \end{eqnarray}%
Using the identity $1-\tanh ^{2}(-x/2)=2/(1+\cosh x)$, this
correlation function can be expressed in terms of the bath
spectral density function
[Eq. (\ref{eq:J})] as follows:%
\begin{eqnarray*}
Q_{2} &=&\sum_{n=1}^{N}g_{n}^{2}(1-\beta _{n}^{2}) \\
&=&\int_{-\infty }^{\infty }\delta (\Omega -\Omega
_{n})|g_{n}|^{2}(1-\tanh
^{2}(-\frac{\Omega }{2kT}))\mathrm{d}\Omega \\
&=&\int_{-\infty }^{\infty }\frac{2J(\Omega )\mathrm{d}\Omega }{1+\cosh (%
\frac{\Omega }{kT})}.
\end{eqnarray*}

Higher order correlation functions are computed analogously.

\section*{4.7 \hspace{2pt} Appendix B: Cumulants for the NZ and TCL master equations}

\label{app:B}

\addcontentsline{toc}{section}{4.7 \hspace{0.15cm} Appendix B:
Cumulants for the NZ and TCL master equations}

We calculate the explicit expressions for the cumulants appearing in Eq. (%
\ref{eq:cumulants}), needed to find the NZ and TCL perturbation
expansions up to fourth order.

Second order:

\begin{eqnarray}
\langle \mathcal{L}^{2}\rangle \rho
&=&-\mathrm{Tr}_{B}\{[H_{I},[H_{I},\rho
]]\}\otimes \rho _{B}  \notag \\
&=&-\mathrm{Tr}_{B}\{H_{I}^{2}\rho -2H_{I}\rho H_{I}+\rho
H_{I}^{2}\}\otimes
\rho _{B}  \notag \\
&=&-2Q_{2}(\rho _{S}-\sigma _{z}\rho _{S}\sigma _{z})\otimes \rho
_{B}
\notag \\
&\equiv &\rho ^{\prime },
\end{eqnarray}%
\begin{eqnarray}
\langle \mathcal{L}^{2}\rangle ^{2}\rho  &=&\mathcal{P}\mathcal{L}^{2}%
\mathcal{P}\mathcal{P}\mathcal{L}^{2}\mathcal{P}\rho   \notag \\
&=&\mathcal{P}\mathcal{L}^{2}\mathcal{P}\rho ^{\prime }  \notag \\
&=&-2Q_{2}(\rho _{S}^{\prime }-\sigma _{z}\rho _{S}^{\prime
}\sigma _{z})\otimes \rho _{B},  \notag
\end{eqnarray}%
where $\rho _{S}^{\prime }=\mathrm{Tr}_{B}{\rho }^{\prime
}=-2Q_{2}(\rho _{S}-\sigma _{z}\rho _{S}\sigma _{z})$. Therefore
\begin{eqnarray}
\langle \mathcal{L}^{2}\rangle ^{2}\rho  &=&-2Q_{2}\{(-2Q_{2}(\rho
_{S}-\sigma _{z}\rho _{S}\sigma _{z}))-\sigma _{z}(-2Q_{2}(\rho
_{S}-\sigma
_{z}\rho _{S}\sigma _{z}))\sigma _{z}\}\otimes \rho _{B}  \notag \\
&=&8Q_{2}^{2}(\rho _{S}-\sigma _{z}\rho _{S}\sigma _{z})\otimes
\rho _{B}.
\end{eqnarray}%
Third order:
\begin{eqnarray}
\langle \mathcal{L}^{3}\rangle \rho  &=&i\mathrm{Tr}_{B}%
\{[H_{I},[H_{I},[H_{I},\rho ]]]\}\otimes \rho _{B}  \notag \\
&=&i\mathrm{Tr}_{B}\{H_{I}^{3}\rho -3H_{I}^{2}\rho
H_{I}+3H_{I}\rho
H_{I}^{2}-\rho H_{I}^{3}\}\otimes \rho _{B}  \notag \\
&=&4iQ_{3}(\sigma _{z}\rho _{S}-\rho _{S}\sigma _{z})\otimes \rho
_{B}.
\end{eqnarray}%
Fourth order:
\begin{eqnarray}
\langle \mathcal{L}^{4}\rangle \rho  &=&\mathrm{Tr}_{B}%
\{[H_{I},[H_{I},[H_{I},[H_{I},\rho ]]]]\}\otimes \rho _{B}  \notag \\
&=&\mathrm{Tr}_{B}\{H_{I}^{4}\rho -4H_{I}^{3}\rho
H_{I}+6H_{I}^{2}\rho
H_{I}^{2}-4H_{I}\rho H_{I}^{3}+\rho H_{I}^{4}\}\otimes \rho _{B}  \notag \\
&=&8Q_{4}(\rho _{S}-\sigma _{z}\rho _{S}\sigma _{z})\otimes \rho
_{B}.
\end{eqnarray}

\chapter*{Chapter 5: \hspace{1pt} Continuous quantum error correction for non-Markovian decoherence}
\addcontentsline{toc}{chapter}{Chapter 5:\hspace{0.15cm}
Continuous quantum error correction for non-Markovian decoherence}

In this chapter we continue our exploration of non-Markovian
decoherence. This time, we compare Markovian and non-Markovian
error models in light of the performance of continuous quantum
error correction. We consider again an Ising decoherence model of
the type we studied in the previous chapter, but in a much simpler
version---when the environment consists of only a single qubit.
This allows us to solve exactly the evolution of a multi-qubit
code in which each qubit is coupled to an independent bath when
the code is subject to continuous error correction. The
conclusions we obtain, however, extend beyond this model and apply
for general non-Markovian decoherence.

\section*{5.1 \hspace{2pt} Preliminaries}
\addcontentsline{toc}{section}{5.1 \hspace{0.15cm} Preliminaries}

\subsection*{5.1.1 \hspace{2pt} Continuous quantum error correction}
\addcontentsline{toc}{subsection}{5.1.1 \hspace{0.15cm} Continuous
quantum error correction}

In general, error probabilities increase with time. No matter how
complicated a code or how many levels of concatenation are
involved, the probability of uncorrectable errors is never truly
zero, and if the system is exposed to noise for a sufficiently
long time, the weight of uncorrectable errors can accumulate. To
combat this, error correction must be applied repeatedly and
sufficiently often. If one assumes that the time for an
error-correcting operation is small compared to other relevant
time scales of the system, error-correcting operations can be
considered instantaneous. Then the scenario of repeated error
correction leads to a discrete evolution which often may be
difficult to describe.  To study the evolution of a system in the
limit of frequently applied instantaneous error correction, Paz
and Zurek proposed to describe error correction as a continuous
quantum jump process \cite{PZ98}. In this model, the infinitesimal
error-correcting transformation that the density matrix of the
encoded system undergoes during a time step $dt$ is
\begin{equation}\label{basicequation}
\rho\rightarrow (1-\kappa dt)\rho + \kappa dt \Phi(\rho),
\end{equation}
where $\Phi(\rho)$ is the completely positive trace-preserving
(CPTP) map describing a full error-correcting operation, and
$\kappa$ is the error-correction rate. The full error-correcting
operation $\Phi(\rho)$ consists of a syndrome detection, followed
(if necessary) by a unitary correction operation conditioned on
the syndrome.

Consider, for example, the three-qubit bit-flip code whose purpose
is to protect an unknown qubit state from bit-flip (Pauli $X$)
errors. The code space is spanned by $|\overline{0}\rangle =
|000\rangle$ and $|\overline{1}\rangle = |111\rangle$, and the
stabilizer generators are $ZZI$ and $IZZ$ (see Section 8.3). Here
by $X=\sigma^x$, $Y=\sigma^y$, $Z=\sigma^z$ and $I$ we denote the
usual Pauli operators and the identity, respectively, and a string
of three operators represents the tensor product of operators on
each of the three qubits. The standard error-correction procedure
involves a measurement of the stabilizer generators, which
projects the state onto one of the subspaces spanned by
$|000\rangle$ and $|111\rangle$, $|100\rangle$ and $|011\rangle$,
$|010\rangle$ and $|101\rangle$, or $|001\rangle$ and
$|110\rangle$; the outcome of these measurements is the error
syndrome. Assuming that the probability for two- or three-qubit
errors is negligible, then with high probability the result of
this measurement is either the original state with no errors, or
with a single $X$ error on the first, the second, or the third
qubit. Depending on the outcome, one then applies an $X$ gate to
the erroneous qubit and transforms the state back to the original
one. The CPTP map $\Phi(\rho)$ for this code can be written
explicitly as
\begin{equation}
\begin{split}
\Phi(\rho) = \left(|000\rangle \langle000| + |111\rangle
\langle111| \right) \rho
\left(|000\rangle \langle000| + |111\rangle \langle111| \right) \\
+ \left(|000\rangle \langle100| + |111\rangle \langle011| \right)
\rho
\left(|100\rangle \langle000| + |011\rangle \langle111| \right) \\
+ \left(|000\rangle \langle010| + |111\rangle \langle101|
\right)\rho
\left(|010\rangle \langle000| + |101\rangle \langle111| \right) \\
+ \left(|000\rangle\langle001| + |111\rangle \langle110|
\right)\rho \left(|001\rangle \langle000| + |110\rangle
\langle111| \right). \label{strongmap}
\end{split}
\end{equation}

The quantum-jump process \eqref{basicequation} can be viewed as a
smoothed version of the discrete scenario of repeated error
correction, in which instantaneous full error-correcting
operations are applied at random times with rate $\kappa$. It can
also be looked upon as arising from a continuous sequence of
infinitesimal CPTP maps of the type \eqref{basicequation}. In
practice, such a weak map is never truly infinitesimal, but rather
has the form
\begin{equation} \rho
\rightarrow (1-\epsilon^2)\rho + \epsilon^2 \Phi(\rho),\label{wm}
\end{equation}
where $\epsilon \ll 1$ is a small but finite parameter, and the
weak operation takes a small but nonzero time $\tau_c$. For times
$t$ much greater than $\tau_c$ ($\tau_c\ll t$), the weak
error-correcting map (\ref{wm}) is well approximated by the
infinitesimal form \eqref{basicequation}, where the rate of error
correction is
\begin{equation}
\kappa = \epsilon^2 /\tau_c. \label{tauc}
\end{equation}
A weak map of the form \eqref{wm} could be implemented, for
example, by a weak coupling between the system and an ancilla via
an appropriate Hamiltonian, followed by discarding the ancilla. A
closely related scenario, where the ancilla is continuously cooled
in order to reset it to its initial state, was studied in
\cite{SarMil05}.

Another way of implementing the weak map \eqref{wm} is via weak
measurements followed by weak unitaries dependent on the outcome. In
the appendix at the end of this chapter, we give an example of such
an implementation for the case of the bit-flip code---when
$\Phi(\rho)$ is given by \eqref{strongmap}. We also present a scheme
in terms of weak measurements for codes that correct arbitrary
single-qubit errors. In the latter case, the resulting weak map is
not the same as \eqref{wm}, but also yields the strong
error-correcting map $\Phi(\rho)$ when exponentiated. We point out
that the weak measurements used in these schemes are not weak
versions of the strong measurements for syndrome detection---they
are in a different basis. They can be regarded as weak versions of a
different set of strong measurements which, when followed by an
appropriate unitary, yield the same map $\Phi(\rho)$ on average.
Thus, the workings of continuous error correction, when it is driven
by weak measurements, does not translate directly into the error
syndrome detection and correction of the standard paradigm. In this
sense, the continuous approach can be regarded as a different
paradigm for error correction---one based on weak measurements and
weak unitary operations. The idea of using continuous weak
measurements and unitary operations for error correction has been
explored in the context of different heuristic schemes \cite{ADL02,
SarMil05g}, some of which are based on a direct ``continuization''
of the syndrome measurements. In this study we consider continuous
error correction of the type given by Eq.~\eqref{basicequation}.

\subsection*{5.1.2 \hspace{2pt} Markovian decoherence}
\addcontentsline{toc}{subsection}{5.1.2 \hspace{0.15cm} Markovian
decoherence}

So far, continuous quantum error correction has been studied only
for Markovian error models. As we discussed in the previous
chapter, the Markovian approximation describes situations where
the bath-correlation times are much shorter than any
characteristic time scale of the system \cite{BrePet02}. In this
limit, the dynamics can be described by a semi-group master
equation in the Lindblad form \cite{Lin76}:
\begin{equation}
\frac{d\rho}{dt}=L(\rho)\equiv-i[H,\rho]+\frac{1}{2}\underset{j}{\sum}\lambda_j(2L_j\rho
L_j^{\dagger}-L_j^{\dagger}L_j\rho-\rho
L_j^{\dagger}L_j).\label{firstLindblad}
\end{equation}
Here $H$ is the system Hamiltonian and the $\{L_j\}$ are suitably
normalized Lindblad operators describing different error channels
with decoherence rates $\lambda_j$. For example, the Liouvillian
\begin{equation}
L(\rho)= \underset{j}{\sum}\lambda_j(X_j\rho X_j - \rho),
\label{Lbitflip}
\end{equation}
where $X_j$ denotes a local bit-flip operator acting on the $j$-th
qubit, describes independent Markovian bit-flip errors.

For a system undergoing Markovian decoherence and error correction
of the type \eqref{basicequation}, the evolution is given by the
equation
\begin{equation}
\frac{d\rho}{dt}=L(\rho)+\kappa\Gamma(\rho),\label{errorcorrectionequation}
\end{equation}
where $\Gamma(\rho)=\Phi(\rho)-\rho$. In \cite{PZ98}, Paz and
Zurek showed that if the set of errors $\{L_j\}$ are correctable
by the code, in the limit of infinite error-correction rate
(strong error-correcting operations applied continuously often)
the state of the system freezes and is protected from errors at
all times. The effect of freezing can be understood by noticing
that the transformation arising from decoherence during a short
time step $\Delta t$, is
\begin{equation}
\rho\rightarrow \rho + L(\rho)\Delta t +\textit{O}(\Delta t^2),
\end{equation}
i.e., the weight of correctable errors emerging during this time
interval is proportional to $\Delta t$, whereas uncorrectable errors
(e.g., multi-qubit bit flips in the case of the three-qubit bit-flip
code) are of order $\textit{O}(\Delta t^2)$. Thus, if errors are
constantly corrected, in the limit $\Delta t \rightarrow 0$
uncorrectable errors cannot accumulate, and the evolution stops.

\subsection*{5.1.3 \hspace{2pt} The Zeno effect. Error correction versus error prevention}
\addcontentsline{toc}{subsection}{5.1.3 \hspace{0.15cm} The Zeno
effect. Error correction versus error prevention}

The effect of ``freezing'' in continuous error correction strongly
resembles the quantum Zeno effect \cite{MisSud77}, in which
frequent measurements slow down the evolution of a system,
freezing the state in the limit where they are applied
continuously. The Zeno effect arises when the system and its
environment are initially decoupled and they undergo a
Hamiltonian-driven evolution, which leads to a quadratic change
with time of the state during the initial moments \cite{NNP96}
(the so called Zeno regime). Let the initial state of the system
plus the bath be $\rho_{SB}(0)=|0\rangle \langle
0|_S\otimes\rho_B(0)$. For small times, the fidelity of the
system's density matrix with the initial state
$\alpha(t)=\textrm{Tr}\left\{\left(|0\rangle\langle0|_S\otimes
I_B\right)\rho_{SB}(t)\right\}$ can be approximated as
\begin{equation}
\alpha(t)= 1-C t^2+\textit{O}(t^3).\label{Zeno}
\end{equation}
In terms of the Hamiltonian $H_{SB}$ acting on the entire system,
the coefficient $C$ is
\begin{equation}
C =
\textrm{Tr}\left\{H_{SB}^2\left(|0\rangle\langle0|_S\otimes\rho_B(0)\right)\right\}
- \textrm{Tr}\left\{H_{SB}\left(|0\rangle\langle0|_S\otimes
I_B\right) H_{SB} \left(|0\rangle\langle0|_S\otimes
\rho_B(0)\right)\right\}. \label{C}
\end{equation}
According to Eq. \eqref{Zeno}, if after a short time step $\Delta
t$ the system is measured in an orthogonal basis which includes
the initial state $|0\rangle$, the probability to find the system
in a state other than the initial state is of order
$\textit{O}(\Delta t^2)$. Thus if the state is continuously
measured ($\Delta t \rightarrow 0$), this prevents the system from
evolving.

It has been proposed to utilize the quantum Zeno effect in schemes
for error prevention \cite{Zur84, BBDEJM97, VGW96}, in which an
unknown encoded state is prevented from errors simply by frequent
measurements which keep it inside the code space. The approach is
similar to error correction in that the errors for which the code
is designed send a codeword to a space orthogonal to the code
space. The difference is that different errors need not be
distinguishable, since the procedure does not involve {\it
correction} of errors, but their prevention. In \cite{VGW96} it
was shown that with this approach it is possible to use codes of
smaller redundancy than those needed for error correction and a
four-qubit encoding of a qubit was proposed, which is capable of
preventing arbitrary independent errors arising from Hamiltonian
interactions. The possibility of this approach implicitly assumes
the existence of a Zeno regime, and fails if we assume Markovian
decoherence for all times. This is because the probability of
errors emerging during a time step $dt$ in a Markovian model is
proportional to $dt$  (rather than $dt^2$), and hence errors will
accumulate with time if not corrected.

From the above observations we see that error {\it correction} is
capable of achieving results in noise regimes where error {\it
prevention} fails. Of course, this advantage is at the expense of
a more complicated procedure---in addition to the measurements
used in error prevention, error correction involves unitary
correction operations, and in general requires codes with higher
redundancy. At the same time, we see that in the Zeno regime it is
possible to reduce decoherence using weaker resources than those
needed in the case of Markovian noise. This suggests that in this
regime error correction may exhibit higher performance than it
does for Markovian decoherence.

\subsection*{5.1.4 \hspace{2pt} Non-Markovian decoherence}
\addcontentsline{toc}{subsection}{5.1.4 \hspace{0.15cm}
Non-Markovian decoherence}

Markovian decoherence is an approximation valid for times much
larger than the memory of the environment. As we saw in the
previous chapter, however, in many situations of practical
significance the memory of the environment cannot be neglected and
the evolution is highly non-Markovian. Furthermore, no evolution
is strictly Markovian, and for a system initially decoupled from
its environment a Zeno regime is always present, short though it
may be \cite{NNP96}. If the time resolution of error-correcting
operations is high enough so that they ``see'' the Zeno regime,
this could give rise to different behavior.

The existence of a Zeno regime is not the only interesting feature
of non-Markovian decoherence. The mechanism by which errors
accumulate in a general Hamiltonian interaction with the
environment may differ significantly from the Markovian case,
since the system may develop nontrivial correlations with the
environment. For example, imagine that some time after the initial
encoding of a system, a strong error-correcting operation is
applied. This brings the state inside the code space, but the
state contains a nonzero portion of errors non-distinguishable by
the code. Thus the new state is mixed and is generally correlated
with the environment. A subsequent error-correcting operation can
only aim at correcting errors arising after this point, since the
errors already present inside the code space are in principle
uncorrectable. Subsequent errors on the density matrix, however,
may not be completely positive due to the correlations with the
environment.

Nevertheless, it follows from a result in \cite{ShaLid07} that an
error-correction procedure which is capable of correcting a
certain class of completely positive (CP) maps, can also correct
any linear noise map whose operator elements can be expressed as
linear combinations of the operator elements in a correctable CP
map. This implies, in particular, that an error-correction
procedure that can correct arbitrary single-qubit CP maps can
correct arbitrary single-qubit linear maps. In this context, we
note that the effects of system-environment correlations in
non-Markovian error models have also been studied from the
perspective of fault tolerance, and it has been shown that the
threshold theorem can be extended to various types of
non-Markovian noise \cite{TB05, AGP06, AKP06}.

Another important difference from the Markovian case is that error
correction and the effective noise on the reduced density matrix
of the system cannot be treated as independent processes. One
could derive an equation for the effective evolution of the system
alone subject to interaction with the environment, like the
Nakajima-Zwanzig \cite{Nak58, Zwa60} or the time-convolutionless
(TCL) \cite{Shibata77, ShiAri80} master equations, but the
generator of transformations at a given moment in general will
depend (implicitly or explicitly) on the entire history up to this
moment. Therefore, adding error correction can nontrivially affect
the effective error model. This means that in studying the
performance of continuous error correction one either has to
derive an equation for the effective evolution of the encoded
system, taking into account error correction from the very
beginning, or one has to look at the evolution of the entire
system---including the bath---where the error generator and the
generator of error correction can be considered independent. In
the latter case, for sufficiently small $\tau_c$, the evolution of
the entire system including the bath can be described by
\begin{equation}
\frac{d \rho}{dt}=-i[H, \rho] + \kappa \Gamma(\rho),
\label{NMerrorcorrectionequation}
\end{equation}
where $\rho$ is the density matrix of the system plus bath, $H$ is
the total Hamiltonian, and the error-correction generator $\Gamma$
acts locally on the encoded system. In this study, we take this
approach for a sufficiently simple bath model which allows us to
find a solution for the evolution of the entire system.

\subsection*{5.1.5 \hspace{2pt} Plan of this chapter}
\addcontentsline{toc}{subsection}{5.1.5 \hspace{0.15cm} Plan of
this chapter}

The rest of the chapter is organized as follows. To develop
understanding of the workings of continuous error correction, in
Section 5.2 we look at a simple example:  an error-correction code
consisting of only one qubit which aims at protecting a known
state. We discuss the difference in performance for Markovian and
non-Markovian decoherence, and argue the implications it has for
the case of multi-qubit codes. In Section 5.3, we study the
three-qubit bit-flip code. We first review the performance of
continuous error correction in the case of Markovian bit-flip
decoherence, which was first studied in \cite{PZ98}. We then
consider a non-Markovian model, where each qubit in the code is
coupled to an independent bath qubit. This model is a simple
version of the one studied in the previous chapter, and it allows
us to solve for its evolution analytically.  In the limit of large
error-correction rates, the effective evolution approaches the
evolution of a single qubit without error correction, but the
coupling strength is now decreased by a factor which scales
quadratically with the error-correction rate. This is opposed to
the case of Markovian decoherence, where the same factor scales
linearly with the rate of error-correction. In Section 5.4, we
show that the quadratic enhancement in the performance over the
case of Markovian noise can be attributed to the presence of a
Zeno regime and argue that for general stabilizer codes and
independent errors, the performance of continuous error correction
would exhibit the same qualitative characteristics. In Section
5.5, we conclude. In the Appendix (Section 5.6), we present an
implementation of the quantum-jump error correcting model that
uses weak measurements and weak unitary operations.

\section*{5.2 \hspace{2pt} The single-qubit code}
\addcontentsline{toc}{section}{5.2 \hspace{0.15cm} The
single-qubit code}

Consider the problem of protecting a qubit in state $|0\rangle$
from bit-flip errors. This problem can be regarded as a trivial
example of a stabilizer code, where the code space is spanned by
$|0\rangle$ and its stabilizer is $Z$. Let us consider the
Markovian bit-flip model first. The evolution of the state subject
to bit-flip errors and error correction is described by Eq.
\eqref{errorcorrectionequation} with
\begin{equation}
L(\rho)=\lambda( X \rho X - \rho), \label{bitflipgen}
\end{equation}
and
\begin{equation}
\Gamma(\rho)=|0\rangle \langle 0| \rho |0\rangle \langle0| +
|0\rangle \langle 1|\rho |1\rangle \langle 0| - \rho.
\label{ECgen}
\end{equation}
If the state lies on the z-axis of the Bloch sphere, it will never
leave it, since both the noise generator \eqref{bitflipgen} and
the error-correction generator \eqref{ECgen} keep it on the axis.
We will take the qubit to be initially in the desired state
$|0\rangle$, and therefore at any later moment it will have the
form $\rho (t) = \alpha(t) |0\rangle\langle
0|+(1-\alpha(t))|1\rangle\langle 1|$, $\alpha(t) \in [0,1]$. The
coefficient $\alpha(t)$ has the interpretation of a fidelity with
the trivial code space spanned by $|0\rangle$. For an
infinitesimal time step $dt$, the effect of the noise is to
decrease $\alpha(t)$ by the amount $\lambda (2\alpha(t)-1) dt$ and
that of the correcting operation is to increase it by $\kappa
(1-\alpha(t)) dt$. The net evolution is then described by
\begin{equation}
\label{equation1}
\frac{d\alpha(t)}{dt}=-(\kappa+2\lambda)\alpha(t)+(\kappa +
\lambda).
\end{equation}
The solution is
\begin{equation}
\alpha(t)=(1-\alpha_*^{\rm
M})e^{-(\kappa+2\lambda)t}+\alpha_*^{\rm M}, \label{MSQS}
\end{equation}
where
\begin{equation}
\alpha_*^{\rm M}=1-\frac{1}{2+r}, \label{attractor}
\end{equation}
and $r=\kappa/\lambda$ is the ratio between the rate of error
correction and the rate of decoherence. We see that the fidelity
decays, but it is confined above its asymptotic value
$\alpha_*^{\rm M}$, which can be made arbitrarily close to 1 for a
sufficiently large $r$.

Now let us consider a non-Markovian error model. We choose the
simple scenario where the system is coupled to a single bath qubit
via the Hamiltonian
\begin{equation}
H=\gamma X\otimes X,
\end{equation}
where $\gamma$ is the coupling strength. This is the Ising
Hamiltonian \eqref{eq:HI} for the case of a single bath qubit, but
in the basis $|+\rangle=\frac{|0\rangle+|1\rangle}{\sqrt{2}}$,
$|-\rangle=\frac{|0\rangle-|1\rangle}{\sqrt{2}}$. As we noted in
Chapter 4 (Section 4.1), the model of a single bath qubit can be a
good approximation for situations in which the coupling to a
single spin from the bath dominates over other interactions.

We will assume that the bath qubit is initially in the maximally
mixed state, which can be thought of as an equilibrium state at
high temperature. From Eq. \eqref{NMerrorcorrectionequation} one
can verify that if the system is initially in the state
$|0\rangle$, the state of the system plus the bath at any moment
will have the form
\begin{eqnarray}
\rho(t) = \left(\alpha (t) |0\rangle \langle 0| +
(1-\alpha(t))|1\rangle \langle 1|\right)\otimes \frac{I}{2} -
\beta(t)Y \otimes \frac{X}{2}.
\end{eqnarray}
In the tensor product, the first operator belongs to the Hilbert
space of the system and the second to the Hilbert space of the
bath. We have $\alpha(t) \in [0,1]$, and
$|\beta(t)|\le\sqrt{\alpha(t)(1-\alpha(t))}, \beta(t)\in R$. The
reduced density matrix of the system has the same form as the one
for the Markovian case. The traceless term proportional to
$\beta(t)$ can be thought of as a ``hidden'' part, which
nevertheless plays an important role in the error-creation
process, since errors can be thought of as being transferred to
the ``visible'' part from the ``hidden'' part (and vice versa).
This can be seen from the fact that during an infinitesimal time
step $dt$, the Hamiltonian changes the parameters $\alpha$ and
$\beta$ as follows:
\begin{gather}
\alpha\rightarrow \alpha-2\beta \gamma dt ,\notag\\
\beta \rightarrow \beta +(2\alpha-1)\gamma dt .\label{sqe}
\end{gather}
The effect of an infinitesimal error-correcting operation is
\begin{gather}
\alpha \rightarrow \alpha + (1-\alpha)\kappa dt,\notag\\
\beta\rightarrow \beta-\beta\kappa dt.
\end{gather}
Note that the hidden part is also being acted upon. Putting it all
together, we get the system of equations
\begin{gather}
\frac{d \alpha(t)}{dt}=\kappa(1-\alpha(t))-2\gamma \beta(t),\notag\\
\frac{d
\beta(t)}{dt}=\gamma(2\alpha-1)-\kappa\beta(t)\label{equation2}.
\end{gather}
The solution for the fidelity $\alpha(t)$ is
\begin{gather}
\alpha(t) = \frac{2\gamma^2 + \kappa^2}{4\gamma^2+\kappa^2} +
e^{-\kappa t}\left(\frac{\kappa\gamma}{4\gamma^2+\kappa^2}
\sin{2\gamma t} + \frac{2\gamma^2}{4\gamma^2+\kappa^2}\cos{2\gamma
t}\right). \label{singlequbitsolution}
\end{gather}
We see that as time increases, the fidelity stabilizes at the
value
\begin{equation}
\alpha_*^{\rm NM}= \frac{2+R^2}{4+R^2}=1-\frac{2}{4+R^2},
\end{equation}
where $R=\kappa/\gamma$ is the ratio between the error-correction
rate and the coupling strength. In Fig. 1 we have plotted the
fidelity as a function of the dimensionless parameter $\gamma t$
for three different values of $R$. For error-correction rates
comparable to the coupling strength ($R=1$), the fidelity
undergoes a few partial recurrences before it stabilizes close to
$\alpha_*^{\rm NM}$. For larger $R=2$, however, the oscillations
are already heavily damped and for $R=5$ the fidelity seems
confined above $\alpha_*^{\rm NM}$. As $R$ increases, the
evolution becomes closer to a decay like the one in the Markovian
case.

\begin{figure}[h]
\begin{center}
\includegraphics[width=4in]{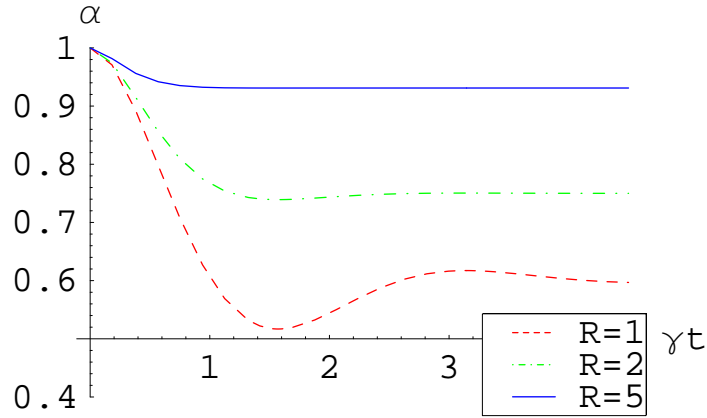}
\caption{Fidelity of the single-qubit code with continuous bit-flip
errors and error correction, as a function of dimensionless time
$\gamma t$, for three different values of the ratio
$R=\kappa/\gamma$.} \label{fig1}
\end{center}
\end{figure}

A remarkable difference, however, is that the asymptotic weight
outside the code space ($1-\alpha_*^{\rm NM}$) decreases with
$\kappa$ as $1/\kappa^2$, whereas in the Markovian case the same
quantity decreases as $1/\kappa$. The asymptotic value can be
obtained as an equilibrium point at which the infinitesimal weight
flowing out of the code space during a time step $dt$ is equal to
the weight flowing into it. The latter corresponds to vanishing
right-hand sides in Eqs. \eqref{equation1} and \eqref{equation2}.
In Section 5.4, we will show that the difference in the
equilibrium code-space fidelity for the two different types of
decoherence arises from the difference in the corresponding
evolutions during initial times.

For multi-qubit codes, error correction cannot preserve a high
fidelity with the initial codeword for all times, because there
will be multi-qubit errors that can lead to errors within the code
space itself. But it is natural to expect that the code-space
fidelity can be kept above a certain value, since the effect of
the error-correcting map \eqref{basicequation} is to oppose its
decrease. If similarly to the single-qubit code there is a
quadratic difference in the code-space fidelity for the cases of
Markovian and non-Markovian decoherence, this could lead to a
different performance of the error-correction scheme with respect
to the rate of accumulation of uncorrectable errors inside the
code space. This is because multi-qubit errors that can lead to
transformations entirely within the code space during a time step
$dt$ are of order $\textit{O}(dt^2)$. This means that if the state
is kept constantly inside the code space (as in the limit of an
infinite error-correction rate), uncorrectable errors will never
develop. But if there is a finite nonzero portion of correctable
errors, by the error mechanism it will give rise to errors not
distinguishable or misinterpreted by the code. Therefore, the
weight outside the code space can be thought of as responsible for
the accumulation of uncorrectable errors, and consequently a
difference in its magnitude may lead to a difference in the
overall performance. In the following sections we will see that
this is indeed the case.

\section*{5.3 \hspace{2pt} The three-qubit bit-flip code}
\addcontentsline{toc}{section}{5.3 \hspace{0.15cm} The three-qubit
bit-flip code}

\subsection*{5.3.1 \hspace{2pt} A Markovian error model}
\addcontentsline{toc}{subsection}{5.3.1 \hspace{0.15cm} A
Markovian error model}

Even though the three-qubit bit-flip code can correct only
bit-flip errors, it captures most of the important characteristics
of nontrivial stabilizer codes. Before we look at a non-Markovian
model, we will review the Markovian case which was studied in
\cite{PZ98}. Let the system decohere through identical independent
bit-flip channels, i.e., $L(\rho)$ is of the form \eqref{Lbitflip}
with $\lambda_1=\lambda_2=\lambda_3=\lambda$. Then one can verify
that the density matrix at any moment can be written as
\begin{equation}
\rho(t) = a(t)\rho(0)+b(t)\rho_{1}+c(t)\rho_{2}+d(t)\rho_{3},
\label{rhooft}
\end{equation}
where
\begin{gather}
\rho_{1}=\frac{1}{3}(X_1\rho(0)X_1+X_2\rho(0)X_2 +
X_3\rho(0)X_3),\notag\\
\rho_{2}= \frac{1}{3}(X_1X_2\rho(0)X_1X_2+
X_2X_3\rho(0)X_2X_3+X_1X_3\rho(0)X_1X_3),\\
\rho_{3} = X_1X_2X_3\rho(0) X_1X_2X_3,\notag
\end{gather}
are equally-weighted mixtures of single-qubit, two-qubit and
three-qubit errors on the original state.

The effect of decoherence for a single time step $dt$ is
equivalent to the following transformation of the coefficients in
Eq. \eqref{rhooft}:
\begin{equation}
\begin{split}
a\rightarrow a - 3a \lambda dt + b \lambda dt,\\
b\rightarrow b + 3a \lambda dt - 3 b \lambda dt + 2 c \lambda dt,\\
c\rightarrow c +2b \lambda dt - 3c\lambda dt+3d\lambda dt,\\
d\rightarrow d +c \lambda dt -3d \lambda dt.
\label{decohtransform}
\end{split}
\end{equation}
If the system is initially inside the code space, combining Eq.
\eqref{decohtransform} with the effect of the weak
error-correcting map $\rho\rightarrow (1-\kappa dt)\rho + \kappa
dt \Phi(\rho)$, where $\Phi(\rho)$ is given in Eq.
\eqref{strongmap}, yields the following system of first-order
linear differential equations for the evolution of the system
subject to decoherence plus error correction:
\begin{equation}
\begin{split}
\frac{da(t)}{dt} = -3\lambda a(t) + (\lambda+\kappa)b(t),\\
\frac{db(t)}{dt} = 3\lambda a(t) - (3\lambda+\kappa)b(t) + 2 \lambda c(t),\\
\frac{dc(t)}{dt} = 2\lambda b(t) - (3\lambda+\kappa)c(t) + 3 \lambda d(t),\\
\frac{dd(t)}{dt} = (\lambda+\kappa)c(t)-3\lambda d(t).
\label{equations}
\end{split}
\end{equation}
The exact solution has been found in \cite{PZ98}. Here we just
note that for the initial conditions $a(0)=1, b(0)=c(0)=d(0)=0$,
the exact solution for the weight outside the code space is
\begin{equation}
b(t)+c(t)=\frac{3}{4+r}(1-e^{-(4+r)\lambda t}),
\end{equation}
where $r=\kappa/\lambda$. We see that similarly to what we
obtained for the trivial code in the previous section, the weight
outside the code space quickly decays to its asymptotic value
$\frac{3}{4+r}$ which scales as $1/r$. But note that here the
asymptotic value is roughly three times greater than that for the
single-qubit model. This corresponds to the fact that there are
three single-qubit channels. More precisely, it can be verified
that if for a given $\kappa$ the uncorrected weight by the
single-qubit scheme is small, then the uncorrected weight by a
multi-qubit code using the same $\kappa$ and the same kind of
decoherence for each qubit scales approximately linearly with the
number of qubits. Similarly, the ratio $r$ required to preserve a
given overlap with the code space scales linearly with the number
of qubits in the code.

The most important difference from the single-qubit model is that
in this model there are uncorrectable errors that cause a decay of
the state's fidelity {\it inside} the code space. Due to the
finiteness of the resources employed by our scheme, there always
remains a nonzero portion of the state outside the code space,
which gives rise to uncorrectable three-qubit errors. To
understand how the state decays inside the code space, we ignore
the terms of the order of the weight outside the code space in the
exact solution. We obtain:
\begin{equation}
a(t)\approx \frac{1+e^{-\frac{6}{r}2\lambda t}}{2} \approx 1 -
d(t),
\end{equation}
\begin{equation}
b(t) \approx c(t) \approx 0.
\end{equation}
Comparing this solution to the expression for the fidelity of a
single decaying qubit without error correction---which can be seen
from Eq. \eqref{MSQS} for $\kappa=0$---we see that the encoded
qubit decays roughly as if subject to bit-flip decoherence with
rate $6\lambda/r$. Therefore, for large $r$ this error-correction
scheme can reduce the rate of decoherence approximately $r/6$
times. In the limit $r \rightarrow \infty$, it leads to perfect
protection of the state for all times.

\subsection*{5.3.2 \hspace{2pt} A non-Markovian error model}
\addcontentsline{toc}{subsection}{5.3.2 \hspace{0.15cm} A
non-Markovian error model}

We consider a model where each qubit independently undergoes the
same kind of non-Markovian decoherence as the one we studied for
the single-qubit code. Here the system we look at consists of six
qubits---three for the codeword and three for the environment. We
assume that all system qubits are coupled to their corresponding
environment qubits with the same coupling strength, i.e., the
Hamiltonian is
\begin{equation}
H=\gamma\overset{3}{\underset{i=1}{\sum}}X^S_i\otimes
X^B_i,\label{Hamiltonian}
\end{equation}
where the operators $X^S$ act on the system qubits and $X^B$ act
on the corresponding bath qubits. The subscripts label the
particular qubit on which they act. Obviously, the types of
effective single-qubit errors on the density matrix of the system
that can result from this Hamiltonian at any time, whether they
are CP or not, will have operator elements which are linear
combinations of $I$ and $X^S$, i.e., they are correctable by the
procedure according to \cite{ShaLid07}. Considering the forms of
the Hamiltonian \eqref{Hamiltonian} and the error-correcting map
\eqref{strongmap}, one can see that the density matrix of the
entire system at any moment is a linear combination of terms of
the following type:
\begin{equation}
\varrho_{lmn,pqr}\equiv X_1^lX_2^mX_3^n\rho(0)
X_1^pX_2^qX_3^r\otimes \frac{X_1^{l+p}}{2}\otimes
\frac{X_2^{m+q}}{2} \otimes \frac{X_3^{n+r}}{2}.
\end{equation}
Here the first term in the tensor product refers to the Hilbert
space of the system, and the following three refer to the Hilbert
spaces of the bath qubits that couple to the first, second and
third qubits from the code, respectively. The powers $l,m,n,p,q,r$
take values $0$ and $1$ in all possible combinations, and $X^1=X$,
$X^0=X^2=I$.  Note that $\varrho_{lmn,pqr}$ should not be mistaken
for the components of the density matrix in the computational
basis. Collecting these together, we can write the density matrix
in the form
\begin{eqnarray}
\rho(t)&=&\underset{l,m,n,p,q,r}{\sum}(-i)^{l+m+n}(i)^{p+q+r}C_{lmn,pqr}(t)\times
\varrho_{lmn,pqr},\label{fullDM}
\end{eqnarray}
where the coefficients $C_{lmn,pqr}(t)$ are real. The coefficient
$C_{000,000}$ is less than or equal to the codeword fidelity (with
equality when $\rho(0)=|\bar{0}\rangle\langle \bar{0}|$ or
$\rho(0)=|\bar{1}\rangle\langle \bar{1}|$). Since the scheme is
intended to protect an unknown codeword, we are interested in its
worst-case performance; we will therefore use $C_{000,000}$ as a
lower bound on the codeword fidelity.

Using the symmetry with respect to permutations of the different
system-bath pairs of qubits and the Hermiticity of the density
matrix, we can reduce the description of the evolution to a system
of equations for only $13$ of the $64$ coefficients.  (In fact,
$12$ coefficients are sufficient if we invoke the normalization
condition $\textrm{Tr}\rho=1$, but we have found it more
convenient to work with $13$.) The equations are linear, and we
write them as a single 13-dimensional vector equation:
\begin{equation}
{\tiny \setcounter{MaxMatrixCols}{13} \frac{d}{dt}\begin{bmatrix}
C_{000,000}\\
C_{100,000}\\
C_{110,000}\\
C_{100,010}\\
C_{100,100}\\
C_{110,001}\\
C_{111,000}\\
C_{110,100}\\
C_{110,110}\\
C_{110,011}\\
C_{111,100}\\
C_{111,110}\\
C_{111,111}
\end{bmatrix}=\gamma
\begin{bmatrix}
0&-6&0&0&3R&0&0&0&0&0&0&0&0\\
1&-R&-2&-2&-1&0&0&0&0&0&0&0&0\\
0&2&-R&0&0&-1&-1&-2&0&0&0&0&0\\
0&2&0&-R&0&-2&0&-2&0&0&0&0&0\\
0&2&0&0&-R&0&0&-4&0&0&0&0&0\\
0&0&1&2&0&-R&0&0&0&-2&-1&0&0\\
0&0&3&0&0&-3R&0&0&0&0&-3&0&0\\
0&0&1&1&1&0&0&-R&-1&-1&-1&0&0\\
0&0&0&0&0&0&0&4&-R&0&0&-2&0\\
0&0&0&0&0&2&0&2&0&-R&0&-2&0\\
0&0&0&0&0&1&1&2&0&0&-R&-2&0\\
0&0&0&0&0&0&0&0&1&2&2&-R&-1\\
0&0&0&0&0&0&0&0&3R&0&0&6&0
\end{bmatrix}\cdot
\begin{bmatrix}
C_{000,000}\\
C_{100,000}\\
C_{110,000}\\
C_{100,010}\\
C_{100,100}\\
C_{110,001}\\
C_{111,000}\\
C_{110,100}\\
C_{110,110}\\
C_{110,011}\\
C_{111,100}\\
C_{111,110}\\
C_{111,111}
\end{bmatrix}}
\label{NMsystem1}
\end{equation}
where $R=\kappa/\gamma$.  Each nonzero component in this matrix
represents an allowed transition process for the quantum states;
these transitions can be driven either by the decoherence process
or the continuous error-correction process.  We plot these allowed
transitions in Fig.~2.

\begin{figure}[htbp]
\includegraphics[width=5.6in]{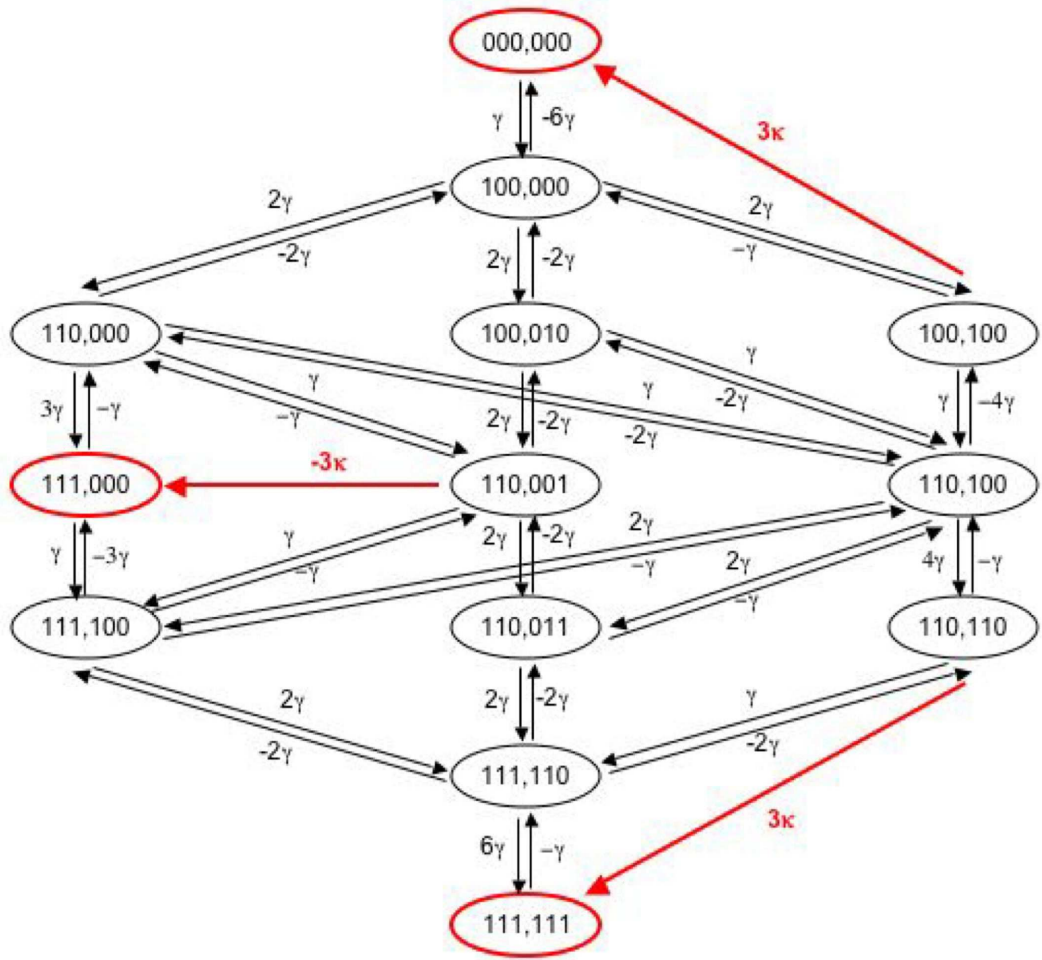}
\caption{These are the allowed transitions between the different
components of the system (\ref{NMsystem1}) and their rates,
arising from both the decoherence (bit-flip) process (with rate
$\gamma$) and the continuous error-correction process (with rate
$\kappa$).  Online, the transitions due to decoherence are black,
and the transitions due to error correction are red.} \label{fig2}
\end{figure}

We can use the symmetries of the process to recover the 64
coefficients of the full state.  Each of the 13 coefficients
represents a set of coefficients having the same number of $1$s on
the left and the same number of $1$s on the right, as well as the
same number of places which have $1$ on both sides.  All such
coefficients are equal at all times. For example, the coefficient
$C_{110,011}$ is equal to all coefficients with two $1$s on the
left, two $1$s on the right and exactly one place with $1$ on both
sides; there are exactly six such coefficients:
\[
C_{110,011} = C_{110,101} = C_{101,011} = C_{101,110} =
C_{011,110} = C_{011,101} .
\]
In determining the transfer rate from one coefficient to another
in Fig.~2, one has to take into account the number of different
coefficients of the first type which can make a transition to a
coefficient of the second type of order $dt$ according to Eq.
\eqref{NMerrorcorrectionequation}. The sign of the flow is
determined from the phases in front of the coefficients in Eq.
\eqref{fullDM}.

The eigenvalues of the matrix in Eq. \eqref{NMsystem1} up to the
first two lowest orders in $1/\kappa$ are presented in Table I.

\begin{table}[htdp]
\caption{Eigenvalues of the matrix}
\begin{center}
\begin{tabular}{|c|c|}
\hline Eigenvalues \\ \hline $\lambda_0 = 0$   \\ \hline
$\lambda_{1,2} = -\kappa$ \\ \hline $\lambda_{3,4} =  - \kappa\pm
i 2\gamma$ \\ \hline $\lambda_{5,6} =  - \kappa \pm i 4\gamma$ \\
\hline $\lambda_{7,8} = -\kappa\pm
i(\sqrt{13}+3)\gamma+\textit{O}(1/\kappa)$ \\ \hline
$\lambda_{9,10} = -\kappa\pm
i(\sqrt{13}-3)\gamma+\textit{O}(1/\kappa)$
\\ \hline $\lambda_{11,12} = \pm i(24/R^2)\gamma  - (144/R^3)
\gamma + \textit{O}(1/\kappa^4)$ \\ \hline
\end{tabular}
\end{center}
\label{eigenvalue_table}
\end{table}%
Obviously all eigenvalues except the first one and the last two
describe fast decays with rates $\sim \kappa$. They correspond to
terms in the solution which will vanish quickly after the
beginning of the evolution. The eigenvalue $0$ corresponds to the
asymptotic ($t\rightarrow \infty$) solution, since all other terms
will eventually decay. The last two eigenvalues are those that
play the main role in the evolution on a time scale
$t\gg\frac{1}{\kappa}$. We see that on such a time scale, the
solution will contain an oscillation with an angular frequency
approximately equal to $(24/R^2)\gamma$ which is damped by a decay
factor with a rate of approximately $(144/R^3)\gamma$.  In Fig.~3
we have plotted the codeword fidelity $C_{000,000}(t)$ as a
function of the dimensionless parameter $\gamma t$ for $R=100$.
The graph indeed represents this type of behavior, except for very
short times after the beginning ($\gamma t \sim 0.1$), where one
can see a fast but small in magnitude decay (Fig. 4). The maximum
magnitude of this quickly decaying term obviously decreases with
$R$, since in the limit of $R\rightarrow \infty$ the fidelity
should remain constantly equal to $1$.

\begin{figure}[h]
\begin{center}
\includegraphics[width=4in]{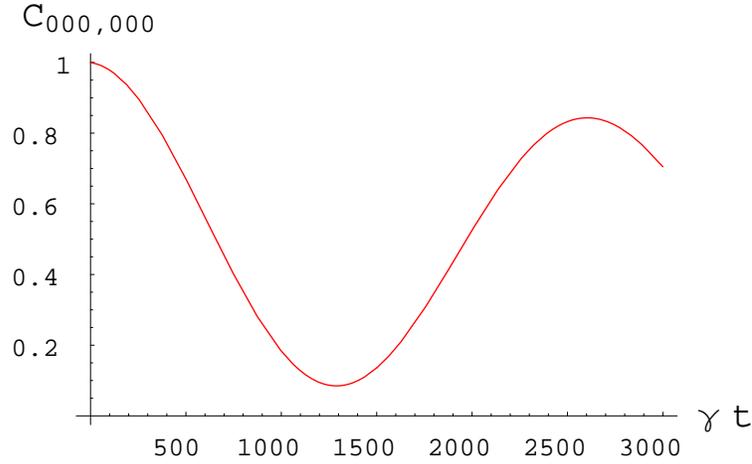}
\caption{Long-time behavior of the three-qubit system with bit-flip
noise and continuous error correction.  The ratio of correction rate
to decoherence rate is $R=\kappa/\gamma=100$.} \label{fig3}
\end{center}
\end{figure}

\begin{figure}[h]
\begin{center}
\includegraphics[width=4in]{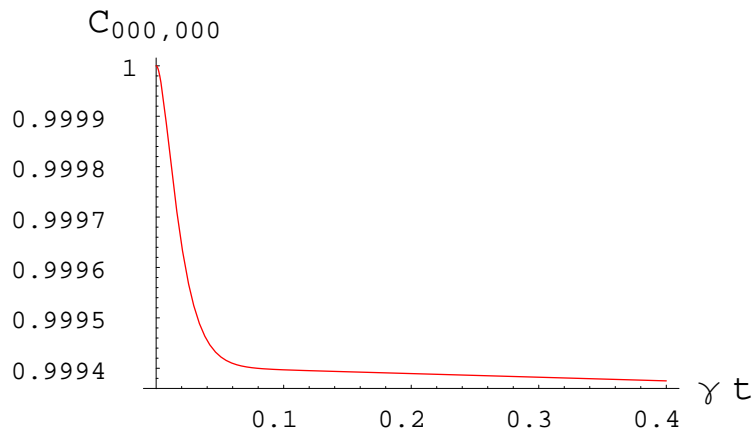}
\caption{Short-time behavior of the three-qubit system with bit-flip
noise and continuous error correction.  The ratio of correction rate
to decoherence rate is $R=\kappa/\gamma=100$.} \label{fig4}
\end{center}
\end{figure}

From the form of the eigenvalues one can see that as $R$
increases, the frequency of the main oscillation decreases as
$1/R^2$ while the rate of decay decreases faster, as $1/R^3$. Thus
in the limit $R\rightarrow \infty$, the evolution approaches an
oscillation with an angular frequency $(24/R^2)\gamma$. (We
formulate this statement more rigorously below.) This is the same
type of evolution as that of a single qubit interacting with its
environment, but the coupling constant is effectively reduced by a
factor of $R^2/12$.

While the coupling constant serves to characterize the decoherence
process in this particular case, this is not valid in general. To
handle the more general situation, we propose to use the
instantaneous rate of decrease of the codeword fidelity $F_{cw}$
as a measure of the effect of decoherence:
\begin{equation}
\Lambda(F_{cw}(t)) = -\frac{dF_{cw}(t)}{dt}. \label{errorrate}
\end{equation}
(In the present case, $F_{cw}=C_{000,000}$.) This quantity does
not coincide with the decoherence rate in the Markovian case
(which can be defined naturally from the Lindblad equation), but
it is a good estimate of the rate of loss of fidelity and can be
used for any decoherence model. From now on we will refer to it
simply as an error rate, but we note that there are other possible
definitions of instantaneous error rate suitable for non-Markovian
decoherence, which in general may depend on the kind of errors
they describe. Since the goal of error correction is to preserve
the codeword fidelity, the quantity \eqref{errorrate} is a useful
indicator for the performance of a given scheme. Note that
$\Lambda(F_{cw})$ is a function of the codeword fidelity and
therefore it makes sense to use it for a comparison between
different cases only for identical values of $F_{cw}$. For our
example, the fact that the coupling constant is effectively
reduced approximately $R^2/12$ times implies that the error rate
for a given value of $F_{cw}$ is also reduced $R^2/12$ times.
Similarly, the reduction of $\lambda$ by the factor $r/6$ in the
Markovian case implies a reduction of $\Lambda$ by the same
factor. We see that the effective reduction of the error rate
increases quadratically with $\kappa^2$ in the non-Markovian case,
whereas it increases only linearly with $\kappa$ in the Markovian
case.

Now let us rigorously derive the approximate solution to this
model of non-Markovian decoherence with continuous error
correction. Assuming that $\gamma \ll \kappa$ (or equivalently,
$R\gg1$), the superoperator driving the evolution of the system
during a time step $\delta t$ can be written as
\begin{eqnarray}
e^{\mathcal{L}\delta t}&=&e^{\mathcal{L}_{\kappa}\delta
t}+\overset{\delta t}{\underset{0}{\int}}dt'
e^{\mathcal{L}_{\kappa}(\delta
t-t')}\mathcal{L}_{\gamma}e^{\mathcal{L}_{\kappa}t'}
+\overset{\delta t}{\underset{0}{\int}}dt'\overset{\delta
t}{\underset{t'}{\int}}dt''e^{\mathcal{L}_{\kappa}(\delta
t-t'')}\mathcal{L}_{\gamma}e^{\mathcal{L}_{\kappa}(t''-t')}\mathcal{L}_{\gamma}e^{\mathcal{L}_{\kappa}t'}+\notag\\
&+&\overset{\delta t}{\underset{0}{\int}}dt'\overset{\delta
t}{\underset{t'}{\int}}dt''\overset{\delta
t}{\underset{t''}{\int}}dt'''e^{\mathcal{L}_{\kappa}(\delta
t-t''')}\mathcal{L}_{\gamma}e^{\mathcal{L}_{\kappa}(t'''-t'')}
\mathcal{L}_{\gamma}e^{\mathcal{L}_{\kappa}(t''-t')}
\mathcal{L}_{\gamma}e^{\mathcal{L}_{\kappa}t'}+...
\label{perturbation}
\end{eqnarray}
We have denoted the Liouvillian by
$\mathcal{L}=\mathcal{L}_{\gamma}+\mathcal{L}_{\kappa}$, where
$\mathcal{L}_{\kappa}\rho=\kappa\Gamma(\rho)$, and
$\mathcal{L}_{\gamma}\rho=-i[H,\rho]$.

Let $\gamma \delta t \ll 1 \ll \kappa \delta t $. We will derive
an approximate differential equation for the evolution of
$\rho(t)$ by looking at the terms of order $\delta t$ in the
change of $\rho$ according to Eq. \eqref{perturbation}. When
$\kappa=0$, we have $d\rho/dt = \mathcal{L}_{\gamma}\rho$, so the
effect of $\mathcal{L}_{\gamma}$ on the state of the system can be
seen from Eq. \eqref{NMsystem1} with $\kappa$ taken equal to $0$.
By the action of $\exp({\mathcal{L}_{\kappa} t})$, the different
terms of the density matrix transform as follows:
$\varrho_{000,000},\varrho_{111,000},\varrho_{111,111}$ remain
unchanged, $\varrho_{100,100}\rightarrow e^{-\kappa
t}\varrho_{100,100}+(1-e^{-\kappa t})\varrho_{000,000}$,
$\varrho_{110,110}\rightarrow e^{-\kappa
t}\varrho_{110,110}+(1-e^{-\kappa t})\varrho_{111,111}$,
$\varrho_{110,001}\rightarrow e^{-\kappa
t}\varrho_{110,001}-(1-e^{-\kappa t})\varrho_{111,000}$, and all
other terms are changed as $\varrho\rightarrow e^{-\kappa t}
\varrho$. Since $\kappa \delta t \gg 1$, we will ignore terms of
order $e^{-\kappa \delta t}$. But from Eq. \eqref{perturbation} it
can be seen that all terms except
$\varrho_{000,000},\varrho_{111,000},\varrho_{000,111},\varrho_{111,111}$
will get multiplied by the factor $e^{-\kappa \delta t}$ by the
action of $\exp({\mathcal{L}_{\kappa}\delta t})$ in Eq.
\eqref{perturbation}. The integrals in Eq. \eqref{perturbation}
also yield negligible factors, since every integral either gives
rise to a factor of order $\delta t$ when the integration variable
is trivially integrated, or a factor of $1/\kappa$ when the
variable participates nontrivially in the exponent. Therefore, in
the above approximation these terms of the density matrix can be
neglected, which amounts to an effective evolution entirely within
the code space. According to Eq. \eqref{NMsystem1}, the terms
$\varrho_{000,000},\varrho_{111,000},\varrho_{111,111}$ can couple
to each other only by a triple or higher application of
$\mathcal{L}_{\gamma}$. This means that if we consider the
expansion up to the lowest nontrivial order in $\gamma$, we only
need to look at the triple integral in Eq. \eqref{perturbation}.

Let us consider the effect of $\exp({\mathcal{L}\delta t})$ on
$C_{000,000}$. Any change can come directly only from
$\varrho_{111,000}$ and $\varrho_{000,111}$. The first exponent
$e^{\mathcal{L}_{\kappa}t'}$ acts on these terms as the identity.
Under the action of the first operator $\mathcal{L}_{\gamma}$ each
of these two terms can transform to six terms that can eventually
be transformed to $\varrho_{000,000}$. They are
$\varrho_{110,000}$, $\varrho_{101,000}$, $\varrho_{011,000}$,
$\varrho_{111,100}$, $\varrho_{111,010}$, $\varrho_{111,001}$, and
$\varrho_{000,110}$, $\varrho_{000,101}$, $\varrho_{000,011}$,
$\varrho_{100,111}$, $\varrho_{010,111}$, $\varrho_{001,111}$,
with appropriate factors. The action of the second exponent is to
multiply each of these new terms by $e^{-\kappa(t''-t')}$. After
the action of the second $\mathcal{L}_{\gamma}$, the action of the
third exponent on the relevant resultant terms will be again to
multiply them by a factor $e^{-\kappa(t'''-t'')}$. Thus the second
and the third exponents yield a net factor of
$e^{-\kappa(t'''-t')}$. After the second and the third
$\mathcal{L}_{\gamma}$, the relevant terms that we get are
$\varrho_{000,000}$ and $\varrho_{100,100}$, $\varrho_{010,010}$,
$\varrho_{001,001}$, each with a corresponding factor. Finally,
the last exponent acts as the identity on $\varrho_{000,000}$ and
transforms each of the terms $\varrho_{100,100}$,
$\varrho_{010,010}$, $\varrho_{001,001}$ into
$(1-e^{-\kappa(\delta t - t''')})\varrho_{000,000}$. Counting the
number of different terms that arise at each step, and taking into
account the factors that accompany them, we obtain:
\begin{eqnarray}
C_{000,000} &\rightarrow& C_{000,000}+\overset{\delta
t}{\underset{0}{\int}}dt'\overset{\delta
t}{\underset{t'}{\int}}dt''\overset{\delta
t}{\underset{t''}{\int}}dt'''
(24e^{-\kappa(t'''-t')}-36e^{-\kappa(\delta
t-t')})C_{111,000}+\cdots\notag \\
&\approx& C_{000,000}+C_{111,000}\frac{24}{R^2}\gamma \delta
t+\textit{O}(\delta t^2).
\end{eqnarray}
Using that $C_{000,000}+C_{111,111}\approx 1$, in a similar way
one obtains
\begin{equation}
C_{111,000}\rightarrow
C_{111,000}-(2C_{000,000}-1)\frac{12}{R^2}\gamma \delta
t+\textit{O}(\delta t^2).
\end{equation}
For times much larger than $\delta t$, we can write the
approximate differential equations
\begin{gather}
\frac{d C_{000,000}}{dt}=\frac{24}{R^2}\gamma C_{111,000},\notag\\
\frac{d C_{111,000}}{dt}=-\frac{12}{R^2}\gamma
(2C_{000,000}-1).\label{approxeqn}
\end{gather}
Comparing with Eq. \eqref{sqe}, we see that the encoded qubit
undergoes approximately the same type of evolution as that of a
single qubit without error correction, but the coupling constant
is effectively decreased $R^2/12$ times. The solution of Eq.
\eqref{approxeqn} yields for the codeword fidelity
\begin{equation}
C_{000,000}(t)=\frac{1+\cos (\frac{24}{R^2}\gamma t)}{2}
\label{firstapproxsoln}.
\end{equation}
This solution is valid only with precision $\textit{O}(1/R)$ for
times $\gamma t \ll R^3$. This is because we ignored terms whose
magnitudes are always of order $\textit{O}(1/R)$ and ignored
changes of order $\textit{O}(\gamma\delta t/R^3)$ per time step
$\delta t$ in the other terms. The latter changes could accumulate
with time and become of the order of unity for times $\gamma
t\approx R^3$, which is why the approximate solution is invalid
for such times. In fact, if one carries out the expansion
\eqref{perturbation} to fourth order in $\gamma$, one obtains the
approximate equations
\begin{gather}
\frac{d C_{000,000}}{dt}=\frac{24}{R^2}\gamma C_{111,000}-\frac{72}{R^3}\gamma (2 C_{000,000}-1),\notag\\
\frac{d C_{111,000}}{dt}=-\frac{12}{R^2}\gamma (2C_{000,000}-1)
-\frac{144}{R^3}\gamma C_{111,000},\label{approxeqn2}
\end{gather}
which yield for the fidelity
\begin{equation}
C_{000,000}(t)=\frac{1+e^{-144\gamma t/R^3}\cos (24\gamma
t/R^2)}{2}.
\end{equation}
We see that in addition to the effective error process which is of
the same type as that of a single qubit, there is an extra
Markovian bit-flip process with rate $72\gamma/R^3$. This
Markovian behavior is due to the Markovian character of our
error-correcting procedure which, at this level of approximation,
is responsible for the direct transfer of weight between
$\varrho_{000,000}$ and $\varrho_{111,111}$, and between
$\varrho_{111,000}$ and $\varrho_{000,111}$. The exponential
factor explicitly reveals the range of applicability of the
solution \eqref{firstapproxsoln}: with precision
$\textit{O}(1/R)$, it is valid only for times $\gamma t$ of up to
order $R^2$. For times of the order of $R^3$, the decay becomes
significant and cannot be neglected. The exponential factor may
also play an important role for short times of up to order $R$,
where its contribution is bigger than that of the cosine. But in
the latter regime the difference between the cosine and the
exponent is of order $\textit{O}(1/R^2)$, which is negligible for
the precision that we consider.

In general, the effective evolution that one obtains in the limit
of high error-correction rate does not have to approach a form
identical to that of a single decohering qubit. The reason we
obtain such behavior here is that for this particular model the
lowest order of uncorrectable errors that transform the state
within the code space is 3, and three-qubit errors have the form
of an encoded $X$ operation. Furthermore, the symmetry of the
problem ensured an identical evolution of the three qubits in the
code. For general stabilizer codes, the errors that a single qubit
can undergo are not limited to bit flips only. Therefore,
different combinations of single-qubit errors may lead to
different types of lowest-order uncorrectable errors inside the
code space, none of which in principle has to represent an encoded
version of the single-qubit operations that compose it. In
addition, if the noise is different for the different qubits,
there is no unique single-qubit error model to compare to.
Nevertheless, we will show that with regard to the effective
decrease in the error-correction rate, general stabilizer codes
will exhibit the same qualitative performance.

\section*{5.4 \hspace{2pt} Relation to the Zeno regime}
\addcontentsline{toc}{section}{5.4 \hspace{0.15cm} Relation to the
Zeno regime}

The effective continuous evolution \eqref{approxeqn} was derived
under the assumption that $\gamma \delta t \ll 1 \ll \kappa \delta
t $. The first inequality implies that $\delta t$ can be
considered within the Zeno time scale of the system's evolution
without error correction. On the other hand, from the relation
between $\kappa$ and $\tau_c$ in \eqref{tauc} we see that
$\tau_c\ll\delta t$. Therefore, the time for implementing a weak
error-correcting operation has to be sufficiently small so that on
the Zeno time scale the error-correction procedure can be
described approximately as a continuous Markovian process. This
suggests a way of understanding the quadratic enhancement in the
non-Markovian case based on the properties of the Zeno regime.

Let us consider again the single-qubit code from Section 5.2, but
this time let the error model be any Hamiltonian-driven process.
We assume that the qubit is initially in the state $|0\rangle$,
i.e., the state of the system including the bath has the form
$\rho(0)=|0\rangle \langle 0|\otimes\rho_B(0)$. For times smaller
than the Zeno time $\delta t_Z$, the evolution of the fidelity
without error correction can be described by Eq. \eqref{Zeno}.
Equation \eqref{Zeno} naturally defines the Zeno regime in terms
of $\alpha$ itself:
\begin{equation}
\alpha\geq \alpha_Z \equiv 1-C\delta t_Z^2.
\end{equation}
For a single time step $\Delta t \ll \delta t_Z$, the change in
the fidelity is
\begin{equation}
\alpha\rightarrow \alpha-2\sqrt{C}\sqrt{1-\alpha}\Delta
t+\textit{O}(\Delta t^2).\label{singlestepdecoh}
\end{equation}
On the other hand, the effect of error correction during a time
step $\Delta t$ is
\begin{equation}
\alpha \rightarrow \alpha+\kappa (1-\alpha)\Delta t
+\textit{O}(\Delta t^2),\label{singlestepcorr}
\end{equation}
i.e., it tends to oppose the effect of decoherence.  If both
processes happen simultaneously, the effect of decoherence will
still be of the form \eqref{singlestepdecoh}, but the coefficient
$C$ may vary with time. This is because the presence of
error-correction opposes the decrease of the fidelity and
consequently can lead to an increase in the time for which the
fidelity remains within the Zeno range. If this time is
sufficiently long, the state of the environment could change
significantly under the action of the Hamiltonian, thus giving
rise to a different value for $C$ in Eq. \eqref{singlestepdecoh}
according to Eq. \eqref{C}.

Note that the strength of the Hamiltonian puts a limit on $C$, and
therefore this constant can vary only within a certain range. The
equilibrium fidelity $\alpha_*^{\rm NM}$ that we obtained for the
error model in Section 5.2, can be thought of as the point at
which the effects of error and error correction cancel out. For a
general model, where the coefficient $C$ may vary with time, this
leads to a quasi-stationary equilibrium. From Eqs.
\eqref{singlestepdecoh} and \eqref{singlestepcorr}, one obtains
the equilibrium fidelity
\begin{equation}
\alpha_*^{\rm NM}\approx 1-\frac{4C}{\kappa^2}.
\end{equation}
In agreement with what we obtained in Section 5.2, the equilibrium
fidelity differs from $1$ by a quantity proportional to
$1/\kappa^2$. This quantity is generally quasi-stationary and can
vary within a limited range. If one assumes a Markovian error
model, for short times the fidelity changes linearly with time
which leads to $1-\alpha_*^{\rm M}\propto 1/\kappa$. Thus the
difference can be attributed to the existence of a Zeno regime in
the non-Markovian case.

But what happens in the case of non-trivial codes? As we saw,
there the state decays inside the code space and therefore can be
highly correlated with the environment. Can we talk about a Zeno
regime then? It turns out that the answer is positive. Assuming
that each qubit undergoes an independent error process, then up to
first order in $\Delta t$ the Hamiltonian cannot map terms in the
code space to other terms without detectable errors. (This
includes both terms in the code space and terms from the hidden
part, like $\varrho_{111,000}$ in the example of the bit-flip
code.) It can only transform terms from the code space into
traceless terms from the hidden part which correspond to
single-qubit errors (like $\varrho_{100,000}$ in the same
example). Let $|\bar{0}\rangle$, $|\bar{1}\rangle$ be the two
logical codewords and $|\psi_i \rangle$ be an orthonormal basis
that spans the space of all single-qubit errors. Then in the basis
$|\bar{0}\rangle$, $|\bar{1}\rangle$, $|\psi_i \rangle$, all the
terms that can be coupled directly to terms inside the code space
are $|\bar{0}\rangle \langle \psi_i|$, $|\psi_i\rangle \langle
\bar{0}|$, $|\bar{1}\rangle \langle \psi_i|$, $|\psi_i\rangle
\langle \bar{1}|$. From the condition of positivity of the density
matrix, one can show that the coefficients in front of these terms
are at most $\sqrt{\alpha(1-\alpha)}$ in magnitude, where $\alpha$
is the code-space fidelity. This implies that for small enough
$1-\alpha$, the change in the code-space fidelity is of the type
\eqref{singlestepdecoh}, which is Zeno-like behavior. Then using
only the properties of the Zeno behavior as we did above, we can
conclude that the weight outside the code space will be kept at a
quasi-stationary value of order $1/\kappa^2$. Since uncorrectable
errors enter the code space through the action of the
error-correction procedure, which misinterprets some multi-qubit
errors in the error space, the effective error rate will be
limited by a factor proportional to the weight in the error space.
That is, this will lead to an effective decrease of the error rate
at least by a factor proportional to $1/\kappa^2$.

The accumulation of uncorrectable errors in the Markovian case is
similar, except that in this case there is a direct transfer of
errors between the code space and the visible part of the error
space. In both cases, the error rate is effectively reduced by a
factor which is roughly proportional to the inverse of the weight
in the error space, and therefore the difference in the
performance comes from the difference in this weight. The
quasi-stationary equilibrium value of the code-space fidelity
establishes a quasi-stationary flow between the code space and the
error space. One can think that this flow effectively takes
non-erroneous weight from the code space, transports it through
the error space where it accumulates uncorrectable errors, and
brings it back into the code space. Thus by minimizing the weight
outside the code space, error correction creates a ``bottleneck''
which reduces the rate at which uncorrectable errors accumulate.

Finally, a brief remark about the resources needed for quadratic
reduction of the error rate. As pointed out above, two conditions
are involved:  one concerns the rate of error correction; the
other concerns the time resolution of the weak error-correcting
operations. Both of these quantities must be sufficiently large.
There is, however, an interplay between the two, which involves
the strength of the interaction required to implement the weak
error-correcting map \eqref{wm}. Let us imagine that the weak map
is implemented by making the system interact weakly with an
ancilla in a given state, after which the ancilla is discarded.
The error-correction procedure consists of a sequence of such
interactions, and can be thought of as a cooling process which
takes away the entropy accumulated in the system as a result of
correctable errors. If the time for which a single ancilla
interacts with the system is $\tau_c$, one can verify that the
parameter $\epsilon$ in Eq. \eqref{wm} would be proportional to
$g^2\tau_c^2$, where $g$ is the coupling strength between the
system and the ancilla. From Eq. \eqref{tauc} we then obtain that
\begin{equation}
\kappa \propto g^2\tau_c.
\end{equation}
The two parameters that can be controlled are the interaction time
and the interaction strength, and they determine the
error-correction rate. Thus if $g$ is kept constant, a decrease in
the interaction time $\tau_c$ leads to a proportional decrease in
$\kappa$, which may be undesirable. In order to achieve a good
working regime, one may need to adjust both $\tau_c$ and $g$. But
it has to be pointed out that in some situations decreasing
$\tau_c$ alone can prove advantageous, if it leads to a time
resolution revealing the non-Markovian character of an error model
which was previously described as Markovian. The quadratic
enhancement of the performance as a function of $\kappa$ may
compensate the decrease in $\kappa$, thus leading to a seemingly
paradoxical result:  better performance with a lower
error-correction rate.

\section*{5.5 \hspace{2pt} Summary and outlook}
\addcontentsline{toc}{section}{5.5 \hspace{0.15cm} Summary and
outlook}

In this chapter we studied the performance of a particular
continuous quantum error-correction scheme for non-Markovian
errors. We analyzed the evolution of the single-qubit code and the
three-qubit bit-flip code in the presence of continuous error
correction for a simple non-Markovian bit-flip error model. This
enabled us to understand the workings of the error-correction
scheme, and the mechanism whereby uncorrectable errors accumulate.
The fidelity of the state with the code space in both examples
quickly reaches an equilibrium value, which can be made
arbitrarily close to $1$ by a sufficiently high rate of error
correction. The weight of the density matrix outside the code
space scales as $1/\kappa$ in the Markovian case, while it scales
as $1/\kappa^2$ in the non-Markovian case. Correspondingly, the
rate at which uncorrectable errors accumulate in the three-qubit
code is proportional to $1/\kappa$ in the Markovian case, and to
$1/\kappa^2$ in the non-Markovian case. These differences have the
same cause, since the equilibrium weight in the error space is
closely related to the rate of uncorrectable error accumulation.

The quadratic difference in the error weight between the Markovian
and non-Markovian cases can be attributed to the existence of a
Zeno regime in the non-Markovian case. Regardless of the
correlations between the density matrix inside the code space and
the environment, if the lowest-order errors are correctable by the
code, there exists a Zeno regime in the evolution of the
code-space fidelity. The effective reduction of the error rate
with the rate of error correction for non-Markovian error models
depends crucially on the assumption that the time resolution of
the continuous error correction is much shorter than the Zeno time
scale of the evolution {\it without} error correction. This
suggests that decreasing the time for a single (infinitesimal)
error-correcting operation can lead to an increase in the
performance of the scheme, even if the average error-correction
rate goes down.

While here we have only considered codes for the correction of
single-qubit errors, our results can be extended to other types of
codes and errors as well. As long as the error process only
produces errors correctable by the code to lowest order, an
argument analogous to the one given here shows that a Zeno regime
will exist, which leads to an enhancement in the error-correction
performance. Unfortunately, it is very difficult to describe the
evolution of a system with a continuous correction protocol, based
on a general error-correction code and subject to general
non-Markovian interactions with the environment. This is
especially true if one must include the evolution of a complicated
environment in the description, as would be necessary in general.
A more practical step in this direction might be to find an
effective description for the evolution of the reduced density
matrix of the system subject to decoherence plus error correction,
using projection techniques like the Nakajima-Zwanzig or the TCL
master equations. Since one is usually interested in the evolution
during initial times before the codeword fidelity decreases
significantly, a perturbation approach could be useful.  This is a
subject for further research.

\section*{5.6 \hspace{2pt} Appendix: Implementation of the quantum-jump error-correcting process via weak measurements and weak unitary operations}
\addcontentsline{toc}{section}{5.6 \hspace{0.15cm} Appendix:
Implementation of the quantum-jump error-correcting process via
weak measurements and weak unitary operations}

Here we show how the weak CPTP map \eqref{wm} for the bit-flip
code can be implemented using weak measurements and weak unitary
operations. We also present a similar scheme for codes that
correct arbitrary-single qubit errors, which yields a weak map
different from \eqref{wm} but one that also results in the strong
error-correcting map $\Phi(\rho)$ when exponentiated. To introduce
our construction, we start again with the single-qubit code with
stabilizer $\langle Z \rangle$.

\subsection*{5.6.1 \hspace{2pt} The single-qubit model}
\addcontentsline{toc}{subsection}{5.6.1 \hspace{0.15cm} The
single-qubit model}

Consider the completely positive map corresponding to the strong
error-correcting operation for the single-qubit code:
\begin{equation}
\Phi(\rho)= X |1\rangle \langle 1| \rho |1\rangle \langle 1| X +
|0\rangle \langle 0| \rho |0\rangle \langle 0| =|0\rangle \langle
1| \rho |1\rangle \langle 0|+|0\rangle \langle 0| \rho |0\rangle
\langle 0|.\label{singlequbitstrongmap}
\end{equation}
Observe that this transformation can also be written as
\begin{equation}
\Phi(\rho)= |0\rangle \langle +| \rho |+\rangle \langle
0|+|0\rangle \langle -| \rho |-\rangle \langle 0|= ZR |+\rangle
\langle +| \rho |+\rangle \langle +|RZ + XR|-\rangle \langle -|
\rho |-\rangle \langle -|RX,
\end{equation}
where $|\pm\rangle = (|0\rangle\pm |1\rangle)/\sqrt{2}$ and
\begin{equation}
R=\frac{1}{\sqrt{2}}
\begin{pmatrix}
1&1\\
1&-1\\
\end{pmatrix}\label{Hadamard}
\end{equation}
is the Hadamard gate. Therefore the same error-correcting
operation can be implemented as a measurement in the $|\pm\rangle$
basis (measurement of the operator $X$), followed by a unitary
conditioned on the outcome: if the outcome is '+', we apply $ZR$;
if the outcome is '-', we apply $XR$. This choice of unitaries is
not unique---for example, we could apply just $R$ instead of $ZR$
after outcome '+'. But this particular choice has a convenient
geometric interpretation---the unitary $ZR$ corresponds to a
rotation around the Y-axis by an angle $\pi/2$: $ZR =
e^{i\frac{\pi}{2}\frac{Y}{2}}$, and $XR$ corresponds to a rotation
around the same axis by an angle $-\pi/2$: $ZR =
e^{-i\frac{\pi}{2}\frac{Y}{2}}$.

A weak version of the above error-correcting operation can be
constructed by taking the corresponding weak measurement of the
operator $X$, followed by a weak rotation around the Y-axis, whose
direction is conditioned on the outcome:
\begin{equation}
\begin{split}
 \rho \rightarrow \frac{I+i\epsilon'Y}{\sqrt
{1+{\epsilon'}^2}}\sqrt{\frac{I+\epsilon
X}{2}}\rho\sqrt{\frac{I+\epsilon
X}{2}}\frac{I-i\epsilon'Y}{\sqrt {1+{\epsilon'}^2}}+ \\
 +\frac{I-i\epsilon'Y}{\sqrt
{1+{\epsilon'}^2}}\sqrt{\frac{I-\epsilon
X}{2}}\rho\sqrt{\frac{I-\epsilon X}{2}}\frac{I+i\epsilon'Y}{\sqrt
{1+{\epsilon'}^2}}.\label{singlequbitweakmap}
\end{split}
\end{equation}

Here $\epsilon$ and $\epsilon'$ are small parameters. From the
symmetry of this map it can be seen that if the map is applied to a
state which lies on the Z-axis, the resultant state will still lie
on the Z-axis. Whether the state will move towards $|0\rangle\langle
0|$ or towards $|1\rangle\langle 1|$, depends on the relation
between $\epsilon$ and $\epsilon'$. Since our goal is to protect the
state from drifting away from $|0\rangle\langle 0|$ due to bit-flip
decoherence, we will assume that the state lies on the Z-axis in the
northern hemisphere (although the transformation we will obtain
works for any kind of decoherence where the state need nor remain on
the Z-axis). We would like, if possible, to choose the relation
between the parameters $\epsilon$ and $\epsilon'$ in such a way that
the effect of this map on any state on the Z-axis to be to move the
state towards $|0\rangle\langle 0|$.

In order to calculate the effect of this map on a given state, it
is convenient to write the state in the $|\pm\rangle$ basis. For a
state on the Z-axis, $\rho = \alpha |0\rangle\langle 0|+(1-\alpha)
|1\rangle\langle 1|$, we have
\begin{equation}
\rho = \frac{1}{2}|+\rangle\langle +|+ \frac{1}{2}|-\rangle\langle
-| + (2\alpha -1)\left(\frac{1}{2}|+\rangle\langle -|+
\frac{1}{2}|-\rangle\langle +|\right).\label{rhopm}
\end{equation}
For the action of our map on the state \eqref{rhopm} we obtain:
\begin{equation}
\rho \rightarrow \frac{1}{2}|+\rangle\langle +|+
\frac{1}{2}|-\rangle\langle -| +
\frac{(1-{\epsilon'}^2)\sqrt{1-\epsilon^2}(2\alpha -1) +
2\epsilon\epsilon'}{1+{\epsilon'}^2}\left(\frac{1}{2}|+\rangle\langle
-|+ \frac{1}{2}|-\rangle\langle +|\right).\label{transf}
\end{equation}
Thus we can think that upon this transformation the parameter
$\alpha$ transforms to $\alpha'$, where
\begin{equation}
2\alpha'-1 =\frac{(1-{\epsilon'}^2)\sqrt{1-\epsilon^2}(2\alpha -1)
+ 2\epsilon\epsilon'}{1+{\epsilon'}^2}.\label{alpha'}
\end{equation}
If it is possible to choose the relation between $\epsilon$ and
$\epsilon'$ in such a way that $\alpha'\geq \alpha$ for every
$0\leq \alpha \leq 1$, then clearly the state must remain
invariant when $\alpha = 1$. Imposing this requirement, we obtain
\begin{equation}
\epsilon = \frac{2\epsilon'}{1+{\epsilon'}^2},
\end{equation}
or equivalently
\begin{equation}
\epsilon'=\frac{1-\sqrt{1-\epsilon^2}}{\epsilon}.\label{varepsilon'}
\end{equation}
Substituting back in \eqref{alpha'}, we can express
\begin{equation}
\alpha'-\alpha =
\frac{4{\epsilon'}^2}{(1+{\epsilon'}^2)^2}(1-\alpha)\geq 0.
\end{equation}
We see that the coefficient $\alpha$ (which is the fidelity of our
state with $|0\rangle\langle 0|$) indeed increases after every
application of our weak completely positive map (Fig.1). The
amount by which it increases for fixed $\epsilon'$ depends on
$\alpha$ and becomes smaller as $\alpha$ approaches 1.

Since we will be taking the limit $\epsilon\rightarrow 0$, we can
write Eq.~\eqref{varepsilon'} as
\begin{equation}
\epsilon'=\frac{\epsilon}{2}+\textit{O}(\epsilon^3).
\end{equation}
If we define the relation between the time step $\tau_c$ and
$\epsilon$ as in Eq.~\eqref{tauc}, for the effect of the CPTP map
\eqref{singlequbitweakmap} on an arbitrary state of the form $\rho
= \alpha|0\rangle \langle 0 | +\beta |0\rangle \langle 1| +
\beta^* |1\rangle \langle 0| + (1-\alpha)|1\rangle \langle 1|$,
$\alpha\in R$, $\beta\in C$, we obtain
\begin{gather}
\alpha \rightarrow \alpha + (1-\alpha) \kappa \tau_c,\\
\beta \rightarrow \sqrt{1-\kappa \tau_c} \beta = \beta -
\frac{1}{2}\kappa \beta \tau_c + O({\tau_c}^2).
\end{gather}
This is exactly the map \eqref{wm} for $\Phi(\rho)$ given by
Eq.~\eqref{singlequbitstrongmap}.

\subsection*{5.6.2 \hspace{2pt} The bit-flip code}
\addcontentsline{toc}{subsection}{5.6.2 \hspace{0.15cm} The
bit-flip code}

While in the toy model from the previous section we had to protect
a given state from errors, here we have to protect the whole
subspace spanned by $|\overline{0}\rangle$ and
$|\overline{1}\rangle$. This makes a geometric visualization of
the problem significantly more difficult than in the previous
case, which is why we will take a different approach.

In the single-qubit model we saw how to protect a qubit in state
$|0\rangle$ from bit-flip errors. Similarly we could protect a
qubit in state $|1\rangle$; the only difference is that the weak
unitaries following the two outcomes of the weak measurement of
$X$ have to be exchanged. For the three-qubit bit-flip code, every
block of the code lies in the subspace spanned by the codewords
$|000\rangle$ and $|111\rangle$, i.e., each qubit is in state
$|0\rangle$ when the other two qubits are in state $|00\rangle$,
or in state $|1\rangle$ when the other qubits are in state
$|11\rangle$. This correlation is what makes it possible for the
code to correct single-qubit bit-flip errors without ever
acquiring information about the actual state of the system. We
propose to utilize this correlation in a three-qubit scheme which
protects each qubit by applying to it the corresponding
single-qubit scheme for either $|0\rangle$ or $|1\rangle$
depending on the value of the other two qubits. This, of course,
has to be done without acquiring information about the encoded
state.

Just as in the single-qubit case, the scheme consist of weak
measurements followed by weak unitaries conditioned on the
outcomes of the measurements. For error correction on the first
qubit, we propose the weak measurement with measurement operators
\begin{equation}
M^1_{\pm} = \sqrt{\frac{I{\pm}\epsilon X}{2}}\otimes (|00\rangle
\langle 00| + |11\rangle \langle 11| )+\frac{I}{\sqrt{2}}
\otimes(|01\rangle \langle 01| +|10\rangle \langle 10|),
\end{equation}
where $\sqrt{\frac{I{\pm}\epsilon X}{2}}$ are the same weak
measurement operators that we used in \eqref{singlequbitweakmap},
acting on the first qubit. This measurement can be thought of as a
weak measurement of the operator $X\otimes(|00\rangle\langle
00|+|11\rangle\langle 11|)$. In order to understand its effect
better, consider the expansion of the density matrix of our system
in the computational basis of the three qubits in a given block of
the code. Assuming that the state begins inside the code space and
that the system decoheres through single-qubit bit-flip channels,
the density matrix at any time can be written as a linear
combination of the following terms: $|000\rangle \langle
000|$,$|000\rangle \langle 111|$, $|111\rangle \langle 000|$,
$|111\rangle \langle 111|$, $|100\rangle \langle 100|$,
$|100\rangle \langle 011|$, $|011\rangle \langle 100|$,
$|011\rangle \langle 011|$, $|010\rangle \langle 010|$,
$|010\rangle \langle 101|$, $|101\rangle \langle 010|$,
$|101\rangle \langle 101|$, $|001\rangle \langle 001|$,
$|001\rangle \langle 110|$,\\ $|110\rangle \langle 001|$,
$|110\rangle \langle 110|$.  For those terms in the expansion for
which the second and third qubits are in the subspace spanned by
$|00\rangle$ and $|11\rangle$, the effect of this measurement will
be the same as the effect of a weak single-qubit measurement of
$X$ on the first qubit. Those terms in which the second and third
qubits are in the subspace spanned by $|01\rangle$ and
$|10\rangle$ will not be affected by the measurement. This is
because the three-qubit bit-flip code cannot distinguish
multi-qubit errors from single-qubit errors; the subspaces
corresponding to two- and three-qubit errors are the same as the
subspaces corresponding to single-qubit or no errors. This is why,
if the second and third qubits have different values, the
error-correction scheme will assume that an error has occurred on
one of these two qubits and will not apply any correction on the
first qubit.

The unitary operation conditioned on the outcome of the
measurement is
\begin{equation}
U^1_{\pm} =\frac{I\pm i\epsilon'Y}{\sqrt
{1+{\epsilon'}^2}}\otimes|00\rangle \langle 00| +\frac{I\mp
i\epsilon'Y}{\sqrt {1+{\epsilon'}^2}}\otimes|11\rangle \langle
11|+I\otimes (|01\rangle\langle 01| +|10\rangle\langle 10|).
\end{equation}
This is a weak unitary driven by the Hamiltonian $\pm
Y\otimes(|00\rangle\langle00|-|11\rangle\langle 11|)$. Again, it
is designed in such a way that those components of the density
matrix which correspond to an error on the second or third qubits
will undergo no transformation, while the terms for which the
second and third qubits have the same value (these are the same
terms that have undergone non-trivial transformation during the
measurement) will undergo a rotation of the first qubit analogous
to that from the single-qubit model. One can verify that the only
terms that undergo non-trivial transformation after the completely
positive map $\rho \rightarrow U^1_+M^1_+ \rho
M^1_+{U^1_+}^{\dagger} +U^1_-M^1_- \rho M^1_-{U^1_-}^{\dagger}$
are:
\begin{equation}
\begin{split}
|100\rangle\langle 100| \rightarrow (1-\kappa \tau_c)
|100\rangle\langle 100| + \kappa \tau_c |000\rangle\langle 000|,\\
|100\rangle\langle 011| \rightarrow (1-\kappa \tau_c)
|100\rangle\langle 011| + \kappa \tau_c |000\rangle\langle 111|,\\
|011\rangle\langle 100| \rightarrow (1-\kappa \tau_c)
|011\rangle\langle 100| + \kappa \tau_c |111\rangle\langle 000|,\\
|011\rangle\langle 011| \rightarrow (1-\kappa \tau_c)
|011\rangle\langle 011| + \kappa \tau_c |111\rangle\langle 111|.\\
|100\rangle\langle \overline{\phi}| \rightarrow
(1-\frac{1}{2}\kappa \tau_c) |100\rangle\langle \overline{\phi}|
,\hspace{0.2cm}|\overline{\phi}\rangle\langle 100| \rightarrow
(1-\frac{1}{2}\kappa \tau_c)
|\overline{\phi}\rangle\langle 100|\\
|011\rangle\langle \overline{\phi}| \rightarrow
(1-\frac{1}{2}\kappa \tau_c) |011\rangle\langle
\overline{\phi}|,\hspace{0.2cm} |\overline{\phi}\rangle\langle
011| \rightarrow (1-\frac{1}{2}\kappa \tau_c)
|\overline{\phi}\rangle\langle 011|
\end{split}
\end{equation}
where $|\overline{\phi}\rangle$ is any state orthogonal to the
subspace spanned by $|100\rangle$ and $|011\rangle$. We see that
the effect of this operation on the terms that correspond to bit
flip on the first qubit is to correct these terms by the same
amount as in the single-qubit error-correction scheme. All other
terms remain unchanged. If we write the state of the system at a
given moment as
\begin{gather}
\rho = a\rho(0) + b_1X_1\rho(0)X_1 + b_2X_2\rho(0)X_2
+ b_3X_3\rho(0)X_3+\\
 +c_1X_2X_3\rho(0)X_2X_3
+c_2X_1X_3\rho(0)X_1X_3+c_3X_1X_2\rho(0)X_1X_2 + d
X_1X_2X_3\rho(0)X_1X_2X_3, \nonumber
\end{gather}
where $\rho(0)$ is the initial state, then the effect of the above
completely positive map is:
\begin{equation}
\begin{split}
a \rightarrow  a + b_1 4\kappa \tau_c,\hspace{0.2cm} b_1
\rightarrow  b_1 - b_1 4\kappa \tau_c,\hspace{0.2cm} b_2
\rightarrow b_2,\hspace{0.2cm}
b_3 \rightarrow b_3,\\
c_1 \rightarrow c_1 - c_1 4\kappa \tau_c,\hspace{0.2cm} c_2
\rightarrow c_2,\hspace{0.2cm} c_3 \rightarrow c_3,\hspace{0.2cm}
d \rightarrow d + c_1 4\kappa \tau_c.
\end{split}
\end{equation}

We apply the same correction ($\rho \rightarrow U^i_+M^i_+ \rho
M^i_+{U^i_+}^{\dagger} +U^i_-M^i_- \rho M^i_-{U^i_-}^{\dagger}$)
to each of the other two qubits ($i=2,3$) as well. One can easily
see that the effect of all three corrections (up to first order in
$\Delta t$) is equivalent to the map \eqref{wm} with $\Phi(\rho)$
given in Eq.~\eqref{strongmap}.

\subsection*{5.6.3 \hspace{2pt} General single-error-correcting stabilizer codes}
\addcontentsline{toc}{subsection}{5.6.3 \hspace{0.15cm} General
single-error-correcting stabilizer codes}

We now proceed to generalizing this scheme to error-correcting
codes that correct arbitrary single-qubit errors. A stabilizer
code which is able to correct arbitrary single-qubit errors, has
the property that a single-qubit $X$, $Y$ or $Z$ error on a state
inside the code space, sends that state to a subspace orthogonal
to the code space \cite{Got97}. One can verify that this implies
that any two orthogonal codewords can be written as
\begin{gather}
|\overline{0}\rangle = \frac{1}{\sqrt{2}}|0\rangle
|\psi^0_0\rangle +\frac{1}{\sqrt{2}}|1\rangle
|\psi^0_1\rangle,\nonumber\\
|\overline{1}\rangle = \frac{1}{\sqrt{2}}|0\rangle
|\psi^1_0\rangle +\frac{1}{\sqrt{2}}|1\rangle
|\psi^1_1\rangle,\label{codeform}
\end{gather}
where $|\psi^i_j\rangle$, $i,j=0,1$ form an orthonormal set. Here we
have expanded the codewords in the computational basis (the
eigenbasis of $Z$) of the first qubit, but the same can be done with
respect to any qubit in the code. Note that an $X$, $Y$, or $Z$
error on one of the other qubits sends each of the vectors
$|\psi^i_j\rangle$ to a subspace orthogonal to the subspace spanned
by $|\psi^i_j\rangle$, $i,j=0,1$. This can be shown to follow from
the fact that different single-qubit errors send the code space to
different orthogonal subspaces. An exception is the case of
degenerate codes where the error in question has the same effect on
a codeword as an error on the first qubit. In such a case, however,
we can assume that the error has occurred on the first qubit. The
weak operation for correcting bit flips on a given qubit (say the
first one) is therefore constructed similarly to that for the
bit-flip code. We first apply the weak measurement
\begin{gather}
M^1_{\pm} = \sqrt{\frac{I{\pm}\epsilon X}{2}}\otimes
(|\psi^0_0\rangle \langle \psi^0_0| + |\psi^0_1\rangle \langle
\psi^0_1| +|\psi^1_0\rangle \langle \psi^1_0| +|\psi^1_1\rangle
\langle \psi^1_1|)+\nonumber\\
+\frac{I}{\sqrt{2}} \otimes(I^{n-1}-|\psi^0_0\rangle \langle
\psi^0_0| - |\psi^0_1\rangle \langle \psi^0_1| -|\psi^1_0\rangle
\langle \psi^1_0| -|\psi^1_1\rangle \langle \psi^1_1|),
\label{weakmeas}
\end{gather}
where $I^{n-1}$ is the identity on the space of all qubits in the
code except the first one. This can be thought of as a weak
measurement of the operator $X(|\psi^0_0\rangle \langle \psi^0_0|
+ |\psi^0_1\rangle \langle \psi^0_1| +|\psi^1_0\rangle \langle
\psi^1_0| +|\psi^1_1\rangle \langle \psi^1_1|)$. The measurement
is followed by the unitary
\begin{gather}
U^1_{\pm} =\frac{I\pm i\epsilon'Y}{\sqrt
{1+{\epsilon'}^2}}\otimes(|\psi^0_0\rangle \langle
\psi^0_0|+|\psi^1_0\rangle \langle \psi^1_0|) +\frac{I\mp
i\epsilon'Y}{\sqrt {1+{\epsilon'}^2}}\otimes (|\psi^0_1\rangle
\langle \psi^0_1|+|\psi^1_1\rangle \langle
\psi^1_1|)+\nonumber\\
+I\otimes (I^{n-1}-|\psi^0_0\rangle \langle \psi^0_0| -
|\psi^0_1\rangle \langle \psi^0_1| -|\psi^1_0\rangle \langle
\psi^1_0| -|\psi^1_1\rangle \langle \psi^1_1|) \label{weakunit}
\end{gather}
conditioned on the outcome. The Hamiltonian driving this unitary
is $\pm Y(|\psi^0_0\rangle \langle \psi^0_0|+|\psi^1_0\rangle
\langle \psi^1_0| - |\psi^0_1\rangle \langle
\psi^0_1|-|\psi^1_1\rangle \langle \psi^1_1|)$. It is easy to
verify that the effect of the corresponding completely positive
map is analogous to that for the bit-flip code. The action of each
of the operators $U^1_+M^1_+$ and $U^1_-M^1_-$ can be summarized
as follows:
\begin{gather}
U^1_{\pm}M^1_{\pm}|i\rangle |\phi\rangle =\frac{1}{\sqrt{2}}
|i\rangle |\phi\rangle, \hspace{0.3 cm}\textrm{for}\hspace{0.2 cm}
|\phi\rangle \in
I^{n-1}-\underset{j,k}{\sum}|\psi^j_k\rangle \langle \psi^j_k|,\label{actionofUM1}\\
U^1_{\pm}M^1_{\pm} |j\rangle |\psi^i_k\rangle =
\sqrt{\frac{1-\kappa \tau_c}{2}}|j\rangle |\psi^i_k\rangle \pm
\sqrt{\frac{\kappa \tau_c}{2}}|k\rangle
|\psi^i_k\rangle,\hspace{0.3 cm}\textrm{for}\hspace{0.2 cm} j\neq
k,\\
U^1_{\pm}M^1_{\pm} |j\rangle |\psi^i_k\rangle = |j\rangle
|\psi^i_k\rangle,\hspace{0.3 cm}\textrm{for}\hspace{0.2 cm} j =
k.\label{actionofUM3}
\end{gather}

This implies that the effect of the map $\sigma \rightarrow
U^1_+M^1_+\sigma M^1_+ {U^1_+}^{\dagger} + U^1_-M^1_-\sigma M^1_-
{U^1_-}^{\dagger}$ on a bit-flip error on the first qubit of a
codeword $\rho$ is:
\begin{gather}
X_1\rho X_1 \rightarrow (1-\kappa \tau_c)X_1\rho X_1 + \kappa
\tau_c \rho,\\
X_1\rho \rightarrow (1-\frac{1}{2}\kappa \tau_c)X_1\rho,\\
\rho X_1 \rightarrow (1-\frac{1}{2}\kappa \tau_c)\rho X_1.
\end{gather}
Just like in the bit-flip code, the error-correcting procedure for
the case where each qubit decoheres through an independent
bit-flip channel consists of simultaneous corrections of all
qubits ($i=1,2,...,n$) by continuous application of the maps
$\sigma \rightarrow U^i_+M^i_+\sigma M^i_+ {U^i_+}^{\dagger} +
U^i_-M^i_-\sigma M^i_- {U^i_-}^{\dagger}$.

From \eqref{codeform} it can be seen that the codewords have
analogous forms when expanded in the eigenbasis of another Pauli
operator ($X$ or $Y$) acting on a given qubit:
\begin{gather}
|\overline{0}\rangle = \frac{1}{\sqrt{2}}|x_+\rangle
|\psi^0_{x_+}\rangle +\frac{1}{\sqrt{2}}|x_-\rangle
|\psi^0_{x_-}\rangle = \frac{1}{\sqrt{2}}|y_+\rangle
|\psi^0_{y_+}\rangle +\frac{1}{\sqrt{2}}|y_-\rangle
|\psi^0_{y_-}\rangle,\nonumber\\
|\overline{1}\rangle = \frac{1}{\sqrt{2}}|x_+\rangle
|\psi^1_{x_+}\rangle +\frac{1}{\sqrt{2}}|x_-\rangle
|\psi^1_{x_-}\rangle = \frac{1}{\sqrt{2}}|y_+\rangle
|\psi^1_{y_+}\rangle +\frac{1}{\sqrt{2}}|y_-\rangle
|\psi^1_{y_-}\rangle.\label{codeform2}
\end{gather}
Here
\begin{equation}
|x_{\pm}\rangle = (-i)^{\frac{1\mp 1}{2}}\frac{|0\rangle \pm
|1\rangle}{\sqrt{2}}
\end{equation}
and
\begin{equation}
|y_{\pm}\rangle = \frac{|0\rangle \pm i|1\rangle}{\sqrt{2}}
\end{equation}
are eigenbases of $X$ and $Y$ respectively, and
\begin{equation}
|\psi^i_{x_{\pm}}\rangle=i^{\frac{1\mp
1}{2}}\frac{|\psi^i_0\rangle \pm |\psi^i_1\rangle}{\sqrt{2}},
\hspace{0.1cm}i=0,1
\end{equation}
and
\begin{equation}
|\psi^i_{y_{\pm}}\rangle=\frac{|\psi^i_0\rangle \mp
i|\psi^i_1\rangle}{\sqrt{2}},\hspace{0.1cm}i=0,1
\end{equation}
are orthonormal sets. The reason why we have chosen these
particular overall phases in the definition of the eigenvectors of
$X$ and $Y$, is that we want to have our expressions explicitly
symmetric with respect to cyclic permutations of $X$, $Y$ and $Z$.
More precisely, the expansions of the operators $X$, $Y$, $Z$ in
the $|0,1\rangle$ basis are the same as the expansions of $Y$,
$Z$, $X$ in the $|x_{\pm}\rangle$ basis, and the same as the
expansions of $Z$, $X$, $Y$ in the $|y_{\pm}\rangle$ basis. This
means that $Y$ and $Z$ errors in the computational basis can be
treated as $X$ errors in the bases $|x_{\pm}\rangle$ and
$|y_{\pm}\rangle$, and therefore can be corrected accordingly. The
weak measurement and unitary for the correction of $Y$ errors on
the first qubit (let's call them $M^1_{y\pm}$ and $U^1_{y\pm}$)
are obtained from \eqref{weakmeas} and \eqref{weakunit} by making
the substitutions $X\rightarrow Y$, $Y\rightarrow Z$, $|0,1\rangle
\rightarrow |x_{\pm}\rangle$, $|\psi^i_{0,1}\rangle \rightarrow
|\psi^i_{x_{\pm}}\rangle$. The operations for the correction of
$Z$ errors ($M^1_{z\pm}$ and $U^1_{z\pm}$) are obtained from
\eqref{weakmeas} and \eqref{weakunit} by $X\rightarrow Z$,
$Y\rightarrow X$, $|0,1\rangle \rightarrow |y_{\pm}\rangle$,
$|\psi^i_{0,1}\rangle \rightarrow |\psi^i_{y_{\pm}}\rangle$. The
operations for correction of $Y$ and $Z$ errors on any qubit
($M^i_{y\pm}$, $U^i_{y\pm}$, and $M^i_{z\pm}$, $U^i_{z\pm}$,
$i=1,2,...,n$) are defined analogously.

To prove that the weak error-correcting map resulting from the
application of the described weak measurements and unitary
operations is equal to Eq.~\eqref{wm}, we are going to look at its
effect on different components of the density matrix. Any density
matrix can be written as a linear combination of terms of the type
$|\phi\rangle \langle \chi |$, where each of the vectors
$|\phi\rangle$ and $|\chi\rangle$ belongs to one of the orthogonal
subspaces on which a state gets projected if we measure the
stabilizer generators of the code. Let us denote the code space by
$C$ and the subspaces corresponding to different single-qubit
errors by $C_{X_i}$, $C_{Y_i}$, and $C_{Z_i}$, where the subscript
refers to the type of error ($X$, $Y$, or $Z$) and the number of
the qubit on which it occurred. The code space and the subspaces
corresponding to single-qubit errors in general do not cover the
whole Hilbert space. Some of the outcomes of the measurement of
the stabilizer generators may project the state onto subspaces
corresponding to multi-qubit errors. We are going to denote the
direct sum of these subspaces by $C_M$. Our weak error-correcting
operation consists of a simultaneous application of the weak maps
$\rho \rightarrow U^i_+M^i_+\rho M^i_+ {U^i_+}^{\dagger} +
U^i_-M^i_-\rho M^i_- {U^i_-}^{\dagger}$, $\rho \rightarrow
U^i_{y+}M^i_{y+}\rho M^i_{y+} {U^i_{y+}}^{\dagger} +
U^i_{y-}M^i_{y-}\rho M^i_{y-} {U^i_{y-}}^{\dagger}$, $\rho
\rightarrow U^i_{z+}M^i_{z+}\rho M^i_{z+} {U^i_{z+}}^{\dagger} +
U^i_{z-}M^i_{z-}\rho M^i_{z-} {U^i_{z-}}^{\dagger}$,
$i=1,2,...,n$. The order of application is irrelevant since we
consider only contributions of up to first order in $\Delta t$.
Using \eqref{actionofUM1}-\eqref{actionofUM3} and the symmetry
under cyclic permutations of $X$, $Y$ and $Z$, one can show that
this map has the following effect:
\begin{gather}
|\phi\rangle \langle \chi | \rightarrow |\phi\rangle \langle \chi
|, \hspace{0.2 cm} \textrm{if} \hspace{0.2 cm} |\phi\rangle , |
\chi \rangle \in C\oplus C_M, \label{actionofweakmap1}\\
|\phi\rangle \langle \chi | \rightarrow
(1-2\kappa\tau_c)|\phi\rangle \langle \chi | + \kappa \tau_c
X_i|\phi\rangle \langle \chi |X_i + \kappa \tau_c Z_i|\phi\rangle
\langle \chi |Z_i, \hspace{0.2 cm} \textrm{if} \hspace{0.2 cm}
|\phi\rangle , |
\chi \rangle \in C_{X_i},\label{actionofweakmap2}\\
|\phi\rangle \langle \chi | \rightarrow
(1-2\kappa\tau_c)|\phi\rangle \langle \chi | + \kappa \tau_c
Y_i|\phi\rangle \langle \chi |Y_i + \kappa \tau_c X_i|\phi\rangle
\langle \chi |X_i, \hspace{0.2 cm} \textrm{if} \hspace{0.2 cm}
|\phi\rangle , |
\chi \rangle \in C_{Y_i},\label{actionofweakmap3}\\
|\phi\rangle \langle \chi | \rightarrow (1-2\kappa\tau_c
t)|\phi\rangle \langle \chi | + \kappa \tau_c Z_i|\phi\rangle
\langle \chi |Z_i + \kappa \tau_c Y_i|\phi\rangle \langle \chi
|Y_i, \hspace{0.2 cm} \textrm{if} \hspace{0.2 cm} |\phi\rangle , |
\chi \rangle \in C_{Z_i},\label{actionofweakmap4}\\
|\phi\rangle \langle \chi | \rightarrow
(1-\kappa\tau_c)|\phi\rangle \langle \chi |, \hspace{0.2 cm}
\textrm{if} \hspace{0.2 cm} |\phi\rangle \in C_{X_i}\oplus C_{Y_i}
\oplus C_{Z_i}, \hspace{0.2 cm}|
\chi \rangle \in C\oplus C_M,\label{actionofweakmap5}\\
|\phi\rangle \langle \chi | \rightarrow
(1-2\kappa\tau_c)|\phi\rangle \langle \chi | +\kappa \tau_c
X_i|\phi\rangle \langle \chi |X_i, \hspace{0.2 cm} \textrm{if}
\hspace{0.2 cm} |\phi\rangle \in C_{X_i},
|\chi \rangle \in C_{Y_i},\label{actionofweakmap6}\\
|\phi\rangle \langle \chi | \rightarrow
(1-2\kappa\tau_c)|\phi\rangle \langle \chi | +\kappa \tau_c
Y_i|\phi\rangle \langle \chi |Y_i, \hspace{0.2 cm} \textrm{if}
\hspace{0.2 cm} |\phi\rangle \in C_{Y_i},
|\chi \rangle \in C_{Z_i},\label{actionofweakmap7}\\
|\phi\rangle \langle \chi | \rightarrow
(1-2\kappa\tau_c)|\phi\rangle \langle \chi | +\kappa \tau_c
Z_i|\phi\rangle \langle \chi |Z_i, \hspace{0.2 cm} \textrm{if}
\hspace{0.2 cm} |\phi\rangle \in C_{Z_i},
|\chi \rangle \in C_{X_i},\label{actionofweakmap8}\\
|\phi\rangle \langle \chi | \rightarrow
(1-2\kappa\tau_c)|\phi\rangle \langle \chi |, \hspace{0.2 cm}
\textrm{if} \hspace{0.2 cm} |\phi\rangle \in C_{X_i},\oplus
C_{Y_i} \oplus C_{Z_i}, | \chi \rangle \in C_{X_j}\oplus C_{Y_j}
\oplus C_{Z_j}, \hspace{0.2 cm} i\neq j \label{actionofweakmapN}.
\end{gather}

This is sufficient to determine the effect of the error-correcting
map on any density matrix. One can easily see that this map is not
equal to the map \eqref{wm} because of the last terms on the
right-hand sides of
Eqs.~\eqref{actionofweakmap2}-\eqref{actionofweakmap4} and
\eqref{actionofweakmap6}-\eqref{actionofweakmap8}. These terms
appear because the operation we proposed for correcting $X$ errors,
for example, cannot distinguish between $X$ and $Y$ errors and
corrects both. This gives rise to the last terms in
Eqs.~\eqref{actionofweakmap3} and \eqref{actionofweakmap6}. The same
holds for the operations we proposed for correcting $Y$ and $Z$
errors. Nevertheless, this map is also a weak error-correcting map
in the sense that in the limit of infinitely many applications, it
corrects single-qubit errors fully, i.e., it results in the strong
error-correcting map $\Phi(\rho)$.

To see this, consider all possible single-qubit errors on a
density matrix $\rho \in C$. The most general form of a
single-qubit error on the $i^{\textrm{th}}$ qubit is
\begin{equation}
\rho_i=\overset{4}{\underset{j=1}{\sum}}M_{i,j}\rho
M_{i,j}^{\dagger},\label{singlequbiterror}
\end{equation}
where the Kraus operators $M_{i,j}$ are complex linear
combinations of $I$, $X_i$, $Y_i$ and $Z_i$ that satisfy
$\overset{4}{\underset{j=1}{\sum}}M_{i,j}^{\dagger}M_{i,j}=I$.
Observe that $\rho_i$ is a real superposition of the following
terms: $\rho$, $X_i\rho X_i$, $Y_i\rho Y_i$, $Z_i\rho Z_i$,
$i(X_i\rho- \rho X_i)$, $i(Y_i\rho- \rho Y_i)$, $i(Z_i\rho- \rho
Z_i)$, $X_i\rho Y_i + Y_i \rho X_i$, $Y_i\rho Z_i + Z_i \rho Y_i$,
$X_i\rho Z_i + Z_i \rho X_i$. Each of the first four terms has
trace $1$ and the rest of the terms are traceless. From
\eqref{actionofweakmap1}-\eqref{actionofweakmapN} one can see that
the weak map does not couple the first four terms with the rest.
Therefore, their evolution under continuous application of the map
(without decoherence) can be treated separately. If we write the
single-qubit error \eqref{singlequbiterror} as
\begin{equation}
\rho_i = a \rho + b X_i\rho X_i + c Y_i\rho Y_i +d Z_i\rho Z_i +
\textrm{traceless terms} \label{generalsinglequbiterror},
\end{equation}
a single application of the weak map causes the transformation
\begin{gather}
a \rightarrow a + (b + c + d)\kappa \tau_c,\label{ageneralcode}\\
b \rightarrow b - b 2\kappa \tau_c + c \kappa \tau_c,\\
c \rightarrow c - c 2\kappa \tau_c + d \kappa \tau_c,\\
d \rightarrow d - d 2\kappa \tau_c + b \kappa \tau_c.
\end{gather}
Using that at any moment $a+b+c+d=1$ and taking the limit $\tau_c
\rightarrow 0$, from \eqref{ageneralcode} we obtain that the
evolution of $a$ is described by
\begin{equation}
\frac{d a(t)}{dt}=\kappa (1-a(t)).
\end{equation}
The solution is
\begin{equation}
a(t) = 1-(1-a(0))e^{-\kappa t},
\end{equation}
i.e., in the limit of $t\rightarrow \infty$ we obtain
$a(t)\rightarrow 1$ (and therefore $b,c,d \rightarrow 0$). We
don't need to look at the evolution of the traceless terms in
$\rho_i$ because our map is completely positive and therefore the
transformed $\rho_i$ is also a density matrix, which implies that
if $a=1, b=0,c=0,d=0$, all traceless terms have to vanish. This
completes the proof that in the limit of infinitely many
applications, our weak error-correcting map is able to correct
arbitrary single-qubit errors.

It is interesting whether a similar implementation in terms of
weak measurements and weak unitary operations can be found for the
map \eqref{wm} for general codes. One way to approach this problem
might be to look at the error-correcting operations in the decoded
basis. Another interesting question is whether the scheme we
presented can be modified to include feedback which depends more
generally on the history of measurement outcomes and not only on
the outcome of the last measurement. It is natural to expect that
using fully the available information about the state could lead
to a better performance. These questions are left open for future
investigation.

\chapter*{Chapter 6: \hspace{1pt} Correctable subsystems under continuous decoherence}
\addcontentsline{toc}{chapter}{Chapter 6:\hspace{0.15cm}
Correctable subsystems under continuous decoherence}

In the previous chapter, we were concerned with a situation in
which the information stored in an error-correcting code was only
approximately correctable. For the model we considered, there were
non-correctable multi-qubit errors that accumulated with time,
albeit with a slower rate. This is, in practice, the general
situation---the probability for non-correctable errors is never
truly zero and in order to deal with higher-order terms we need to
use concatenation and fault-tolerant techniques (see Section 8.3).
But as we saw in the previous chapter, the idea of perfect error
correction can be crucial for understanding the approximate
process. In view of this, in this chapter we ask the question of
the conditions under which a code is perfectly correctable during
an entire time interval of continuous decoherence. We consider the
most general form of quantum codes---\textit{operator}, or
\textit{subsystem} codes.

\section*{6.1 \hspace{2pt} Preliminaries}
\addcontentsline{toc}{section}{6.1 \hspace{0.15cm} Preliminaries}

Operator quantum error correction (OQEC) \cite{KLP05, KLPL06, BKK07}
is a unified approach to error correction which uses the most
general encoding for the protection of information---encoding in
subsystems \cite{Knill06, VKL01} (see also \cite{BKNPV07}). This
approach contains as special cases the standard quantum
error-correction method \cite{Shor95, Ste96, Bennett96c, KL96} as
well as the methods of decoherence-free subspaces \cite{DG98, ZR97,
LCW98, LBKW01} and subsystems \cite{KLV00, DeF00, KBLW01, YGB01}. In
the OQEC formalism, noise is represented by a completely positive
trace-preserving (CPTP) linear map or a noise channel, and
correctability is defined with respect to such channels. In
practice, however, noise is a continuous process and if it can be
represented by a CPTP map, that map is generally a function of time.
Correctability is therefore a time-dependent property. Furthermore,
the evolution of an open system is completely positive if the system
and the environment are initially uncorrelated, and necessary and
sufficient conditions for CPTP dynamics are not known. As pointed
out in the previous chapter, for more general cases one might need a
notion of correctability that can capture non-CP transformations
\cite{ShaLid07}.

Whether completely positive or not, the noise map is a result of
the action of the generator driving the evolution and possibly of
the initial state of the system and the environment. Therefore,
our goal will be to understand the conditions for correctability
in terms of the generator that drives the evolution. We will
consider conditions on the system-environment Hamiltonian, or in
the case of Markovian evolution---on the Lindbladian.

Conditions on the generator of evolution have been derived for
decoherence-free subsystems (DFSs) \cite{ShaLid05}, which are a
special type of operator codes. DFSs are \textit{fixed} subsystems
of the system's Hilbert space, inside which all states evolve
unitarily. One generalization of this concept are the so called
unitarily correctable subsystems \cite{KLPL06}. These are
subsystems, all states inside of which can be corrected via a
unitary operation up to an arbitrary transformation inside the
gauge subsystem. Unlike DFSs, the unitary evolution followed by
states in a unitarily correctable code are not restricted to the
initial subsystem. An even more general concept is that of
\textit{unitarily recoverable} subsystems \cite{KLP05, KLPL06},
for which states can be recovered by a unitary transformation up
to an expansion of the gauge subsystem. It was shown that any
correctable subsystem is in fact a unitarily recoverable subsystem
\cite{KS06}. This reflects the so called subsystem principle
\cite{Knill06, VKL01}, according to which protected information is
always contained in a subsystem of the system's Hilbert space. The
connection between DFSs and unitarily recoverable subsystems
suggests that similar conditions on the generators of evolution to
those for DFSs can be derived in the case of general correctable
subsystems. This is the subject of the present study.

The chapter is organized as follows. In Section 6.2 we review the
definitions of correctable subsystems and unitarily recoverable
subsystems. In Section 6.3, we discuss the necessary and sufficient
conditions for such subsystems to exist in the case of CPTP maps. In
Section 6.4, we derive conditions for the case of Markovian
decoherence. The conditions for general correctability in this case
are essentially the same as those for unitary correctability except
that the dimension of the gauge subsystem is allowed to suddenly
increase. For the case when the evolution is non-correctable, we
conjecture a procedure for tracking the subsystem which contains the
optimal amount of undissipated information and discuss its possible
implications for the problem of optimal error correction. In Section
6.5, we derive conditions on the system-environment Hamiltonian. In
this case, the conditions for unitary correctability concern only
the effect of the Hamiltonian on the system, whereas the conditions
for general correctability concern the entire system-environment
Hamiltonian. In the latter case, the state of the noisy subsystem
plus environment belongs to a particular subspace which plays an
important role in the conditions. We extend the conditions to the
case where the environment is initialized inside a particular
subspace. In Section 6.6, we conclude.

\section*{6.2 \hspace{2pt} Correctable subsystems}
\addcontentsline{toc}{section}{6.2 \hspace{0.15cm} Correctable
subsystems}

For simplicity, we consider the case where information is stored
in only one subsystem. Then there is a corresponding decomposition
of the Hilbert space of the system,
\begin{equation}
\mathcal{H^S}=\mathcal{H^A}\otimes\mathcal{H}^B\oplus
\mathcal{K},\label{decomposition}
\end{equation}
where the subsystem $\mathcal{H}^A$ is used for encoding of the
protected information. The subsystem $\mathcal{H}^B$ is referred
to as the gauge subsystem, and $\mathcal{K}$ denotes the rest of
the Hilbert space. In the formulation of OQEC \cite{KLP05,
KLPL06}, the noise process is a completely positive
trace-preserving (CPTP) linear map
$\mathcal{E}:\mathcal{B}(\mathcal{H}^S)\rightarrow
\mathcal{B}(\mathcal{H}^S)$, where $\mathcal{B}(\mathcal{H})$
denotes the set of linear operators on a finite-dimensional
Hilbert space $\mathcal{H}$. Let the operator-sum representation
of the map $\mathcal{E}$ be
\begin{equation}
\mathcal{E}(\sigma)=\underset{i}{\sum}M_i\sigma M_i^{\dagger},
\hspace{0.2cm} \textrm{for all } \sigma\in
\mathcal{B}(\mathcal{H}^S),\label{Kraus1}
\end{equation}
where the Kraus operators
$\{M_i\}\subseteq\mathcal{B}(\mathcal{H}^S)$ satisfy
\begin{equation}
\underset{i}{\sum}M_i^{\dagger}M_i=I^S.\label{completeness}
\end{equation}

The subsystem $\mathcal{H}^A$ in Eq.~\eqref{decomposition} is
called \textit{noiseless} with respect to the noise process
$\mathcal{E}$, if
\begin{gather}
\textrm{Tr}_B\{(\mathcal{P}^{AB}\circ\mathcal{E})(\sigma)\}=\textrm{Tr}_B\{\sigma\},\label{noiselesssystem}\\
\hspace{0.1cm} \textrm{for all }\sigma\in
\mathcal{B}(\mathcal{H}^S) \textrm{ such that }
\sigma=\mathcal{P}^{AB}(\sigma)\hspace{0.1cm} ,\notag
\end{gather}
where
\begin{equation}
\mathcal{P}^{AB}(\cdot)=P^{AB}(\cdot)P^{AB}
\end{equation}
with $P^{AB}$ being the projector of $\mathcal{H}^S$ onto
$\mathcal{H}^A\otimes \mathcal{H}^B$,
\begin{equation}
P^{AB}\mathcal{H}^S=\mathcal{H}^A\otimes \mathcal{H}^B.
\end{equation}

Similarly, a \textit{correctable} subsystem is one for which there
exists a correcting CPTP map
$\mathcal{R}:\mathcal{B}(\mathcal{H}^S)\rightarrow
\mathcal{B}(\mathcal{H}^S)$, such that the subsystem is noiseless
with respect to the map $\mathcal{R}\circ \mathcal{E}$:
\begin{gather}
\textrm{Tr}_B\{(\mathcal{P}^{AB}\circ\mathcal{R}\circ\mathcal{E})(\sigma)\}=\textrm{Tr}_B\{\sigma\},\label{correctablesystem}\\
\hspace{0.1cm} \textrm{for all }\sigma\in
\mathcal{B}(\mathcal{H}^S) \textrm{ such that }
\sigma=\mathcal{P}^{AB}(\sigma)\hspace{0.1cm} .\notag
\end{gather}

When the correcting map $\mathcal{R}$ is unitary,
$\mathcal{R}=\mathcal{U}$, the subsystem is called
\textit{unitarily correctable}:
\begin{gather}
\textrm{Tr}_B\{(\mathcal{P}^{AB}\circ\mathcal{U}\circ\mathcal{E})(\sigma)\}=\textrm{Tr}_B\{\sigma\},\label{unitarilycorrectable}\\
\hspace{0.1cm} \textrm{for all }\sigma\in
\mathcal{B}(\mathcal{H}^S) \textrm{ such that }
\sigma=\mathcal{P}^{AB}(\sigma)\hspace{0.1cm} .\notag
\end{gather}

A similar but more general notion is that of a \textit{unitarily
recoverable} subsystem, for which the unitary $\mathcal{U}$ need
not bring the erroneous state back to the original subspace
$\mathcal{H}^A\otimes \mathcal{H}^B$ but can bring it in a
subspace $\mathcal{H}^A\otimes \mathcal{H}^{B'}$ such that
\begin{gather}
\textrm{Tr}_{B'}\{(\mathcal{P}^{AB'}\circ\mathcal{U}\circ\mathcal{E})(\sigma)\}=\textrm{Tr}_B\{\sigma\},\label{unitarilyrecoverable}
\hspace{0.1cm} \textrm{for all }\sigma\in
\mathcal{B}(\mathcal{H}^S) \textrm{ such that }
\sigma=\mathcal{P}^{AB}(\sigma)\hspace{0.1cm} .\notag
\end{gather}

Obviously, if $\mathcal{H}^A$ is unitarily recoverable, it is also
correctable, since one can always apply a local CPTP map
$\mathcal{E}^{B'\rightarrow B}:
\mathcal{B}(\mathcal{H}^{B'})\rightarrow
\mathcal{B}(\mathcal{H}^{B})$ which brings all states from
$\mathcal{H}^{B'}$ to $\mathcal{H}^{B}$. (In fact, if the
dimension of $\mathcal{H}^{B'}$ is smaller or equal to that of
$\mathcal{H}^{B}$, this can always be done by a unitary map, i.e.,
$\mathcal{H}^A$ is unitarily correctable.) In Ref.~\cite{KS06} it
was shown that the reverse is also true---if $\mathcal{H}^A$ is
correctable, it is unitarily recoverable. This equivalence will
provide the basis for our derivation of correctability conditions
for continuous decoherence.

Before we proceed with our discussion, we point out that condition
\eqref{unitarilyrecoverable} can be equivalently written as
\cite{KLP05, KLPL06}
\begin{gather}
\mathcal{U}\circ\mathcal{E}(\rho\otimes\tau)=\rho\otimes\tau',\hspace{0.4cm}
\tau'\in\mathcal{B}(\mathcal{H}^{B'}),
\label{unitarilyrecoverable2} \hspace{0.2cm} \textrm{for all
}\rho\in \mathcal{B}(\mathcal{H}^A), \hspace{0.1cm} \tau\in
\mathcal{B}(\mathcal{H}^B)\hspace{0.1cm} .
\end{gather}

\section*{6.3 \hspace{2pt} Completely positive linear maps}
\addcontentsline{toc}{section}{6.3 \hspace{0.15cm} Completely
positive linear maps}

Let $\mathcal{H}^S$ and $\mathcal{H}^E$ denote the Hilbert spaces
of a system and its environment, and let
$\mathcal{H}=\mathcal{H}^S\otimes\mathcal{H}^E$ be the total
Hilbert space. As we pointed out earlier, a common example of a CP
map is the transformation that the state of a system undergoes if
the system is initially decoupled from its environment,
$\rho(0)=\rho^S(0)\otimes \rho^E(0)$, and both the system and
environment evolve according to the Schr\"{o}dinger equation:
\begin{equation}
\frac{d\rho(t)}{dt}=-i[H(t),\rho(t)].\label{Schrodinger}
\end{equation}
Equation \eqref{Schrodinger} gives rise to the unitary
transformation
\begin{equation}
\rho(t)=V(t)\rho(0)V^{\dagger}(t),
\end{equation}
with
\begin{equation}
V(t)=\mathcal{T}\textrm{exp}(-i\int_0^t H(\tau)d\tau),
\end{equation}
where $\mathcal{T}$ denotes time ordering. Under the assumption of
an initially-decoupled state of the system and the environment,
the transformation of the state of the system is described by the
time-dependent CPTP map
\begin{equation}
\rho^S(0)\rightarrow
\rho^S(t)\equiv\textrm{Tr}_E(\rho(t))=\underset{i}{\sum}M_{i}(t)\rho^S(0)M_{i}^{\dagger}(t),\notag
\end{equation}
with Kraus operators
\begin{equation}
M_{i}(t)=\sqrt{\lambda_{\nu}}\langle \mu | V(t)|\nu\rangle
,\hspace{0.3cm}i=(\mu,\nu)\label{Krausoperator}
\end{equation}
where $\{|\mu\rangle \}$ is a basis in which the initial
environment density matrix is diagonal,
$\rho^B(0)=\underset{\mu}{\sum}\lambda_{\mu}|\mu\rangle\langle
\mu|$. We already saw one example of such a map in Chapter 4 where
we studied the evolution of a qubit coupled to a spin bath
(Eq.~\eqref{KrausForm}).

The Kraus representation \eqref{Kraus1} applies to any CP linear map
which need not necessarily arise from evolution of the type
\eqref{Schrodinger}. This is why in the following theorem we derive
conditions for discrete CP maps. For correctability under continuous
decoherence, the same conditions must apply at any moment of time,
i.e., one can think that the quantities $M_i$, $U$, $C_i$, as well
as the subsystem $\mathcal{H}^{B'}$ in the theorem are implicitly
time-dependent.

\textbf{Theorem 1:} \textit{The subsystem $\mathcal{H}^A$ in the
decomposition \eqref{decomposition} is correctable under a CP
linear map in the form \eqref{Kraus1}, if and only if there exists
a unitary operator $U\in\mathcal{B}(\mathcal{H}^S)$ such that the
Kraus operators satisfy
\begin{gather}
M_{i}P^{AB}=U^{\dagger}I^{A}\otimes C^{B\rightarrow
B'}_{i},\hspace{0.2cm}C^{B\rightarrow B'}_{i}:
\mathcal{H}^B\rightarrow \mathcal{H}^{B'}, \hspace{0.2cm} \forall
i. \label{conditionKraus}
\end{gather}}

\textbf{Proof:} The sufficiency of condition
\eqref{conditionKraus} is obvious---using that $\rho\otimes\tau$
in Eq.~\eqref{unitarilyrecoverable2} satisfies $\rho\otimes\tau=
P^{AB}\rho\otimes\tau P^{AB}$, it can be immediately verified that
Eq.~\eqref{conditionKraus} implies
Eq.~\eqref{unitarilyrecoverable2} with
$\mathcal{U}=U(\cdot)U^{\dagger}$. Now assume that $\mathcal{H}^A$
is unitarily recoverable and the recovery map is
$\mathcal{U}=U(\cdot)U^{\dagger}$. The map $\mathcal U\circ
\mathcal{E}$ in Eq.~\eqref{unitarilyrecoverable2} can then be
thought of as having Kraus operators $U M_{i}$. In particular,
condition \eqref{unitarilyrecoverable2} has to be satisfied for
$\rho=|\psi\rangle \langle \psi|$, $\tau=|\phi\rangle \langle
\phi|$ where $|\psi\rangle\in \mathcal{H}^A$ and $|\phi\rangle\in
\mathcal{H}^B$ are pure states. Notice that the image of
$|\psi\rangle\langle\psi|\otimes|\phi\rangle\langle\phi| $ under
the map $\mathcal{U}\circ \mathcal{E}$ would be of the form
$|\psi\rangle\langle\psi|\otimes \tau'$, only if all terms in
Eq.~\eqref{Kraus1} are of the form
\begin{gather}
UM_{i}|\psi\rangle\langle\psi|\otimes |\phi\rangle\langle\phi|
M_{i}^{\dagger}U^{\dagger}=|g_{i}(\psi)|^2|\psi\rangle\langle\psi|\otimes
|\phi'_{i}(\psi)\rangle\langle \phi'_{i}(\psi)|,
\hspace{0.2cm}g_{i}(\psi)\in C,
\end{gather}
where for now we assume that $g_{i}$ and $|\phi'_{i}\rangle$ may
depend on $|\psi\rangle$. In other words,
\begin{equation}
UM_{i}|\psi\rangle |\phi\rangle
=g_{i}(\psi)|\psi\rangle|\phi'_{i}(\psi)\rangle
,\hspace{0.2cm}g_{i}(\psi)\in C, \hspace{0.2cm} \forall
i.\label{edno'}
\end{equation}

But if we impose \eqref{edno'} on a linear superposition
$|\psi\rangle=a|\psi_1\rangle+b|\psi_2\rangle$, ($a,b\neq 0$), we
obtain $g_{i}(\psi_1)=g_{i}(\psi_2)$ and
$|\phi'_{i}(\psi_1)\rangle = |\phi'_{i}(\psi_2)\rangle$ i.e.,
\begin{equation}
g_{i}(\psi)\equiv g_{i}, \hspace{0.2cm}|\phi'_{i}(\psi)\rangle
\equiv |\phi'_{i}\rangle, \hspace{0.2cm} \forall |\psi\rangle \in
\mathcal{H}^A, \hspace{0.2cm} \forall i.
\end{equation}
Since Eq.~\eqref{edno'} has to be satisfied for all $|\psi\rangle
\in \mathcal{H}^A$ and all $|\phi\rangle \in \mathcal{H}^B$, we
obtain
\begin{gather}
UM_{i}P^{AB}=I^{A}\otimes C_{i}^{B\rightarrow
B'},\hspace{0.2cm}C_{i}^{B\rightarrow B'}:
\mathcal{H}^B\rightarrow \mathcal{H}^{B'}, \hspace{0.2cm} \forall
i.
\end{gather}
Applying $U^{\dagger}$ from the left yields condition
\eqref{conditionKraus}.

We remark that condition \eqref{conditionKraus} is equivalent to
the conditions obtained in Ref.~\cite{KLPL06}.

\section*{6.4 \hspace{2pt} Markovian dynamics}
\addcontentsline{toc}{section}{6.4 \hspace{0.15cm} Markovian
dynamics}

The most general continuous completely positive time-local
evolution of the state of a quantum system is described by a
semi-group master equation in the form \eqref{firstLindblad} but
with time dependent coefficients,
\begin{eqnarray}
\frac{d\rho(t)}{dt}=-i[H(t),\rho(t)]-\frac{1}{2}\underset{j}{\sum}(2L_j(t)\rho(t)
L_j^{\dagger}(t)\notag\\
-L_j^{\dagger}(t)L_j(t)\rho(t)-\rho(t)
L_j^{\dagger}(t)L_j(t))\equiv
\mathcal{L}(t)\rho(t).\label{Lindblad}
\end{eqnarray}
(For a discussion of the situations in which such time-dependent
Markovian evolution can arise, see, e.g., Ref.~\cite{Len86}.) Here
$H(t)$ is a system Hamiltonian, $L_j(t)$ are Lindblad operators, and
$\mathcal{L}(t)$ is the Liouvillian superoperator corresponding to
this dynamics. (The decoherence rates $\lambda_j$ that appear in
Eq.~\eqref{firstLindblad}, here have been absorbed in $L_j(t)$.) The
general evolution of a state is given by
\begin{equation}
\rho(t_2)=\mathcal{T}\textrm{exp}(\int_{t_1}^{t_2}
\mathcal{L}(\tau)d \tau)\rho(t_1), \hspace{0.3 cm}
t_2>t_1.\label{evolutionLindblad}
\end{equation}

We will first derive necessary and sufficient conditions for
unitarily correctable subsystems under the dynamics
\eqref{Lindblad}, and then will extend them to the case of
unitarily recoverable subsystems.

In the case of continuous dynamics, the error map $\mathcal{E}$
and the error-correcting map $\mathcal{U}$ in
Eq.~\eqref{unitarilycorrectable} are generally time dependent. If
we set $t=0$ as the initial time at which the system is prepared,
the error map resulting from the dynamics \eqref{Lindblad} is
\begin{equation}
\mathcal{E}(t)(\cdot)=\mathcal{T}\textrm{exp}\left(\int_{0}^{t}
\mathcal{L}(\tau)d \tau\right)(\cdot).
\end{equation}
Let the $\mathcal{U}(t)=U(t)(\cdot)U^{\dagger}(t)$ be the unitary
error-correcting map in Eq.~\eqref{unitarilycorrectable}. We can
define the rotating frame corresponding to $U^{\dagger}(t)$ as the
transformation of each operator as
\begin{equation}
O(t)\rightarrow
\widetilde{O}(t)=U(t)O(t)U^{\dagger}(t).\label{rotatingframe}
\end{equation}
In this frame, the Lindblad equation \eqref{Lindblad} can be
written as
\begin{gather}
\frac{d\widetilde{\rho}(t)}{dt}=-i[\widetilde{H}(t)+{H}'(t),\widetilde{\rho}(t)]-\frac{1}{2}\underset{j}{\sum}(2\widetilde{L}_j(t)\widetilde{\rho}(t)
\widetilde{L}_j^{\dagger}(t)\notag\\
-\widetilde{L}_j^{\dagger}(t)\widetilde{L}_j(t)\widetilde{\rho}(t)-\widetilde{\rho}(t)
\widetilde{L}_j^{\dagger}(t)\widetilde{L}_j(t))\equiv
\widetilde{\mathcal{L}}(t)\widetilde{\rho}(t),\label{Lindbladrot}
\end{gather}
where $H'(t)$ is defined through
\begin{equation}
i\frac{d U(t)}{dt}=H'(t) U(t),\label{defineU}
\end{equation}
i.e.,
\begin{equation}
U(t)=\mathcal{T}\textrm{exp}\left(-i\int_0^t H'(\tau)d\tau\right).
\end{equation}
The CPTP map resulting from the dynamics \eqref{Lindbladrot} is
\begin{equation}
\widetilde{\mathcal{E}}(t)(\cdot)=\mathcal{T}\textrm{exp}\left(\int_{0}^{t}
\widetilde{\mathcal{L}}(\tau)d \tau\right)(\cdot).
\end{equation}

\textbf{Theorem 2:} \textit{Let $\widetilde{H}(t)$ and
$\widetilde{L}_j(t)$ be the Hamiltonian and the Lindblad operators
in the rotating frame \eqref{rotatingframe} with $U(t)$ given by
Eq.~\eqref{defineU}. Then the subsystem $\mathcal{H}^A$ in the
decomposition \eqref{decomposition} is correctable by $U(t)$
during the evolution \eqref{Lindblad}, if and only if
\begin{gather}
\widetilde{L}_{j}(t)P^{AB}=I^{A}\otimes C^B_j(t),\hspace{0.2cm}\
C^B_j(t)\in\mathcal{B}(\mathcal{H}^B),\hspace{0.2cm}\forall
j\label{Markov1}
\end{gather}
and
\begin{gather}
\mathcal{P}^{AB}(\widetilde{H}(t)+H'(t))=I^A\otimes
D^B(t),\hspace{0.2cm} D^B(t)\in
\mathcal{B}(\mathcal{H}^B)\label{Markov2}
\end{gather}
and
\begin{gather}
P^{AB}(\widetilde{H}(t)+H'(t)+\frac{i}{2}\underset{j}{\sum}\widetilde{L}_j^{\dagger}(t)\widetilde{L}_j(t))P_{\mathcal{K}}=0\label{Markov3}
\end{gather}
for all $t$, where $P_{\mathcal{K}}$ denotes the projector on
$\mathcal{K}$. }

\textbf{Proof:} Since by definition $U(t)$ is an error-correcting
map for subsystem $\mathcal{H}^A$, if
$\mathcal{P}^{AB}(\rho(0))=\rho(0)$, we have
$\textrm{Tr}_B\{\mathcal{P}^{AB}\circ\widetilde{\mathcal{E}}(\widetilde{\rho}(0))\}=\textrm{Tr}_B\{\mathcal{P}^{AB}(\widetilde{\rho}(t))\}=
\textrm{Tr}_B\{\mathcal{P}^{AB}\circ\mathcal{U}(t)\circ
\mathcal{E}(t)(\rho(0))\}=\textrm{Tr}_B\{\rho(0)\}=\textrm{Tr}_B\{\tilde{\rho}(0)\}$,
i.e, $\mathcal{H}^A$ is a noiseless subsystem under the evolution
in the rotating frame \eqref{Lindbladrot}. Then the theorem
follows from Eq.~\eqref{Lindbladrot} and the conditions for
noiseless subsystems under Markovian decoherence obtained in
\cite{ShaLid05}.

\textbf{Comment:} Conditions \eqref{Markov2} and \eqref{Markov3}
can be used to obtain the operator $H'(t)$ (and hence $U(t)$) if
the initial decomposition \eqref{decomposition} is known. Note
that there is a freedom in the definition of $H'(t)$. For example,
$D^B(t)$ in Eq.~\eqref{Markov2} can be any Hermitian operator. In
particular, we can choose $D^B(t)=0$. Also, the term
$P_{\mathcal{K}}H'(t)P_{\mathcal{K}}$ does not play a role and can
be chosen arbitrary. Using that $P_{\mathcal{K}}=I-P^{AB}$, we can
choose
\begin{gather}
H'(t)=-\widetilde{H}(t)-\frac{i}{2}P^{AB}\left(\underset{j}{\sum}\widetilde{L}_j^{\dagger}(t)\widetilde{L}_j(t)\right)
+\frac{i}{2}\left(\underset{j}{\sum}\widetilde{L}_j^{\dagger}(t)\widetilde{L}_j(t)\right)P^{AB},\label{defineH'}
\end{gather}
which satisfies Eq.~\eqref{Markov2} and Eq.~\eqref{Markov3}. Using
Eq.~\eqref{rotatingframe}, Eq.~\eqref{defineU} and
Eq.~\eqref{defineH'}, we obtain the following first-order
differential equation for $U(t)$:
\begin{gather}
i\frac{dU(t)}{dt}=-U(t)H(t)-\frac{i}{2}P^{AB}U(t)\left(\underset{j}{\sum}{L}_j^{\dagger}(t){L}_j(t)\right)\notag\\
+\frac{i}{2}U(t)\left(\underset{j}{\sum}{L}_j^{\dagger}(t){L}_j(t)\right)U^{\dagger}(t)P^{AB}U(t).
\end{gather}
This equation can be used to solve for $U(t)$ starting from
$U(0)=I$.

Notice that since $\mathcal{H}^A$ is unitarily correctable by
$U(t)$, at time $t$ the initially encoded information can be
thought of as contained in the subsystem $\mathcal{H}^A(t)$
defined through
\begin{equation}
\mathcal{H}^A(t)\otimes\mathcal{H}^B(t)\equiv
U^{\dagger}(t)\mathcal{H}^A\otimes\mathcal{H}^B,
\end{equation}
i.e., this subsystem is obtained from $\mathcal{H}^A$ in
Eq.~\eqref{decomposition} via the unitary transformation
$U^{\dagger}(t)$. One can easily verify that the fact that the
right-hand side of Eq.~\eqref{Markov1} acts trivially on
$\mathcal{H}^{A}$ together with Eq.~\eqref{Markov2} are necessary
and sufficient conditions for an arbitrary state encoded in
subsystem $\mathcal{H}^A(t)$ to undergo trivial dynamics at time
$t$. Therefore, these conditions can be thought of as the
conditions for lack of noise in the instantaneous subsystem that
contains the protected information. On the other hand, the fact
that the right-hand side of Eq.~\eqref{Markov1} maps states from
$\mathcal{H}^A\otimes\mathcal{H}^B$ to
$\mathcal{H}^A\otimes\mathcal{H}^B$ together with
Eq.~\eqref{Markov3} are necessary and sufficient conditions for
states inside the time-dependent subspace
$U^{\dagger}(t)\mathcal{H}^{AB}$ not to leave this subspace during
the evolution. Thus the conditions of the theorem can be thought
off as describing a time-varying noiseless subsystem
$\mathcal{H}^A(t)$.

We now extend the above conditions to the case of unitarily
recoverable subsystems. As we pointed out earlier, the difference
between a unitarily correctable and a unitarily recoverable
subsystem is that in the latter the dimension of the gauge
subsystem may increase. Since the dimension of the gauge subsystem
is an integer, this increase can happen only in a jump-like
fashion at particular moments. Between these moments, the
evolution is unitarily correctable. Therefore, we can state the
following

\textbf{Theorem 3:} \textit{The subsystem $\mathcal{H}^A$ in
Eq.~\eqref{decomposition} is correctable during the evolution
\eqref{Lindblad}, if and only if there exist times $t_i$, $i=0,1,
2,...$, $t_0=0$, $t_i<t_{i+1}$, such that for each interval
between $t_{i-1}$ and $t_i$ there exists a decomposition
\begin{equation}
\mathcal{H}^S=\mathcal{H}^A\otimes\mathcal{H}^B_i\oplus
\mathcal{K}_i, \hspace{0.4cm}
\mathcal{H}^B_{i}\ni\mathcal{H}^B_{i-1},
\end{equation}
with respect to which the evolution during this interval is
unitarily correctable. }

\textbf{Remark:} An increase of the gauge subsystem at time $t_i$
happens if the operator $C_j(t)$ in Eq.~\eqref{Markov1} obtains
non-zero components that map states from $\mathcal{H}^{B}_i$ to
$\mathcal{H}^{B}_{i+1}$. From that moment on, $t_i\leq t \leq
t_{i+1}$, Eq.~\eqref{Markov1} must hold for the new decomposition
$\mathcal{H}^S=\mathcal{H}^A\otimes\mathcal{H}^B_{i+1}\oplus
\mathcal{K}_{i+1}$. The unitary $U(t)$ is determined from
Eq.~\eqref{Markov2} and Eq.~\eqref{Markov3} as described earlier.

The conditions derived in this section provide insights into the
mechanism of information preservation under Markovian dynamics, and
thus could have implications for the problem of error correction
when perfect correctability is not possible \cite{Schum02, Klesse07,
RW05, YHT05, FSW07, KSL08}. For example, it is possible that the
unitary operation constructed according to Eq.~\eqref{defineU} with
the appropriate modification for the case of increasing gauge
subsystem, may be useful for error-correction also when the
conditions of the theorems are only approximately satisfied. Notice
that the generator driving the effective evolution of the subspace
$U^{\dagger}(t)\mathcal{H}^A\otimes\mathcal{H}^B$ whose projector we
denote by $P^{AB}(t)\equiv U^{\dagger}(t)P^{AB}U(t)$, can be written
as
\begin{equation}
\mathcal{L}(t)(\cdot)=-i[H_{\textrm{eff}}(t),
\cdot]+\mathcal{D}(t)(\cdot) + \mathcal{S}(t)(\cdot),
\end{equation}
where
\begin{gather}
H_{\textrm{eff}}(t)=H(t)+\frac{i}{2}P^{AB}(t)\left(
\underset{j}{\sum}{L}_j^{\dagger}(t){L}_j(t)\right)
-\frac{i}{2}\left(\underset{j}{\sum}{L}_j^{\dagger}(t){L}_j(t)\right)P^{AB}(t)
\end{gather}
is an effective Hamiltonian,
\begin{gather}
\mathcal{D}(t)(\cdot)=\underset{j}{\sum}L_j(t)(\cdot)L_j^{\dagger}(t)
\end{gather}
is a dissipator, and
\begin{gather}
\mathcal{S}(t)(\cdot)= -\frac{1}{2}P^{AB}(t) \left(
\underset{j}{\sum}{L}_j^{\dagger}(t){L}_j(t)\right)
P^{AB}(t)(\cdot)\notag\\
-\frac{1}{2}(\cdot)P^{AB}(t) \left(
\underset{j}{\sum}{L}_j^{\dagger}(t){L}_j(t)\right) P^{AB}(t)
\end{gather}
is a superoperator acting on
$\mathcal{B}(U^{\dagger}(t)\mathcal{H}^{AB})$. The dissipator most
generally causes an irreversible loss of the information contained
in the current subspace, which may involve loss of the information
stored in subsystem $\mathcal{H}^A(t)$ as well as an increase of
the gauge subsystem. The superoperator $\mathcal{S}(t)(\cdot)$
gives rise to a transformation solely inside the current subspace.
In the case when the evolution is correctable, this operator acts
locally on the gauge subsystem, but in the general case it may act
non-trivially on $\mathcal{H}^A(t)$. The role of the effective
Hamiltonian is to rotate the current subspace by an infinitesimal
amount. If one could argue that the information lost under the
action of $\mathcal{D}(t)$ and $\mathcal{S}(t)$ is in principle
irretrievable, then heuristically one could expect that after a
single time step $dt$, the corresponding factor of the
infinitesimally rotated (possibly expanded) subspace will contain
the maximal amount of the remaining encoded information. Note that
to keep track of the increase of the gauge subsystem one would
need to determine the operator $C_j$ on the right-hand side of
Eq.~\eqref{Markov1} that optimally approximates the left-hand
side. Of course, since the dissipator generally causes leakage of
states outside of the current subspace, the error-correcting map
at the end would have to involve more than just a unitary recovery
followed by a CPTP map on the gauge subsystem. In order to
maximize the fidelity \cite{Ore08} of the encoded information with
a perfectly encoded state, one would have to bring the state of
the system fully inside the subspace
$\mathcal{H}^A\otimes\mathcal{H}^B$. These heuristic arguments,
however, require a rigorous analysis. It is possible that the
action of the superoperators $\mathcal{D}(t)$ and $\mathcal{S}(t)$
may be partially correctable and thus one may have to modify the
unitary \eqref{defineU} in order to optimally track the
retrievable information. We leave this as a problem for future
investigation.

\section*{6.5 \hspace{2pt} Conditions on the system-environment Hamiltonian}
\addcontentsline{toc}{section}{6.5 \hspace{0.15cm} Conditions on
the system-environment Hamiltonian}

We now derive conditions for correctability of a subsystem when the
dynamics of the system and the environment is described by the
Schr\"{o}dinger equation \eqref{Schrodinger}. While the CP-map
conditions can account for such dynamics when the states of the
system and the environment are initially disentangled, they depend
on the initial state of the environment. Below, we will first derive
conditions on the system-environment Hamiltonian that hold for any
state of the environment, and then extend them to the case when the
environment is initialized inside a particular subspace.

We point out that the equivalence between unitary recoverable
subsystems and correctable subsystems has been proven for CPTP
maps. Here, we could have a non-CP evolution since the initial
state of the system and the environment may be entangled.
Nevertheless, since correctability must hold for the case when the
initial state of the system and the environment is separable, the
conditions we obtain are necessary. They are obviously also
sufficient since unitary recoverability implies correctability.

Let us write the system-environment Hamiltonian as
\begin{equation}
H_{SE}(t)=H_S(t)\otimes I_E + I_E\otimes
H_E(t)+H_I(t),\label{Hamiltonian}
\end{equation}
where $H_S(t)$ and $H_E(t)$ are the system and the environment
Hamiltonians respectively, and
\begin{equation}
H_I(t)=\underset{j}{\sum}S_{j}(t)\otimes E_{j}(t),\label{HI}
\end{equation}
is the interaction Hamiltonian.

From the point of view of the Hilbert space of the system plus
environment, the decomposition \eqref{decomposition} reads
\begin{equation}
\mathcal{H}=(\mathcal{H}^A\otimes\mathcal{H}^B\oplus\mathcal{K})\otimes
\mathcal{H}^E=\mathcal{H}^A\otimes\mathcal{H}^B\otimes\mathcal{H}^E\oplus
\mathcal{K}\otimes\mathcal{H}^E \label{decompositionfull}.
\end{equation}

\subsection*{6.5.1 \hspace{2pt} Conditions independent of the state of the
environment} \addcontentsline{toc}{subsection}{6.5.1
\hspace{0.15cm} Conditions independent of the state of the
environment}

We will consider again conditions for unitary correctability
first, and then conditions for general correctability. In the
rotating frame \eqref{rotatingframe}, the Schr\"{o}dinger equation
\eqref{Schrodinger} becomes
\begin{equation}
\frac{d\widetilde{\rho}(t)}{dt}=-i[\widetilde{H}_{SE}(t)+H'(t),\widetilde{\rho}(t)].\label{Schrodingerrot}
\end{equation}

Since in this picture a unitarily-correctable subsystem is
noiseless, we can state the following

\textbf{Theorem 4:} \textit{Consider the evolution
\eqref{Schrodinger} driven by the Hamiltonian \eqref{Hamiltonian}.
Let $\widetilde{H}_S(t)$ and $\widetilde{S}_j(t)$ be the system
Hamiltonian and the interaction operators \eqref{HI} in the
rotating frame \eqref{rotatingframe} with $U(t)$ given by
Eq.~\eqref{defineU}. Then the subsystem $\mathcal{H}^A$ in the
decomposition \eqref{decomposition} is correctable by $U(t)$
during this evolution, if and only if
\begin{gather}
\widetilde{S}_{j}(t)P^{AB}=I^{A}\otimes C^B_j(t),\hspace{0.2cm}\
C^B_j(t)\in\mathcal{B}(\mathcal{H}^B),\hspace{0.2cm}\forall
j\label{Hamilton1}
\end{gather}
and
\begin{gather}
(\widetilde{H}_S(t)+H'(t))P^{AB}=I^A\otimes D^B(t),\hspace{0.2cm}
D^B(t)\in \mathcal{B}(\mathcal{H}^B).\label{Hamilton2}
\end{gather}}

\textbf{Proof:} With respect to the evolution in the rotating
frame \eqref{rotatingframe}, the subsystem $\mathcal{H}^A$ is
noiseless. The theorem follows from the conditions for noiseless
subsystems under Hamiltonian dynamics \cite{ShaLid05} applied to
the Hamiltonian in the rotating frame. Note that the fact that the
operator on the right-hand side of Eq.~\eqref{Hamilton2} sends
states from $\mathcal{H}^A\otimes\mathcal{H}^B$ to
$\mathcal{H}^A\otimes\mathcal{H}^B$ implies that the off-diagonal
terms of $\widetilde{H}_S(t)+H'(t)$ in the block basis
corresponding to the decomposition \eqref{decomposition} vanish,
i.e., $P^{AB}(\widetilde{H}_S(t)+H'(t))P_{\mathcal{K}}=0$.

\textbf{Comment:} The Hamiltonian $H'(t)$ can be obtained from
conditions \eqref{Hamilton1} and \eqref{Hamilton2}. We can choose
$D^B(t)=0$ and define $H'(t)=-\widetilde{H}_S(t)$, which together
with Eq.~\eqref{defineU} yields
\begin{equation}
i\frac{dU(t)}{dt}=-U(t)H_S(t),
\end{equation}
i.e.,
\begin{equation}
U^{\dagger}(t)=\mathcal{T}\textrm{exp}\left(-i\int_0^t H_S(\tau)
d\tau \right).
\end{equation}
This simply means that the evolution of the subspace that contains
the encoded information is driven by the system Hamiltonian.

The conditions again can be separated into two parts. The fact
that the right-hand sides of Eq.~\eqref{Hamilton1} and
Eq.~\eqref{Hamilton2} act trivially on $\mathcal{H}^A$ is
necessary and sufficient for the information stored in the
instantaneous subsystem $\mathcal{H}^A(t)$ to undergo trivial
dynamics at time $t$. The fact that the right-hand-sides of these
equations do not take states outside of
$\mathcal{H}^A\otimes\mathcal{H}^B$ is necessary and sufficient
for states not to leave the subspace
$U^{\dagger}(t)\mathcal{H}^A\otimes\mathcal{H}^B$ as it evolves.

The conditions for general correctability, however, are not
obtained directly from Theorem 4 in analogy to the case of
Markovian decoherence. Such conditions would certainly be
sufficient, but it turns out that they are not necessary. This is
because after applying the unitary recovery operation, the state
of the gauge subsystem $\mathcal{H}^{B'}$ (which is generally
larger than the initial gauge subsystem $\mathcal{H}^B$) plus the
environment would generally belong to a proper subspace of
$\mathcal{H}^{B'}\otimes\mathcal{H}^E$ which cannot be factored
into a subsystem belonging to $\mathcal{H}^S$ and a subsystem
belonging to $\mathcal{H}^E$. Thus it is not necessary that the
Hamiltonian acts trivially on the factor $\mathcal{H}^A$ in
$\mathcal{H}^A\otimes\mathcal{H}^{B'}\otimes\mathcal{H}^E$, but
only on the factor $\mathcal{H}^A$ in $\mathcal{H}^A\otimes
\widetilde{\mathcal{H}}^{BE}$, where
$\widetilde{\mathcal{H}}^{BE}$ is the proper subspace in question.
In the case of unitary correctability, tracing out the environment
provides necessary conditions because $\mathcal{H}^{B'}=
\mathcal{H}^{B}$, and hence $\mathcal{H}^{B}\otimes \mathcal{H}^E$
is fully occupied.

Let
\begin{equation}
\mathcal{H}^S=\mathcal{H}^A\otimes \mathcal{H}^{B'}\oplus
\mathcal{K'}\label{maximaldecomposition}
\end{equation}
be a decomposition of the Hilbert space of the system such that the
factor $\mathcal{H}^{B'}\supset \mathcal{H}^{B}$ has the largest
possible dimension. Since the evolution of the state of the system
plus the environment is unitary, at time $t$ the initial subspace
$\mathcal{H}^A\otimes\mathcal{H}^B\otimes\mathcal{H}^E$ will be
transformed to some other subspace of
$\mathcal{H}^S\otimes\mathcal{H}^E$, which is unitarily related to
the initial one. Applying the unitary recovery operation $U(t)$
returns this subspace to the form
$\mathcal{H}^A\otimes\widetilde{\mathcal{H}}^{BE}(t)$, where
$\widetilde{\mathcal{H}}^{BE}(t)$ is a subspace of
$\mathcal{H}^{B'}\otimes\mathcal{H}^E$. Clearly, there exists a
unitary operator $W_0(t):
\mathcal{H}^{B'}\otimes\mathcal{H}^E\rightarrow
\mathcal{H}^{B'}\otimes\mathcal{H}^E$ that maps this subspace to the
initial subspace $\mathcal{H}^B\otimes \mathcal{H}^E$:
\begin{equation}
W_0(t) \widetilde{P}^{BE}(t) W^{\dagger}_0(t)=P^{BE}.\label{defineW}
\end{equation}
(Here $\widetilde{P}^{BE}(t)$ denotes the projector on
$\widetilde{\mathcal{H}}^{BE}(t)$.) Note that as an operator on the
entire Hilbert space, this unitary has the form $W_0(t)\equiv
I^A\otimes W_0^{B'E}(t)\oplus I_{\mathcal{K'}}\otimes I^E$. Let us
define the frame
\begin{equation}
\widehat{O}(t)=W(t)O(t)W^{\dagger}(t),\label{rotatingframe2}
\end{equation}
where
\begin{equation}
i\frac{dW(t)}{dt}=H''(t)W(t).\label{defineW2}
\end{equation}
Then the evolution driven by a Hamiltonian $G(t)$, in this frame
will be driven by $\widehat{G}(t)+H''(t)$.

\textbf{Theorem 5:} \textit{Let $\widetilde{O}(t)$ denote the
image of an operator $O(t)\in \mathcal{B}(\mathcal{H})$ under the
transformation \eqref{rotatingframe} with
$U(t)\in\mathcal{B}(\mathcal{H}^S)$ given by Eq.~\eqref{defineU}
($H'(t)\in \mathcal{B}(\mathcal{H}^S)$), and let $\widehat{O}(t)$
denote the image of $O(t)$ under the transformation
\eqref{rotatingframe2} with $W(t)$ given by Eq.~\eqref{defineW2}.
Let $P^{ABE}$ be the projector on
$\mathcal{H}^A\otimes\mathcal{H}^B\otimes\mathcal{H}^E$. The
subsystem $\mathcal{H}^A$ in the decomposition
\eqref{decompositionfull} is recoverable by $U(t)$ during the
evolution driven by the system-environment Hamiltonian
$H_{SE}(t)$, if and only if there exists $H''(t)\in
\mathcal{B}(\mathcal{H}^{B'}\otimes\mathcal{H}^E)$, where
$\mathcal{H}^{B'}$ was defined in \eqref{maximaldecomposition},
such that
\begin{gather}
(\widehat{\widetilde{H}}_{SE}(t)+\widehat{H}'(t)+H''(t))P^{ABE}=I^A\otimes
D^{BE}(t),\label{Hamcond}\\
\hspace{0.2cm} D^{BE}(t)\in
\mathcal{B}(\mathcal{H}^{B}\otimes\mathcal{H}^E),
\hspace{0.2cm}\forall t.\notag
\end{gather}}

\textbf{Proof:} Assume that the information encoded in
$\mathcal{H}^A$ is unitarily recoverable by $U(t)$. Consider the
evolution in the frame defined through the unitary operation
$W(t)U(t)$, where $W(t)=W_0(t)$ for some differentiable $W_0(t)$
that satisfies the property \eqref{defineW}. In this frame, which
can be obtained by consecutively applying the transformations
\eqref{rotatingframe} and \eqref{rotatingframe2}, the Hamiltonian is
$\widehat{\widetilde{H}}_{SE}(t)+\widehat{H}'(t)+H''(t)$. Under this
Hamiltonian, the subsystem $\mathcal{H}^A$ must be noiseless and no
states should leave the subspace
$\mathcal{H}^A\otimes\mathcal{H}^B\otimes\mathcal{H}^E$. It is
straightforward to see that the first requirement means that
$\mathcal{H}^A$ must be acted upon trivially by all terms of the
Hamiltonian, hence the factor $I^A$ on the right-hand side of
Eq.~\eqref{Hamcond}. At the same time, the subspace
$\mathcal{\mathcal{H}^B\otimes\mathcal{H}^E}$ must be preserved by
the action of the Hamiltonian, which implies that the factor
$D^{BE}(t)$ on the right-hand side of Eq.~\eqref{Hamcond} must send
states from $\mathcal{H}^B\otimes\mathcal{H}^E$ to
$\mathcal{H}^B\otimes\mathcal{H}^E$. Note that this implies that the
off-diagonal terms of the Hamiltonian in the block form
corresponding to the decomposition \eqref{decompositionfull} must
vanish, i.e.,
$P^{ABE}(\widehat{\widetilde{H}}_{SE}(t)+\widehat{H}'(t)+H''(t))P^{ABE}_{\perp}=0$,
where $P^{ABE}_{\perp}$ denoted the projector on
$\mathcal{K}\otimes\mathcal{H}^E$. Obviously, these conditions are
also sufficient, since they ensure that in the frame defined by the
unitary transformation $W(t)U(t)$, the evolution of $\mathcal{H}^A$
is trivial and states inside the subspace
$\mathcal{H}^B\otimes\mathcal{H}^E$ evolve unitarily under the
action of the Hamiltonian $D^{BE}(t)$. Since $W(t)$ acts on
$\mathcal{H}^{B'}\otimes\mathcal{H}^E$, subsystem $\mathcal{H}^A$ is
invariant also in the rotating frame \eqref{rotatingframe}. This
means that $\mathcal{H}^A$ is recoverable by the unitary $U(t)$.

\textbf{Comment:} Similarly to the previous cases, the unitary
operators $U(t)$ and $W(t)$ can be obtained iteratively from
Eq.~\eqref{Hamcond} if the decomposition \eqref{decomposition} is
given. Since $H''(t)$ acts on
$\mathcal{H}^{B'}\otimes\mathcal{H}^E$, from Eq.~\eqref{Hamcond}
it follows that the operator
$\widehat{\widetilde{H}}_{SE}(t)+\widehat{H}'(t)$ must satisfy
\begin{gather}
(\widehat{\widetilde{H}}_{SE}(t)+\widehat{H}'(t))P^{ABE}=I^A\otimes
F^{B'E}(t), \hspace{0.4cm} F^{B'E}(t)\in
\mathcal{B}(\mathcal{H}^{B'}\otimes\mathcal{H}^E).
\label{determineH'}
\end{gather}
At the same time, we can choose $H''(t)$ so that $D^{BE}(t)=0$.
This corresponds to
\begin{equation}
W(t)\widetilde{\mathcal{H}}^{BE}(t)=
\mathcal{H}^B\otimes\mathcal{H}^E,
\end{equation}
where $\widetilde{\mathcal{H}}^{BE}(t)$ was defined in the
discussion before Theorem 5. To ensure $D^{BE}(t)=0$, we can
choose
\begin{equation}
H''(t)=-\widehat{\widetilde{H}}_{SE}(t)-\widehat{H}'(t)+\mathcal{P}^{ABE}_{\perp}\left(\widehat{\widetilde{H}}_{SE}(t)+\widehat{H}'(t)\right),\label{determineH''}
\end{equation}
where
$\mathcal{P}^{ABE}_{\perp}(\cdot)=P^{ABE}_{\perp}(\cdot)P^{ABE}_{\perp}$.
For $t=0$ ($U(0)=I$, $W(0)=I$), we can find a solution for
$\widehat{H}'(0)={H}'(0)$ from Eq.~\eqref{determineH'}, given the
Hamiltonian $\widehat{\widetilde{H}}_{SE}(0)=H_{SE}(0)$. Plugging
the solution in Eq.~\eqref{determineH''}, we can obtain $H''(0)$.
For the unitaries after a single time step $dt$ we then have
\begin{equation}
U(dt)=I-iH'(0)dt+\textit{O}(dt^2),
\end{equation}
\begin{equation}
W(dt)=I-iH''(0)dt+\textit{O}(dt^2).
\end{equation}
Using $U(dt)$ and $W(dt)$ we can calculate
$\widehat{\widetilde{H}}_{SE}(dt)$ according to
Eq.~\eqref{rotatingframe} and Eq.~\eqref{rotatingframe2}. Then we
can solve Eq.~\eqref{determineH'} for
$\widehat{H}'(dt)=W(dt)H'(dt)W^{\dagger}(dt)$, which we can use in
Eq.~\eqref{determineH''} to find $H''(dt)$, and so on. Note that
here we cannot specify a simple expression for $\widehat{H}'(t)$
in terms of $\widehat{\widetilde{H}}_{SE}(t)$, since we do not
have the freedom to choose fully $F^{B'E}(t)$ in
Eq.~\eqref{determineH'} due to the restriction that $H'(t)$ acts
locally on $\mathcal{H}^S$.

We point out that condition \eqref{Hamcond} again can be
understood as consisting of two parts---the fact that the
right-hand side acts trivially on $\mathcal{H}^A$ is necessary and
sufficient for the instantaneous dynamics undergone by the
subsystem $U^{\dagger}(t)W^{\dagger}(t)\mathcal{H}^A$ at time $t$
to be trivial, while the fact that it preserves
$\mathcal{H}^A\otimes\mathcal{H}^{B}\otimes\mathcal{H}^E$ is
necessary and sufficient for states not to leave
$U^{\dagger}(t)W^{\dagger}(t)
\mathcal{H}^A\otimes\mathcal{H}^{B}\otimes\mathcal{H}^E$ as it
evolves.

It is tempting to perform an argument similar to the one we
presented for the Markovian case about the possible relation of
the specified recovery unitary operation $U(t)$ and the optimal
error-correcting map in the case of approximate error correction.
If the encoded information is not perfectly preserved, we can
construct the unitary operation $U(t)$ as explained in the comment
after Theorem 5 by optimally approximating Eq.~\eqref{determineH'}
and Eq.~\eqref{determineH''}. However, in this case the evolution
is not irreversible and the information that leaks out of the
system may return back to it. Thus we cannot argue that the
unitary map specified in this manner would optimally track the
remaining encoded information.

\subsection*{6.5.2 \hspace{2pt} Conditions depending on the initial state of the
environment} \addcontentsline{toc}{subsection}{6.5.2
\hspace{0.15cm} Conditions depending on the initial state of the
environment}

We can easily extend Theorem 5 to the case when the initial state
of the environment belongs to a particular subspace
$\mathcal{H}^{E_0}\in\mathcal{H}^E$. The only modification is that
instead of $P^{ABE}$ in Eq.~\eqref{Hamcond}, we must have
$P^{ABE_0}$, where $P^{ABE_0}$ is the projector on
$\mathcal{H}^A\otimes\mathcal{H}^B\otimes \mathcal{H}^{E_0}$, and
on the right-hand side must have $D^{BE_0}(t)\in
\mathcal{B}(\mathcal{H}^{B}\otimes\mathcal{H}^{E_0})$.

The following two theorems follow by arguments analogous to those
for Theorem 5. We assume the same definitions as in Theorem 5
(Eq.~\eqref{rotatingframe}, Eq.~\eqref{defineU},
Eq.~\eqref{rotatingframe2}, Eq.~\eqref{defineW2} ), except that in
the second theorem we restrict the definition of $H''(t)$.

\textbf{Theorem 6:} \textit{Let $P^{ABE_0}$ be the projector on
$\mathcal{H}^A\otimes\mathcal{H}^B\otimes\mathcal{H}^{E_0}$, where
$\mathcal{H}^{E_0}\in \mathcal{H}^{E}$. The subsystem
$\mathcal{H}^A$ in the decomposition \eqref{decompositionfull} is
recoverable by $U(t)\in\mathcal{B}(\mathcal{H}^S)$ during the
evolution driven by the system-environment Hamiltonian $H_{SE}(t)$
when the state of the environment is initialized inside
$\mathcal{H}^{E_0}$, if and only if there exists $H''(t)\in
\mathcal{B}(\mathcal{H}^{B'}\otimes\mathcal{H}^E)$ such that
\begin{gather}
(\widehat{\widetilde{H}}_{SE}(t)+\widehat{H}'(t)+H''(t))P^{ABE_0}=I^A\otimes
D^{BE_0}(t),\label{Hamcondano}\\
\hspace{0.2cm} D^{BE_0}(t)\in
\mathcal{B}(\mathcal{H}^{B}\otimes\mathcal{H}^{E_0}),
\hspace{0.2cm}\forall t.\notag
\end{gather}}

The conditions for unitary correctability in this case require the
additional restriction that $W(t)$ acts on $\mathcal{H}^B\otimes
\mathcal{H}^E$ and not on $\mathcal{H}^{B'}\otimes \mathcal{H}^E$,
since in this case $U(t)$ brings the state inside
$\mathcal{H}^A\otimes\mathcal{H}^B\otimes\mathcal{H}^E$. Notice
that when the state of the environment is initialized in a
particular subspace, we cannot use conditions for unitary
correctability similar to those in Theorem 4. This is because
after the correction $U(t)$, the state of the gauge subsystem plus
environment may belong to a proper subspace of
$\mathcal{H}^B\otimes \mathcal{H}^E$ and tracing out the
environment would not yield necessary conditions.

\textbf{Theorem 7:} \textit{Let $P^{ABE_0}$ be the projector on
$\mathcal{H}^A\otimes\mathcal{H}^B\otimes\mathcal{H}^{E_0}$, where
$\mathcal{H}^{E_0}\in \mathcal{H}^{E}$. The subsystem
$\mathcal{H}^A$ in the decomposition \eqref{decompositionfull} is
correctable by $U(t)\in\mathcal{B}(\mathcal{H}^S)$ during the
evolution driven by the system-environment Hamiltonian $H_{SE}(t)$
when the state of the environment is initialized inside
$\mathcal{H}^{E_0}$, if and only if there exists $H''(t)\in
\mathcal{B}(\mathcal{H}^{B}\otimes\mathcal{H}^E)$ such that
\begin{gather}
(\widehat{\widetilde{H}}_{SE}(t)+\widehat{H}'(t)+H''(t))P^{ABE_0}=I^A\otimes
D^{BE_0}(t),\label{Hamcondanother}\\
\hspace{0.2cm} D^{BE_0}(t)\in
\mathcal{B}(\mathcal{H}^{B}\otimes\mathcal{H}^{E_0}),
\hspace{0.2cm}\forall t.\notag
\end{gather}}

Notice that the conditions of Theorem 6 and Theorem 7 do not
depend on the particular initial state of the environment but only
on the subspace to which it belongs. This can be understood by
noticing that different environment states inside the same
subspace give rise to Kraus operators \eqref{Krausoperator} which
are linear combinations of each other. The discretization of
errors in operator quantum error correction \cite{KLP05, KLPL06}
implies that all such maps will be correctable.

The conditions for correctable dynamics dependent on the state of
the environment could be useful if we are able to prepare the
state of the environment in the necessary subspace. The
environment, however, is generally outside of the experimenter's
control. Nevertheless, it is conceivable that the experimenter may
have some control over the environment (for example, by varying
its temperature), which for certain Hamiltonians could bring the
environment state close to a subspace for which the evolution of
the system is correctable. It is important to point out that
according to the result we derive in the next chapter, the error
due to imperfect initialization of the bath will not increase
under the evolution.

\section*{6.6 \hspace{2pt} Summary and outlook}
\addcontentsline{toc}{section}{6.6 \hspace{0.15cm} Summary and
outlook}

We have derived conditions for correctability of subsystems under
continuous decoherence. We first presented conditions for the case
when the evolution can be described by a CPTP linear map. These
conditions are equivalent to those known for operator codes
\cite{KLP05, KLPL06} except that we consider them for
time-dependent noise processes. We then derived condition for the
case of Markovian decoherence and general Hamiltonian evolution of
the system and the environment. We derived conditions for both
unitary correctability and general correctability, using the fact
that correctable subsystems are unitarily recoverable \cite{KS06}.

The conditions for correctability in both Markovian and
Hamiltonian evolution can be understood as consisting of two
parts---the first is necessary and sufficient for lack of noise
inside the instantaneous subsystem that contains the information,
and the second is necessary and sufficient for states not to leave
the subsystem as it evolves with time. In this sense, the new
conditions can be thought of a generalizations of the conditions
for noiseless subsystems to the case where the subsystem is
time-dependent.

In the Hamiltonian case, the conditions for unitary correctability
concern only the action of the Hamiltonian on the system, whereas
the conditions for general correctability concern the entire
system-bath Hamiltonian. The reason for this is that the state of
the gauge subsystem plus the environment generally belongs to a
particular subspace, which does not factor into sectors belonging
separately to the system and the environment. We also derived
conditions in the Hamiltonian case that depend on the initial
state of the environment. These conditions could be useful, in
principle, since errors due to imperfect initialization of the
environment do not increase under the evolution. Furthermore,
these conditions could provide a better understanding of
correctability under CPTP maps, since a CPTP map that results from
Hamiltonian evolution depends on both the Hamiltonian and the
initial state of the environment. An interesting generalization of
this work would be to derive similar condition for the case of the
Nakajima-Zwanzig or the TCL master equations.

We discussed possible implications of the conditions we derived
for the problem of optimal recovery in the case of imperfectly
preserved information. We hope that the results obtained in this
study will provide insight into the mechanisms of information flow
under decoherence that could be useful in the area of approximate
error correction as well.

\chapter*{Chapter 7: \hspace{1pt} Robustness of operator quantum error correction against initialization errors}
\addcontentsline{toc}{chapter}{Chapter 7:\hspace{0.15cm}
Robustness of operator quantum error correction against
initialization errors}

The conditions we derived in the previous chapter, as well as the
standard OQEC conditions for discrete errors, depend on the
assumption that states are perfectly initialized inside the
subspace factored by the correctable subsystem. In practice,
however, perfect initialization of the state may not be easy to
achieve. Hence, it is important to understand to what extent the
preparation requirement can be relaxed. In this chapter, we
examine the performance of OQEC in the case of imperfect encoding.

\section*{7.1 \hspace{2pt} Preliminaries}
\addcontentsline{toc}{section}{7.1 \hspace{0.15cm} Preliminaries}

As can be seen from the definitions \eqref{noiselesssystem} and
\eqref{correctablesystem}, the concept of noiseless subsystem is a
cornerstone in the theory of OQEC; it serves as a basis for the
definition of correctable subsystem and error correction in
general. As shown in Ref.~\cite{ShaLid05}, in order to ensure
perfect noiselessness of a subsystem in the case of imperfect
initialization, the noise process has to satisfy more restrictive
conditions than those required in the case of perfect
initialization. It was believed that these conditions are
necessary if a noiseless (or more generally decoherence-free)
subsystem is to be robust against arbitrarily large initialization
errors. The fundamental relation between a noiseless subsystem and
a correctable subsystem implies that in the case of imperfect
initialization, more restrictive conditions would be needed for
OQEC codes as well.

In this chapter we show that with respect to the ability of a code
to protect from errors, more restrictive conditions are not
necessary. For this purpose, we define a measure of the fidelity
between the encoded information in two states for the case of
subsystem encoding. We first give an intuitive motivation for the
definition, and then study the properties of the measure. We then
show that the effective noise that can arise inside the code due
to imperfect initialization under the standard conditions, is such
that it can only increase the fidelity of the encoded information
with the information encoded in a perfectly prepared state. This
robustness against initialization errors is shown to hold also
when the state is subject to encoded operations.

\section*{7.2 \hspace{2pt} Review of the noiseless-subsystem conditions on the Kraus operators}
\addcontentsline{toc}{section}{7.2 \hspace{0.15cm} Review of the
noiseless-subsystem conditions on the Kraus operators}

For simplicity, we consider again the case where information is
stored in only one subsystem, i.e., we consider the decomposition
\eqref{decomposition}. The definition of noiseless subsystem
\eqref{noiselesssystem} implies that the information encoded in
$\mathcal{B}(\mathcal{H}^A)$ remains invariant after the process
$\mathcal{E}$, if the initial density operator of the system
$\rho(0)$ belongs to
$\mathcal{B}(\mathcal{H}^A\otimes\mathcal{H}^B)$. If, however, one
allows imperfect initialization,
$\rho(0)\neq\mathcal{P}^{AB}(\rho(0))$, this need not be the case.
Consider the ``initialization-free" analogue of the definition
\eqref{noiselesssystem}:
\begin{gather}
\textrm{Tr}_B\{(\mathcal{P}^{AB}\circ\mathcal{E})(\sigma)\}=\textrm{Tr}_B\{\mathcal{P}^{AB}(\sigma)\},\label{IFnoiselesssystem}
\hspace{0.4cm} \textrm{for all } \sigma\in
\mathcal{B}(\mathcal{H}^S).\notag
\end{gather}
Obviously Eq.~\eqref{IFnoiselesssystem} implies
Eq.~\eqref{noiselesssystem}, but the reverse is not true. As shown
in \cite{ShaLid05}, the definition \eqref{IFnoiselesssystem}
imposes more restrictive conditions on the channel $\mathcal{E}$
than those imposed by \eqref{noiselesssystem}. To see this,
consider the form of the Kraus operators $M_i$ of $\mathcal{E}$
(\eqref{Kraus1}) in the block basis corresponding to the
decomposition \eqref{decomposition}. From a result derived in
\cite{ShaLid05} it follows that the subsystem $\mathcal{H}^A$ is
noiseless in the sense of Eq.~\eqref{noiselesssystem}, if and only
if the Kraus operators have the form
\begin{equation}
M_i=\begin{bmatrix} I^A\otimes C_i^B&D_i\\
0&G_i
\end{bmatrix},\label{Krausoperatorsblock}
\end{equation}
where the upper left block corresponds to the subspace
$\mathcal{H}^A\otimes\mathcal{H}^B$, and the lower right block
corresponds to $\mathcal{K}$. The completeness relation
\eqref{completeness} implies the following conditions on the
operators $C^B_i$, $D_i$, and $G_i$:
\begin{gather}
\underset{i}{\sum}C_i^{\dagger B}C_i^B=I^B,\label{one}\\
\underset{i}{\sum}I^A\otimes C_i^{\dagger B}D_i=0,\label{two}\\
\underset{i}{\sum}(D_i^{\dagger}D_i+G_i^{\dagger}G_i)=I_{\mathcal{K}}\label{three}.
\end{gather}

In the same block basis, a perfectly initialized state $\rho$ and
its image under the map \eqref{Krausoperatorsblock} have the form
\begin{gather}
{\rho}=\begin{bmatrix} {\rho}_1&0\\
0&0
\end{bmatrix}, \hspace{0.2cm} \mathcal{E}(\rho)=\begin{bmatrix}
\rho_1'&0\\
0&0
\end{bmatrix},\label{perfectini}
\end{gather}
where $\rho_1'=\underset{i}{\sum}I^A\otimes C_i^B\rho_1 I^A\otimes
{C_i^{\dagger B}}$. Using the linearity and cyclic invariance of
the trace together with Eq.~\eqref{one}, we obtain
\begin{gather}
\textrm{Tr}_B\{(\mathcal{P}^{AB}\circ\mathcal{E})(\rho)
\}=\textrm{Tr}_B\{\underset{i}{\sum}I^A\otimes C_i^B\rho_1
I^A\otimes {C_i^{\dagger B}}\}\notag\\=\textrm{Tr}_B\{\rho_1
\underbrace{\underset{i}{\sum}I^A\otimes C_i^{\dagger
B}C_i^B}_{I^A\otimes
I^B}\}=\textrm{Tr}_B\{\mathcal{P}^{AB}(\rho)\},\label{reducedrho}
\end{gather}
i.e., the reduced operator on $\mathcal{H}^A$ remains invariant.

On the other hand, an imperfectly initialized state $\tilde{\rho}$
and its image have the form
\begin{equation}
\tilde{\rho}=\begin{bmatrix} \tilde{\rho}_1&\tilde{\rho}_2\\
\tilde{\rho}_2^{\dagger}&\tilde{\rho}_3
\end{bmatrix}, \hspace{0.2cm} \mathcal{E}(\tilde{\rho})=\begin{bmatrix}
\tilde{\rho}_1'&\tilde{\rho}_2'\\
\tilde{\rho}_2'^{\dagger}&\tilde{\rho}_3'
\end{bmatrix}.\label{imperfect}
\end{equation}
Here $\tilde{\rho}_2$ and/or $\tilde{\rho}_3$ are non-vanishing,
and
\begin{gather}
\tilde{\rho}_1'=\underset{i}{\sum}(I^A\otimes C_i^B\tilde{\rho}_1
I^A\otimes {C_i^{\dagger B}}+D_i\tilde{\rho}_2^{\dagger}I^A\otimes
{C^{\dagger B}_i}+I^A\otimes C^B_i\tilde{\rho}_2
D_i^{\dagger}+D_i\tilde{\rho}_3D_i^{\dagger}),\\
\tilde{\rho}_2'=\underset{i}{\sum}(I^A\otimes C_i^B\tilde{\rho}_2
G_i^{\dagger}+D_i\tilde{\rho}_3G_i^{\dagger}),\\
\tilde{\rho}_3'=\underset{i}{\sum}G_i\tilde{\rho}_3G_i^{\dagger}.
\end{gather}
In this case, using the linearity and cyclic invariance of the
trace together with Eq.~\eqref{one} and Eq.~\eqref{two}, we obtain
\begin{eqnarray}
\textrm{Tr}_B\{(\mathcal{P}^{AB}\circ\mathcal{E})(\tilde{\rho})
\}&=&\textrm{Tr}_B\{\underset{i}{\sum}(I^A\otimes
C_i^B\tilde{\rho}_1 I^A\otimes {C_i^{\dagger B}}
+D_i\tilde{\rho}_2^{\dagger}I^A\otimes {C^{\dagger
B}_i}\notag\\
&& +I^A\otimes C^B_i\tilde{\rho}_2
D_i^{\dagger}+D_i\tilde{\rho}_3D_i^{\dagger})\}\label{reducedrhotilde}
\\&=&\textrm{Tr}_B\{\tilde{\rho}_1
\underbrace{\underset{i}{\sum}I^A\otimes C_i^{\dagger
B}C_i^B}_{I^A\otimes
I^B}\}+\textrm{Tr}_B\{(\underbrace{\underset{i}{\sum}{I^A\otimes
C^{\dagger B}_i}D_i}_{0})\tilde{\rho}_2^{\dagger}
\}\notag\\
&&
+\textrm{Tr}_B\{\tilde{\rho}_2(\underbrace{\underset{i}{\sum}{I^A\otimes
C^{\dagger
B}_i}D_i}_{0})^{\dagger}\}+\textrm{Tr}_B\{\underset{i}{\sum}D_i\tilde{\rho}_3D_i^{\dagger}
\}\notag\\
&=&\textrm{Tr}_B\tilde{\rho}_1+\textrm{Tr}_B\{\underset{i}{\sum}D_i\tilde{\rho}_3D_i^{\dagger}
\}\neq
\textrm{Tr}_B\tilde{\rho}_1\equiv\textrm{Tr}_B\{\mathcal{P}^{AB}(\tilde{\rho})\},\notag
\end{eqnarray}
i.e., the reduced operator on $\mathcal{H}^A$ is not preserved. It
is easy to see that the reduced operator would be preserved for
every imperfectly initialized state if and only if we impose the
additional condition
\begin{equation}
D_i=0, \hspace{0.2cm} \textrm{for all } i.\label{extraconstraint}
\end{equation}
This further restriction to the form of the Kraus operators is
equivalent to the requirement that there are no transitions from
the subspace $\mathcal{K}$ to the subspace
$\mathcal{H}^A\otimes\mathcal{H}^B$ under the process
$\mathcal{E}$. This is in addition to the requirement that no
states leave $\mathcal{H}^A\otimes\mathcal{H}^B$, which is ensured
by the vanishing lower left blocks of the Kraus operators
\eqref{Krausoperatorsblock}. Condition \eqref{extraconstraint}
automatically imposes an additional restriction on the
error-correction conditions, since if $\mathcal{R}$ is an
error-correcting map in this ``initialization-free" sense, the map
$\mathcal{R}\circ \mathcal{E}$ would have to satisfy
Eq.~\eqref{extraconstraint}. But is this constraint necessary from
the point of view of the ability of the code to correct further
errors?

Notice that since $\tilde{\rho}$ is a positive operator,
$\tilde{\rho_3}$ is positive, and hence
$\textrm{Tr}_B\{\underset{i}{\sum}D_i\tilde{\rho}_3D_i^{\dagger}
\}$ is positive. The reduced operator on subsystem
$\mathcal{H}^A$, although unnormalized, can be regarded as a
(partial) probability mixture of states on $\mathcal{H}^A$. The
noise process modifies the original mixture
($\textrm{Tr}_B\tilde{\rho}_1$) by \textit{adding} to it another
partial mixture (the positive operator
$\textrm{Tr}_B\{\underset{i}{\sum}D_i\tilde{\rho}_3D_i^{\dagger}
\}$). Since the weight of any state already present in the mixture
can only increase by this process, this should not worsen the
faithfulness with which information is encoded in $\tilde{\rho}$.
In order to make this argument rigorous, however, we need a
measure that quantifies the faithfulness of the encoding.

\section*{7.3 \hspace{2pt} Fidelity between the encoded information
in two states} \addcontentsline{toc}{section}{7.3 \hspace{0.15cm}
Fidelity between the encoded information in two states}

\subsection*{7.3.1 \hspace{2pt} Motivating the definition}
\addcontentsline{toc}{subsection}{7.3.1 \hspace{0.15cm} Motivating
the definition}

If we consider two states with density operators $\tau$ and
$\upsilon$, a good measure of the faithfulness with which one
state represents the other is given by the fidelity between the
states:
\begin{equation}
F(\tau,
\upsilon)=\textrm{Tr}\sqrt{\sqrt{\tau}\upsilon\sqrt{\tau}}.\label{fidelity}
\end{equation}
This quantity can be thought of as a square root of a generalized
``transition probability" between the two states $\tau$ and
$\upsilon$ as defined by Uhlmann \cite{Uhl76}. Another
interpretation due to Fuchs \cite{Fuchs96} gives an operational
meaning of the fidelity as the minimal overlap between the
probability distributions generated by all possible generalized
measurements on the states:
\begin{equation}
F(\tau,
\upsilon)=\underset{\{M_i\}}{\textrm{min}}\underset{i}{\sum}\sqrt{\textrm{Tr}\{
M_i\tau\}}\sqrt{\textrm{Tr}\{ M_i\upsilon\}}.\label{fidelityoper}
\end{equation}
Here, minimum is taken over all positive operators $\{M_i\}$ that
form a positive operator valued measure (POVM) \cite{Kraus83},
$\underset{i}{\sum}M_i=I^S$.

In our case, we need a quantity that compares the \textit{encoded}
information in two states. Clearly, the standard fidelity between
the states will not do since it measures the similarity between
the states on the entire Hilbert space. The encoded information,
however, concerns only the reduced operators on subsystem
$\mathcal{H}^A$. In view of this, we propose the following

\textbf{Definition 1:} Let $\tau$ and $\upsilon$ be two density
operators on a Hilbert space $\mathcal{H}^S$ with decomposition
\eqref{decomposition}. The fidelity between the information
encoded in subsystem $\mathcal{H}^A$ in the two states is given
by:
\begin{equation}
F^A(\tau, \upsilon)=\underset{\tau',
\upsilon'}{\textrm{max}}F(\tau',\upsilon'),\label{measure}
\end{equation}
where maximum is taken over all density operators $\tau'$ and
$\upsilon'$ that have the same reduced operators on
$\mathcal{H}^A$ as $\tau$ and $\upsilon$: $\textrm{Tr}_B\{
\mathcal{P}^{AB}(\tau') \} = \textrm{Tr}_B\{
\mathcal{P}^{AB}(\tau) \}$, $\textrm{Tr}_B\{
\mathcal{P}^{AB}(\upsilon') \} = \textrm{Tr}_B\{
\mathcal{P}^{AB}(\upsilon) \}$.

The intuition behind this definition is that by maximizing over
all states that have the same reduced operators on $\mathcal{H}^A$
as the states being compared, we ensure that the measure does not
penalize for differences between the states that are not due
specifically to differences between the reduced operators.

\subsection*{7.3.2 \hspace{2pt} Properties of the measure}
\addcontentsline{toc}{subsection}{7.3.2 \hspace{0.15cm} Properties
of the measure}

\textbf{Property 1 (Symmetry):} Since the fidelity is symmetric
with respect to its inputs, it is obvious from Eq.~\eqref{measure}
that $F^A$ is also symmetric:
\begin{equation}
F^A(\tau, \upsilon)=F^A(\upsilon, \tau).
\end{equation}

Although intuitive, the definition \eqref{measure} does not allow
for a simple calculation of $F^A$. We now derive an equivalent
form for $F^A$, which is simple and easy to compute. Let
$\mathcal{P}_{\mathcal{K}}(\cdot)=P_\mathcal{K}(\cdot)P_\mathcal{K}$
denote the superoperator projector on $\mathcal{B}(\mathcal{K})$,
and let
\begin{gather}
\rho^A\equiv\textrm{Tr}_B\{\mathcal{P}^{AB}(\rho)\}/\textrm{Tr}\{
\mathcal{P}^{AB}(\rho)\}
\end{gather}
denote the \textit{normalized} reduced operator of $\rho$ on
$\mathcal{H}^A$.

\textbf{Theorem 1:} The definition \eqref{measure} is equivalent
to
\begin{eqnarray}
F^A(\tau, \upsilon)=f^A(\tau,\upsilon)\label{easyform}
+\sqrt{\textrm{Tr}\{
\mathcal{P}_{\mathcal{K}}(\tau)\}\textrm{Tr}\{
\mathcal{P}_{\mathcal{K}}(\upsilon)\}},
\end{eqnarray}
where
\begin{gather}
f^A(\tau,\upsilon)=\sqrt{\textrm{Tr}\{
\mathcal{P}^{AB}(\tau)\}\textrm{Tr}\{
\mathcal{P}^{AB}(\upsilon)\}}F(\tau^A,
\upsilon^A).\label{intermsofF}
\end{gather}

\textbf{Proof:} Let $\tau^*$ and $\upsilon^*$ be two states for
which the maximum on the right-hand side of Eq.~\eqref{measure} is
attained. From the monotonicity of the standard fidelity under
CPTP maps \cite{BCF96} it follows that
\begin{gather}
F^A(\tau, \upsilon)=F(\tau^*, \upsilon^*)\leq F(\Pi(\tau^*),
\Pi(\upsilon^*)),
\end{gather}
where
$\Pi(\cdot)=\mathcal{P}^{AB}(\cdot)+\mathcal{P}_{\mathcal{K}}(\cdot)$.
But the states $\Pi(\tau^*)$ and $\Pi(\upsilon^*)$ satisfy
\begin{gather}
\textrm{Tr}_B\{ \mathcal{P}^{AB}(\Pi(\tau^*)) \} = \textrm{Tr}_B\{
\mathcal{P}^{AB}(\tau) \},\label{edno}\\
 \textrm{Tr}_B\{
\mathcal{P}^{AB}(\Pi(\tau^*))\} = \textrm{Tr}_B\{
\mathcal{P}^{AB}(\upsilon) \}\label{dve},
\end{gather}
i.e., they are among those states over which the maximum in
Eq.~\eqref{measure} is taken. Therefore,
\begin{gather}
F^A(\tau, \upsilon)= F(\Pi(\tau^*),
\Pi(\upsilon^*)).\label{pistar}
\end{gather}
Using Eq.~\eqref{fidelity} and the fact that in the block basis
corresponding to the decomposition \eqref{decomposition} the
states $\Pi(\tau^*)$ and $\Pi(\upsilon^*)$ have block-diagonal
forms, it is easy to see that
\begin{gather}
F(\Pi(\tau^*), \Pi(\upsilon^*))=
\check{F}(\mathcal{P}^{AB}(\tau^*), \mathcal{P}^{AB}(\upsilon^*))
+ \check{F}(\mathcal{P}_{\mathcal{K}}(\tau^*),
\mathcal{P}_{\mathcal{K}}(\upsilon^*)),\label{intermediate}
\end{gather}
where $\check{F}$ is a function that has the same expression as
the fidelity \eqref{fidelity}, but is defined over all positive
operators. From Eq.~\eqref{edno} and Eq.~\eqref{dve} it can be
seen that $\textrm{Tr}\{ \mathcal{P}^{AB}(\tau^*)\}=\textrm{Tr}\{
\mathcal{P}^{AB}(\tau)\}$, $\textrm{Tr}\{
\mathcal{P}^{AB}(\upsilon^*)\}=\textrm{Tr}\{
\mathcal{P}^{AB}(\upsilon)\}$, which also implies that
$\textrm{Tr}\{ \mathcal{P}_{\mathcal{K}}(\tau^*)\}=\textrm{Tr}\{
\mathcal{P}_{\mathcal{K}}(\tau)\}=1-\textrm{Tr}\{
\mathcal{P}^{AB}(\tau)\}$, $\textrm{Tr}\{
\mathcal{P}_{\mathcal{K}}(\upsilon^*)\}=\textrm{Tr}\{
\mathcal{P}_{\mathcal{K}}(\upsilon)\}=1-\textrm{Tr}\{
\mathcal{P}^{AB}(\upsilon)\}$. The two terms on the right-hand
side of Eq.~\eqref{intermediate} can therefore be written as
\begin{gather}
\check{F}(\mathcal{P}^{AB}(\tau^*), \mathcal{P}^{AB}(\upsilon^*))=
\sqrt{\textrm{Tr}\{ \mathcal{P}^{AB}(\tau)\}\textrm{Tr}\{
\mathcal{P}^{AB}(\upsilon)\}}\notag\\\times
F\left(\frac{\mathcal{P}^{AB}(\tau^*)}{\textrm{Tr}\{
\mathcal{P}^{AB}(\tau)\}},
\frac{\mathcal{P}^{AB}(\upsilon^*)}{\textrm{Tr}\{
\mathcal{P}^{AB}(\upsilon)\}}\right),\label{firstterm}
\end{gather}
\begin{gather}
\check{F}(\mathcal{P}_{\mathcal{K}}(\tau^*),
\mathcal{P}_{\mathcal{K}}(\upsilon^*))= \sqrt{\textrm{Tr}\{
\mathcal{P}_{\mathcal{K}}(\tau)\}\textrm{Tr}\{
\mathcal{P}_{\mathcal{K}}(\upsilon)\}}\notag\\\times
F\left(\frac{\mathcal{P}_{\mathcal{K}}(\tau^*)}{\textrm{Tr}\{
\mathcal{P}_{\mathcal{K}}(\tau)\}},
\frac{\mathcal{P}_{\mathcal{K}}(\upsilon^*)}{\textrm{Tr}\{
\mathcal{P}_{\mathcal{K}}(\upsilon)\}}\right).\label{secondterm}
\end{gather}
Since $\tau^*$ and $\sigma^*$ should maximize the right-hand side
of Eq.~\eqref{intermediate}, and the only restriction on
$\mathcal{P}_{\mathcal{K}}(\tau^*)$ and
$\mathcal{P}_{\mathcal{K}}(\upsilon^*)$ is $\textrm{Tr}\{
\mathcal{P}_{\mathcal{K}}(\tau^*)\}=\textrm{Tr}\{
\mathcal{P}_{\mathcal{K}}(\tau)\}$, $\textrm{Tr}\{
\mathcal{P}_{\mathcal{K}}(\upsilon^*)\}=\textrm{Tr}\{
\mathcal{P}_{\mathcal{K}}(\upsilon)\}$, we must have
\begin{gather}
F\left(\frac{\mathcal{P}_{\mathcal{K}}(\tau^*)}{\textrm{Tr}\{
\mathcal{P}_{\mathcal{K}}(\tau)\}},
\frac{\mathcal{P}_{\mathcal{K}}(\upsilon^*)}{\textrm{Tr}\{
\mathcal{P}_{\mathcal{K}}(\upsilon)\}}\right)=1,
\end{gather}
i.e.,
\begin{gather}
\frac{\mathcal{P}_{\mathcal{K}}(\tau^*)}{\textrm{Tr}\{
\mathcal{P}_{\mathcal{K}}(\tau)\}} =
\frac{\mathcal{P}_{\mathcal{K}}(\upsilon^*)}{\textrm{Tr}\{
\mathcal{P}_{\mathcal{K}}(\upsilon)\}}.\label{tauupsstar}
\end{gather}
Thus we obtain
\begin{eqnarray}
\check{F}(\mathcal{P}_{\mathcal{K}}(\tau^*),
\mathcal{P}_{\mathcal{K}}(\upsilon^*))= \sqrt{\textrm{Tr}\{
\mathcal{P}_{\mathcal{K}}(\tau)\}\textrm{Tr}\{
\mathcal{P}_{\mathcal{K}}(\upsilon)\}}.
\end{eqnarray}
The term \eqref{firstterm} also must be maximized. Applying again
the monotonicity of the fidelity under CPTP maps for the map
$\Gamma(\rho^{AB})=\textrm{Tr}_B\{\rho^{AB}\}\otimes
|0^B\rangle\langle 0^B|$ defined on operators over
$\mathcal{H}^A\otimes\mathcal{H}^B$, where $|0^B\rangle$ is some
state in $\mathcal{H}^B$, we see that the term \eqref{firstterm}
must be equal to
\begin{gather}
\check{F}(\mathcal{P}^{AB}(\tau^*),
\mathcal{P}^{AB}(\upsilon^*))=\sqrt{\textrm{Tr}\{
\mathcal{P}^{AB}(\tau)\}\textrm{Tr}\{
\mathcal{P}^{AB}(\upsilon)\}} F(\tau^A,\upsilon^A)\equiv f^A(\tau,
\upsilon).\label{tauupsstar2}
\end{gather}
This completes the proof.

We next provide an operational interpretation of the measure
$F^A$. For this we need the following

\textbf{Lemma:} The function $f^A(\tau,\upsilon)$ defined in
Eq.~\eqref{intermsofF} equals the minimum overlap between the
statistical distributions generated by all local measurements on
subsystem $\mathcal{H}^A$:
\begin{equation}
f^A(\tau,
\upsilon)=\underset{\{M_i\}}{\textrm{min}}\underset{i}{\sum}\sqrt{\textrm{Tr}\{
M_i\tau\}}\sqrt{\textrm{Tr}\{ M_i\upsilon\}},\label{fA}
\end{equation}
where $M_i=M_i^A\otimes I^B$, $\underset{i}{\sum}M_i=I^A\otimes
I^B$, $M^A_i>0$, for all $i$.

Note that since the operators $M_i$ do not form a complete POVM on
the entire Hilbert space, the probability distributions
$p_{\tau}(i)=\textrm{Tr}\{ M_i \tau\}$ and
$p_{\upsilon}(i)=\textrm{Tr}\{ M_i \upsilon\}$ generated by such
measurements generally do not sum up to 1. This reflects the fact
that a measurement on subsystem $\mathcal{H}^A$ requires a
projection onto the subspace $\mathcal{H}^A\otimes \mathcal{H}^B$,
i.e., it is realized through post-selection.

\textbf{Proof:} Using that
\begin{gather}
\textrm{Tr}\{ M_i \tau\} =\textrm{Tr}\{M^A_i\otimes I^B
\mathcal{P}^{AB}(\tau)\}=\textrm{Tr}\{
\mathcal{P}^{AB}(\tau)\}\textrm{Tr}\{M^A_i\otimes I^B
\frac{\mathcal{P}^{AB}(\tau)}{\textrm{Tr}\{
\mathcal{P}^{AB}(\tau)\}}\}\notag\\
=\textrm{Tr}\{ \mathcal{P}^{AB}(\tau)\}\textrm{Tr}\{M^A_i\tau^A\},
\end{gather}
we can write Eq.~\eqref{fA} in the form
\begin{gather}
f^A(\tau,\upsilon)=\sqrt{\textrm{Tr}\{
\mathcal{P}^{AB}(\tau)\}\textrm{Tr}\{
\mathcal{P}^{AB}(\upsilon)\}}
\underset{\{M^A_i\}}{\textrm{min}}\underset{i}{\sum}\sqrt{\textrm{Tr}\{
M^A_i\tau^A\}}\sqrt{\textrm{Tr}\{
M^A_i\upsilon^A\}}.\label{interm}
\end{gather}
From Eq.~\eqref{fidelityoper}, we see that \eqref{interm} is
equivalent to \eqref{intermsofF}.

\textbf{Theorem 2:} $F^A(\tau,\upsilon)$ equals the minimum
overlap
\begin{equation}
F^A(\tau,
\upsilon)=\underset{\{M_i\}}{\textrm{min}}\underset{i\geq
0}{\sum}\sqrt{\textrm{Tr}\{ M_i\tau\}}\sqrt{\textrm{Tr}\{
M_i\upsilon\}}\label{operational}
\end{equation}
between the statistical distributions generated by all possible
measurements of the form $M_0=P_\mathcal{K}$, $M_i=M_i^A\otimes
I^B$ for $i\geq 1$, $\underset{i\geq 0}{\sum}M_i=I^S$.

\textbf{Proof:} The proof follows from Eq.~\eqref{easyform} and
Eq.~\eqref{fA}.

Note that the measure $F^A$ compares the information stored in
subsystem $\mathcal{H}^A$, which is the information extractable
through local measurements on $\mathcal{H}^A$. The last result
reflects the intuition that extracting information encoded in
$\mathcal{H}^A$ involves a measurement that projects on the
subspaces $\mathcal{H}^A\otimes\mathcal{H}^B$ or $\mathcal{K}$.

\textbf{Property 2 (Normalization):} From the definition
\eqref{measure} it is obvious that
\begin{equation} F^A(\tau,
\upsilon)\leq F^A(\tau, \tau)=1 , \hspace{0.2cm} \tau \neq
\upsilon.
\end{equation}
From Eq.~\eqref{easyform} we can now see that
\begin{gather}
F^A(\tau, \upsilon)=1, \textrm{  iff
}\hspace{0.1cm}\textrm{Tr}_B\{ \mathcal{P}^{AB}(\tau) \} =
\textrm{Tr}_B\{ \mathcal{P}^{AB}(\upsilon) \},
\end{gather}
as one would expect from a measure that compares only the encoded
information in $\mathcal{H}^A$.

\textbf{Proposition:} Using that the maximum in
Eq.~\eqref{measure} is attained for states of the form
$\Pi(\tau^*)$ and $\Pi(\upsilon^*)$ (Eq.~\eqref{pistar}) where
$\tau^*$ and $\upsilon^*$ satisfy Eq.~\eqref{tauupsstar} and
Eq.~\eqref{tauupsstar2}, without loss of generality we can assume
that for all $\tau$ and $\upsilon$,
\begin{equation}
F^A(\tau, \upsilon)=F(\tau^*, \upsilon^*),\label{useful}
\end{equation}
where
\begin{eqnarray}
\tau^*=\textrm{Tr}_B\{ \mathcal{P}^{AB}(\tau)\}\otimes
|0^B\rangle\langle 0^B| +\textrm{Tr}\{
\mathcal{P}_{\mathcal{K}}(\tau)\}|0_{\mathcal{K}}\rangle\langle
0_{\mathcal{K}}|,\label{taustar1}
\end{eqnarray}
\begin{eqnarray}
\upsilon^*=\textrm{Tr}_B\{ \mathcal{P}^{AB}(\upsilon)\}\otimes
|0^B\rangle\langle 0^B|+ \textrm{Tr}\{
\mathcal{P}_{\mathcal{K}}(\upsilon)\}
|0_{\mathcal{K}}\rangle\langle 0_{\mathcal{K}}|,\label{upsstar1}
\end{eqnarray}
with $|0^B\rangle$ and $|0_{\mathcal{K}}\rangle$ being some fixed
states in $\mathcal{H}^B$ and $\mathcal{K}$, respectively.

\textbf{Property 3 (Strong concavity and concavity of the square
of $F^A$):} The form of $F^A$ given by
Eqs.~\eqref{useful}--\eqref{upsstar1} can be used for deriving
various useful properties of $F^A$ from the properties of the
standard fidelity. For example, it implies that for all mixtures
$\underset{i}{\sum}p_i\tau_i$ and
$\underset{i}{\sum}q_i\upsilon_i$ we have
\begin{gather}
F^A(\underset{i}{\sum}p_i\tau_i,
\underset{i}{\sum}q_i\upsilon_i)=F(\underset{i}{\sum}p_i\tau^*_i,
\underset{i}{\sum}q_i\upsilon^*_i).
\end{gather}
This means that the property of \textit{strong concavity} of the
fidelity \cite{NieChu00} (and all weaker concavity properties that
follow from it) as well as the \textit{concavity of the square of
the fidelity} \cite{Uhl76}, are automatically satisfied by the
measure $F^A$.

\textbf{Definition 2:} Similarly to the concept of angle between
two states \cite{NieChu00} which can be defined from the standard
fidelity, we can define an \textit{angle between the encoded
information in two states}:
\begin{equation}
\Lambda^A(\tau, \upsilon) \equiv \arccos F^A(\tau, \upsilon).
\end{equation}

\textbf{Property 4 (Triangle inequality):} From
Eqs.~\eqref{useful}--\eqref{upsstar1} it follows that just as the
angle between states satisfies the triangle inequality, so does
the angle between the encoded information:
\begin{equation}
\Lambda^A(\tau, \upsilon) \leq \Lambda^A(\tau, \phi) +
\Lambda^A(\phi, \upsilon).
\end{equation}

\textbf{Property 5 (Monotonicity of $F^A$ under local CPTP maps):}
We point out that the monotonicity under CPTP maps of the standard
fidelity does not translate directly to the measure $F^A$. Rather,
as can be seen from Eq.~\eqref{easyform}, $F^A$ satisfies
monotonicity under local CPTP maps on $\mathcal{H}^A$:
\begin{equation}
F^A(\mathcal{E}(\tau),\mathcal{E}(\upsilon) )\geq
F^A(\tau,\upsilon)
\end{equation}
for
\begin{gather}
\mathcal{E}=\mathcal{E}^A\otimes \mathcal{E}^B \oplus
\mathcal{E}_{\mathcal{K}},\label{localCPTPmaps}
\end{gather}
where $\mathcal{E}^A$, $\mathcal{E}^B$ and
$\mathcal{E}_{\mathcal{K}}$ are CPTP maps on operators over
$\mathcal{H}^A$, $\mathcal{H}^B$ and $\mathcal{K}$, respectively.

\textbf{Remark:} There exist other maps under which $F^A$ is also
non-decreasing. Such are the maps which take states from
$\mathcal{H}^A\otimes\mathcal{H}^B$ to $\mathcal{K}$ without
transfer in the opposite direction. But in general, maps which
couple states in $\mathcal{H}^{A}\otimes \mathcal{H}^B$ with
states in $\mathcal{K}$, or states in $\mathcal{H}^A$ with states
in $\mathcal{H}^B$, do not obey this property. For example, a
unitary map which swaps the states in $\mathcal{H}^A$ and
$\mathcal{H}^B$ (assuming both subsystems are of the same
dimension) could both increase or decrease the measure depending
on the states in $\mathcal{H}^B$. Similarly, a unitary map
exchanging states between $\mathcal{H}^{A}\otimes \mathcal{H}^B$
and $\mathcal{K}$ could give rise to both increase or decrease of
the measure depending on the states in $\mathcal{K}$.

Finally, the monotonicity of $F^A$ under local CPTP maps implies

\textbf{Property 6 (Contractivity of the angle under local CPTP
maps):} For CPTP maps of the form \eqref{localCPTPmaps},
$\Lambda^A$ satisfies
\begin{equation}
\Lambda^A(\mathcal{E}(\tau),\mathcal{E}(\upsilon)
)\leq\Lambda^A(\tau,\upsilon).
\end{equation}

\section*{7.4 \hspace{2pt} Robustness of OQEC with respect to initialization
errors} \addcontentsline{toc}{section}{7.4 \hspace{0.15cm}
Robustness of OQEC with respect to initialization errors}

Let us now consider the fidelity between the encoded information
in an ideally prepared state \eqref{perfectini} and in a state
which is not perfectly initialized \eqref{imperfect}:
\begin{gather}
F^A(\rho,\tilde{\rho})=\sqrt{\textrm{Tr}\rho_1}
\sqrt{\textrm{Tr}\tilde{\rho}_1}
F(\rho^A,\tilde{\rho}^A)+0\\
=\textrm{Tr}\sqrt{\sqrt{\textrm{Tr}_B\rho_1}\textrm{Tr}_B\tilde{\rho}_1\sqrt{\textrm{Tr}_B\rho_1}}\equiv
\check{F}(\textrm{Tr}_B\rho_1, \textrm{Tr}_B\tilde{\rho}_1).\notag
\end{gather}
After the noise process $\mathcal{E}$ with Kraus operators
\eqref{Krausoperatorsblock}, the imperfectly encoded state
transforms to $\mathcal{E}(\tilde{\rho})$. Its fidelity with the
perfectly encoded state becomes
\begin{gather}
F^A(\rho,\mathcal{E}(\tilde{\rho}))=\check{F}(\textrm{Tr}_B\rho_1,
\textrm{Tr}_B\tilde{\rho}'_1) =\check{F}(\textrm{Tr}_B\rho_1,
\textrm{Tr}_B\tilde{\rho}_1+\textrm{Tr}_B\{\underset{i}{\sum}D_i\tilde{\rho}_3D_i^{\dagger}\}),
\end{gather}
where we have used the expressions for $\textrm{Tr}_B\rho_1'$ and
$\textrm{Tr}_B\tilde{\rho}_1'$ obtained in Eq.~\eqref{reducedrho}
and Eq.~\eqref{reducedrhotilde}. As we pointed out earlier, the
operator
$\textrm{Tr}_B\{\underset{i}{\sum}D_i\tilde{\rho}_3D_i^{\dagger}
\}$ is positive. Then from the concavity of the \textit{square} of
the fidelity \cite{Uhl76}, it follows that
\begin{gather}
\check{F}^2\left(\textrm{Tr}_B\rho_1,\textrm{Tr}_B\tilde{\rho}_1+\textrm{Tr}_B\{\underset{i}{\sum}D_i\tilde{\rho}_3D_i^{\dagger}
\}\right)=\textrm{Tr}\rho_1\textrm{Tr}\{\tilde{\rho}_1+
\underset{i}{\sum}D_i\tilde{\rho}_3D_i^{\dagger} \}\label{argumentfidelity}\\
\times F^2\left(\rho^A,
\frac{\textrm{Tr}\tilde{\rho}_1}{\textrm{Tr}\{\tilde{\rho}_1+
\underset{i}{\sum}D_i\tilde{\rho}_3D_i^{\dagger} \}}
\tilde{\rho}^A+
\frac{\textrm{Tr}\{\underset{i}{\sum}D_i\tilde{\rho}_3D_i^{\dagger}
\}}{\textrm{Tr}\{\tilde{\rho}_1
+\underset{i}{\sum}D_i\tilde{\rho}_3D_i^{\dagger}\}}\frac{\textrm{Tr}_B\{\underset{i}{\sum}D_i\tilde{\rho}_3D_i^{\dagger}\}
}
{\textrm{Tr}\{\underset{i}{\sum}D_i\tilde{\rho}_3D_i^{\dagger}\}}
\right)\notag\\\geq
 \textrm{Tr}\rho_1\textrm{Tr}\tilde{\rho}_1
F^2(\rho^A,\tilde{\rho}^A)+\textrm{Tr}\rho_1\textrm{Tr}\{\underset{i}{\sum}D_i\tilde{\rho}_3D_i^{\dagger}
\}
 F^2\left(\rho^A,
\frac{\textrm{Tr}_B\{\underset{i}{\sum}D_i\tilde{\rho}_3D_i^{\dagger}\}
}{\textrm{Tr}\{\underset{i}{\sum}D_i\tilde{\rho}_3D_i^{\dagger}\}}\right)\notag\\=\check{F}^2(\textrm{Tr}_B\rho_1,
\textrm{Tr}_B\tilde{\rho}_1)+ \check{F}^2(\textrm{Tr}_B\rho_1,
\textrm{Tr}_B\{\underset{i}{\sum}D_i\tilde{\rho}_3D_i^{\dagger}\}
)\geq \check{F}^2(\textrm{Tr}_B\rho_1,
\textrm{Tr}_B\tilde{\rho}_1).\notag
\end{gather}
(Here, the transition from the first to the second line is
obtained by pulling out the normalization factors of the operators
in $\check{F}$ so that the latter can be expressed in terms of the
fidelity $F$. The transition form the second to the third line is
by using the concavity of the square of the fidelity. The last
line is obtained by expressing the quantities again in terms of
$\check{F}$). Therefore, we can state the following

\textbf{Theorem 3:} The fidelity between the encoded information
in a perfectly initialized state \eqref{perfectini} and an
imperfectly initialized state \eqref{imperfect} does not decrease
under CPTP maps $\mathcal{E}$ with Kraus operators of the form
\eqref{Krausoperatorsblock}:
\begin{equation}
F^A(\rho,\mathcal{E}(\tilde{\rho}))\geq F^A(\rho,\tilde{\rho}).
\end{equation}

We see that even if the ``initialization-free" constraint
\eqref{extraconstraint} is not satisfied, no further decrease in
the fidelity occurs as a result of the process. The effective
noise (the term
$\textrm{Tr}_B\{\underset{i}{\sum}D_i\tilde{\rho}_3D_i^{\dagger}
\}$) that arises due to violation of that constraint, can only
decrease the initialization error.

The above result can be generalized to include the possibility for
information processing on the subsystem. Imagine that we want to
perform a computational task which ideally corresponds to applying
the CPTP map $\mathcal{C}^A$ on the encoded state. In general, the
subsystem $\mathcal{H}^A$ may consist of many subsystems encoding
separate information units (e.g., qubits), and the computational
process may involve many applications of error correction. The noise
process itself generally acts continuously during the computation.
Let us assume that all operations following the initialization are
performed fault-tolerantly \cite{Sho96,ABO98,Kit97,KLZ98,Got97} so
that the overall transformation $\mathcal{C}$ on a \textit{perfectly
initialized} state succeeds with an arbitrarily high probability
(for a model of fault-tolerant quantum computation on subsystems,
see, e.g., Ref.~\cite{AC07}). This means that the effect of
$\mathcal{C}$ on the reduced operator of a perfectly initialized
state is
\begin{equation}
\textrm{tr}_B\rho_1\rightarrow
\mathcal{C}^A(\textrm{Tr}_B\rho_1)\label{FTcomputation}
\end{equation}
up to an arbitrarily small error.

\textbf{Theorem 4:} Let $\mathcal{C}$ be a CPTP map whose effect
on reduced operator of every perfectly initialized state
\eqref{perfectini} is given by Eq.~\eqref{FTcomputation} with
$\mathcal{C}^A$ being a CPTP map on $\mathcal{B}(\mathcal{H}^A)$.
Then the fidelity between the encoded information in a perfectly
initialized state \eqref{perfectini} and an imperfectly
initialized state \eqref{imperfect} does not decrease under
$\mathcal{C}$:
\begin{equation}
F^A(\mathcal{C}(\rho),\mathcal{C}(\tilde{\rho}))\geq
F^A(\rho,\tilde{\rho}).
\end{equation}

\textbf{Proof:} From Eq.~\eqref{FTcomputation} it follows that the
map $\mathcal{C}$ has Kraus operators with vanishing lower left
blocks, similarly to \eqref{Krausoperatorsblock}. If the state is
not perfectly initialized, an argument similar to the one
performed earlier shows that the reduced operator on the subsystem
transforms as $\textrm{Tr}_B\tilde{\rho_1}\rightarrow
\mathcal{C}^A(\textrm{Tr}_B\tilde{\rho}_1)+
\tilde{\rho}^A_{\textrm{err}}$, where
$\tilde{\rho}^A_{\textrm{err}}$ is a positive operator which
appears as a result of the possibly non-vanishing upper right
blocks of the Kraus operators. Using an argument analogous to
\eqref{argumentfidelity} and the monotonicity of the fidelity
under CPTP maps \cite{BCF96}, we obtain
\begin{gather}
F^A(\mathcal{C}(\rho),\mathcal{C}(\tilde{\rho}))=
\check{F}(\mathcal{C}^A(\textrm{Tr}_B\rho_1),\mathcal{C}^A(\textrm{Tr}_B\tilde{\rho}_1)+
\tilde{\rho}^A_{\textrm{err}})+0\\
\geq
\check{F}(\mathcal{C}^A(\textrm{Tr}_B\rho_1),\mathcal{C}^A(\textrm{Tr}_B\tilde{\rho}_1))
=\sqrt{\textrm{Tr}\rho_1}\sqrt{\textrm{Tr}\tilde{\rho}_1}
F(\mathcal{C}^A(\rho^A),\mathcal{C}^A(\tilde{\rho}^A)\notag\\
\geq\sqrt{\textrm{Tr}\rho_1}\sqrt{\textrm{Tr}\tilde{\rho}_1}
F(\rho^A,\tilde{\rho}^A)=\check{F}(\textrm{Tr}_B\rho_1,
\textrm{Tr}_B\tilde{\rho}_1)=F^A(\rho,\tilde{\rho}).\notag
\end{gather}

Again, the preparation error is not amplified by the process. The
problem of how to deal with preparation errors has been discussed in
the context of fault-tolerant computation on standard
error-correction codes, e.g., in Ref.~\cite{Pre99}. The situation
for general OQEC is similar---if the initial state is known, the
error can be eliminated by repeating the encoding. If the state to
be encoded is unknown, the preparation error generally cannot be
corrected. Nevertheless, encoding would still be worthwhile as long
as the initialization error is smaller than the error which would
result from leaving the state unprotected.

\section*{7.5 \hspace{2pt} Summary and outlook}
\addcontentsline{toc}{section}{7.5 \hspace{0.15cm} Summary and
outlook}

In summary, we have shown that a noiseless subsystem is robust
against initialization errors without the need for modification of
the noiseless subsystem conditions. Similarly, we have argued that
general OQEC codes are robust with respect to imperfect
preparation in their standard form. This property is compatible
with fault-tolerant methods of computation, which is essential for
reliable quantum information processing. In order to rigorously
prove our result, we introduced a measure of the fidelity
$F^A(\tau, \upsilon)$ between the encoded information in two
states. The measure is defined as the maximum of the fidelity
between all possible states which have the same reduced operators
on the subsystem code as the states being compared. We derived a
simple form of the measure and discussed many of its properties.
We also gave an operational interpretation of the quantity.

Since the concept of encoded information is central to quantum
information science, the fidelity measure introduced in this study
may find various applications. It provides a natural means for
extending key concepts such as the fidelity of a quantum channel
\cite{KL96} or the entanglement fidelity \cite{Sch96b} to the case
of subsystem codes.

\chapter*{Chapter 8: \hspace{1pt} A fault-tolerant scheme for holonomic quantum
computation} \addcontentsline{toc}{chapter}{Chapter
8:\hspace{0.15cm} A fault-tolerant scheme for holonomic quantum
computation}

\section*{8.1 \hspace{2pt} Preliminaries}
\addcontentsline{toc}{section}{8.1 \hspace{0.15cm} Preliminaries}

There are two main sources of errors in quantum
computers---environment-induced decoherence and imperfect control.
According to the theory of fault tolerance \cite{Sho96, DVS96,
ABO98, Kit97, KLZ98, Got97', Got97, Pre99}, if the errors of each
type are sufficiently uncorrelated and their rates are bellow a
certain threshold, it is possible to implement reliably an
arbitrarily long computational task with an efficient overhead of
resources. Quantum error correction thus provides a universal
software strategy to combat noise in quantum computers.

In addition to the software approach, there have also been
proposals to deal with the effects of noise by hardware methods
that provide robustness through their inherent properties. One
such method is holonomic quantum computation (HQC)\cite{ZR99,
PZR99}---an adiabatic, all-geometric method of computation which
uses non-Abelian generalizations \cite{WZ84} of the Berry phase
\cite{Berry84}. It has been shown that due to its geometric
nature, this approach is robust against various types of errors in
the control parameters driving the evolution \cite{CGSV03, CP03,
SZZ04, FGL05, ZZ05}, and thus provides a degree of built-in
resilience at the hardware level.

In Ref.~\cite{WZL05} HQC was combined with the method of
decoherence-free subspaces (DFSs) \cite{DG98, ZR97, LCW98,
LBKW01}, which was the first step towards systematic error
protection in conjunction with the holonomic approach. DFSs
provide \textit{passive} protection against certain types of
correlated noise; however, they cannot protect against independent
errors. The standard tool to deal with the latter is
\textit{active} error correction \cite{Shor95, Ste96}. Active
error correction is also the basis of quantum fault tolerance,
which is necessary for the scalability of any method of
computation. Even if the system is perfectly isolated from its
environment, when the size of the circuit increases, errors due to
imperfect operations would accumulate detrimentally unless they
are corrected. Therefore, scalability of HQC requires combining
the holonomic approach with active error correction.

In this chapter, we presented a scheme which combines HQC with the
techniques for fault-tolerant computation on stabilizer codes. This
demonstrates that HQC is a scalable method of computation
\cite{OBL08}. The scheme uses Hamiltonians which are elements of the
stabilizer, or in the case of subsystem codes---elements of the
gauge group. Gates are implemented by slowly varying the
Hamiltonians along suitable paths in parameter space, such that the
resulting geometric transformation in each eigenspace of the
Hamiltonian is transversal. On certain codes such as the 9-qubit
Shor code \cite{Shor95} or its subsystem generalizations
\cite{Bac06, BC06}, universal computation according to our scheme
can be implemented with Hamiltonians of weight 2 and 3.

\section*{8.2 \hspace{2pt} Holonomic quantum computation}
\addcontentsline{toc}{section}{8.2 \hspace{0.15cm} Holonomic quantum
computation}

Let $\{H_{\lambda}\}$ be a family of Hamiltonians on an
$N$-dimensional Hilbert space, which is continuously parameterized
by a point $\lambda$ in a control-parameter manifold
$\mathcal{M}$. Assume that the family has the same degeneracy
structure, i.e., there are no level crossings. The Hamiltonians
can then be written as
$H_{\lambda}=\sum_{n=1}^{R}\varepsilon_n(\lambda)\Pi_n(\lambda)$,
where $\{\varepsilon_n(\lambda)\}_{n=1}^{R}$ are the $R$ different
$d_n$-fold degenerate eigenvalues of $H_\lambda$, ($\sum_{n=1}^{R}
d_n=N$), and $\Pi_n(\lambda)$ are the projectors on the
corresponding eigenspaces. If the parameter $\lambda$ is changed
adiabatically, a state which initially belongs to an eigenspace of
the Hamiltonian will remain in the corresponding eigenspace as the
Hamiltonian evolves. The unitary evolution that results from the
action of the Hamiltonian $H(t):=H_{\lambda(t)}$ is
\begin{gather}
U(t)=\mathcal{T}\textrm{exp}(-i\int_0^t d\tau H(\tau)) =
\oplus_{n=1}^R e^{i\omega_n(t)}U^{\lambda}_{A_n}(t),
\label{adiabaticevolution}
\end{gather}
where $\omega_n(t)=-\int_0^td\tau\varepsilon_n(\lambda(\tau))$ is
a dynamical phase, and $U^{\lambda}_{A_n}(t)$ is given by the
following path-ordered exponent:
\begin{equation}
U^{\lambda}_{A_n}(t)=\mathcal{P}\textrm{exp}(\int_{\lambda(0)}^{\lambda(t)}A_n).\label{openpathholonomy}
\end{equation}
Here $A_n$ is the Wilczek-Zee connection \cite{WZ84},
$A_n=\sum_\mu A_{n,\mu} d\lambda^\mu$, where $A_{n,\mu}$ has
matrix elements \cite{WZ84}
\begin{equation}
(A_{n,\mu})_{\alpha\beta}=\langle n\alpha;
\lambda|\frac{\partial}{\partial
\lambda^\mu}|n\beta;\lambda\rangle.\label{matrixelementsA}
\end{equation}
The parameters $\lambda^\mu$ are local coordinates on
$\mathcal{M}$ ($1\leq\mu\leq \textrm{dim}\mathcal{M}$) and
$\{|n\alpha; \lambda\rangle\}_{\alpha=1}^{d_n}$ is an orthonormal
basis of the $n^{\textrm{th}}$ eigenspace of the Hamiltonian at
the point $\lambda$.

When the path $\lambda(t)$ forms a loop $\gamma(t)$,
$\gamma(0)=\gamma(T)= \lambda_0$, the unitary matrix
\begin{equation}
U_n^{\gamma}\equiv
U^{\lambda}_{A_n}(T)=\mathcal{P}\textrm{exp}(\oint_{\gamma}A_n)\label{holonomy}
\end{equation}
is called the holonomy associated with the loop. In the case when
the $n^{\textrm{th}}$ energy level is non-degenerate ($d_n=1$),
the corresponding holonomy reduces to the Berry phase
\cite{Berry84}. The set $\textrm{Hol}(A)=\{U_\gamma/ \gamma\in
L_{\lambda_0}(\mathcal{M})\}$, where $L_{\lambda_0}(\mathcal{M})=
\{\gamma: [0,T]\rightarrow \mathcal{M}
/\gamma(0)=\gamma(T)=\lambda_0\}$ is the space of all loops based
on $\lambda_0$, is a subgroup of $U(d_n)$ called the holonomy
group.

In Refs.~\cite{ZR99, PZR99} it was shown that if the dimension of
the control manifold is sufficiently large, quantum holonomies can
be used as a means of universal quantum computation. In the
proposed approach, logical states are encoded in the degenerate
eigenspace of a Hamiltonian and gates are implemented by
adiabatically varying the Hamiltonian along suitable loops in the
parameter manifold (for a construction of a universal set of
gates, see also Ref.~\cite{NNS03}).

\section*{8.3 \hspace{2pt} Stabilizer codes and fault tolerant computation}
\addcontentsline{toc}{section}{8.3 \hspace{0.15cm} Stabilizer
codes and fault tolerant computation}

A large class of quantum error-correcting codes can be described
by the so called stabilizer formalism \cite{Got96, CRSS96,
CRSS96'}. A stabilizer $S$ is an Abelian subgroup of the Pauli
group $\mathcal{G}_n$ on $n$ qubits, which does not contain the
element $-I$ \cite{NieChu00}. The Pauli group consists of all
possible $n$-fold tensor products of the Pauli matrices $X$, $Y$,
$Z$ together with the multiplicative factors $\pm1$, $\pm i$. The
stabilizer code corresponding to $S$ is the subspace of all states
$|\psi\rangle$ which are left invariant under the action of every
operator in $S$ ($G|\psi\rangle=|\psi\rangle$, $\forall G \in S$).
It is easy to see that the stabilizer of a code encoding $k$
qubits into $n$ has $n-k$ generators. For the case of operator
codes, the stabilizer leaves the subspace $\mathcal{H}^A\otimes
\mathcal{H}^B$ in the decomposition \eqref{decomposition}
invariant but the encoded information is invariant also under
operations that act on the gauge subsystem. An operator stabilizer
code encoding $k$ qubits into $n$ with $r$ gauge qubits, has
$n-r-k$ stabilizer generators, while the gauge group has $2r$
generators \cite{Pou05}. According to the error-correction
condition for stabilizer codes \cite{NieChu00, Pou05}, a set of
errors $\{E_i\}$ in $\mathcal{G}_n$ (which without loss of
generality are assumed to be Hermitian) is correctable by the code
if and only if, for all $i$ and $j$, $E_iE_j$ anticommutes with at
least one element of $S$, or otherwise belongs to $S$ or to the
gauge group. In this chapter we will be concerned with stabilizer
codes for the correction of single-qubit errors and the techniques
for fault-tolerant computation \cite{Sho96, DVS96, ABO98, Kit97,
KLZ98, Got97', Got97, Pre99} on such codes.

A quantum information processing scheme is called fault-tolerant
if a single error occurring during the implementation of any given
operation introduces at most one error per block of the code
\cite{Got97'}. This property has to apply for unitary gates as
well as measurements, including those that constitute the
error-correcting operations themselves. Fault-tolerant schemes for
computation on stabilizer codes generally depend on the code being
used---some codes, like the Bacon-Shor subsystem codes
\cite{Bac06, BC06} for example, are better suited for
fault-tolerant computation than others \cite{AC07}. In spite of
these differences, however, it has been shown that fault-tolerant
information processing is possible on any stabilizer code
\cite{Got97', Got97}. The general procedure can be described
briefly as follows. DiVincenzo and Shor \cite{DVS96} demonstrated
how to perform fault-tolerant measurements of the stabilizer for
any stabilizer code. Their method makes use of an approach
introduced by Shor \cite{Sho96}, which involves the ``cat'' state
$(|0...0\rangle + |1...1\rangle)/\sqrt{2}$ which can be prepared
and verified fault-tolerantly. As pointed out by Gottesman
\cite{Got97'}, by the same method one can measure any operator in
the Pauli group. Since the encoded $X$, $Y$ and $Z$ operators
belong to the Pauli group for any stabilizer code \cite{Got97'},
one can prepare fault-tolerantly various superpositions of the
logical basis states $|\overline{0}\rangle$ and
$|\overline{1}\rangle$, like
$|\overline{+}\rangle=(|\overline{0}\rangle
+|\overline{1}\rangle)/\sqrt{2}$ for example. The latter can be
used to implement fault-tolerantly the encoded Phase and Hadamard
gates, as long as a fault-tolerant C-NOT gate is available
\cite{Got97'}. Gottesman showed how the C-NOT gate can be
implemented fault-tolerantly by first applying a transversal
operation on four encoded qubits and then measuring the encoded
$X$ operator on two of them. Finally, for universal computation
one needs a gate outside of the Clifford group, e.g., the Toffoli
gate. The Toffoli construction was demonstrated first by Shor in
\cite{Sho96} for a specific type of codes---those obtained from
doubly-even self-dual classical codes by the CSS construction
\cite{CS96, St96b}. Gottesman showed \cite{Got97} that a
transversal implementation of the same procedure exists for any
stabilizer code.

Note that the described method for universal fault-tolerant
computation on stabilizer codes uses almost exclusively transversal
operations---these are operations for which each qubit in a block
interacts only with the corresponding qubit from another block or
from a special ancillary state such as Shor's ``cat" state (see also
Steane's \cite{Ste97} and Knill's \cite{Kni05} methods). Since
single-qubit unitaries together with the C-NOT gate form a universal
set of gates, fault-tolerant computation can be realized entirely in
terms of single-qubit operations and C-NOT operations between qubits
from different blocks, assuming that the ``cat'' state can be
prepared reliably. Hence, our goal will be to construct holonomic
realizations of these operations as well as of the preparation of
the ``cat'' state. It is not evident that by doing so we will obtain
a fault-tolerant construction, because the geometric approach
requires that we use degenerate Hamiltonians which generally couple
qubits within the same block. Nevertheless, we will see that it is
possible to design the scheme so that single-qubit errors do not
propagate.

\section*{8.4 \hspace{2pt} The scheme}
\addcontentsline{toc}{section}{8.4 \hspace{0.15cm} The scheme}

Consider an $[[n,1,r,3]]$ stabilizer code. This is a code that
encodes $1$ qubit into $n$, has $r$ gauge qubits, and can correct
arbitrary single-qubit errors. In order to perform a holonomic
operation on this code, we need a nontrivial starting Hamiltonian
which leaves the code space invariant. It is easy to verify that
the only Hamiltonians that satisfy this property are linear
combinations of the elements of the stabilizer, or in the case of
subsystem codes---elements of the gauge group. Note that the
stabilizer and the gauge group transform during the course of the
computation under the operations being applied. At any stage when
we complete an encoded operation, they return to their initial
forms. During the implementation of a standard encoded gate, the
Pauli group $\mathcal{G}_n$ on a given codeword may spread over
other codewords, but it can be verified that this spreading can be
limited to at most $4$ other codewords counting the ``cat" state.
This is because the encoded C-NOT gate can be implemented
fault-tolerantly on any stabilizer code by a transversal operation
on $4$ encoded qubits \cite{Got97}, and any encoded Clifford gate
can be realized using only the encoded C-NOT provided that we are
able to do fault-tolerant measurements (the encoded Clifford group
is generated by the encoded Hadamard, Phase and C-NOT gates).
Encoded gates outside of the Clifford group, such as the encoded
$\pi/8$ or Toffoli gates, can be implemented fault-tolerantly
using encoded C-NOT gates conditioned on the qubits in a ``cat"
state, so they may require transversal operations on a total of
$5$ blocks. More precisely, the fault-tolerant implementation of
the Toffoli gate requires the preparation of a special state of
three encoded qubits \cite{Sho96}, which involves a sequence of
conditional encoded Phase operations and conditional encoded C-NOT
operations with conditioning on the qubits in a ``cat" state
\cite{Got97}. But the encoded Phase gate has a universal
implementation using an encoded C-NOT between the qubit and an
ancilla, so the conditional Phase gate may require applying a
conditional encoded C-NOT. The procedure for implementing an
encoded $\pi/8$ gate involves applying an encoded $SX$ gate
conditioned on the qubits in a ``cat" state \cite{BMPRV99}, where
\begin{equation}
S=\begin{pmatrix} 1&0\\
0&i
\end{pmatrix},
\end{equation}
is the Phase gate, but the encoded $S$ gate generally also
involves an encoded C-NOT on the qubit and an ancilla, so it may
also require the interaction of $4$ blocks. For CSS codes,
however, the spreading of the Pauli group of one block during the
implementation of a basic encoded operation can be limited to a
total of $3$ blocks, since the encoded C-NOT gate has a
transversal implementation \cite{Got97}.

It also has to be pointed out that fault-tolerant encoded Clifford
operations can be implemented using only Clifford gates on the
physical qubits \cite{Got97}. These operations transform the
stabilizer and the gauge group into subgroups of the Pauli group,
and their elements remain in the form of tensor products of Pauli
matrices. The fault-tolerant implementation of encoded gates
outside of the Clifford group, however, involves operations that
take these groups outside of the Pauli group. We will, therefore,
consider separately two cases---encoded operations in the Clifford
group, and encoded operations outside of the Clifford group.

\subsection*{8.4.1 \hspace{2pt} Encoded operations in the Clifford group}
\addcontentsline{toc}{subsection}{8.4.1 \hspace{0.15cm} Encoded
operations in the Clifford group}

In Ref.~\cite{Got97} it was shown that every encoded operation in
the Clifford group can be implemented fault-tolerantly using
Clifford gates on physical qubits. The Clifford group is generated
by the Hadamard, Phase and C-NOT gates, but in addition to these
gates, we will also demonstrate the holonomic implementation of
the $X$ and $Z$ gates which are standard for quantum computation.
We will restrict our attention to implementing single-qubit
unitaries on the first qubit in a block, as well as C-NOT
operations between the first qubits in two blocks. The operations
on the rest of the qubits can be obtained analogously.

\subsubsection*{8.4.1.1 \hspace{2pt} Single-qubit unitary operations}
\addcontentsline{toc}{subsubsection}{8.4.1.1 \hspace{0.15cm}
Single-qubit unitary operations}

In order to implement a single-qubit holonomic operation on the
first qubit in a block, we will choose as a starting Hamiltonian
an element of the stabilizer (with a minus sign) or an element of
the gauge group that acts non-trivially on that qubit. Since we
are considering codes that can correct arbitrary single-qubit
errors, one can always find an element of the initial stabilizer
or the initial gauge group that has a factor $\sigma^0=I$,
$\sigma^1=X$, $\sigma^2=Y$ or $\sigma^3=Z$ acting on the first
qubit, i.e.,
\begin{equation}
\widehat{G}=\sigma^i\otimes \widetilde{G},\hspace{0.4cm}i=0,1,2,3
\label{stabelementgen}
\end{equation}
where $\widetilde{G}$ is a tensor product of Pauli matrices and
the identity on the rest $n-1$ qubits. It can be verified that
under Clifford gates the stabilizer and the gauge group transform
in such a way that this is always the case except that the factor
$\widetilde{G}$ may spread on qubits in other blocks. From now on,
we will use ``hat" to denote operators on all these qubits and
``tilde" to denote operators on all the qubits excluding the first
one.

Without loss of generality we will assume that the chosen
stabilizer or gauge-group element for that qubit has the form
\begin{equation}
\widehat{G}=Z\otimes \widetilde{G}.\label{stabelement}
\end{equation}
As initial Hamiltonian, we will take the operator
\begin{equation}
\widehat{H}(0)=-\widehat{G}=-Z\otimes \widetilde{G}.\label{H(0)}
\end{equation}
Thus, if $\widehat{G}$ is an element of the stabilizer, the code
space will belong to the ground space of $\widehat{H}(0)$. Our
goal is to find paths in parameter space such that when the
Hamiltonian is varied adiabatically along these paths, it gives
rise to single-qubit transformations on the first qubit in each of
its eigenspaces.

\textbf{Proposition:} If the initial Hamiltonian \eqref{H(0)} is
varied adiabatically so that only the factor acting on the first
qubit changes,
\begin{equation}
\widehat{H}(t)=-H(t)\otimes \widetilde{G},\label{Ham1}
\end{equation}
where
\begin{equation}
\textrm{Tr}\{H(t)\}=0,
\end{equation}
the geometric transformation resulting in each of the eigenspaces
of this Hamiltonian will be equivalent to a local unitary on the
first qubit, $\widehat{U}(t)\approx U(t)\otimes \widetilde{I}$.\\

\textbf{Proof.} Observe that \eqref{Ham1} can be written as
\begin{equation}
\widehat{H}(t)=H(t)\otimes \widetilde{P}_0- H(t)\otimes
\widetilde{P}_1,\label{Ham2}
\end{equation}
where
\begin{equation}
\widetilde{P}_{0,1}=\frac{\widetilde{I}\pm
\widetilde{G}}{2}\label{Projectors}
\end{equation}
are orthogonal complementary projectors. The evolution driven by
$\widehat{H}(t)$ is therefore
\begin{equation}
\widehat{U}(t) = U_0(t)\otimes \widetilde{P}_0 + U_1(t) \otimes
\widetilde{P}_1,\label{overallunitary}
\end{equation}
where
\begin{equation}
U_{0,1}(t)=\mathcal{T}\textrm{exp}(-i\overset{t}{\underset{0}{\int}}\pm
H(\tau)d\tau).\label{unitaries}
\end{equation}

Let $|\phi_{0}(t)\rangle$ and $|\phi_{1}(t)\rangle$ be the
instantaneous ground and excited states of $H(t)$ with eigenvalues
$E_{0,1}(t)=\mp E(t)$ ($E(t)>0$). Using
Eq.~\eqref{adiabaticevolution} for the expressions
\eqref{unitaries}, we obtain that in the adiabatic limit
\begin{equation}
U_{0,1}(t)= e^{i\omega(t)}U_{A_{0,1}}(t)\oplus
e^{-i\omega(t)}U_{A_{1,0}}(t),\label{unita}
\end{equation}
where $\omega(t)= \int_0^t d\tau E(\tau)$ and
\begin{gather}
U_{A_{0,1}}(t)=e^{\int_0^t d\tau
\langle\phi_{0,1}(\tau)|\frac{d}{d\tau}|\phi_{0,1}(\tau)\rangle
}|\phi_{0,1}(t)\rangle \langle \phi_{0,1}(0)|.\label{holon}
\end{gather}

The projectors on the instantaneous ground and excited eigenspaces
of $\widehat{H}(t)$ are
\begin{equation}
\widehat{P}_0=|\phi_0(t)\rangle\langle\phi_0(t)|\otimes
\widetilde{P}_0 + |\phi_1(t)\rangle\langle\phi_1(t)|\otimes
\widetilde{P}_1
\end{equation}
and
\begin{equation}
\widehat{P}_1=|\phi_1(t)\rangle\langle\phi_1(t)|\otimes
\widetilde{P}_0 + |\phi_0(t)\rangle\langle\phi_0(t)|\otimes
\widetilde{P}_1,
\end{equation}
respectively. Using Eq.~\eqref{unita} and Eq.~\eqref{holon}, one
can see that the effect of the unitary \eqref{overallunitary} on
each of these projectors is
\begin{equation}
\widehat{U}(t)\widehat{P}_0=e^{i\omega(t)}(U_{A_{0}}(t)\oplus
U_{A_{1}}(t))\otimes\widetilde{I} \hspace{0.1cm}\widehat{P}_0,
\end{equation}
\begin{equation}
\widehat{U}(t)\widehat{P}_1=e^{-i\omega(t)}(U_{A_{0}}(t)\oplus
U_{A_{1}}(t))\otimes\widetilde{I} \hspace{0.1cm}\widehat{P}_1,
\end{equation}
i.e, up to an overall dynamical phase its effect on each of the
eigenspaces is the same as that of the unitary
\begin{equation}
\widehat{U}(t)=U(t)\otimes\widetilde{I},
\end{equation}
where
\begin{equation}
U(t)=U_{A_{0}}(t)\oplus U_{A_{1}}(t)\label{finalU}.
\end{equation}

We next show how by suitably choosing $H(t)$ we can implement all
necessary single-qubit gates. We will identify a set of points in
parameter space, such that by interpolating between these points
we can draw various paths resulting in the desired
transformations. We remark that if a path does not form a loop,
the resulting geometric transformation \eqref{finalU} is an
open-path holonomy \cite{KAS06}.

Consider the single-qubit unitary operator
\begin{equation}
V_{\theta\pm}=\frac{1}{\sqrt{2}}\begin{pmatrix} 1& \mp e^{-i\theta}\\
\pm e^{i\theta}&1
\end{pmatrix},
\end{equation}
where $\theta$ is a real parameter (note that $V_{\theta
-}=V^{\dagger}_{\theta +}$). Define the following single-qubit
Hamiltonian:
\begin{equation}
H_{\theta\pm}\equiv V_{\theta\pm}ZV^{\dagger}_{\theta\pm}.
\end{equation}
Let $H(t)$ in Eq.~\eqref{Ham1} be a Hamiltonian which interpolates
between $H(0)=Z$ and $H(T)=H_{\theta\pm}$ (up to a factor) as
follows:
\begin{equation}
H(t)=f(t)Z+g(t) H_{\theta\pm}\equiv H_{\theta\pm;f,g}(t),
\label{interpolation}
\end{equation}
where $f(0),g(T)>0$, $f(T)=g(0)=0$. To simplify our notations, we
will drop the indices $f$ and $g$ of the Hamiltonian, since the
exact form of these functions is not important for our analysis as
long as they are sufficiently smooth (see discussion below). This
Hamiltonian has eigenvalues $\pm \sqrt{f(t)^2+g(t)^2}$ and its
energy gap is non-zero unless the entire Hamiltonian vanishes. We
will show that in the adiabatic limit, the Hamiltonian
\eqref{Ham1} with $H(t)=H_{\theta\pm}(t)$ gives rise to the
geometric transformation
\begin{equation}
\widehat{U}_{\theta\pm}(T)=V_{\theta\pm}\otimes \widetilde{I}.
\end{equation}

To prove this, observe that
\begin{equation}
-H_{\theta\pm}(t)=W_{\theta}H_{\theta\pm}(t)W_{\theta},
\end{equation}
where $W_{\theta}$ is the Hermitian unitary
\begin{equation}
W_{\theta}=\begin{pmatrix} 0&ie^{-i\theta}\\-ie^{i\theta}&0
\end{pmatrix}.
\end{equation}
The unitaries $U_{\theta\pm{0,1}}$, given by Eq.~\eqref{unitaries}
for $H(t)=H_{\theta\pm}(t)$, are then related by
\begin{equation}
U_{{\theta\pm}0}=W_{\theta} U_{{\theta\pm}1}
W_{\theta}.\label{U01}
\end{equation}
Using that $W_{\theta}|0\rangle=-ie^{i\theta}|1\rangle$,
$W_{\theta}|1\rangle=ie^{-i\theta}|0\rangle$, from
Eq.~\eqref{unita} and Eq.~\eqref{holon} one can see that
Eq.~\eqref{U01} implies
\begin{equation}
U_{{\theta\pm}A_0}=W_{\theta}U_{{\theta\pm}A_1}W_{\theta}.
\label{wrelation}
\end{equation}
Let us define the eigenstates of $H_{\theta\pm}(t)$ at time $T$ as
$|\phi_{{\theta\pm}0}(T)\rangle = V_{\theta\pm}|0\rangle$ and
$|\phi_{{\theta\pm}1}(T)\rangle = V_{\theta\pm}|1\rangle$.
Expression \eqref{holon} can then be written as
\begin{gather}
U_{{\theta\pm}A_{0}}(T)=e^{i\alpha_{{\theta\pm}0}}
V_{\theta\pm}|0\rangle \langle 0|,\notag\\
U_{{\theta\pm}A_{1}}(T)=e^{i\alpha_{{\theta\pm}1}}
V_{\theta\pm}|1\rangle \langle 1|,\label{UA01}
\end{gather}
where $\alpha_{{\theta\pm}0}$ and $\alpha_{{\theta\pm}1}$ are
geometric phases. Without explicitly calculating the geometric
phases, from Eq.~\eqref{UA01} and Eq.~\eqref{wrelation} we obtain
\begin{equation}
e^{i\alpha_{{\theta\pm}0}}=e^{i\alpha_{{\theta\pm}1}}.
\end{equation}
Therefore, up to a global phase, Eq.~\eqref{finalU} yields
\begin{equation}
U_{\theta\pm}(T)\sim V_{\theta\pm}.
\end{equation}

We will use this result, to construct a set of standard gates by
sequences of operations of the form $V_{\theta\pm}$, which can be
generated by interpolations of the type \eqref{interpolation} run
forward or backward. For single-qubit gates in the Clifford group,
we will only need three values of the parameter $\theta$: $0$,
$\pi/2$ and $\pi/4$. For completeness, however, we will also
demonstrate how to implement the $\pi/8$ gate, which together with
the Hadamard gate is sufficient to generate any single-qubit
unitary transformation \cite{BMPRV99}. For this we will need
$\theta=\pi/8$. Note that
\begin{equation}
H_{\theta\pm}= \pm (\cos{\theta}X+\sin{\theta}Y),
\end{equation}
so for these values of $\theta$ we have $H_{0\pm}=\pm X$,
$H_{\pi/2\pm}=\pm Y$, $H_{\pi/4\pm}=\pm
(\frac{1}{\sqrt{2}}X+\frac{1}{\sqrt{2}}Y)$, $H_{\pi/8\pm}=\pm
(\cos{\frac{\pi}{8}}X+\sin{\frac{\pi}{8}}Y)$.

Consider the adiabatic interpolations between the following
Hamiltonians:
\begin{equation}
-Z\otimes\widetilde{G} \rightarrow -Y\otimes \widetilde{G}
\rightarrow Z\otimes\widetilde{G}.
\end{equation}
According to the above result, the first interpolation yields the
transformation $V_{\pi/2+}$. The second interpolation can be
regarded as the inverse of $Z\otimes\widetilde{G}\rightarrow
-Y\otimes\widetilde{G}$ which is equivalent to
$-Z\otimes\widetilde{G}\rightarrow Y\otimes\widetilde{G}$ since
$\widehat{H}(t)$ and $-\widehat{H}(t)$ yield the same geometric
transformations. Thus the second interpolation results in
$V_{\pi/2-}^{\dagger}=V_{\pi/2+}$. The net result is therefore
$V_{\pi/2+}V_{\pi/2+}=iX$. We see that up to a global phase the
above sequence results in a geometric implementation of the $X$
gate.

Similarly, one can verify that the $Z$ gate can be realized via
the loop
\begin{equation}
-Z\otimes\widetilde{G} \rightarrow -X\otimes
\widetilde{G}\rightarrow Z\otimes\widetilde{G} \rightarrow
Y\otimes \widetilde{G}\rightarrow
-Z\otimes\widetilde{G}.\label{Zgate}
\end{equation}

The Phase gate can be realized by applying
\begin{equation}
-Z\otimes\widetilde{G}\rightarrow
-(\frac{1}{\sqrt{2}}X+\frac{1}{\sqrt{2}}Y)\otimes\widetilde{G}\rightarrow
Z\otimes\widetilde{G},
\end{equation}
followed by the $X$ gate.

The Hadamard gate can be realized by first applying $Z$, followed
by
\begin{equation}
-Z\otimes\widetilde{G} \rightarrow -X\otimes \widetilde{G}.
\end{equation}

Finally, the $\pi/8$ gate can be implemented by first applying
$XZ$, followed by
\begin{equation}
Z\otimes\widetilde{G}\rightarrow
-(\cos{\frac{\pi}{8}}X+\sin{\frac{\pi}{8}}Y)\otimes\widetilde{G}\rightarrow
-Z\otimes\widetilde{G}.
\end{equation}

\subsubsection*{8.4.1.2 \hspace{2pt} A note on the adiabatic condition}
\addcontentsline{toc}{subsubsection}{8.4.1.2 \hspace{0.15cm} A
note on the adiabatic condition}

Before we show how to implement the C-NOT gate, let us comment on
the conditions under which the adiabatic approximation assumed in
the above operations is satisfied. Because of the form
\eqref{overallunitary} of the overall unitary, the adiabatic
approximation depends on the extent to which each of the unitaries
\eqref{unitaries} approximate the expressions \eqref{unita}. The
latter depends only on the properties of the single-qubit
Hamiltonian $H(t)$, for which the adiabatic condition \cite{mes}
reads
\begin{equation}
\frac{\varepsilon}{{\Delta}^2}\ll 1, \label{adi}
\end{equation}
where
\begin{equation} \varepsilon =
\operatornamewithlimits{max}_{0\leq t\leq T}|\langle
\phi_1(t)|\frac{dH(t)}{dt}|\phi_0(t)\rangle|, \label{eps}
\end{equation}
and
\begin{equation}
\Delta = \operatornamewithlimits{min}_{0\leq t\leq
T}(E_1(t)-E_0(t))= \operatornamewithlimits{min}_{0\leq t\leq
T}2E(t) \label{Delt}
\end{equation}
is the minimum energy gap of $H(t)$.

Along the segments of the parameter paths we described, the
Hamiltonian is of the form \eqref{interpolation} and its
derivative is
\begin{equation}
\frac{dH_{\theta\pm}(t)}{dt}=\frac{df(t)}{dt}Z+\frac{dg(t)}{dt}H_{\theta\pm},
\hspace{0.6cm} 0<t<T.
\end{equation}
This derivative is well defined as long as $\frac{df(t)}{dt}$ and
$\frac{dg(t)}{dt}$ are well defined. The curves we described,
however, may not be differentiable at the points connecting two
segments. In order for the Hamiltonians \eqref{interpolation} that
interpolate between these points to be differentiable, the
functions $f(t)$ and $g(t)$ have to satisfy $\frac{df(T)}{dt}=0$
and $\frac{dg(0)}{dt}=0$. This means that the change of the
Hamiltonian slows down to zero at the end of each segment (except
for a possible change in its strength), and increases again from
zero along the next segment. We point out that when the
Hamiltonian stops changing, we can turn it off completely by
decreasing its strength. This can be done arbitrarily fast and it
would not affect a state which belongs to an eigenspace of the
Hamiltonian. Similarly, we can turn on another Hamiltonian for the
implementation of a different operation.

The above condition guarantees that the adiabatic approximation is
satisfied with precision
$\textit{O}((\frac{\varepsilon}{{\Delta}^2})^2)$. It is known,
however, that under certain conditions on the Hamiltonian, we can
obtain better results \cite{HJ02}. Let us write the
Schr\"{o}dinger equation as
\begin{equation}
i\frac{d}{dt}|\psi(t)\rangle = H(t) |\psi(t)\rangle\equiv
\frac{1}{\epsilon} \bar{H}(t) |\psi(t)\rangle,
\end{equation}
where $\epsilon>0$ is small. If $\bar{H}(t)$ is smooth and all its
derivatives vanish at the end points $t=0$ and $t=T$, the error
would scale super-polynomially with $\epsilon$, i.e., it will
decrease with $\epsilon$ faster than $\textit{O}(\epsilon^N)$ for
any $N$. (Notice that $\frac{\varepsilon}{{\Delta}^2}\propto
\epsilon$, i.e., the error according to the standard adiabatic
approximation is of order $\textit{O}(\epsilon^2)$.)

In our case, the smoothness condition translates directly to the
functions $f(t)$ and $g(t)$. For any choice of these functions which
satisfies the standard adiabatic condition, we can ensure that the
stronger condition is satisfied by the reparameterization
$f(t)\rightarrow f(y(t))$, $g(t)\rightarrow g(y(t))$ where $y(t)$ is
a smooth function of $t$ which satisfies $y(0)=0$, $y(T)=T$, and has
vanishing derivatives at $t=0$ and $t=T$. Then by slowing down the
change of the Hamiltonian by a constant factor $\epsilon$, which
amounts to an increase of the total time $T$ by a factor
$1/\epsilon$, we can decrease the error super-polynomially in
$\epsilon$. We will use this result to obtain a low-error
interpolation in Section 8.5 where we estimate the time needed to
implement a holonomic gate with certain precision.

\subsubsection*{8.4.1.3 \hspace{2pt} The C-NOT gate}
\addcontentsline{toc}{subsubsection}{8.4.1.3 \hspace{0.15cm} The
C-NOT gate}

The stabilizer or the gauge group on multiple blocks of the code
is a direct product of the stabilizers or the gauge groups of the
individual blocks. Therefore, from Eq.~\eqref{stabelementgen} it
follows that one can always find an element of the initial
stabilizer or gauge group on multiple blocks which has any desired
combination of factors $\sigma^i$, $i=0,1,2,3$ on the first qubits
in these blocks. It can be verified that applying transversal
Clifford operations on the blocks does not change this property.
Therefore, we can assume that for implementing a C-NOT gate we can
find an element of the stabilizer or the gauge group which has the
form \eqref{stabelement} where the factor $Z$ acts on the target
qubit and $\widetilde{G}$ acts trivially on the control qubit.

Then it is straightforward to verify that the C-NOT gate can be
implemented by first applying the inverse of the Phase gate
($S^{\dagger}$) on the control qubit, as well as the
transformation $V_{\pi/2+}$ on the target qubit, followed by the
transformation
\begin{equation}
-I^c\otimes Y\otimes \widetilde{G} \rightarrow -Z^c\otimes
Z\otimes \widetilde{G},\label{CNOTint}
\end{equation}
where the superscript $c$ denotes the control qubit. The
interpolation \eqref{CNOTint} is understood as in
Eq.~\eqref{interpolation}. To see that this yields the desired
transformation, observe that the Hamiltonian corresponding to
Eq.~\eqref{CNOTint} can be written in the form
\begin{equation}
\widehat{\widehat{H}}(t)= |0\rangle\langle 0|^c \otimes
{H}_{\pi/2+}(T-t)\otimes\widetilde{G}+|1\rangle\langle 1|^c\otimes
{H}_{\pi/2-}(T-t)\otimes\widetilde{G}.\label{HCNOT}
\end{equation}
The application of this Hamiltonian from time $t=0$ to time $t=T$
results in the unitary
\begin{equation}
\widehat{\widehat{U}}(T)=|0\rangle\langle 0|^c \otimes
\widehat{U}^{\dagger}_{\pi/2+}(T)+|1\rangle\langle 1|^c\otimes
\widehat{U}^{\dagger}_{\pi/2-}(T),\label{doublehatU}
\end{equation}
where
\begin{equation}
\widehat{U}_{\pi/2\pm}(T)=\mathcal{T}\textrm{exp}(-i\overset{T}{\underset{0}{\int}}d\tau
H_{\pi/2\pm}(\tau)\otimes\widetilde{G}).
\end{equation}
But the Hamiltonians ${H}_{\pi/2+}(T-t)\otimes \widetilde{G}$ and
${H}_{\pi/2-}(T-t)\otimes\widetilde{G}$ have the same
instantaneous spectrum, and Eq.~\eqref{adiabaticevolution} implies
that up to a dynamical phase, each of the eigenspaces of
$\widehat{\widehat{H}}$ will undergo the geometric transformation
\begin{equation}
\widehat{\widehat{U}}_g(T)=|0\rangle\langle 0|^c \otimes
V_{\pi/2+}^{\dagger}\otimes \widetilde{I}+|1\rangle\langle
1|^c\otimes V_{\pi/2-}^{\dagger}\otimes \widetilde{I},
\end{equation}
where $V_{\pi/2\pm}^{\dagger}\otimes \widetilde{I}$ are the
geometric transformations generated by
${H}_{\pi/2\pm}(T-t)\otimes\widetilde{G}$ as shown earlier. This
transformation was preceded by the operation $S^{\dagger c}\otimes
V_{\pi/2+}\otimes \widetilde{I}$, which means that the net result
is
\begin{gather}
\widehat{\widehat{U}}_g(T) S^{\dagger c}\otimes V_{\pi/2+}\otimes
\widetilde{I}=|0\rangle\langle 0|^c \otimes
V_{\pi/2+}^{\dagger}V_{\pi/2+}\otimes
\widetilde{I}-i|1\rangle\langle 1|^c\otimes
V_{\pi/2-}^{\dagger}V_{\pi/2+}\otimes \widetilde{I}\notag\\
=|0\rangle\langle 0|^c \otimes I\otimes
\widetilde{I}+|1\rangle\langle 1|^c\otimes X\otimes \widetilde{I}.
\end{gather}
This is exactly the C-NOT transformation. Note that because of the
form \eqref{HCNOT} of $\widehat{\widehat{H}}(t)$, the extent to
which the adiabatic approximation is satisfied during this
transformation depends only on the adiabatic properties of the
single-qubit Hamiltonians $H_{\pi/2\pm}(T-t)$ which we discussed
in the previous subsection.

Our construction allowed us to prove the resulting geometric
transformations without explicitly calculating the holonomies
\eqref{holonomy}. It may be instructive, however, to demonstrate
this calculation for at least one of the gates we described. In
the appendix at the end of this chapter we present an explicit
calculation of the geometric transformation for the $Z$ gate for
the following two cases: $f(t)=1-\frac{t}{T}$, $g(t)=\frac{t}{T}$
(linear interpolation); $f(t)=\cos{\frac{\pi t}{2 T}}$, $g(t)=\sin
{\frac{\pi t}{2 T}}$ (unitary interpolation).

\subsection*{8.4.2 \hspace{2pt} Encoded operations outside of the Clifford group}
\addcontentsline{toc}{subsection}{8.4.2 \hspace{0.15cm} Encoded
operations outside of the Clifford group}

For universal fault-tolerant computation we also need at least one
encoded gate outside of the Clifford group. The fault-tolerant
implementation of such gates is based on the preparation of a
special encoded state \cite{Sho96,KLZ98,Got97,BMPRV99,ZLC00} which
involves a measurement of an encoded operator in the Clifford group.
For example, the $\pi/8$ gate requires the preparation of the state
$\frac{|0\rangle+\textrm{exp}(i\pi/4)|1\rangle}{\sqrt{2}}$, which
can be realized by measuring the operator $e^{-i\pi/4}SX$
\cite{BMPRV99}. Equivalently, the state can be obtained by applying
the operation $RS^{\dagger}$, where $R$ denotes the Hadamard gate,
on the state
$\frac{\cos(\pi/8)|0\rangle+\sin(\pi/8)|1\rangle}{\sqrt{2}}$ which
can be prepared by measuring the Hadamard gate \cite{KLZ98}. The
Toffoli gate requires the preparation of the three-qubit encoded
state $\frac{|000\rangle+|010\rangle+|100\rangle+|111\rangle}{2}$
and involves a similar procedure \cite{ZLC00}. In all these
instances, the measurement of the Clifford operator is realized by
applying transversally the operator conditioned on the qubits in a
``cat" state.

We now show a general method that can be used to implement
holonomically any conditional transversal Clifford operation with
conditioning on the ``cat" state. Let $O$ be a Clifford gate
acting on the first qubits from some set of blocks. As we
discussed in the previous section, under this unitary the
stabilizer and the gauge group transform in such a way that we can
always find an element with an arbitrary combination of Pauli
matrices on the first qubits. If we write this element in the form
\begin{equation}
\widehat{G}=G_1\otimes G_{2,...,n},
\end{equation}
where $G_1$ is a tensor product of Pauli matrices acting on the
first qubits from the blocks, and $G_{2,...,n}$ is an operator on
the rest of the qubits, then applying $O$ conditioned on the first
qubit in a ``cat" state transforms this stabilizer or gauge-group
element as follows:
\begin{gather}
I^c\otimes G_1\otimes G_{2,...,n}=|0\rangle\langle 0|^c\otimes
G_1\otimes G_{2,...,n}+|1\rangle\langle 1|^c\otimes G_1\otimes
G_{2,...,n} \notag\\
\rightarrow |0\rangle\langle 0|^c\otimes
G_1\otimes G_{2,...,n}+|1\rangle\langle 1|^c\otimes
OG_1O^{\dagger}\otimes G_{2,...,n},
\end{gather}
where the superscript $c$ denotes the control qubit from the
``cat" state. We can implement this operation by choosing the
factor $G_1$ the same as the one we would use if we wanted to
implement the operation $O$ according to the previously described
procedure. Then we can apply the following Hamiltonian:
\begin{equation}
\widehat{\widehat{H}}_{C(O)}(t)=-|0\rangle\langle 0|^c\otimes
G_1\otimes G_{2,...,n}-\alpha(t)|1\rangle\langle 1|^c\otimes
H_O(t)\otimes G_{2,...,n},\label{HamA}
\end{equation}
where $H_O(t)\otimes G_{2,...,n}$ is the Hamiltonian that we would
use for the implementation of the operation $O$ and $\alpha(t)$ is
a real parameter chosen such that at every moment the operator
$\alpha(t)|1\rangle\langle 1|^c\otimes H_O(t)\otimes G_{2,...,n}$
has the same instantaneous spectrum as the operator
$|0\rangle\langle 0|^c\otimes G_1\otimes G_{2,...,n}$. This
guarantees that the overall Hamiltonian is degenerate and the
geometric transformation in each of its eigenspaces is
\begin{equation}
\widehat{\widehat{U}}_g(t)=|0\rangle\langle 0|^c\otimes I_1\otimes
I_{2,...,n}+|1\rangle\langle 1|^c\otimes U_O(t)\otimes
I_{2,...,n},
\end{equation}
where $U_O(t)$ is the geometric transformation on the first qubits
generated by $H_O(t)\otimes G_{2,...,n}$. Since we presented the
constructions of our basic Clifford operations up to an overall
phase, the operation $U_O(t)$ may differ from the desired
operation by a phase. This phase can be corrected by applying a
suitable gate on the control qubit from the ``cat" state (we
explain how this can be done in the next section). We remark that
a Hamiltonian of the type \eqref{HamA} requires fine tuning of the
parameter $\alpha(t)$ and generally can be complicated. Our goal
in this section is to prove that universal fault-tolerant
holonomic computation is possible in principle. In Section 8.6 we
show that depending on the code one can find more natural
implementations of these operations.

If we want to apply a second conditional Clifford operation $Q$ on
the first qubits in the block, we can do this via the Hamiltonian
\begin{equation}
\widehat{\widehat{H}}_{C(Q)}(t)=-|0\rangle\langle 0|^c\otimes
G_1\otimes G_{2,...,n}-\beta(t)|1\rangle\langle 1|^c\otimes
H_Q(t)\otimes G_{2,...,n}, \label{HamB}
\end{equation}
where $H_Q(t)\otimes G_{2,...,n}$ is now the Hamiltonian we would
use to implement the operation $Q$, had we implemented the
operation $O$ before that. Here again, the factor $\beta(t)$
guarantees that there is no splitting of the energy levels of the
Hamiltonian. Subsequent operations are applied analogously. Using
this general method, we can implement holonomically any
transversal Clifford operation conditioned on the ``cat" state.

\subsection*{8.4.3 \hspace{2pt} Using the ``cat" state}
\addcontentsline{toc}{subsection}{8.4.3 \hspace{0.15cm} Using the
``cat" state}

In addition to transversal operations, a complete fault-tolerant
scheme requires the ability to prepare, verify and use a special
ancillary state such as the ``cat" state
$(|00...0\rangle+|11...1\rangle)/\sqrt{2}$ proposed by Shor
\cite{Sho96}. This can also be done in the spirit of our holonomic
scheme. Since the ``cat" state is known and its construction is
non-fault-tolerant, we can prepare it by simply treating each
initially prepared qubit as a simple code (with $\widetilde{G}$ in
Eq.~\eqref{stabelement} being trivial), and updating the
stabilizer of the code via the applied geometric transformation as
the operation progresses. The stabilizer of the prepared ``cat"
state is generated by $Z_iZ_j$, $i<j$. Transversal unitary
operations between the ``cat" state and other codewords are
applied as described in the previous section.

We also have to be able to measure the parity of the state, which
requires the ability to apply successively C-NOT operations from
two different qubits in the ``cat" state to one and the same
ancillary qubit initially prepared in the state $|0\rangle$. We
can regard the qubit in state $|0\rangle$ as a simple code with stabilizer $%
\langle Z \rangle $, and we can apply the first C-NOT as described
before. Even though after this operation the state of the target
qubit is unknown, the second C-NOT gate can be applied via the
same interaction, since the transformation in each eigenspace of
the Hamiltonian is the same and at the end when we measure the
qubit we project on one of the eigenspaces.

\subsection*{8.4.4 \hspace{2pt} Fault-tolerance of the scheme}
\addcontentsline{toc}{subsection}{8.4.4 \hspace{0.15cm}
Fault-tolerance of the scheme}

We showed how we can generate any transversal operation on the
code space holonomically, assuming that the state is
non-erroneous. But what if an error occurs on one of the qubits?

At any moment, we can distinguish two types of errors---those that
result in transitions between the ground and the excited spaces of
the current Hamiltonian, and those that result in transformations
inside the eigenspaces. Due to the discretization of errors in
QEC, it suffices to prove correctability for each type separately.
The key property of our construction is that in each of the
eigenspaces, the geometric transformation is the same and it is
transversal. Because of this, if we are applying a unitary on the
first qubit, an error on that qubit will remain localized
regardless of whether it causes an excitation or not. If the error
occurs on one of the other qubits, at the end of the
transformation the result would be the desired single-qubit
unitary gate plus the error on the other qubit, which is
correctable.

It is remarkable that even though the Hamiltonian couples qubits
within the same block, single-qubit errors do not propagate. This
is because the coupling between the qubits amounts to a change in
the relative phase between the ground and excited spaces, but the
latter is irrelevant since it is either equivalent to a gauge
transformation, or when we apply a correcting operation we project
on one of the eigenspaces. In the case of the C-NOT gate, an error
can propagate between the control and the target qubits, but it
never results in two errors within the same codeword.

\section*{8.5 \hspace{2pt} Effects on the accuracy threshold for environment noise}
\addcontentsline{toc}{section}{8.5 \hspace{0.15cm} Effects on the
accuracy threshold for environment noise}

Since the method we presented conforms completely to a given
fault-tolerant scheme, it would not affect the error threshold per
operation for that scheme. Some of its features, however, would
affect the threshold for \textit{environment} noise.

First, observe that when applying the Hamiltonian \eqref{Ham1}, we
cannot at the same time apply
operations on the other qubits on which the factor $%
\widetilde{G}$ acts non-trivially. Thus, some operations at the
lowest level of concatenation that would otherwise be implemented
simultaneously might have to be implemented serially. The effect of
this is equivalent to slowing down the circuit by a constant factor.
(Note that we could also vary the factor $\widetilde{G}$
simultaneously with $H(t)$, but in order to obtain the same
precision as that we would achieve by a serial implementation, we
would have to slow down the change of the Hamiltonian by the same
factor.) The slowdown factor resulting from this loss of parallelism
is usually small since this problem occurs only at the lowest level
of concatenation. For example, for the Bacon-Shor code, we can
implement operation on up to 6 out of the 9 qubits in a block
simultaneously. As we show in Section 8.6, when implementing an
encoded single-qubit gate, we can address any two qubits in a row or
column using our method by taking $\widetilde{G}$ in
Eq.~~\eqref{Ham1} to be a single-qubit operator $Z$ or $X$ on the
third qubit in the same row or column. The Hamiltonians used for
applying operations on the two qubits commute with each other at all
times and do not interfere. A similar thing holds for implementation
of the encoded C-NOT and the operations involving the ``cat" state.
Thus for the Bacon-Shor code we have a slowdown due to parallelism
by a factor of $1.5$.

A more significant slowdown results from the fact that the
evolution is adiabatic. In order to obtain a rough estimate of the
slowdown due specifically to the adiabatic requirement, we will
compare the time $T_{h}$ needed for the implementation of a
holonomic gate with precision $1-\delta$ to the time $T_{d}$
needed for a dynamical realization of the same gate with the same
strength of the Hamiltonian. We will consider a realization of the
$X$ gate via the interpolation
\begin{equation}
\widehat{H}(t)=-V_{X}(\tau(t))ZV_{X}^{\dagger}(\tau(t))\otimes
\widetilde{G}, \hspace{0.2cm} V_{X}(\tau(t))=\textrm{exp}
\left(i\tau(t)\frac{\pi }{2T_{h}}X\right),\label{Hamest}
\end{equation}
where $\tau(0)=0$, $\tau(T_h)=T_h$. Thus the energy gap of the
Hamiltonian is always at maximum. The optimal dynamical
implementation of the same gate is via the Hamiltonian $-X$ for
time $T_{d}=\frac{\pi}{2}$.

As we argued in Section 8.4, the accuracy with which the adiabatic
approximation holds for the Hamiltonian \eqref{Hamest} is the same
as that for the Hamiltonian
\begin{equation}
H(t)=V_{X}(\tau(t))ZV_{X}^{\dagger }(\tau(t)).\label{Hamest2}
\end{equation}
We now present estimates for two different choices of the function
$\tau(t)$. The first one is
\begin{equation}
\tau(t)=t.
\end{equation}
In this case the Schr\"{o}dinger equation can be easily solved in
the instantaneous eigenbasis of the Hamiltonian \eqref{Hamest2}.
For the probability that the initial ground state remains a ground
state at the end of the evolution, we obtain
\begin{equation}
p=\frac{1}{1+\varepsilon ^{2}}+\frac{%
\varepsilon ^{2}}{1+\varepsilon ^{2}}\cos ^{2}(\frac{\pi }{4\varepsilon }%
\sqrt{1+\varepsilon ^{2}}),
\end{equation}
where
\begin{equation}
\varepsilon =\frac{T_{d}}{T_{h}}.
\end{equation}
Expanding in powers of $\varepsilon $ and averaging the square of
the cosine whose period is much smaller than $T_{h}$, we obtain
the condition
\begin{equation}
\varepsilon ^{2}\leq 2\delta.
\end{equation}
Assuming, for example, that $\delta \approx 10^{-4}$
(approximately the threshold for the 9-qubit Bacon-Shor
\cite{AC07}), we obtain that the time of evolution for the
holonomic case must be about 70 times longer than that in the
dynamical case.

It is known, however, that if $H(t)$ is smooth and its derivatives
vanish at $t=0$ and $t=T_h$, the adiabatic error decreases
super-polynomially with $T_h$ \cite{HJ02}. To achieve this, we
will choose
\begin{equation}
\tau(t)= \frac{1}{a}\int_0^t dt' e^{-1/\sin(\pi
t'/T_h)},\hspace{0.2cm} a=\int_0^{T_h} dt' e^{-1/\sin(\pi
t'/T_h)}.
\end{equation}
For this interpolation, by a numerical solution we obtain that
when $T_h/T_d\approx 17$ the error is already of the order of
$10^{-6}$, which is well below the threshold values obtained for
the Bacon-Shor codes \cite{AC07}. This is a remarkable improvement
in comparison to the previous interpolation which shows that the
smoothness of the Hamiltonian plays an important role in the
performance of the scheme.

An additional slowdown in comparison to a perfect dynamical scheme
may result from the fact that the constructions for some of the
standard gates we presented involve long sequences of loops. With
more efficient parameter paths, however, it should be possible to
reduce this slowdown to minimum. An approach for finding loops
presented in Ref.~\cite{NNS03} may be useful in this respect.

In comparison to a dynamical implementation, the allowed rate of
environment noise for the holonomic case would decrease by a factor
similar to the slowdown factor. In practice, however, dynamical
gates are not perfect and the holonomic approach may be advantageous
if it gives rise to a better operational precision.

We finally point out that an error in the factor $H(t)$ in the
Hamiltonian \eqref{Ham1} would result in an error on the first qubit
according to Eq.~\eqref{finalU}. Such an error clearly has to be
below the accuracy threshold. More dangerous errors, however, are
also possible. For example, if the degeneracy of the Hamiltonian is
broken, this can result in an unwanted dynamical transformation
affecting all qubits on which the Hamiltonian acts non-trivially.
Such multi-qubit errors have to be of higher order in the threshold,
which imposes more severe restrictions on the Hamiltonian.

\section*{8.6 \hspace{2pt} Fault-tolerant holonomic computation with low-weight
Hamiltonians} \addcontentsline{toc}{section}{8.6 \hspace{0.15cm}
Fault-tolerant holonomic computation with low-weight Hamiltonians}

The weight of the Hamiltonians needed for the scheme we described
depend on the weight of the stabilizer or gauge-group elements.
Remarkably, certain codes possess stabilizer or gauge-group
elements of low weight covering all qubits in the code, which
allows us to perform holonomic computation using low-weight
Hamiltonians. Here we will consider as an example a subsystem
generalization of the 9-qubit Shor code \cite{Shor95}---the
Bacon-Shor code \cite{Bac06, BC06}---which has particularly
favorable properties for fault-tolerant computation \cite{Ali07,
AC07}. In the 9-qubit Bacon-Shor code, the gauge group is
generated by the weight-two operators $Z_{k,j}Z_{k,j+1}$ and
$X_{j,k}X_{j+1,k}$, where the subscripts label the qubits by row
and column when they are arranged in a $3\times 3$ square lattice.
Since the Bacon-Shor code is a CSS code, the C-NOT gate has a
direct transversal implementation. We now show that the C-NOT gate
ca be realized using at most weight-three Hamiltonians.

If we want to apply a C-NOT gate between two qubits each of which
is, say, in the first row and column of its block, we can use as a
starting Hamiltonian $-Z^{t}_{1,1}\otimes Z^t_{1,2}$, where the
superscript $t$ signifies that these are operators in the target
block. We can then apply the C-NOT gate as described in the previous
section. After the operation, however, this gauge-group element will
transform to $-Z^{t}_{1,1}\otimes Z^c_{1,1}\otimes Z^t_{1,2}$. If we
now want to implement a C-NOT gate between the qubits with index
$\{1,2\}$ using as a starting Hamiltonian the operator
$-Z^{t}_{1,1}\otimes Z^c_{1,1}\otimes Z^t_{1,2}$ according to the
same procedure, we will have to use a four-qubit Hamiltonian. Of
course, at this point we can use the starting Hamiltonian
$-Z^t_{1,2}\otimes Z^t_{1,3}$, but if we had also applied a C-NOT
between the qubits labeled $\{1,3\}$, this operator would not be
available---it would have transformed to $-Z^t_{1,2}\otimes
Z^t_{1,3}\otimes Z^c_{1,3}$.

What we can do instead, is to use as a starting Hamiltonian the
operator $ -Z^{t}_{1,1}\otimes Z^t_{1,2}\otimes Z^c_{1,2}$ which
is obtained from the gauge-group element $ Z^{t}_{1,1}\otimes
Z^c_{1,1}\otimes Z^t_{1,2}\otimes Z^c_{1,2}$ after the application
of the C-NOT between the qubits with index $\{1,1\}$. Since the
C-NOT gate is its own inverse, we can regard the factor
$Z^{t}_{1,1}$ as $\widetilde{G}$ in Eq.~\eqref{CNOTint} and use
this starting Hamiltonian to apply our procedure backwards. Thus
we can implement any transversal C-NOT gate using at most
weight-three Hamiltonians.

Since the encoded $X$, $Y$ and $Z$ operations have a bitwise
implementation, we can always apply them according to our procedure
using Hamiltonians of weight 2. For the Bacon-Shor code, the encoded
Hadamard gate can be applied via bitwise Hadamard transformations
followed by a rotation of the grid to a $90$ degree angle
\cite{AC07}. The encoded Phase gate can be implemented using the
encoded C-NOT and an ancilla.

We point out that the preparation and measurement of the ``cat"
state can also be done using Hamiltonians of weight 2. To prepare
the ``cat" state, we prepare first all qubits in the state
$|f_+^0\rangle = (|0\rangle + |1\rangle)/\sqrt{2}$, which can be
done by measuring each of them in the $|0\rangle$, $|1\rangle$
basis (this ability is assumed for any type of computation) and
applying the transformation $-Z\rightarrow -X$ or $Z\rightarrow
-X$ depending on the outcome. To complete the preparation of the
``cat" state, apply a two-qubit transformation between the first
qubit and each of the other qubits ($j>1$) via the transformation
\begin{equation}
-I_1\otimes X_j\rightarrow -Z_1\otimes Z_j .
\end{equation}
Single-qubit transformations on qubits from the ``cat" state can
be applied according to the method described in the previous
section using at most weight-two Hamiltonians.

To measure the parity of the state, we need to apply successively
C-NOT operations from two different qubits in the ``cat" state to
the same ancillary qubit initially prepared in the state
$|0\rangle$. As described in the previous section, this can also
be done according to our method and requires Hamiltonians of
weight 2.

For universal computation with the Bacon-Shor code, we also need to
be able to apply one encoded transformation outside of the Clifford
group. As we mentioned earlier, in order to implement the Toffoli
gate or the $\pi/8$ gate, it is sufficient to be able to implement a
C-NOT gate conditioned on a ``cat" state. For the Bacon-Shor code,
the C-NOT gate has a transversal implementation, so the conditioned
C-NOT gate can be realized by a series of transversal Toffoli
operations between the ``cat" state and the two encoded states. We
now show that the latter can be implemented using at most
three-qubit Hamiltonians.

Ref.~\cite{NieChu00} provides a circuit for implementing the
Toffoli gate as a sequence of one- and two-qubit gates. We will
use the same circuit, except that we flip the control and target
qubits in every C-NOT gate using the identity
\begin{equation}
R_{1} R_{2} C_{1,2}R_{1} R_{2}= C_{2,1},
\end{equation}
where $R_{i}$ denotes a Hadamard gate on the qubit labeled by $i$
and $C_{i,j}$ denotes a C-NOT gate between qubits $i$ and $j$ with
$i$ being the control and $j$ being the target. Let
$\textrm{Toffoli}_{i,j,k}$ denote the Toffoli gate on qubits $i$,
$j$ and $k$ with $i$ and $j$ being the two control qubits and $k$
being the target qubit, and let $S_{i}$ and $T_i$ denote the Phase
and $\pi/8$ gates on qubit $i$, respectively. Then the Toffoli gate
on three qubits (the first one of which we will assume to belong to
the ``cat" state), can be written as:
\begin{gather}
\textrm{Toffoli}_{1,2,3}=R_{2}C_{3,2}R_{3}T_{3}^{\dagger}R_{3}R_{1}C_{3,1}R_{3}T_{3}R_{3}C_{3,2}R_{3}T_{3}^{\dagger}R_{3}
C_{3,1}\times\notag\\R_{3}T_{3}R_{3}R_{2}T_{2}^{\dagger}R_{2}
C_{2,1}R_{2}T_{2}^{\dagger}R_{2}
C_{2,1}R_{2}S_{2}R_{1}T_{1}.\label{Toffoli}
\end{gather}
To show that each of the above gates can be implemented
holonomically using Hamiltonians of weight at most 3, we will need
an implementation of the C-NOT gate which is suitable for the case
when we have a stabilizer or gauge-group element of the form
\begin{equation}
\widehat{G}=X\otimes \widetilde{G},\label{Xgenerator}
\end{equation}
where the factor $X$ acts on the target qubit and $\widetilde{G}$
acts trivially on the control qubit. By a similar argument to the
one in Section 8.4, one can verify that in this case the C-NOT
gate can be implemented as follows: apply the operation
$S^{\dagger}$ on the control qubit (we describe how to do this for
our particular case below) together with the transformation
\begin{equation}
-X\otimes\widetilde{G}\rightarrow - Z\otimes
\widetilde{G}\rightarrow X\otimes\widetilde{G}\label{CNOT1}
\end{equation}
on the target qubit, followed by the transformation
\begin{equation}
I^c\otimes X\otimes \widetilde{G}\rightarrow -(|0\rangle\langle
0|^c\otimes Z+|1\rangle\langle 1|^c\otimes Y)\otimes
\widetilde{G}\rightarrow-I^c\otimes X\otimes
\widetilde{G}.\label{CNOT2}
\end{equation}

Since the second and the third qubits belong to blocks encoded
with the Bacon-Shor code, there are weight-two elements of the
initial gauge group of the form $Z\otimes Z$ covering all qubits.
The stabilizer generators on the ``cat" state are also of this
type. Following the transformation of these operators according to
the sequence of operations \eqref{Toffoli}, one can see that
before every C-NOT gate in this sequence, there is an element of
the form \eqref{Xgenerator} with $\widetilde{G}=Z$ which can be
used to implement the C-NOT gate as described provided that we can
implement the gate $S^{\dagger}$ on the control qubit. We also
point out that all single-qubit operations on qubit $1$ in this
sequence can be implemented according to the procedure describes
in Section 8.4, since at every step we have a weight-two
stabilizer element on that qubit with a suitable form. Therefore,
all we need to show is how to implement the necessary single-qubit
operations on qubits $2$ and $3$. Due to the complicated
transformation of the gauge-group elements during the sequence of
operations \eqref{Toffoli}, we will introduce a method of applying
a single-qubit operation with a starting Hamiltonian that acts
trivially on the qubit. For implementing single-qubit operations
on qubits $2$ and $3$ we will use as a starting Hamiltonian the
operator
\begin{equation}
\widehat{\widehat{H}}(0)=-I_{i}\otimes X_1\otimes \widetilde{Z},
\hspace{0.4 cm} i=2,3\label{specialcase}
\end{equation}
where the first factor ($I_i$) acts on the qubit on which we want
to apply the operation ($2$ or $3$), and $X_1\otimes
\widetilde{Z}$ is the transformed (after the Hadamard gate $R_1$)
stabilizer element of the ``cat" state that acts non-trivially on
qubit $1$ (the factor $\widetilde{Z}$ acts on some other qubit in
the ``cat" state).

To implement a single-qubit gate on qubit $3$ for example, we
first apply the interpolation
\begin{equation}
-I_{3}\otimes X_1\otimes \widetilde{Z}\rightarrow -Z_{3}\otimes
Z_1\otimes \widetilde{Z}.\label{nexttolast}
\end{equation}
This results in a two-qubit geometric transformation $U_{1,3}$ on
qubits $1$ and $3$. We do not have to calculate this
transformation exactly since we will undo it later, but the fact
that each eigenspace undergoes the same two-qubit geometric
transformation can be verified similarly to the C-NOT gate we
described in Section 8.4.

At this point, the Hamiltonian is of the form \eqref{H(0)} with
respect to qubit 3, and we can apply any single-qubit unitary gate
$V^3$ according to the method described in Section 8.4. This
transforms the Hamiltonian to $-V_3Z_{3}V_3^{\dagger}\otimes
X_1\otimes \widetilde{Z}$. We can now ``undo" the transformation
$U_{1,3}$ by the interpolation
\begin{equation}
-V_3Z_{3}V_{3}^{\dagger}\otimes Z_1\otimes
\widetilde{Z}\rightarrow-I_{3}\otimes X_1\otimes
\widetilde{Z}.\label{lastHamiltonian}
\end{equation}
The latter transformation is the inverse of Eq.~\eqref{nexttolast}
up to the single-qubit unitary transformation $V_3$, i.e., it
results in the transformation $V_3U^{\dagger}_{1,3}V^{\dagger}_{3}$.
Thus the net result is
\begin{equation}
V_3U_{1,3}^{\dagger}V_3^{\dagger}V_3U_{1,3}=V_3,
\end{equation}
which is the desired single-qubit unitary transformation on qubit
$3$. We point out that during this transformation, a single-qubit
error can propagate between qubits $1$ and $3$, but this is not a
problem since we are implementing a transversal Toffoli operation
and such an error would not result in more that one error per
block of the code.

We showed that for the BS code our scheme can be implemented with at
most 3-local Hamiltonians. This is optimal for the construction we
presented, since there are no non-trivial codes with stabilizer or
gauge-group elements of weight smaller than 2 covering all qubits.
One could argue that since the only Hamiltonians that leave the code
space invariant are superpositions of elements of the stabilizer or
the gauge group, one cannot do better than this. However, it may be
possible to approximate the necessary Hamiltonians with sufficient
precision using 2-local interactions. A possible direction to
consider in this respect is the technique introduced in
Ref.~\cite{KKR06} for approximating three-local Hamiltonians by
two-local ones. This is left as a problem for future investigation.

\section*{8.7 \hspace{2pt} Conclusion}
\addcontentsline{toc}{section}{8.7 \hspace{0.15cm} Conclusion}

We described a scheme for fault-tolerant holonomic computation on
stabilizer codes, which demonstrates that HQC is a scalable method
of computation. The scheme opens the possibility of combining the
software protection of error correction with the inherent
robustness of HQC against control imperfections. Our construction
uses Hamiltonians that are elements of the stabilizer or the gauge
group for the code. The Hamiltonians needed for implementing
two-qubit gates are at least 3-local. We have shown that
computation with at most 3-local Hamiltonians is possible with the
Bacon-Shor code.

It is interesting to point out that the adiabatic regime in which
our scheme operates is consistent with the model of Markovian
decoherence. In Ref.~\cite{ALZ06} it was argued that the standard
dynamical paradigm of fault tolerance is based on assumptions that
are in conflict with the rigorous derivation of the Markovian
limit. Although the threshold theorem has been extended to
non-Markovian models \cite{TB05, AGP06, AKP06}, the Markovian
assumption is an accurate approximation for a wide range of
physical scenarios \cite{Car99}. It also allows for a much simpler
description of the evolution in comparison to non-Markovian
models, as we saw in Chapter 5. In Ref.~\cite{ALZ06} it was shown
that the weak-coupling-limit derivation of the Markovian
approximation is consistent with computational methods that employ
slow transformations, such as adiabatic quantum computation
\cite{FGGS00} or HQC. A theory of fault-tolerance for the
adiabatic model of computation at present is not known, although
significant steps in this direction have been undertaken
\cite{JFS06, Lid07}. Our hybrid HQC-QEC scheme provides a solution
for the case of HQC. We point out, however, that it is an open
problem whether the Markovian approximation makes sense for a
fixed value of the adiabatic slowness parameter when the circuit
increases in size.

Applying the present strategy to actual physical systems might
require modifying our abstract construction in accordance with the
available interactions, possibly using superpositions of
stabilizer or gauge-group elements rather than single elements as
the basic Hamiltonians. Given that simple QEC codes and two-qubit
geometric transformations have been realized using NMR \cite{NMR1, NMR2} and ion-trap \cite%
{Tra1, Tra2} techniques, these systems seem particularly suitable
for hybrid HQC-QEC implementations.

We hope that the techniques presented in this study might prove
useful in other areas as well. It is possible that some
combination of transversal adiabatic transformations and active
correction could provide a solution to the problem of fault
tolerance in the adiabatic model of computation.

\section*{8.8 \hspace{2pt} Appendix: Calculating the holonomy for the $Z$ gate}
\addcontentsline{toc}{section}{8.8 \hspace{0.15cm} Appendix:
Calculating the holonomy for the $Z$ gate}

\subsection*{8.8.1 \hspace{2pt} Linear interpolation}
\addcontentsline{toc}{subsection}{8.8.1 \hspace{0.15cm} Linear
interpolation}

We first demonstrate how to calculate the ground-space holonomy for
the $Z$ gate for the case of linear interpolation along each segment
of the path, i.e., when $f(t)$ and $g(t)$ in
Eq.~\eqref{interpolation} are
\begin{equation}
f(t)=1-\frac{t}{T}, \hspace{0.4cm}g(t)=\frac{t}{T}.
\end{equation}

In order to calculate the holonomy \eqref{holonomy} corresponding
to our construction of the $Z$ gate, we need to define a
\textit{single-valued} orthonormal basis of the ground space of
the Hamiltonian along the loop described by Eq.~\eqref{Zgate}.
Since the Hamiltonian has the form \eqref{Ham2} at all times, it
is convenient to choose the basis of the form
\begin{gather}
|j k; \lambda\rangle =
|\chi_j(\lambda)\rangle|\widetilde\psi_{j k}\rangle,\\
\hspace{0.2cm} j=0,1; \hspace{0.2cm}
k=1,...,2^{n-2}\notag\hspace{0.2cm},
\end{gather}
where $|\chi_0(\lambda(t))\rangle$ and
$|\chi_1(\lambda(t))\rangle$ are ground and excited states of
$H(t)$, and $|\widetilde\psi_{0k}\rangle$ and
$|\widetilde\psi_{1k}\rangle$ are fixed orthonormal bases of the
subspaces that support the projectors $\widetilde{P}_0$ and
$\widetilde{P}_1$ defined in Eq.~\eqref{Projectors}, respectively.
The eigenstates $|\chi_0(\lambda(t))\rangle$ and
$|\chi_1(\lambda(t))\rangle$ are defined up to an overall phase,
but we have to chose the phase such that the states are
single-valued along the loop.

Observe that because of this choice of basis, the matrix elements
\eqref{matrixelementsA} become
\begin{gather}
({A_\mu})_{jk, j'k'}=\langle jk; \lambda| \frac{\partial}{\partial
\lambda^{\mu}}|j'k'; \lambda\rangle=\langle
\chi_j(\lambda)|\frac{\partial}{\partial\lambda^\mu}|\chi_{j'}(\lambda)\rangle\notag\\
\times\langle\widetilde{\psi}_{jk}|\widetilde{\psi}_{j'k'}\rangle
=\langle
\chi_j(\lambda)|\frac{\partial}{\partial\lambda^\mu}|\chi_{j'}(\lambda)\rangle
\delta_{jj'}\delta_{kk'},
\end{gather}
i.e., the matrix $A_\mu$ is diagonal. (Since we are looking only
at the ground space, we are not writing the index of the energy
level). We can therefore drop the path-ordering operator. The
resulting unitary matrix $U^{\gamma}_{jk,j'k'}$ acting on the
subspace spanned by $\{|jk;\lambda(0)\rangle\}$ is also diagonal
and its diagonal elements are
\begin{equation}
U^{\gamma}_{jk,jk}=\textrm{exp}\left(\oint_{\gamma}\langle
\chi_j(\lambda)|\frac{\partial}{\partial\lambda^\mu}|\chi_{j}(\lambda)\rangle
d\lambda^\mu\right).
\end{equation}
These are precisely the Berry phases for the loops described by
the states $|\chi_{j}(\lambda\rangle)$. Since the loop in
parameter space consist of four line segments, we can write the
last expression as
\begin{equation}
U^{\gamma}_{jk,jk}=\textrm{exp}\left(\sum_{i=1}^4 \int_{\gamma_i}
\langle
\chi_j(\lambda)|\frac{\partial}{\partial\lambda^\mu}|\chi_{j}(\lambda)\rangle
d\lambda^\mu\right),
\end{equation}
where $\gamma_i$, $i=1,2,3,4$ are the segments indexed in the order
corresponding to Eq.~\eqref{Zgate}. If we parameterize each line
segment by the dimensionless time $0\leq s \leq 1$, we get
\begin{equation}
U^{\gamma}_{jk,jk}=\textrm{exp}\left(\sum_{i=1}^4 \int_0^1 \langle
\chi_j^i(s)|\frac{d}{ds}|\chi_{j}^i(s)\rangle
ds\right),\label{diagonalU}
\end{equation}
where the superscript $i$ in $|\chi_{j}^i(s)\rangle$ indicates the
segment. In the $|0\rangle$, $|1\rangle$ basis, we will write
these states as
\begin{equation}
|\chi_j^i(s)\rangle=\begin{pmatrix}
a^i_j(s)\\
b^i_j(s)
\end{pmatrix}, \hspace{0.2cm}
j=1,2\hspace{0.2cm},\hspace{0.2cm}i=1,2,3,4\hspace{0.2cm},
\end{equation}
where $|a^i_j(s)|^2+|b^i_j(s)|^2=1$.

Along the segment $\gamma_1$, the states $|\chi_0^1(s)\rangle$ and
$|\chi_1^1(s)\rangle$ are the ground and excited states of the
Hamiltonian
\begin{equation}
H_1(t)=(1-s)Z+sX.
\end{equation}
For these states we obtain
\begin{gather}
a^1_0(s)=\frac{(1-s+\sqrt{1-2s+2s^2})e^{i\omega^1_0(s)}}{\sqrt{2-4s+4s^2+(2-2s)\sqrt{1-2s+2s^2}}},\label{a30}\\
b^1_0(s)=\frac{s e^{i\omega^1_0(s)}}{\sqrt{2-4s+4s^2+(2-2s)\sqrt{1-2s+2s^2}}},\label{b30}\\
a^1_1(s)=\frac{(1-s-\sqrt{1-2s+2s^2})e^{i\omega^1_1(s)}}{\sqrt{2-4s+4s^2-(2-2s)\sqrt{1-2s+2s^2}}},\label{a31}\\
b^1_1(s)=\frac{se^{i\omega^1_1(s)}}{\sqrt{2-4s+4s^2-(2-2s)\sqrt{1-2s+2s^2}}},\label{b31}
\end{gather}
where $\omega^1_j(s)$ are arbitrary phases which have to be chosen
so that when we complete the loop, the phases of the corresponding
states will return to their initial values modulo $2\pi$. We will
define the loops as interpolating between the following
intermediate states defined with their overall phases:
\begin{gather}
|\psi_0(\lambda)\rangle: \hspace{0.2cm}|0\rangle \rightarrow
|f_+^0\rangle \rightarrow |1\rangle \rightarrow
|f_-^{\pi/2}\rangle \rightarrow |0\rangle,\\
|\psi_1(\lambda)\rangle: \hspace{0.2cm}|1\rangle \rightarrow
|f_-^0\rangle \rightarrow |0\rangle \rightarrow
|f_+^{\pi/2}\rangle \rightarrow |1\rangle,
\end{gather}
where
\begin{equation}
|f_{\pm}^\theta\rangle =\frac{|0\rangle \pm
e^{i\theta}|1\rangle}{\sqrt{2}}.
\end{equation}
In other words, we impose the conditions
$|\chi_{0,1}^1(0)\rangle=|0,1\rangle$,
$|\chi_{0,1}^1(1)\rangle=|f_{\pm}^0\rangle =
|\chi_{0,1}^2(0)\rangle$, $|\chi_{0,1}^2(1)\rangle =|1,0\rangle
=|\chi_{0,1}^3(0)\rangle$,
$|\chi_{0,1}^3(1)\rangle=|f_{\mp}^{\pi/2}\rangle=|\chi_{0,1}^4(0)\rangle$,
$|\chi_{0,1}^4(1)\rangle=|0,1\rangle$.

From Eq.~\eqref{a30} and Eq.~\eqref{b30} we see that
$a^1_0(0)=e^{i\omega^1_0(0)}$, $b^1_0(0)=0 $ and
$a^1_0(1)=\frac{1}{\sqrt{2}}e^{i\omega^1_0(1)}$,
$b^1_0(1)=\frac{1}{\sqrt{2}}e^{i\omega^1_0(1)}$ , so we can choose
\begin{equation}
\omega^1_0(s)=0,\hspace{0.2cm} \forall s\in [0,1].
\end{equation}
Similarly, from Eq.~\eqref{a31} and Eq.~\eqref{b31} it can be seen
that $a^1_1(0)=0$, $b^1_1(0)=e^{i\omega^1_1(0)} $ and
$a^1_1(1)=-\frac{1}{\sqrt{2}}e^{i\omega^1_1(1)}$,
$b^1_1(1)=\frac{1}{\sqrt{2}}e^{i\omega^1_1(1)}$. This means that
$\omega^1_1(s)$ has to satisfy $e^{i\omega^1_1(0)}=1$,
$e^{i\omega^1_1(1)}=-1$. We can choose any differentiable
$\omega^1_1(s)$ that satisfies
\begin{equation}
\omega^1_1(0)=0, \hspace{0.2cm} \omega^1_1(1)=\pi.
\end{equation}

In order to calculate $\int_0^1 \langle
\chi_j^1(s)|\frac{d}{ds}|\chi_{j}^1(s)\rangle ds$, we also need
\begin{equation}
\frac{d}{ds}|\chi_{j}^1(s)\rangle=
\begin{pmatrix} \frac{d}{ds}a^1_j(s)\\
\frac{d}{ds}b^1_j(s)
\end{pmatrix}.
\end{equation}
Differentiating Eqs.~\eqref{a30}-\eqref{b31} yields
\begin{gather}
\frac{d}{ds}a^1_0(s)=-\frac{s(1-s+\sqrt{1-2s+2s^2})}{2\sqrt{2-4s+4s^2}[1-2s+2s^2+(1-s)\sqrt{1-2s+2s^2}]^{\frac{3}{2}}},\\
\frac{d}{ds}b^1_0(s)=\frac{2-4s+3s^2+(2-2s)\sqrt{1-2s+2s^2}}{2\sqrt{2-4s+4s^2}[1-2s+2s^2+(1-s)\sqrt{1-2s+2s^2}]^{\frac{3}{2}}},\\
\frac{d}{ds}a^1_1(s)=-\frac{s(1-s-\sqrt{1-2s+2s^2})e^{i\omega^1_1(s)}}{2\sqrt{2-4s+4s^2}[1-2s+2s^2-(1-s)\sqrt{1-2s+2s^2}]^{\frac{3}{2}}} +a^1_1(s)i\frac{d}{ds}\omega^1_1(s),\\
\frac{d}{ds}b^1_0(s)=-\frac{(2-4s+3s^2-(2-2s)\sqrt{1-2s+2s^2})e^{i\omega^1_1(s)}}{2\sqrt{2-4s+4s^2}[1-2s+2s^2-(1-s)\sqrt{1-2s+2s^2}]^{\frac{3}{2}}}
+b^1_1(s)i\frac{d}{ds}\omega^1_1(s).
\end{gather}
By a straightforward substitution, we obtain
\begin{eqnarray}
\langle \chi_0^1(s)|\frac{d}{ds}|\chi_{0}^1(s)\rangle &=&a^{1\ast}_0(s)\frac{d}{ds}a^{1}_0(s)+b^{1\ast}_0(s)\frac{d}{ds}b^1_0(s)=0,\\
\langle \chi_1^1(s)|\frac{d}{ds}|\chi_{1}^1(s)\rangle &=&a^{1\ast}_1(s)\frac{d}{ds}a^1_1(s)+b^{1\ast}_1(s)\frac{d}{ds}b^1_1(s)=i\frac{d}{ds}\omega^1_1(s).\\
\end{eqnarray}
Thus the integrals are
\begin{eqnarray}
\int_0^1\langle\chi_0^1(s)|\frac{d}{ds}|\chi_{0}^1(s)\rangle
ds&=&0,\\
\int_0^1\langle\chi_1^1(s)|\frac{d}{ds}|\chi_{1}^1(s)\rangle
ds&=&i\omega^1_1(s)|_0^1=i\pi.
\end{eqnarray}
In the same manner, we calculate the contributions of the other
three line segments. The results are:
\begin{eqnarray}
\int_0^1\langle\chi_0^2(s)|\frac{d}{ds}|\chi_{0}^2(s)\rangle
ds&=&0,\\
\int_0^1\langle\chi_1^2(s)|\frac{d}{ds}|\chi_{1}^2(s)\rangle
ds&=&0,
\end{eqnarray}
\begin{eqnarray}
\int_0^1\langle\chi_0^3(s)|\frac{d}{ds}|\chi_{0}^3(s)\rangle
ds&=&i\frac{\pi}{2},\\
\int_0^1\langle\chi_1^3(s)|\frac{d}{ds}|\chi_{1}^3(s)\rangle
ds&=&0,
\end{eqnarray}
\begin{eqnarray}
\int_0^1\langle\chi_0^4(s)|\frac{d}{ds}|\chi_{0}^4(s)\rangle
ds&=&0,\\
\int_0^1\langle\chi_1^4(s)|\frac{d}{ds}|\chi_{1}^4(s)\rangle
ds&=&i\frac{\pi}{2}.
\end{eqnarray}
Putting everything together, for the diagonal elements of the
holonomy we obtain
\begin{gather}
U^{\gamma}_{0k,0k}=e^{i\frac{\pi}{2}},\notag\\
U^{\gamma}_{1k,1k}=e^{i\frac{3\pi}{2}}.\label{finalholonZ}
\end{gather}
The holonomy transforms any state in the ground space of the
initial Hamiltonian as
\begin{equation}
U^{\gamma}\sum_{jk}\alpha_{jk}|j\rangle|\widetilde{\psi}_{jk}\rangle
=
e^{i\frac{\pi}{2}}\sum_{jk}(-1)^j\alpha_{jk}|j\rangle|\widetilde{\psi}_{jk}\rangle,
\hspace{0.2cm} j=0,1.
\end{equation}
From the point of view of the full Hilbert space, this is
effectively a $Z$ gate on the first qubit up to an overall phase.

We point out that other single-qubit transformations like the
Hadamard or the $X$ gates, which do not form a complete loop in
parameter space, can be obtained in a similar fashion by
calculating the open-path expression \eqref{openpathholonomy}. In
principle, the result of that calculation depends on the choice of
basis $\{|\alpha; \lambda\rangle \}$ which is defined up to a
unitary gauge transformation. However, this ambiguity is removed
by the notion of parallel transport between the initial and the
final subspaces \cite{KAS06}. One can verify that this yields the
correct result for our transformations.

\subsection*{8.8.2 \hspace{2pt} Unitary interpolation}
\addcontentsline{toc}{subsection}{8.8.2 \hspace{0.15cm} Unitary
interpolation}

The calculation is simpler if we choose a unitary interpolation,
\begin{equation}
f(t)=\cos{\frac{\pi t}{2 T}},\hspace{0.4cm} g(t)=\sin {\frac{\pi
t}{2 T}}.
\end{equation}
Such interpolation corresponds to a rotation of the Bloch sphere
around a particular axis for each of the segments of the loop. The
first two segments of the loop \eqref{Zgate} are realized via the
Hamiltonian
\begin{equation}
\widehat{H}_{1,2}(t)=-V_{Y}^{\dagger}(t)ZV_{Y}(t)\otimes
\widetilde{G}, \hspace{0.2cm} V_{Y}(t)=\textrm{exp}
\left(it\frac{\pi }{2T}Y\right),\label{HZ1}
\end{equation}
applied for time $T$, and the third and fourth segments are
realized via the Hamiltonian
\begin{equation}
\widehat{H}_{3,4}(t)=-V_{X}(t)ZV_{X}^{\dagger}(t)\otimes
\widetilde{G}, \hspace{0.2cm} V_{X}(t)=\textrm{exp}
\left(it\frac{\pi }{2T}X\right),\label{HZ1}
\end{equation}
again applied for time $T$. Let us define the eigenstates of the
Hamiltonian along the first two segments as
\begin{equation}
|\chi^{1,2}_0(t)\rangle = V_{X}(t)|0\rangle, \hspace{0.2cm}
|\chi^{1,2}_1(t)\rangle = V_{X}(t)|1\rangle, \hspace{0.2cm}0\leq
t\leq T
\end{equation}
and along the third and forth segments as
\begin{equation}
|\chi^{3,4}_0(t)\rangle = -iV^{\dagger}_Y(t)Y|0\rangle,
\hspace{0.2cm} |\chi^{3,4}_1(t)\rangle =-i
V^{\dagger}_{Y}(t)Y|1\rangle, \hspace{0.2cm}0\leq t\leq T.
\end{equation}
Notice that
\begin{equation}
|\chi^{1,2}_0(T)\rangle=-iY|0\rangle=|\chi^{3,4}_0(0)\rangle,
\hspace{0.2cm}
|\chi^{1,2}_1(T)\rangle=-iY|1\rangle=|\chi^{3,4}_1(0)\rangle,
\end{equation}
but
\begin{equation}
|\chi^{1,2}_0(0)\rangle=|0\rangle \neq |\chi^{3,4}_0(T)\rangle =
-i|0\rangle, \hspace{0.2cm} |\chi^{1,2}_1(0)\rangle=|1\rangle \neq
|\chi^{3,4}_1(T)\rangle = i|1\rangle,
\end{equation}
i.e., this basis is not single-valued. To make it single valued,
we can modify it along the third and fourth segments as
\begin{equation}
|\chi^{3,4}_0(t)\rangle\rightarrow
|\widetilde{\chi}^{3,4}_0(t)\rangle =
e^{i\omega_0(t)}|\chi^{3,4}_0(t)\rangle, \hspace{0.2cm}
|\chi^{3,4}_1(t)\rangle\rightarrow
|\widetilde{\chi}^{3,4}_1(t)\rangle =
e^{i\omega_1(t)}|\chi^{3,4}_0(t)\rangle,
\end{equation}
where
\begin{equation}
\omega_0(0)=0, \hspace{0.2cm} \omega_0(T)=\frac{\pi}{2},
\end{equation}
\begin{equation}
\omega_1(0)=0, \hspace{0.2cm} \omega_1(T)=-\frac{\pi}{2}.
\end{equation}

The expression \eqref{diagonalU} then becomes
\begin{gather}
U^{\gamma}_{jk,jk}=\textrm{exp}\left(\int_0^T \langle
\chi_j^{1,2}(t)|\frac{d}{dt}|\chi_{j}^{1,2}(t)\rangle dt +
\int_0^T \langle
\chi_j^{3,4}(t)|\frac{d}{dt}|\chi_{j}^{3,4}(t)\rangle dt +
(-1)^j\frac{\pi}{2}\right), \notag\\\hspace{0.2cm} j=0,1.
\end{gather}
But
\begin{equation}
\langle \chi_j^{1,2}(t)|\frac{d}{dt}|\chi_{j}^{1,2}(t)\rangle =
-i\frac{\pi }{2T}\langle j|Y|j\rangle =0,
\end{equation}
and
\begin{equation}
\langle \chi_j^{3,4}(t)|\frac{d}{dt}|\chi_{j}^{3,4}(t)\rangle =
i\frac{\pi }{2T}\langle j|YXY|j\rangle =0.
\end{equation}
Therefore, we obtain \eqref{finalholonZ}.

\chapter*{Chapter 9: \hspace{1pt} Conclusion}
\addcontentsline{toc}{chapter}{Chapter 9:\hspace{0.15cm}
Conclusion}

In this thesis we obtained various results in the theory of open
quantum systems and quantum information. These results have opened
interesting questions and suggested promising directions for
future research.

The decomposition into weak measurements presents a practical
prescription for the implementation of any generalized measurement
using weak measurements and feedback control. It also presents a
powerful mathematical tool for the study of measurement processes.
In practice, there may exist limitations on the type of weak
measurements an experimenter can implement, and hence it would be
interesting to look at the inverse problem---given a set of weak
measurements, what are the generalized measurements that one can
generate with them. It might be convenient to recast this problem
in terms of the system-ancilla interactions that are available for
the implementation of such measurements. The decomposition may
prove useful in other problems involving feedback control as well.
One of its interesting features is that the evolution that
corresponds to it is confined on a specific manifold (the
simplex). In that sense, the procedure avoids dissipation into
areas from which the state could drift away from the desired
outcomes. This property could be helpful in designing optimal
feedback-control protocols.

The decomposition into weak measurements furthermore suggests that
it may be possible to find a unified description of measurement
protocols. The operations applied at a given time during the
measurement procedure for generating generalized measurements,
drive the evolution of a stochastic process on the simplex, i.e.,
they can be represented by a stochastic matrix on the coordinate
space. This suggests that there may exist a general coordinate
space, which includes all such simplexes, on which the most
general notion of a measurement protocol can be represented by a
stochastic process. The basic object in such a description would
not be a quantum state but a classical probability distribution on
a space whose coordinates correspond to quantum states. Since
stochastic processes are well understood, such a unified
description could be useful for studies of measurement-driven
schemes.

We also used the decomposition into weak measurements for deriving
necessary and sufficient conditions for entanglement monotones.
These conditions may be useful for proving monotonicity of
conjectured monotones, finding new classes of entanglement
measures, or finding measures with particularly nice properties
such as additivity. Another interesting possibility suggested by
the existence of necessary and sufficient differential conditions
for monotonicity under all types of CPTP transformations, is that
it may be possible to think of all quantum operations as generated
by infinitesimal operations. It is known that CPTP maps cannot be
generated by weak CPTP maps, however, the differential form of the
convexity condition can be thought of as a condition for
monotonicity under infinitesimal loss of information. Therefore,
if we adopt the approach in which the basic objects are ensembles
of states and loss of classical information is a basic operation
on these objects, it may be possible to arrive at a unified
description of the most general form of quantum operations where
every operation can be continuously connected to the identity.

Our investigation of the deterministic evolution of open quantum
systems and the difference between Markovian and non-Markovian
decoherence has also opened various interesting questions. While
we compared the performance of different perturbative master
equations, we have not compared their solutions to the
perturbative expansion of the exact solution. Studying these
equations is important in its own right as it provides
understanding of the actual dynamics driving the effective
evolution. But for the purpose of obtaining an approximation of
the exact solution starting from first principles, it may be more
useful to expand the solution directly. Expanding the exact
solution is justified in the same parameter regime---small $\alpha
t$---and requires computation of the same bath-correlation
functions, but it is significantly simpler since it does not
require deriving an equation and solving it.

As we mentioned in Chapter 5, the TCL or NZ projection techniques
might be useful also for the effective description of the reduced
dynamics of a system subject to non-Markovian decoherence and
continuous error correction. Here too, it would be interesting to
consider expanding the solution directly. We presented a
generalized notion of a Zeno regime applicable for the problem of
error correction and identified the bottle-neck mechanism through
which the performance of the error-correction scheme depends on
this regime. As the Zeno regime plays a central role in the
workings of another error-correction approach---dynamical
decoupling (DD) \cite{VKL99, FLP04}---it might be useful to apply
the insights developed here in the design of hybrid EC-DD schemes.
Another direction for future research is expanding our scheme for
continuous error correction based on weak measurement and weak
unitary operations to include more sophisticated feedback. Since
making full use of the available information about the state can
only help, we expect that this approach would lead to schemes with
better performance.

One of the problems suggested by our study of the conditions for
exact correctability under continuous decoherence, was whether a
similar approach to the one we used could be useful in studies
concerning approximate error correction. A question we raised is
whether the Markovian decoherence process during an infinitesimal
time step can be separated into completely correctable and
non-correctable parts. If this is the case, it could allow us to
formulate conditions for optimal correctability by tracking the
evolution of the maximal information that remains during the
process. As we argued, for non-Markovian decoherence such an
approach cannot be optimal since the information may flow out to
the environment and later return back, but it could nevertheless
be helpful for finding locally optimal solutions. Even if the
answer to this question is negative, the differential approach in
studying information loss certainly seems promising. One
interesting extension of this work would be to derive conditions
for correctability in the context of the TCL or NZ master
equations.

Another promising tool introduced in this thesis is the measure of
fidelity for encoded information that we used to prove the
robustness of operator error correction against imperfect
encoding. As we pointed out, this measure provides a natural means
of extending concepts such as the fidelity of a quantum channel
and the entanglement fidelity to the case of subsystem codes. As
subsystem encoding provides the most general method of encoding,
this measure could also be useful in studies concerning optimal
quantum error correction. Its simple form makes it suitable for
computation which is important in this respect.

Finally, our scheme for fault-tolerant holonomic computation has
also opened a number of interesting questions. We have shown that
for universal computation this scheme requires three-local
Hamiltonians. It may be possible, however, to use perturbative
techniques to approximate three-local Hamiltonians using two-local
ones in a manner similar to the one introduced in
Ref.~\cite{KKR06}. Another direction for future research is
suggested by the fact that the gap of the adiabatic Hamiltonian
provides a natural protection against those types of errors that
lead to excitations. It is interesting whether it is possible to
design more efficient error correction schemes that make use of
this property. Another question is whether the holonomic approach
could provide a solution to the problem of the inconsistency
between the standard fault-tolerance assumptions and the rigorous
derivation of the Markovian limit. Giving a definitive answer to
this question requires a rigorous analysis of the accumulation of
non-Markovian errors due to deviation from perfect adiabaticity.

Our scheme for the Bacon-Shor code uses an approach to holonomic
computation in which the Hamiltonian acts trivially on the
subsystem code and non-trivially on the gauge subsystem. It would
be interesting to formulate this approach as a general method for
holonomic computation on subsystems. The techniques introduced in
this study may also prove useful for the problem of fault
tolerance in the adiabatic model of computation. It is possible
that some combination of transversal adiabatic operations and
active error correction could provide a solution for this case
too.

\bibliographystyle{plain}

\end{document}